\documentclass[final,3p,times]{elsarticle_en}

\usepackage{graphicx}
\usepackage{sidecap}
\usepackage{tikz}

\usepackage{amssymb}
\usepackage{version}

\usepackage[english,french]{babel}

\def\og{\leavevmode\raise.3ex\hbox{$\scriptscriptstyle\langle\!\langle$~}}
\def\fg{\leavevmode\raise.3ex\hbox{~$\!\scriptscriptstyle\,\rangle\!\rangle$}}
\newcommand{\g}[1]{``#1''}

\usepackage[utf8]{inputenc}
\usepackage[T1]{fontenc}
\usepackage{amsmath}
\usepackage{xcolor}
\usepackage{hyperref}
\usepackage{ulem} 
\hypersetup{breaklinks=true}
\hypersetup{colorlinks=true}
\hypersetup{urlcolor=blue}
\hypersetup{citecolor=black}
\hypersetup{linkcolor=black}
\usepackage{cleveref}
\newcommand{\bea}{\begin{eqnarray}}
\newcommand{\eea}{\end{eqnarray}}
\newcommand{\be}{\begin{equation}}
\newcommand{\ee}{\end{equation}}
\newcommand{\rr}{\mathbf{r}}

\newcommand{\kk}{\mathbf{k}}

\newcommand{\KK}{\mathbf{K}}

\newcommand{\qq}{\mathbf{q}}

\newcommand{\vv}{\mathbf{v}}

\newcommand{\mD}{\mathcal{D}}
\newcommand{\mA}{\mathcal{A}}

\newcommand{\mH}{\mathcal{H}}

\newcommand{\mG}{\mathcal{G}}

\newcommand{\zero}{\mathbf{0}}

\newcommand{\veps}{\varepsilon}
\newcommand{\vepsb}{\bar{\varepsilon}}

\newcommand{\qb}{\bar{q}}
\newcommand{\kb}{\bar{k}}
\newcommand{\kpb}{\bar{k}'}
\newcommand{\thetat}{\tilde{\theta}}
\newcommand{\zetat}{\tilde{\zeta}}
\newcommand{\zetac}{\check{\zeta}}

\renewcommand{\vec}{\mathbf}

\newcommand{\ii}{\textrm{i}}
\newcommand{\ff}{\textrm{f}}
\newcommand{\eee}{\textrm{e}}
\newcommand{\dd}{\mathrm{d}}

\DeclareMathOperator\erfi{erfi}

\DeclareMathOperator\re{Re}
\DeclareMathOperator\im{Im}
\DeclareMathOperator\sig{sign}

\newcommand{\yc}{\color{black}}
\newcommand{\ycd}{\color{black}}
\newcommand{\yct}{\color{black}}
\usepackage{float}

\begin{document}

\begin{frontmatter}


\selectlanguage{english}
\title{Phonon damping in a 2D superfluid: insufficiency of Fermi's golden rule at low temperature}


\author{Yvan Castin}
\ead{yvan.castin@lkb.ens.fr}
\author{Alan Serafin}
\ead{a.serafin@protonmail.com}
\author{Alice Sinatra}
\ead{alice.sinatra@lkb.ens.fr}

\address{Laboratoire Kastler Brossel, ENS-Universit\'e PSL, CNRS, Universit\'e Sorbonne and Coll\`ege de France, 24 rue Lhomond, 75231 Paris, France}



\begin{abstract}
It is generally accepted that the phonon gas of a superfluid always enters a weak coupling regime at sufficiently low temperatures, whatever the strength of the interactions between the underlying particles (constitutive of the superfluid). Thus, in this limit, we should always be able to calculate the damping rate of thermal phonons by applying Fermi's golden rule to the Hamiltonian {\yc $H_3$} of cubic phonon-phonon coupling taken from quantum hydrodynamics, at least in the case of a convex acoustic branch and in the collisionless regime (where the eigenfrequency of the considered phonons remains much greater than the gas thermalization rate). Using the many-body Green's function method, we predict that, unexpectedly, this is not true in two dimensions, contrary to the three-dimensional case. We confirm this prediction with classical phonon-field simulations and a non-perturbative theory in $H_3$, where the fourth order is regularized by hand, giving a complex energy to the virtual phonons of the four-phonon collisional processes. For a weakly interacting fluid and a phonon mode in the long-wavelength limit, we predict a damping rate about three times lower than that of the golden rule.

\noindent 
\vskip 0.5\baselineskip
\end{abstract} 
\end{frontmatter}


\selectlanguage{english}
\tableofcontents

\section{Presentation of the problem and general reasoning}
\label{sec0}

In dimension $d\geq 2$, we consider a spatially homogeneous system of particles with short-range interactions, entirely superfluid in its ground state and prepared at a non-zero but arbitrarily low temperature, which allows us to restrict its study to leading order in temperature by means of an effective low-energy theory. Irrespective of the quantum statistics (bosonic or fermionic) of the particles and the strength of their interactions (in gas or liquid phase), this system has an acoustic excitation branch, i.e. with a linear departure in wave number (with no band gap), whose quanta are the phonons. 

The coupling between low-wavenumber phonons (of energy $<\eta k_B T$, where temperature $T\to 0$ then cutoff $\eta\to+\infty$ at the end of the calculations) is deduced from the quantum hydrodynamics of Landau and Khalatnikov \cite{LK}, the effective Hamiltonian taking the form\footnote{\label{note1} The inclusion of higher-order terms $H_5, H_6$, etc., would be meaningless without gradient corrections to the energy functional of quantum hydrodynamics \cite{SonWingate}, corrections which also bring the curvature term (here put by hand) into equation (\ref{eq003}).} 
\begin{equation}\label{eq001}
H=H_2+H_3+H_4
\end{equation} where term $H_n$ is of degree $n$ in terms of $\hat{b}_\kk^\dagger$ and $\hat{b}_\kk$, bosonic creation and annihilation operators of a phonon of wave vector $\kk$ in a quantization box $[0,L]^d$ with periodic boundary conditions. The term $H_2$ represents an ideal gas of phonons of dispersion relation $\veps_\kk$, 
\begin{equation}\label{eq002}
H_2=\sum_{\kk\neq\zero} \veps_\mathbf{k}\hat{b}_\kk^\dagger\hat{b}_\mathbf{k}\end{equation} The precise form of the coupling terms $H_3$ and $H_4$ in equation (\ref{eq001}) is of little importance at this stage and will be given in section \ref{sec1}. The coupling between phonons makes the system acoustically non-linear and causes the damping of sound waves, the subject of the present work. 

We limit our study of damping here (i) to modes of wave vector $\qq$ in the collisionless regime $\veps_\qq\gg\hbar\Gamma_{\rm th}$ where $\Gamma_{\rm th}$ is the thermalization rate of the phonon gas or, what amounts to the same thing, the damping rate of thermal phonons of energy $\veps_\kk\simeq k_B T$, and (ii) to the case of an acoustic branch with a convex start in wavenumber, which can be limited to its cubic approximation at low temperature: 
\begin{equation}\label{eq003}
\veps_\kk=\hbar c k \left[1+\frac{\gamma}{8} \left(\frac{\hbar k}{m c}\right)^2\right]\quad\mbox{with}\quad \gamma>0
\end{equation} where $m$ is the mass of a particle, $c$ is the speed of sound at zero temperature and $\gamma$ is a dimensionless curvature parameter. \footnote{\label{note0} As in quantum electrodynamics, the parameters appearing in the Hamiltonian are in fact the bare values, different from the actual values measured in an experiment, the passage from one to the other being made by a procedure known as \g{renormalisation}. Equation (\ref{eq002}) should therefore refer to a bare dispersion relation $\veps_\kk^{(0)}$ and equation (\ref{eq003}) to a bare speed of sound $c^{(0)}$ and a bare curvature parameter $\gamma^{(0)}$. We do not do so here to lighten the notations, and because this renormalization does not intervene in dimension $d=2$ at the leading order in temperature.}\footnote{The choice $\gamma>0$ excludes superfluids of paired spin $1/2$ fermions in the limit, known as BCS, of a weakly attractive interaction between particles \cite{SDM}.} These assumptions allow us to describe damping in terms of elementary phonon-coupling processes\footnote{In the opposite regime $\veps_\qq\ll\hbar\Gamma_{\rm th}$, the system has time to relax to a juxtaposition of local thermodynamic equilibrium states during an oscillation period of the sound wave of wavevector $\qq$ and damping at non-zero temperature is described through transport or viscosity coefficients, obtained by solving kinetic equations and which enter into partial differential equations on densities, velocities, volume entropy, etc, of a two-fluid model \cite{livreK}.} and, as is generally accepted, to deduce the damping rate from Fermi's golden rule applied to the $H_3$ term of the Hamiltonian.\footnote{In the case $\gamma<0$ of an acoustic branch with a concave start, the three-phonon processes induced by $H_3$ to order one do not conserve momentum-energy, and an extended formulation of the golden rule (taking into account the four-phonon processes induced by $H_3$ to order two and by $H_4$ to order one) must be used to obtain a non-zero damping rate, see references \cite{LK,Annalen} and our section \ref{sec3.2}.} Indeed, it is expected that at sufficiently low temperatures, the phonon gas will always enter the weak coupling regime, even if the constituent particles of the system are strongly interacting (as in liquid helium 4, for example \cite{Maris}). We will show in this section, by simple reasoning, that this is true in dimension $d=3$ but not in dimension $d=2$. Phonon damping in the two-dimensional case will therefore be the subject of further study (beyond the golden rule) in subsequent sections. 

Our simple reasoning is based on many-body Green’s function method \cite{FW}. To fix ideas and the Green's function to be considered, we assume that the gas, initially at thermal equilibrium for the Hamiltonian $H$ (density operator $\hat{\sigma}(0)=\hat{\sigma}_{\rm th}\propto \exp(-H/k_B T)$), is excited by the sudden creation of a coherent state of phonons in mode $\qq$: 
\begin{equation}\label{eq010}
\hat{\sigma}(0^+) = \hat{U}_{\rm exc}\hat{\sigma}_{\rm th} \hat{U}^\dagger_{\rm exc} \quad\mbox{with}\quad \hat{U}_{\rm exc}=\eee^{\alpha\hat{b}_\qq^\dagger-\alpha^*\hat{b}_\qq}
\end{equation} Experimentally, this excitation procedure can be carried out by a short-duration 
Bragg laser pulse \cite{Grynberg,Ketterle,Davidson,Vale,Cartago}. The amplitude $\langle\hat{b}_\qq(t)\rangle$ of the coherent state is then measured over time, for example through the gas density (more precisely, its Fourier components of wave vectors $\pm\qq$). In the limit of very low initial excitation, we enter the linear response regime and can formally expand the unitary excitation operator $\hat{U}_{\rm exc}$ to order one in $\alpha$ in equation (\ref{eq010}) to obtain 
\begin{equation}\label{eq011}
\langle\hat{b}_\qq(t)\rangle\stackrel{t>0}{\underset{\alpha\to 0}{=}} \alpha \langle[\hat{b}_\qq(t),\hat{b}_\qq^\dagger(0)]\rangle_{\rm th}+O(\alpha^2)
\end{equation} where $\hat{b}_\qq(t)$ is written in Heisenberg representation for the evolution unperturbed by $\hat{U}_{\rm exc}$, $\hat{b}_{\qq}(t)=\eee^{\textrm{i}H t/\hbar} \hat{b}_{\qq}(0)\eee^{-\textrm{i}H t/\hbar}$, the mean of the left-hand side is taken in the perturbed state $\hat{\sigma}(0^+)$ and that of the right-hand side in the initial unperturbed thermal state $\hat{\sigma}_{\rm th}$, and we have used the fact that $\langle\hat{b}_{\qq}(0)\hat{b}_{\qq}(t)\rangle_{\rm th}=\langle\hat{b}_{\qq}(t)\hat{b}_{\qq}(0)\rangle_{\rm th}=0$ by non-conservation of the total momentum. Using section 31 of reference \cite{FW}, we relate our signal to the two-point Green's function $\mG$ {\yc of the phonon field} and the associated self-energy $\Sigma$ (noted $\Sigma^{\star}$ in \cite{FW}):\footnote{Recall that, originally, $\mG$ and $\Sigma$ are functions of $(\rr,t)$ and $(\rr',t')$, where $\rr,\rr'$ are two position vectors and $t,t'$ are two times. {\yct By spatial homogeneity and stationarity of the thermal state,} they are reduced to functions of $(\rr-\rr',t-t')$, whose space-time Fourier transforms are functions of $(\kk,\omega)$; the analytic continuation of the latter to complex frequencies $z/\hbar$ leads to functions $\mG(\kk,z)$ and $\Sigma(\kk,z)$ of equation (\ref{eq012}).} 
\begin{equation}\label{eq012}
\boxed{s(t)\equiv \eee^{\ii\veps_\mathbf{q}t/\hbar}\langle[\hat{b}_\qq(t),\hat{b}_\qq^\dagger(0)]\rangle_{\rm th}\stackrel{t>0}{=} \int_{C_+} \frac{\mathrm{d}z}{2\ii\pi} \eee^{-\textrm{i}(z-\veps_\qq)t/\hbar} \mathcal{G}(\qq,z)\quad\mbox{with}\quad \mathcal{G}(\qq,z)=\frac{1}{z-\veps_\qq-\Sigma(\qq,z)}}
\end{equation} where the integration path $C_+$ runs in the upper half {\yc of the complex} plane parallel to the real axis from $+\infty$ to $-\infty$ and where we have, for convenience, removed the free evolution due to the Hamiltonian $H_2$. In the following, it will be advantageous to take as the new complex energy variable in $\Sigma$ the deviation from the free-evolution energy, any confusion being avoided by indexing the wave vector dependence: 
\begin{equation}\label{eq013}
\Sigma(\qq,z)=\Sigma_\qq(\zeta)\quad\mbox{with}\quad\zeta\equiv z-\veps_\mathbf{q}\quad{\ycd\mbox{hence}\quad\mathcal{G}(\qq,z)=\frac{1}{\zeta-\Sigma_\qq(\zeta)}}\end{equation} From this exact formalism, we can recover the exponential damping of Fermi's golden rule $|s(t)|^2\simeq \exp(-\Gamma_q t)$ in two steps. (i) First, the pole approximation \cite{CCTbordeaux} is performed, neglecting the energy dependence of $\Sigma$: 
\begin{equation}\label{eq014}
\Sigma_\qq(\zeta) \approx \Sigma_\qq(\textrm{i}0^+)\quad \mbox{so that}\quad s(t)\simeq s_{\mbox{\scriptsize pole}}(t)=\exp[-\textrm{i}\Sigma_\qq(\textrm{i}0^+)t/\hbar]
\end{equation} (ii) As the pole approximation is generally legitimate only in a weak coupling limit between phonons, it suffices to calculate $\Sigma_\qq$ at the leading order in the coupling, i.e. at second order in $H_3$, but also at first order in $H_4$ if we need to know $\re\Sigma_\qq$. The central question of this section is whether simply taking a low-temperature limit with proper rescaling of the wave number of the mode under consideration,\footnote{It should be pointed out that the strength of the interactions between the constituent particles of the fluid does not vary in the limit (\ref{eq020}), so that the speed of sound $c$ remains constant. On the other hand, limit (\ref{eq020}) automatically brings mode $\qq$ into the collisionless regime $\veps_{\qq}\gg \hbar\Gamma_{\rm th}$ because the rate $\Gamma_{\rm th}$ of gas thermalization, which can be estimated by the damping rate of phonons of wave number $\approx k_B T/\hbar c$, tends to zero faster than linearly in temperature, as we shall see.} 
\begin{equation}\label{eq020}
\epsilon \equiv k_B T/mc^2 \to 0 \quad \mbox{with}\quad \qb\equiv \hbar c q/k_B T \quad \mbox{fixed}
\end{equation} is enough to bring the phonon gas into the weak coupling regime and justify the two steps. We'll see that the answer depends on the dimensionality of the system. 
\paragraph{Step (i)}
 To justify the pole approximation, we rely on the scaling laws describing the behavior of the self-energy at low temperature: 
\begin{equation}\label{eq021}
\Sigma_\qq(\zeta) = k_B T \epsilon^\nu\, \tilde{\Sigma}_{\qb}(\zeta/(k_BT\epsilon^\sigma))
\end{equation} After outputting the energy scale $k_B T$, two exponents $\nu$ and $\sigma$ remain, the first characterizing the order of magnitude of function $\Sigma_\qq(\zeta)$, the second its energy width around $\zeta=0$. It is understood that the new function $\tilde{\Sigma}_{\qb}$ has a finite, non-zero limit when $\epsilon\to 0$. After inserting form (\ref{eq021}) into the signal expression (\ref{eq012}), we {\yct perform} the following scaling on the conjugate variables $\zeta$ and $t$, 
\begin{equation}\label{eq022}
\zeta\equiv k_B T \epsilon^\nu \tilde{\zeta}\quad ; \quad t\equiv \tilde{t}\,\hbar/(k_B T \epsilon^\nu)
\end{equation} to obtain the reduced expression 
\begin{equation}\label{eq023}
\boxed{s(t)=\int_{C_+} \frac{\mathrm{d}\zetat}{2\ii\pi} \frac{\eee^{-\textrm{i}\tilde{\zeta}\tilde{t}}}{\zetat-\tilde{\Sigma}_{\qb}(\epsilon^{\nu-\sigma}\zetat)}}
\end{equation} in which $\epsilon$ still has to tend towards zero at a fixed $\tilde{t}$. We then see that, if $\nu>\sigma$, the argument of function $\tilde{\Sigma}_{\qb}$ tends towards zero by positive imaginary-part values when $\epsilon\to 0$, which means it can be replaced by $\textrm{i}0^+$: we find the pole approximation (\ref{eq014}), which is exact in this limit. 

\paragraph{Step (ii)}
 To find out whether it's sufficient to calculate the self-energy to the leading order in phonon coupling, let's keep only the cubic coupling $H_3$ for simplicity (we can check that the quartic coupling $H_4$ doesn't change anything qualitatively), omit the non-resonant processes it contains (terms in $\hat{b}^\dagger\hat{b}^\dagger\hat{b}^\dagger$ and $\hat{b}\hat{b}\hat{b}$) of sub-dominant contribution in the $\epsilon\to 0$ limit, and estimate the order of magnitude of its contribution of order $2n$ to $\Sigma_\qq{\ycd(\ii 0^+)}$ as follows: \footnote{As $H_3$ changes the parity of the phonon number, the odd-order contributions $2n+1$ are zero.} 
\begin{equation}\label{eq024}
\Sigma^{(2n)}_\qq(\textrm{i}0^+)\approx \int \left(\prod_{i=1}^{n} \dd^d k_i\right) \langle\ |\mathcal{H}_3|\ \rangle \frac{1}{\Delta E_1} \langle\ |\mathcal{H}_3|\ \rangle \ldots \langle\ |\mathcal{H}_3|\ \rangle \frac{1}{\Delta E_{2n-1}} \langle\ |\mathcal{H}_3|\ \rangle
\end{equation} Let's briefly justify form (\ref{eq024}) (the reader is referred to sections \ref{sec2} and \ref{sec3} for a detailed treatment of cases $n=1$ and $n=2$). Order $2n$ involves $2n$ actions of $H_3$ so $2n$ matrix elements $\langle\,|H_3|\,\rangle$, each of which introduces a new independent phonon wave vector $\kk_i$ (Beliaev terms $\hat{b}^\dagger\hat{b}^\dagger\hat{b}$ in $H_3$) or makes one disappear (Landau terms $\hat{b}^\dagger\hat{b}\hat{b}$); the other phonon wave vectors that appear are deduced from $\kk_i$ and $\qq$ by conservation of momentum. Since the $2n$ actions of $H_3$ must conserve the number of phonons as a whole, there must be as many Beliaev processes as Landau processes, leading to exactly $n$ independent wave vectors on which to sum. In the thermodynamic limit, we express $H_3$ in terms of the {\yct volumic} Hamiltonian such that $H_3=\mH_3/L^{d/2}$ and replace the sum over the $\kk_i$ by an integral. Finally, between each matrix element of $H_3$ appears {\yc an energy} denominator representing the free propagation of Hamiltonian $H_2$ and giving the energy difference between {\yct the initial state with a phonon $\qq$ and} an intermediate state. Each $\Delta E_j$ ($1\leq j\leq 2n-1$) can be expressed as the algebraic sum (i.e. with a plus or minus sign) of $j$ elementary energy defects $\veps_\kk+\veps_{\kk'}-\veps_{\kk''}$ corresponding to a $\kk+\kk'\leftrightarrow\kk''$ Beliaev-Landau process. In the low-temperature limit $\epsilon\to 0$, the integral (\ref{eq024}) is in general dominated by \g{small-angle} processes, where {\it all} intermediate phonons (those with wave vectors $\kk_i$ and others) are almost collinear to and in the same direction as $\qq$ \cite{LK,Annalen}; more precisely, we find that the angles $\theta_i$ between vectors $\qq$ and $\kk_i$ must be $O(\gamma^{1/2}\epsilon)$ for $\Delta E_j$ to reach their minimum energy scale {\yc $k_B T\epsilon^2$} and for the integrand to be amplified by the \g{small-denominator} effect, which leads to the expansion of the elementary energy defect 
\begin{equation}\label{eq025}
\veps_\kk+\veps_{\kk'}-\veps_{\kk''=\kk+\kk'}\underset{\epsilon\to 0}{\sim} k_B T\gamma\epsilon^2\left[\frac{\kb\kpb}{\kb+\kpb}\frac{(\theta-\theta')^2}{2\gamma\epsilon^2}-\frac{3}{8}\kb\kpb(\kb+\kpb)\right]
\end{equation} where $\gamma$ is the curvature parameter in the dispersion relation (\ref{eq003}), {\yc angles $\theta=\widehat{(\qq,\kk)}$ and $\theta'=\widehat{(\qq,\kk')}$ are introduced,} and the scaling of $k$ to $\kb$ by its typical value $k_B T/\hbar c$ is based on equation (\ref{eq020}). The $\Delta E_j$ are {\yc thus effectively} of order $k_B T\epsilon^2$ and each $\dd^dk_i$ integration element in (\ref{eq024}) written in polar or spherical coordinates brings out a factor $(k_B T/\hbar c)^d (\gamma^{1/2}\epsilon)^{d-1}$. It remains to give the estimate of the matrix elements of $\mH_3$, which is easily derived from their expression (\ref{eq102}) to come (the coupling constant $\Lambda$ is given by equation (\ref{eq032})): 
\begin{equation}\label{eq027}
\langle\ |\mathcal{H}_3|\ \rangle \approx (k\xi k'\xi k''\xi)^{1/2}(1+\Lambda)\frac{mc^2}{\rho^{1/2}} \approx \epsilon^{3/2}(1+\Lambda)\frac{mc^2 \xi^{d/2}}{(\rho\xi^d)^{1/2}}
\end{equation} where we have introduced the density $\rho$ of the fluid and the so-called relaxation or correlation length of the superfluid, 
\begin{equation}\label{eq026}
\xi=\frac{\hbar}{mc}
\end{equation} We thus end up with the estimate 
\begin{equation}\label{eq028}
\boxed{\Sigma_\qq^{(2n)}(\ii0^+)\underset{\epsilon\to 0}{\approx} k_BT \gamma \epsilon^2 \left[\frac{\epsilon^{2d-4}(1+\Lambda)^2}{\gamma^{(5-d)/2}\rho\xi^d}\right]^n \quad\quad (n>0,d\geq 2)}
\end{equation} which designates the quantity in square brackets as the small parameter of the perturbative expansion of $\Sigma_\qq$ in the coupling. The degree in $\epsilon$ of the leading non-zero order, corresponding to the choice $n=1$ in equation (\ref{eq028}), gives the value of the exponent $\nu$ in equation (\ref{eq021}): 
\begin{equation}\label{eq004}
\nu=2d-2
\end{equation} The value of the second exponent $\sigma$ is obtained by generalizing estimate (\ref{eq024}) to the case $\zeta\neq 0$, which amounts to adding $\zeta$ to the energy of the initial state and therefore to each energy denominator, $\Delta E_j\to \Delta E_j+\zeta$. As we've seen, $\Delta E_j\approx k_B T\epsilon^2$; this is also the width of the self-energy around $z=\epsilon_\qq$ (around $\zeta=0$), so that
\begin{equation}\label{eq005}
\sigma=2
\end{equation} 
irrespective of dimensionality.

\paragraph{Three-dimensional case}
 In dimension $d=3$, where $\nu=4>\sigma=2$ and where $2d-4=2>0$ in the square brackets of equation (\ref{eq028}), steps (i) and (ii) are justified and the damping of mode $\qq$ is described exactly by Fermi's golden rule in the limit (\ref{eq020}). We then find the low-temperature behavior of the imaginary part of $\Sigma_\qq(\textrm{i}0^+)$, 
\begin{multline}
\label{eq030}
\im\Sigma_\qq^{d=3}(\textrm{i}0^+) \stackrel{\qb\,\mbox{\scriptsize fixed}}{\underset{\epsilon\to 0}{\sim}} -\frac{9 k_B T \epsilon^4(1+\Lambda)^2}{32\pi\rho\xi^3}\Bigg[\int_0^{\qb}\dd\kb\, \kb^2(\qb-\kb)^2(\bar{n}^{\rm lin}_k+1/2)+\int_0^{+\infty}\dd\kb\, \kb^2(\kb+\qb)^2(\bar{n}^{\rm lin}_k-\bar{n}^{\rm lin}_{k+q})\Bigg] \\
=-\frac{9 k_B T \epsilon^4(1+\Lambda)^2}{32\pi\rho\xi^3}\Bigg\{\frac{\qb^5}{60}+48[g_5(1)-g_5(\eee^{-\qb})]-24\qb\, g_4(\eee^{-\qb})+4\qb^2[g_3(1)-g_3(\eee^{-\qb})]\Bigg\}
\end{multline}
 corresponding exactly to the $\Gamma_q$ damping rate given in reference \cite{Annalen} (multiplied by the usual $-\hbar/2$ factor). In equation (\ref{eq030}), Bose functions $g_\alpha(z)=\sum_{n=1}^{+\infty} z^n/n^\alpha$ have been introduced and it has been possible to replace the dispersion relation $\veps_\kk$ by its linear approximation $\hbar c k$ in the occupation numbers, which is indicated by the exponent \g{lin}: 
\begin{equation}\label{eq029}
\bar{n}_k^{\rm lin}=\frac{1}{\exp\kb-1}
\end{equation} It is satisfying to note that the prefactor in (\ref{eq030}) is indeed that predicted by (\ref{eq028}) for $n=1$ (and $d=3$); in particular, it is independent of the curvature parameter $\gamma$. Green's function formalism also gives access to the real part, which {\yct was out of reach of} reference \cite{Annalen} and represents the thermal energy shift of phonon $\qq$,\footnote{We omit in (\ref{eq031}) the so-called quantum component, independent of occupation numbers, which is at the leading order in $\epsilon$ the sum of two contributions, (i) an ill-defined contribution, linear combination of monomials $q$, $q^3$ and $q^5$ with coefficients divergently dependent on the ultraviolet cutoff $\eta$ of equation (\ref{eq103}), which must therefore be reabsorbed in $\veps_\qq$, i.e. in a redefinition of the dispersion relation (see also footnote \ref{note0}), and (ii) a universal contribution predicted quantitatively by quantum hydrodynamics, $\re\Sigma_{\qq,{\rm quant\ univ}}^{d=3}(\textrm{i}0^+)=[3(1+\Lambda)^2/(160\pi^2)](\hbar^2 q^5/m\rho)\ln(\hbar q/mc)$. Such a contribution in $q^5\ln q$ was qualitatively expected in reference \cite{SonWingate}. It appears in equation (35) of reference \cite{Barc}, with the correct numerical coefficient but with an extra factor $c$.} 
\begin{multline}
\label{eq031}
\re\Sigma_{\qq,{\rm th}}^{d=3}(\textrm{i}0^+) \stackrel{\qb\,\mbox{\scriptsize fixed}}{\underset{\epsilon\to 0}{\sim}} \frac{k_B T \epsilon^4(1+\Lambda)^2}{2\pi^2\rho\xi^3}
\Bigg\{\qb\zeta(4)\left[-27\ln\frac{\eee(3\gamma)^{1/2}\epsilon}{4}+\frac{2+18\Lambda+3\Lambda_*/2}{(1+\Lambda)^2}\right] \\
+\frac{9}{8} \int_0^{+\infty} \dd\kb\, \kb^2\, \bar{n}^{\rm lin}_k \left[(\kb-\qb)^2\ln|\kb-\qb|-(\kb+\qb)^2\ln(\kb+\qb)\right] \Bigg\}
\end{multline}
 where $\zeta$ is the Riemann zeta function, such {\yct that} $\zeta(s)=g_s(1)$, and $\zeta(4)=\pi^4/90$. For the real part, we obtain a more complicated form than that of (\ref{eq028}) because (a) the angles between $\qq$ and $\kk$ of order $\theta\approx\epsilon^0$ contribute at the same level as the small angles $\theta\approx\epsilon$,\footnote{Indeed, in the angular integration, in $d=3$, $\sin\theta\,\dd\theta/(\Delta E/k_B T)$ is $\approx \epsilon^0$ at large angles and $\approx \epsilon^2/\epsilon^2=\epsilon^0$ at small angles, where $\Delta E$ is an energy denominator. This phenomenon does not occur for the imaginary part, given the energy-conserving Dirac distribution $\delta(\Delta E)$ and the fact that $\Delta E$ only vanishes at small angles for low-temperature thermal phonons. In practice, to obtain (\ref{eq031}), we split the angular integral into two intervals, $\theta\in[0,\alpha]$ and $\theta\in [\alpha,\pi]$, where $\alpha\ll 1$ is fixed. In the first interval, the curvature term in the dispersion relation is kept in the denominator of the integrand, but the numerator is replaced by its leading order in $\theta$ (i.e. $\sin\theta\simeq\theta$). In the second interval, we use the linear dispersion relation, but keep all the dependence in $\theta$ of the trigonometric functions. All calculations done, the lower and upper parts of the integral include a divergence in $\ln\alpha$ which disappears exactly in their sum.} and (b) the quartic Hamiltonian $H_4$ also contributes, which forces us to introduce the second coupling constant $\Lambda_*$ of equation (\ref{eq032}). In particular, we notice a logarithmic correction to the $k_B T \epsilon^4$ law. In the $\qb\to 0$ limit, expression (\ref{eq031}) becomes $\sim \hbar q \delta c_{\rm th}$, where coefficient $\delta c_{\rm th}$ is the thermal correction to the speed of {\ycd collisionless} sound in the fluid. The value of $\delta c_{\rm th}$ derived from Andreev and Khalatnikov's kinetic theory \cite{AK}, see equation (45) in reference \cite{Barc}, is in good agreement with ours,\footnote{Perfect agreement is reached if, in equation (45) of \cite{Barc}, we replace $\ln 27$ by $25/6-2C+\ln(3/2)+2\zeta'(4)/\zeta(4)=\ln(26.852\ldots)$ where $C=0.577\ldots$ is Euler's constant. This small deviation results from an approximation made by Andreev and Khalatnikov in the calculation of an integral, that of equation (13) in \cite{AK}.} but that deduced from the leading order of effective field theory, see equation (33) in reference \cite{Barc}, is in disagreement.

\paragraph{Two-dimensional case}
 In dimension $d=2$, where $\nu=\sigma=2$ and where $2d-4=0$, we fail to justify step (i) and step (ii) at low temperature! In particular, the bracketed expression in equation (\ref{eq028}) does not depend on $\epsilon$ and only becomes a small parameter in the $\rho\xi^2\gg 1$ limit where the underlying superfluid itself (and not just the phonon gas) enters the weakly interacting regime. So, in $d=2$, taking the $\epsilon\to 0$ limit makes neither the pole approximation nor the $\Sigma_\qq(\zeta)$ second-order calculation in phonon coupling accurate. To obtain the correct expression of the signal in this limit, we need to calculate $\tilde{\Sigma}_{\qb}(\zetat)$ for all values of $\zetat$ and at all orders in $H_3$. However, let's take as a reference value the low-temperature behavior of $\Sigma_\qq(\textrm{i}0^+)$ at the leading order in the coupling ($n=1$): 
\begin{multline} 
\label{eq033}
\Sigma_\qq^{(2)}(\textrm{i}0^+)\stackrel{\qb\,\mbox{\scriptsize fixed}}{\underset{\epsilon\to 0}{\sim}} -\ii\frac{9k_B T\epsilon^2(1+\Lambda)^2}{8\pi(3\gamma)^{1/2}\rho\xi^2} \Bigg[\int_0^{\qb}\dd\kb\, \kb(\qb-\kb)(\bar{n}^{\rm lin}_k+1/2)+\int_0^{+\infty}\dd\kb\, \bar{k}(\kb+\qb) (\bar{n}^{\rm lin}_k-\bar{n}^{\rm lin}_{k+q})\Bigg] \\
=\boxed{-\ii\frac{9k_B T\epsilon^2(1+\Lambda)^2}{8\pi(3\gamma)^{1/2}\rho\xi^2} \Bigg[\frac{\qb^3}{12}+2\zeta(2)\qb\Bigg]\equiv -\textrm{i}\frac{\hbar\Gamma_q}{2}}
\end{multline}
 with $\zeta(2)=\pi^2/6$. At sufficiently low temperatures, $\Sigma^{(2)}_\qq(\textrm{i}0^+)$ becomes pure imaginary and reduces to $-\ii\hbar\Gamma_q/2$ where $\Gamma_q$ is the damping rate of the golden rule at the leading order in $\epsilon$ (the quartic Hamiltonian $H_4$ does not show a real part here as it is subleading in $\epsilon$, see section \ref{sec2}). Equation (\ref{eq033}) is, as far as we know, original. In limits $\qb\to 0$ and $\qb\to +\infty$, it reproduces equations (30) and (22) of reference \cite{CB} (times $-\ii\hbar$ given the conventions of \cite{CB}), which deals with the two-dimensional gas of weakly interacting bosons for which $\gamma=1$ (as in the Bogolioubov spectrum) and $\Lambda=0$ (as in the zero-temperature mean-field equation of state $\mu=\rho g$ used in \cite{CB}). In the following, in dimension $d=2$, we go beyond Fermi's golden rule with classical field numerical simulations in section \ref{sec1} and with the Green's function method to higher and higher orders in $H_3$ in later sections: to order two in section \ref{sec2}, to order four in section \ref{sec3} and non-perturbatively in section \ref{sec4}; to our surprise, we are forced to do so by the failure of estimate (\ref{eq028}) for $n\geq 2$. {\ycd In fact, by an explicit calculation of the fourth order contribution in $H_3$ in section \ref{sec3}, we establish a troubling result that disproves (\ref{eq028}) in dimension $d=2$: $\Sigma_{\qq}^{(4)}(\ii 0^+)$ does not tend to zero as $k_B T \epsilon^2$ at low temperature $\epsilon \to 0$, but as $k_B T\epsilon^0$, see equation (\ref{eq296}). This effectively makes the perturbative expansion in powers of $H_3$ unusable in this limit, since the fourth order becomes much larger than the second order for a fixed interaction strength $(\rho\xi^2)^{-1}$ in the superfluid. To obtain an operational description of the damping, section \ref{sec4}, contrary to equation (\ref{eq028}), takes the limit $\epsilon \to 0$ with fixed $\zetat\neq 0$ in $\tilde{\Sigma}_{\qb}^{(4)}(\zetat)$, regularizes the result (divergent at $\zetat=0$) nonperturbatively by assigning finite lifetimes to the intermediate phonons of the four-phonon collisional processes, and calculates the integral over $\zetat$ in equation (\ref{eq023}), finally obtaining a damping rate scaling as $k_B T\epsilon^2$ as expected.} We conclude in section \ref{sec5}.

\section{Numerical experiment in the classical field model}
\label{sec1}

Experimental studies of sound in two-dimensional superfluid ultracold atomic gases have so far concentrated on the hydrodynamic \cite{HM,PCZH} regime or, in the collisionless regime, on the $\epsilon=k_B T/mc^2\geq 1$ \cite{JDJB,Salasnich,Stringari} temperature interval. To verify that in dimension $d=2$, the low-temperature limit (\ref{eq020}) is not sufficient to render accurate the exponential decay predicted by Fermi's golden rule, we therefore carry out in this section a numerical study of phonon damping, at all orders in coupling $H_3$, in the classical field model. We first recall the quantum hydrodynamics Hamiltonian of Landau and Khalatnikov. We then explain how to numerically integrate its classical field version and introduce a correlation function that is more convenient to calculate than signal (\ref{eq012}), but almost equivalent. We then analyze the results obtained for different interaction strengths. 

\subsection{Quantum hydrodynamics Hamiltonian}
\label{sec1.1}

Quantum hydrodynamics is an effective low-energy theory that provides accurate predictions to leading order in temperature, even for a strongly interacting fluid \cite{LK}. The fluid is decomposed into small, but still mesoscopic elements of size $b$ ($b\gg\xi=\hbar/mc$). Each element, with center $\rr$, is (i) sufficiently large that it has a large number of particles ($\rho b^d\gg 1$), allowing it to be assigned a phase operator $\hat{\phi}(\rr)$ canonically conjugated to its particle number operator $\hat{\rho}(\rr)b^d$, $[\hat{\rho}(\rr)b^d,\hat{\phi}({\yct\rr'})]=\ii\delta_{\rr,\rr'},$ and (ii) small enough to be considered both as homogeneous ($b\ll\lambda_{\rm th}$ where $\lambda_{\rm th}=2\pi\hbar c/k_B T$ is the de Broglie wavelength of a thermal phonon in the fluid) and in its ground state ($k_B T\ll\veps_{\rm exc}^{\rm min}$ where $\veps_{\rm exc}^{\rm min}=\hbar c (2\pi/b)$ is the minimum phonon excitation energy in the small element). \footnote{The ultraviolet cutoff to come corresponds more or less to the choice $\pi/b\approx \eta k_B T/\hbar c$. Even so, $b$ remains well below the coherence length of the fluid at sufficiently low temperatures. This is obvious in $d=3$ because of the presence of condensate. It is also true in $d=2$ where the first-order coherence function $g_1(\rr)$ decreases at large distances as $\exp[mk_B T\ln(\xi/r)/(2\pi\hbar^2\rho_s)]$ where $\rho_s\simeq \rho$ is the superfluid density \cite{Bere,Nel,SJ}. This makes the assumption $\hat{\vv}=(\hbar/m)\mathbf{grad}\,\hat{\phi}$ used in reference \cite{LK} and in equation (\ref{eq050}) plausible.} The Hamiltonian is then the sum of the kinetic energies of the center of mass and the internal energy of each small element: 
\begin{equation}\label{eq050}
H_{\rm hydro}=\sum_\mathbf{r}\frac{1}{2} m \hat{\vv}(\rr)\cdot\hat{\rho}(\rr) b^d\hat{\vv}(\rr)+b^d e_0(\hat{\rho}(\rr))
\end{equation} where $\hat{\vv}=(\hbar/m)\mathbf{grad}\,\hat{\phi}$ is the fluid velocity field and $e_0(\rho)$ is the bulk energy density in the ground state at density $\rho$. At low temperatures, the density field $\hat{\rho}(\rr)$ fluctuates little and the phase field $\hat{\phi}(\rr)$ varies slowly; we separate them into their Fourier components of zero wave vector $\hat{\rho}_0=\hat{N}/L^d\simeq \rho$ and $\hat{\phi}_0$, and into spatial fluctuations $\delta\hat{\rho}(\rr)$ and $\delta\hat{\phi}(\rr)$ admitting modal expansions {\yc\cite{broui}}
\begin{equation}\label{eq052}
\delta\hat{\rho}(\rr)=\frac{1}{L^{d/2}} \sum_{\kk\neq\zero} \left(\frac{\hbar\rho k}{2mc}\right)^{1/2}(\hat{b}_\kk+\hat{b}_{-\kk}^\dagger)\,\eee^{\ii\kk\cdot\rr} \quad ; \quad \delta\hat{\phi}(\rr)=\frac{-\ii}{L^{d/2}}  \sum_{\kk\neq\zero} \left(\frac{mc}{2\hbar\rho k}\right)^{1/2} (\hat{b}_\kk-\hat{b}_{-\kk}^\dagger)\,\eee^{\ii\kk\cdot\rr}
\end{equation} where $\hat{b}_\kk^\dagger$ and $\hat{b}_\kk$ are the creation and annihilation operators of a phonon of wave vector $\kk$ in the quantization box $[0,L]^d$, $c$ is the speed of sound at zero temperature, given by $mc^2=\rho(\dd\mu/\dd\rho)$ with $\mu=\mathrm{d}e_0/\dd\rho$ the chemical potential of the fluid in the ground state. As in any effective theory, we need to put an ultraviolet cutoff in the space of wave vectors, which we take here to be isotropic 
\begin{equation}\label{eq103}
\hbar c k < \eta k_B T
\end{equation} and to which the sums (\ref{eq052}) are restricted; we'll let the temperature tend towards zero as in (\ref{eq020}) and then $\eta\to+\infty$ (in the quantum case) without encountering divergences if all goes well. It remains to expand $H_{\rm hydro}$ into powers of $\delta\hat{\rho}$ and $\mathbf{grad}\,\delta\hat{\phi}$. Zero order provides a constant of no interest. Order one is zero by construction.  Order two, diagonalized by the transformation (\ref{eq052}), gives back the Hamiltonian $H_2$ of equation (\ref{eq002}) {\yct with} a linear dispersion relation, the curvature in equation (\ref{eq003}) having to be put in by hand, see references \cite{LK,Annalen} and our footnote \ref{note1}. The third and fourth orders, giving the coupling between phonons in equation (\ref{eq001}), are written in representation $\rr$: 
\begin{equation}\label{eq104}
H_3=b^d\sum_\mathbf{r}\frac{1}{2} m\hat{\vv}(\rr)\cdot\delta\hat{\rho}(\rr)\hat{\vv}(\rr)+\frac{1}{6}\frac{\dd^2\mu}{\dd\rho^2}\delta\hat{\rho}(\rr)^3 \quad ; \quad H_4=b^d\sum_\rr\frac{1}{24} \frac{\dd^3\mu}{\dd\rho^{3}}\delta\hat{\rho}(\rr)^4
\end{equation} knowing that $\mathbf{grad}\,\hat{\phi}=\mathbf{grad}\,\delta\hat{\phi}$. They involve the dimensionless coupling constants 
\begin{equation}\label{eq032}
\Lambda=\rho\frac{\dd^2\mu}{\dd\rho^2}/(3\frac{\dd\mu}{\dd\rho})=\rho^2\frac{\dd^2\mu}{\dd\rho^2}/(3 mc^2)\quad\mbox{and}\quad \Lambda_*=\rho^3\frac{\dd^3\mu}{\dd\rho^3}/(mc^2)
\end{equation} To reveal the physical processes contained in $H_3$ and already mentioned in section \ref{sec0}, we insert in (\ref{eq104}) the modal expansions (\ref{eq052}) and collect together the terms of the same type, Beliaev, Landau and non-resonant respectively: 
\begin{equation}\label{eq100}
H_3 = {\ycd H_3^{(+)} + H_3^{(-)}} + H_3^{\mbox{\yct\scriptsize (non res.)}}
\end{equation} with 
\begin{equation}\label{eq101}
H_3^{(-)}= (H_3^{(+)})^\dagger = \frac{1}{2 L^{d/2}}\sum_{\kk,\kk'} \mA(\kk,\kk';\kk+\kk') \hat{b}_{\kk+\kk'}^\dagger \hat{b}_\mathbf{k}\hat{b}_{\kk'} \quad ; \quad H_3^{\mbox{\yct\scriptsize (non res.)}} = \frac{1}{6 L^{d/2}} \sum_{\kk,\kk'} \mA(\kk,\kk';-(\kk+\kk')) (\hat{b}_{\kk}\hat{b}_{\kk'}\hat{b}_{-(\kk+\kk')}+\mbox{h.c.})
\end{equation} and {\yct the} coupling amplitude\footnote{By choice of notation, our amplitude $\mA$ differs by a factor of two from that of reference \cite{Annalen} {\yct(more precisely by a factor $2mc^2/\rho^{1/2}$)}.} 
\begin{equation}\label{eq102}
\mA(\kk,\kk';\kk'')=mc^2\left(\frac{\hbar^3kk'k''}{8\rho m^3c^3}\right)^{1/2}\left(3\Lambda+\frac{\kk\cdot\kk'}{k k'}+\frac{\kk\cdot\kk''}{k k''}+\frac{\kk'\cdot\kk''}{k' k''}\right)
\end{equation} Of course, all wave vectors $\kk$, $\kk'$, $\pm(\kk+\kk')$ appearing in the Hamiltonian are non-zero and subject to condition (\ref{eq103}). Writing $H_4$ in representation $\hat{b}$ and $\hat{b}^\dagger$ is of little interest here, and we dispense with it.

\subsection{Classical field model}
\label{sec1.2}

The model is built by replacing operators $\hat{b}_\kk$ and $\hat{b}_\kk^\dagger$ in section \ref{sec1.1} with complex amplitudes $b_\kk$ and $b_\kk^*$. This only makes sense for highly populated phonon modes, which imposes a choice of cutoff $\eta\lesssim 1$ in equation (\ref{eq103}); in the simulations, we'll take $\eta=1$. At thermal equilibrium, mode occupation numbers are no longer given by Bose's law, but by equipartition $\bar{n}_\kk^{\rm clas}=k_B T/\veps_\kk$. 

Let's restrict our numerical study to the experimentally relevant case of a {\yct gas of spinless bosons} in the quasi-2D regime, where the chemical potential $\mu$ in the $xOy$ plane is less than, but not very small compared to $\hbar\omega_\perp$, the quantum of excitation of the motion along $Oz$; in this case, the equation of state takes the simple form $\mu=\rho g$ with $g=(8\pi)^{1/2}\hbar^2a_{\rm 3D}/(ma_\perp)$, where $a_{\rm 3D}$ is the 3D scattering length and $a_\perp=(\hbar/m\omega_\perp)^{1/2}$ the size of the harmonic oscillator along $Oz$ \cite{CD,HD}, so that $\Lambda=\Lambda_*=0$ in equation (\ref{eq032}) and $H_4\equiv 0$ in equation {\yc(\ref{eq104})}. \footnote{The Popov equation of state of a strictly two-dimensional Bose gas in the weakly interacting regime corresponds to a chemical potential-dependent pseudo-coupling constant, $g(\mu)=4\pi\hbar^2/[m\ln(\veps_0/\mu)]$ where $\veps_0=4\hbar^2/[ma_{\rm 2D}^2\exp(2C+1)]$, $a_{\rm 2D}$ is the 2D scattering length and $C=0.577\ldots$ is the Euler constant \cite{Popov,Morathese}. In its range of validity \cite{MC2}, it leads to non-zero but small values of $\Lambda$ and $\Lambda_*$, for example {\ycd $\Lambda=mg(\mu)/(12\pi\hbar^2)/[1-mg(\mu)/(4\pi\hbar^2)]^2\simeq mg(\mu)/(12\pi\hbar^2) \ll 1$. The quasi- and strictly 2D regimes are connected by noting that $g=g(\mu=0.105\,965\ldots\hbar\omega_\perp)$ in view of the expression of $a_{\rm 2D}$ in terms of $a_{\rm 3D}$ and $a_\perp$ given in references \cite{Petrov,Olshanii,Ludovic}}.}

The numerical simulation method is described in detail in reference \cite{TheseAlan}. To sample the initial state, $b_\kk$ are drawn according to the (Gaussian) thermal distribution for the $H_2$ Hamiltonian, then the resulting field is evolved with the full Hamiltonian $H$ to reach true thermal equilibrium {\yc by ergodicity}. The time evolution on each $dt$ step is split into the evolution due to $H_2$ during $dt/2$, the evolution due to $H_3$ alone during $dt$ and then again the evolution due to $H_2$ during $dt/2$. The evolution due to $H_2$ is calculated exactly in $\kk$ space, while the evolution due to $H_3$ is calculated in $\rr$\footnote{Since the equations of motion deriving from $H_3$ are local in $\rr$ space (or quasi-local since they involve spatial gradients), the complexity on a step $dt$ is of degree one in the number of points $\mathcal{N}$ on the numerical grid; in $\kk$ space, they are strongly non-local (each amplitude $b_\kk$ is roughly coupled to all the others) resulting in catastrophic complexity $\propto \mathcal{N}^2$.} space by the fourth-order Runge-Kutta method \cite{NumRec}, remembering to filter the field in $\kk$ space as in equation (\ref{eq103}) after each action of $H_3$ in $\rr$ space to eliminate {\yct the cut-off-violating} processes {\yct that $H_3$} has induced.\footnote{The change from $\rr$ to $\kk$ is performed by a fast Fourier transform, in which the wave vectors span the set $\mD\cap(2\pi/L)\mathbb{Z}^d$ with $\mD=[-k_{\rm max},k_{\rm max}[^d$ (each Cartesian component $k_\alpha$ of $\kk$ has a meaning modulo $2 k_{\rm max}$, so that $\mD$ can be regarded as the first Brillouin zone of the spatial lattice $(\pi/k_{\rm max})\mathbb{Z}^d$, which is precisely that spanned by $\rr$ {\yc in the simulations with} restriction to $[0,L[^d$). The naive choice $k_{\rm max}=\eta k_B T/\hbar c$ associated with the cutoff (\ref{eq103}) is not suitable: the processes of $H_3$ bringing phonons out of the corresponding first Brillouin zone are {\yc then} folded back into this zone (Umklapp processes) and can no longer be filtered. At the very least, $k_{\rm max}=2\eta k_B T/\hbar c$ must be used: filtering forces us to double the number of points per dimension.}  The $dt$ step must be small enough for the total energy on a single realization to be conserved to better than one percent of the typical $\langle |H_3|\rangle$ interaction energy, where $\langle\ldots\rangle$ is the average over a large number of realizations, itself a very small fraction of the total energy. 

 It would be numerically inefficient to determine phonon damping by simulating {\yct a} Bragg excitation (\ref{eq010}), as this would require a large number of independent simulations for each new value of wave vector $\qq$, while ensuring that amplitude $\alpha$ is small enough to be in the linear response regime.\footnote{The classical field version $\langle\partial b_\qq(t)/\partial b_\qq(0)\rangle_{\rm th}$ of the correlation function $\langle[\hat{b}_\qq(t),\hat{b}_\qq^\dagger(0)]\rangle_{\rm th}$, which involves a numerical derivative in the particular mode $\qq$, faces the same problems.} We prefer to use the following correlation function, allowing all possible values of $\qq$ to be processed in parallel by simple averaging over the same set of independent realizations: \footnote{The obtention of the {\ycd right-hand} side of equation (\ref{eq110}) {\yc is done by means of the exact relation (31.24) of reference \cite{FW} but is here} approximated as we replaced $\langle\hat{b}_\qq(0)\hat{b}_\qq^\dagger(0)\rangle_{\rm th}$ in the denominator by $1+\bar{n}_{\qq}$, where $\bar{n}_{\qq}$ is the Bose law at energy $\veps_{\qq}$; {\yc this is however} an excellent approximation for our simulations.} 
\begin{equation}\label{eq110}
s_*(t)\equiv \frac{\eee^{\ii\veps_\mathbf{q}t/\hbar}\langle\hat{b}_\qq(t)\hat{b}_\qq^\dagger(0)\rangle_{\rm th}}{\langle\hat{b}_\qq(0)\hat{b}_\qq^\dagger(0)\rangle_{\rm th}}\stackrel{t>0}{\simeq} \lim_{\delta\to 0^+} \int_{-\infty}^{+\infty} 
\frac{\dd\veps}{-2\ii\pi} \eee^{-\ii(\veps-\veps_\qq)t/\hbar} \frac{\mathcal{G}(\qq,\veps+\ii\delta)+\left[\mathcal{G}(\qq,\veps+\ii\delta)-\mathcal{G}(\qq,\veps-\ii\delta)\right]/[\exp(\veps/k_B T)-1]}{1+1/[\exp(\veps_\qq/k_B T)-1]}
\end{equation} {\yct In practice,} this signal is very close to that of equation (\ref{eq012}), which in simulations allows us to identify it with $s(t)$, as we'll do in the figure caption:
\begin{equation}\label{eq111}
s_*(t) \simeq s(t)
\end{equation} Indeed, even for the strongest interaction considered in our simulations, we have, for any value of energy $\veps$, the property $|\im\Sigma(\qq,\veps+\textrm{i}0^+)|\ll \veps_\qq$, so that the difference between the bracketed Green's functions $\mG(\qq,\veps\pm\ii\delta)$ in equation (\ref{eq110}), which can only result from a non-zero imaginary part of the self-energy in the limit $\delta\to 0^+$, is very small except in the width of \g{Lorentzian} $\mathcal{G}(\qq,\veps+\textrm{i}0^+)$ where $|\veps-\veps_\qq|\lesssim |\im\Sigma(\qq,\veps+\textrm{i}0^+)|\ll \veps_\qq, k_B T$. It is therefore an excellent approximation to replace $\veps$ by $\veps_\qq$ in Bose's law in the numerator of the integrand in (\ref{eq110}). By analyticity in the lower complex half-plane, the term proportional to $\mathcal{G}(\qq,z-\ii\delta){\ycd\,\exp(-\ii zt/\hbar)}$ then gives a zero contribution to the integrand {\ycd for $t>0$} (this can be seen by closing the integration path with a semicircle at infinity and using Cauchy's theorem) and we end up with (\ref{eq111}). The same conclusion applies to the classical field model, where the two exponential functions in (\ref{eq110}) are linearized.

\subsection{Numerical results}
\label{sec1.3}
\begin{figure}[t]
\begin{center}
\includegraphics[width=6cm,clip=]{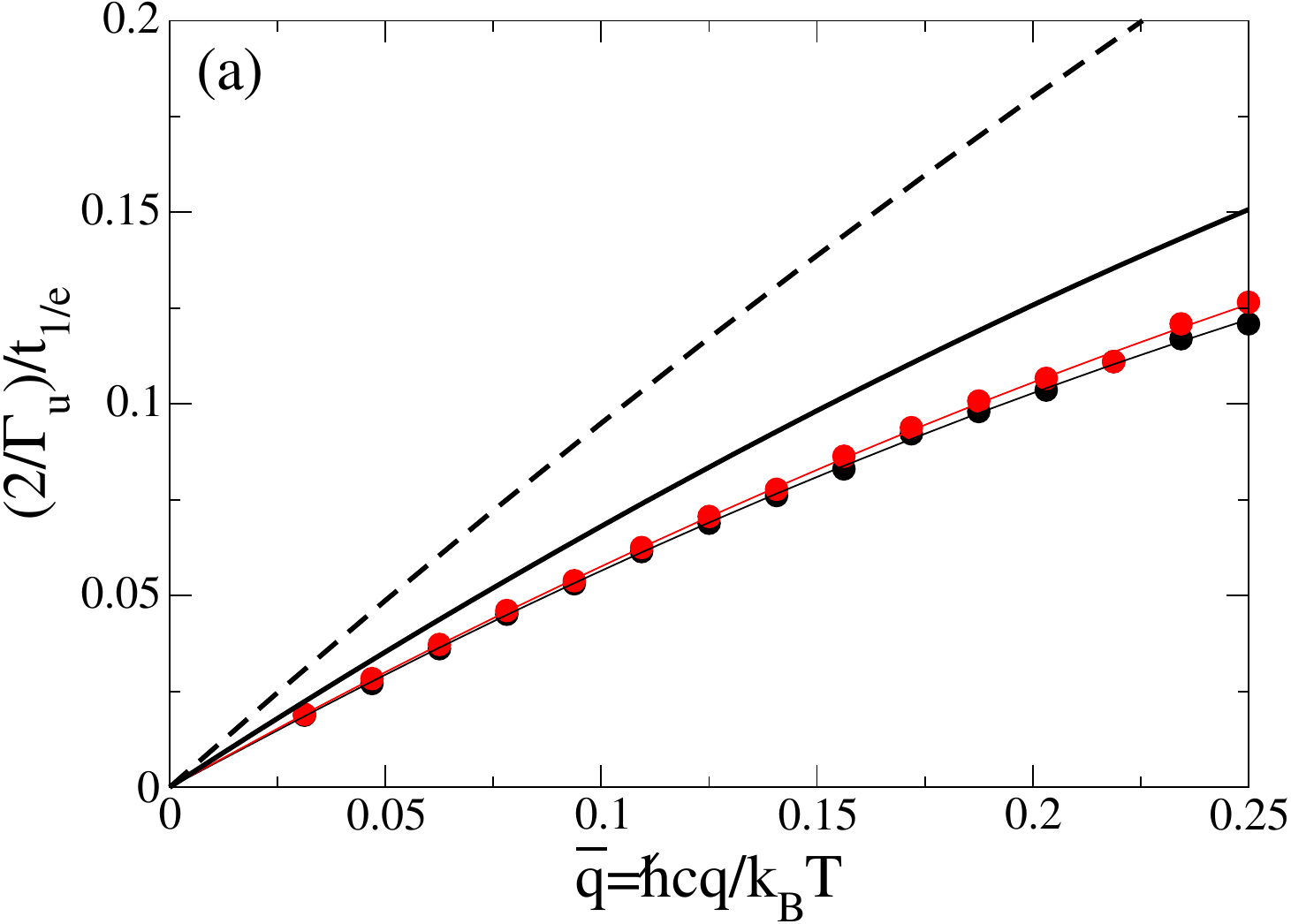} \hspace{1cm} \includegraphics[width=6cm,clip=]{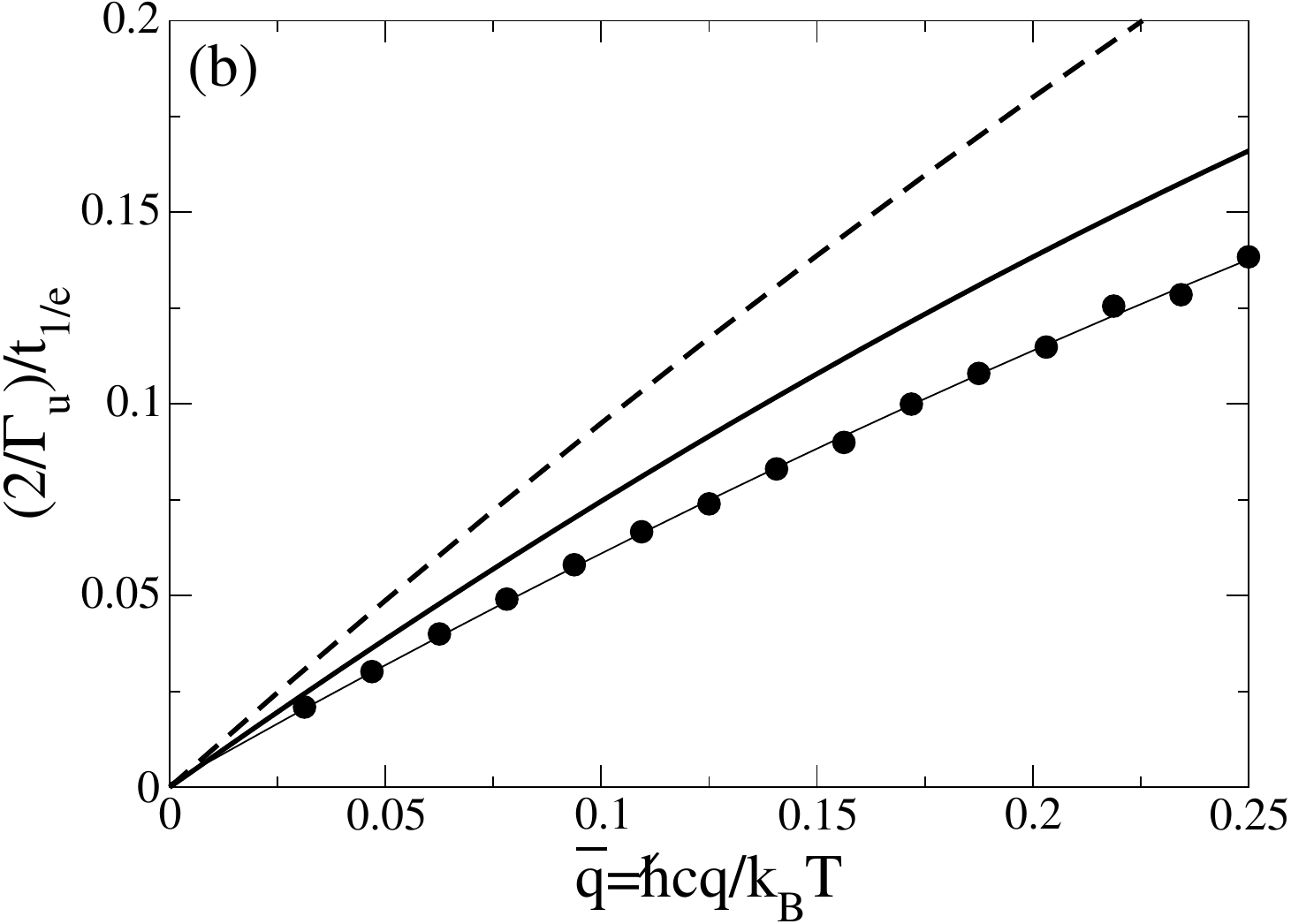} \\
\includegraphics[width=6cm,clip=]{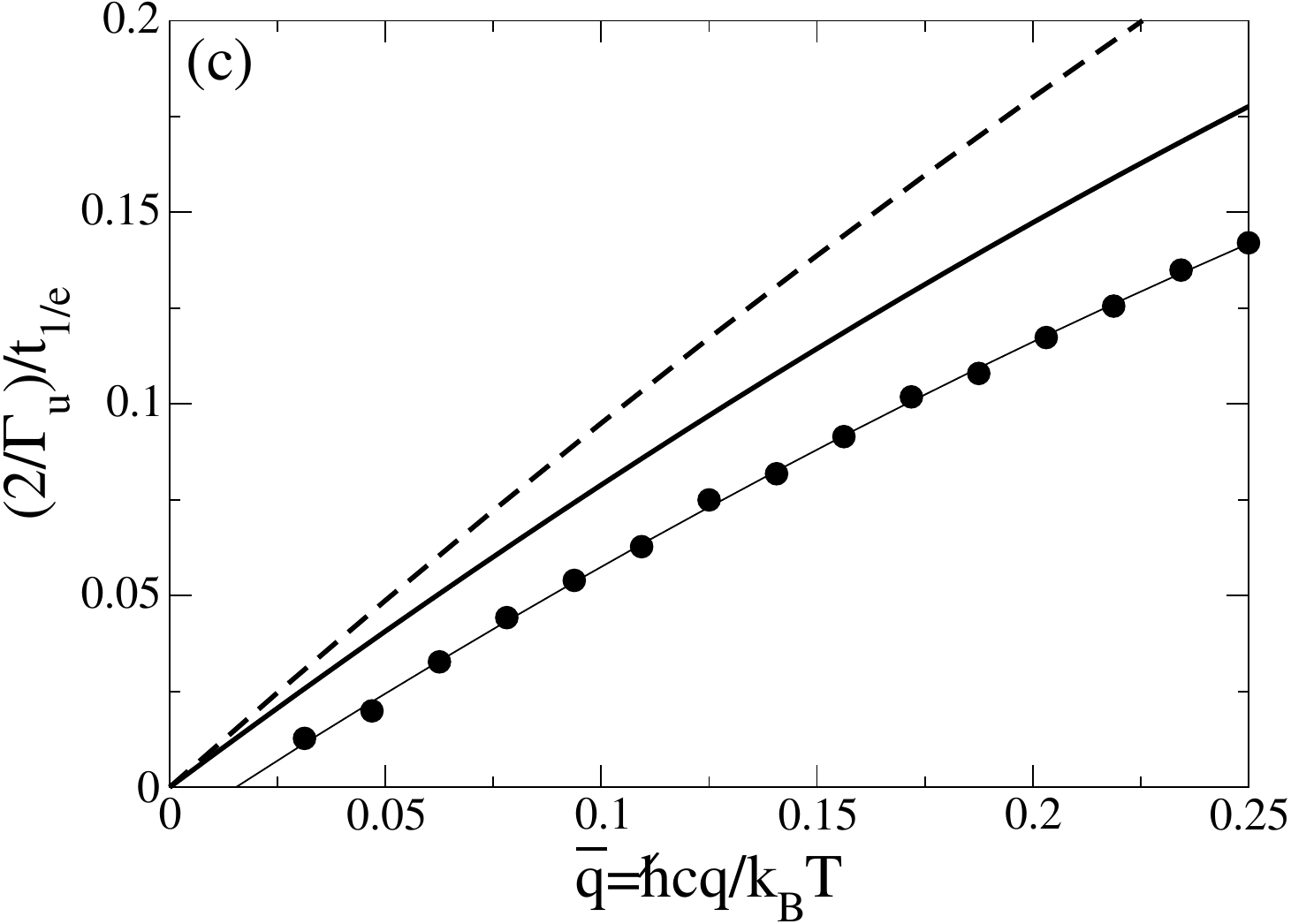} \hspace{1cm} \includegraphics[width=6cm,clip=]{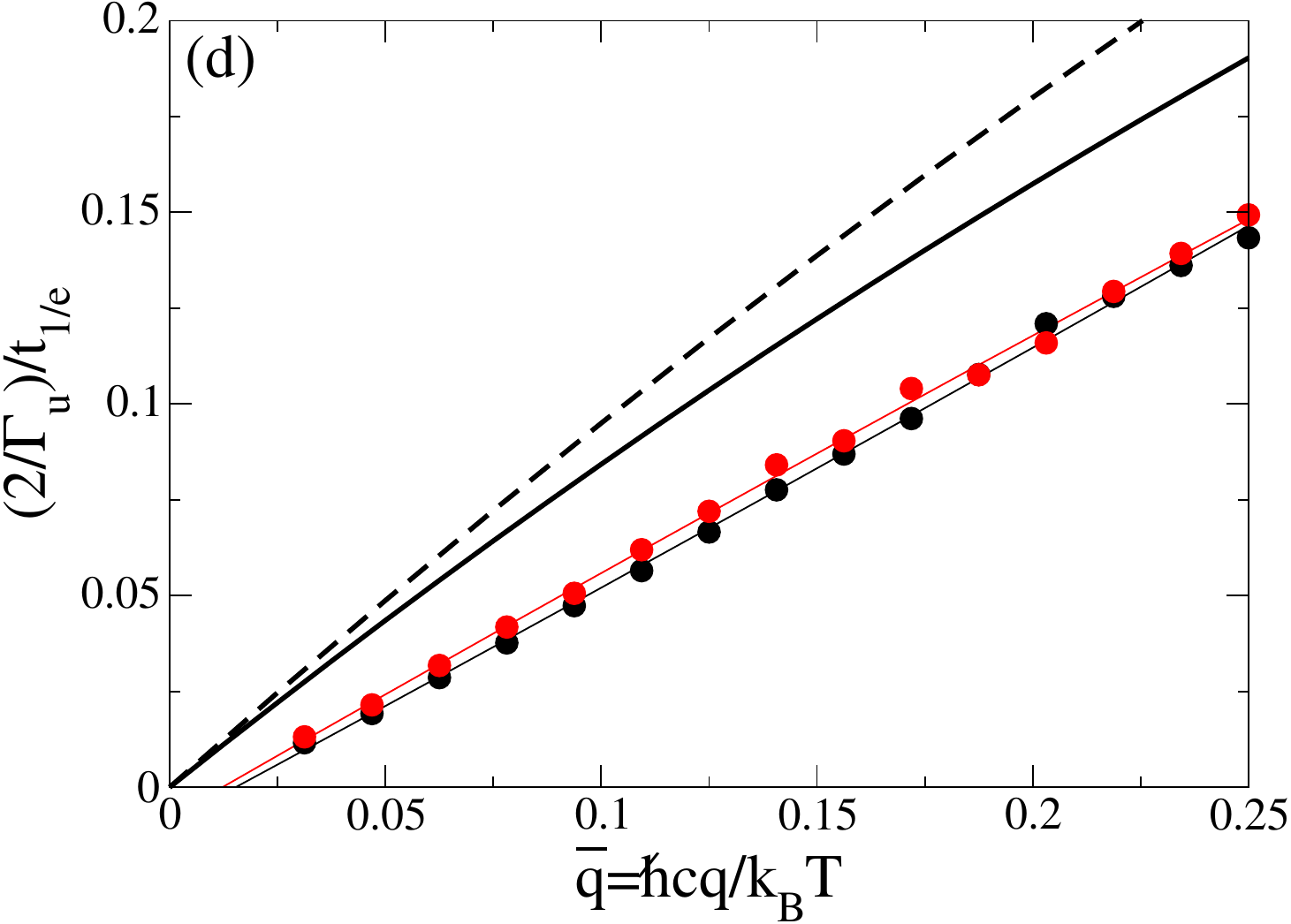}
\end{center}
\caption{Inverse of the decay time at $1/\eee$ of the modulus of signal $s(t)$ (\ref{eq012}) as a function of the wavenumber $q$ of the modes along $Ox$, $\qq=\pm q \mathbf{e}_x$, for different (increasingly weaker) interaction strengths in the {\yct two-dimensional} superfluid {\yct of bosons}: (a) $(\rho\xi^2)^{-1}=0.64$, (b) $(\rho\xi^2)^{-1}=0.46$, (c) $(\rho\xi^2)^{-1}=0.28$, (d) $(\rho\xi^2)^{-1}=0.1$. Discs (superimposed on a thin-line quadratic fit): numerical experiment with $\epsilon=1/2$ (black) or $\epsilon=1/3$ (red). Black dashed line: Fermi's golden rule (\ref{eq113}) for the classical phononic field {\yc in} the $\epsilon\to 0$ limit. Black solid line: many-body Green’s function method with the {\yc$\tilde{\Sigma}^{(2,2)}_{\qb}(\zetat)$} (\ref{eq125}) self-energy (leading order in $H_3$ and $\epsilon$). The other parameters have the following fixed values: $\gamma=1$, $\Lambda=\Lambda_*=0$, $\eta=1$. In the simulations, the system size is large enough for the thermodynamic limit to be reached (a quantization box width of $L=128\pi\hbar c/k_B T$ is sufficient). In the theory, the self-energies are replaced by their classical field versions to enable comparison with numerics. The rate $\Gamma_u$ used as a unit is that of equation (\ref{eq113}).}
\label{fig1}
\end{figure}
 \begin{figure}[t]
\begin{center}
\includegraphics[width=6cm,clip=]{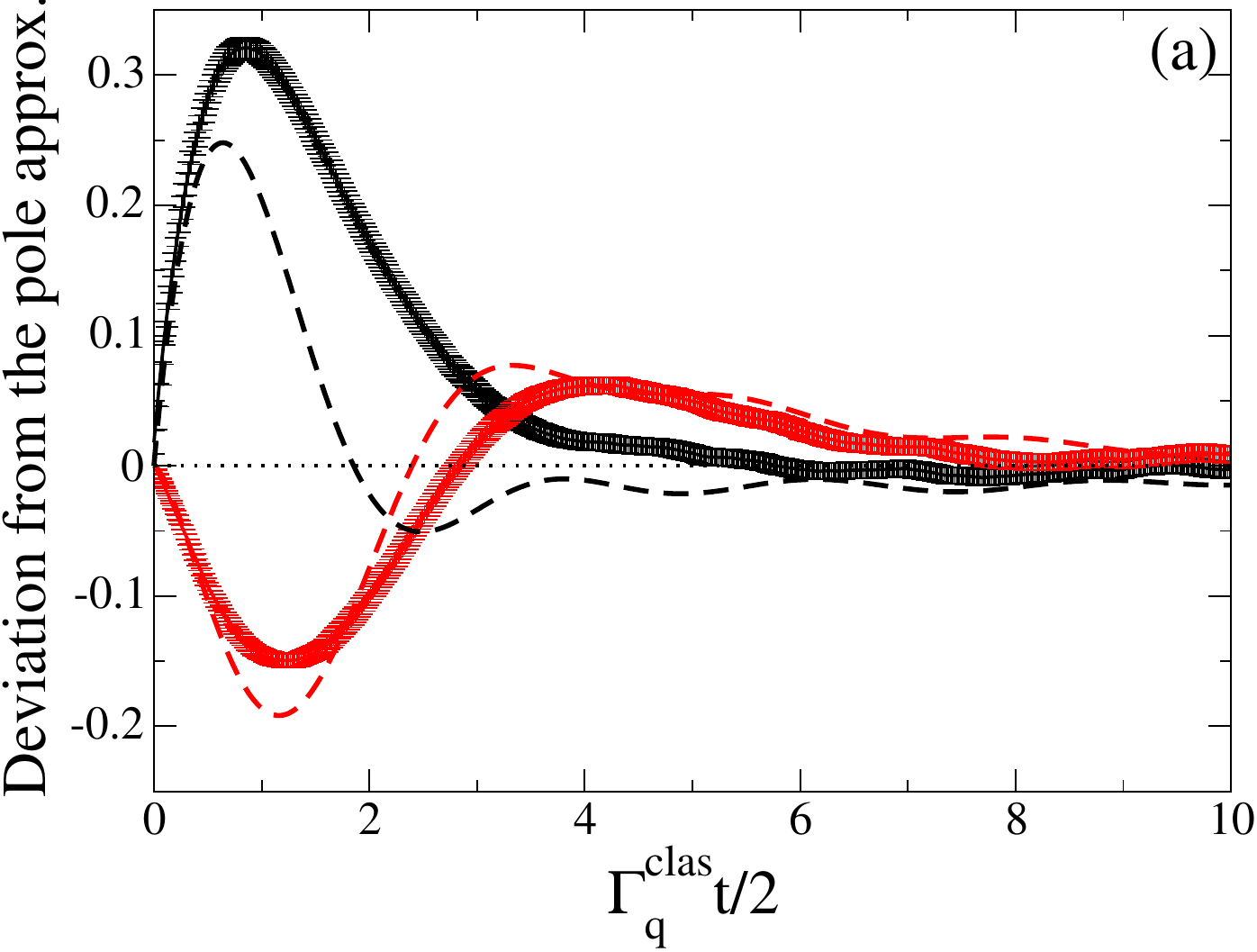}\hspace{1cm} \includegraphics[width=6cm,clip=]{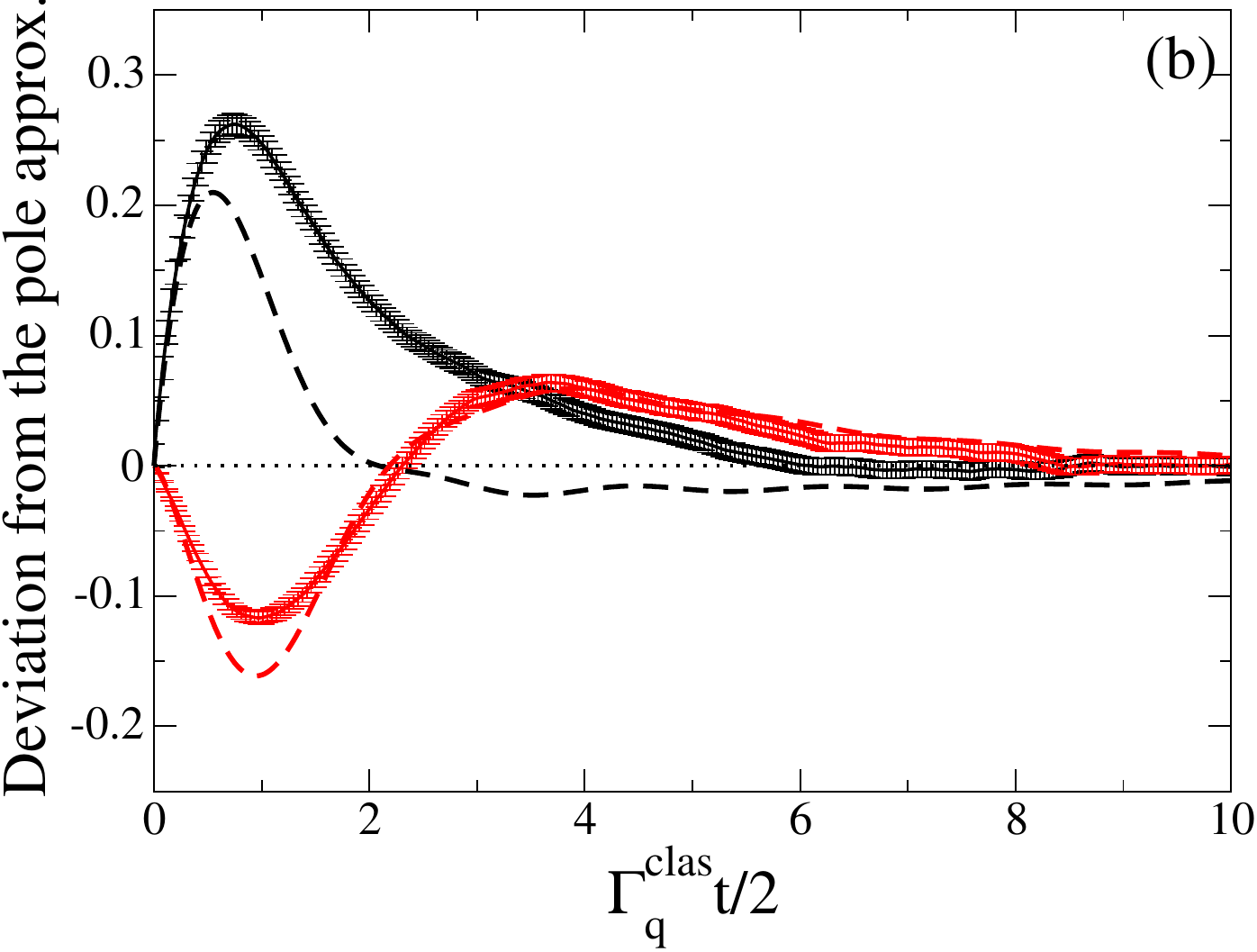} \\
\includegraphics[width=6cm,clip=]{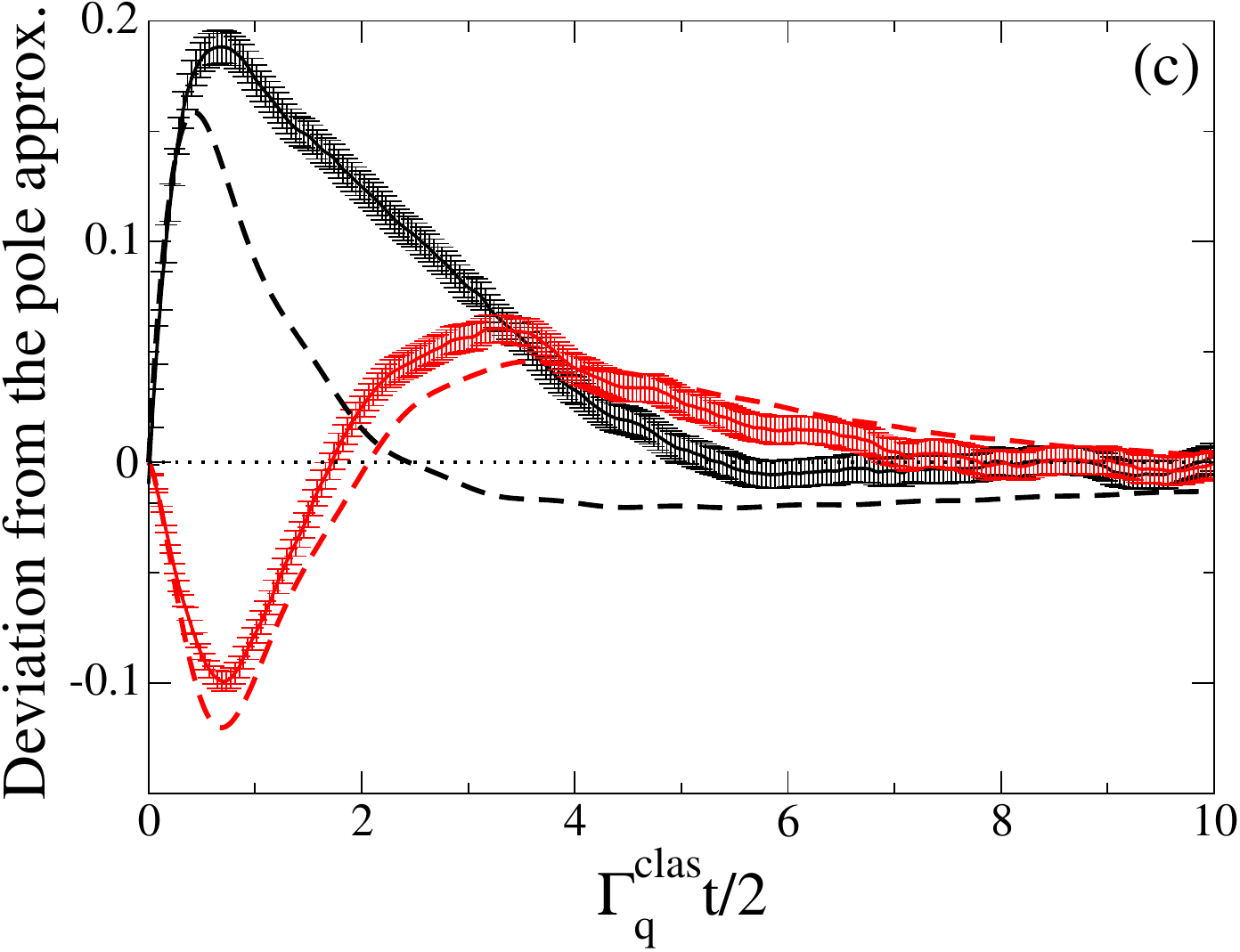}\hspace{1cm} \includegraphics[width=6cm,clip=]{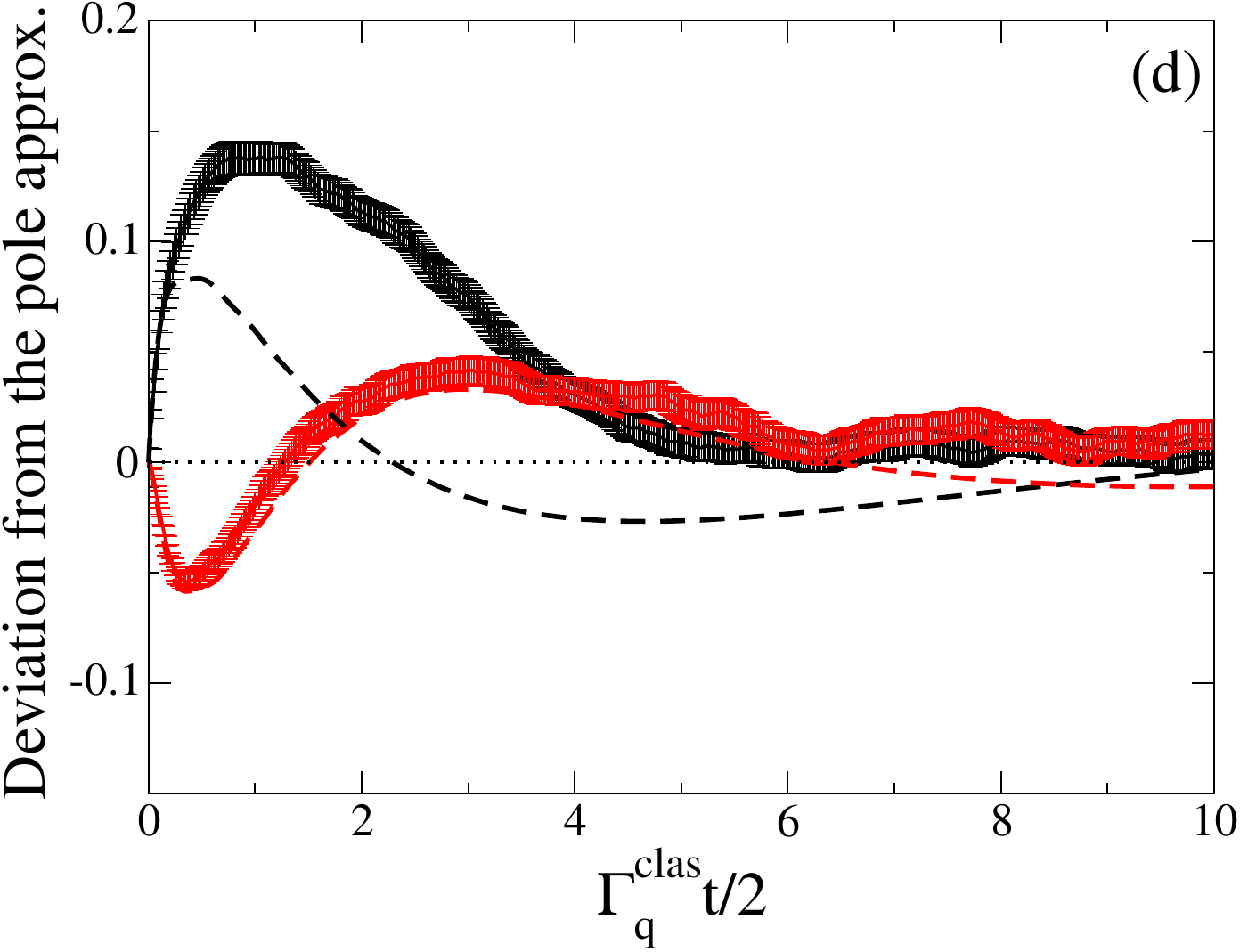}
\end{center}
\caption{Deviation {\yc $s(t)-s_{\mbox{\scriptsize pole}}(t)$} of signal (\ref{eq012}) from pole approximation (\ref{eq014}) as a function of time, for the mode of reduced wavenumber $\qb=1/4$ along $Ox$ and for the same interaction strengths as in Figure \ref{fig1}: (a) $(\rho\xi^2)^{-1}=0.64$, (b) $(\rho\xi^2)^{-1}=0.46$, (c) $(\rho\xi^2)^{-1}=0.28$, (d) $(\rho\xi^2)^{-1}=0.1$. Curves with error bars: results of {\yct the} numerical experiment; dashed line: many-body Green’s function theory with the $\tilde{\Sigma}^{(2,2)}_{\yc\qb}(\zetat)$ (\ref{eq125}) self-energy (leading order in $H_3$ and $\epsilon$) in its classical field version. The real part of the deviation is shown in black, the imaginary part in red. We took $\epsilon=1/2$ in the numerical experiment and for the calculation of $s_{\mbox{\scriptsize pole}}(t)$ (carried out to second order in $H_3$ and to all orders in $\epsilon$) but we checked that we obtain the same results (within the error bars) for $\epsilon=1/3$. The other parameters are fixed at $\gamma=1$, $\Lambda=\Lambda_*=0$, $\eta=1$. In the simulations, the sizes are chosen to ensure that the thermodynamic limit is reached (width $L=128\pi\hbar c/k_B T$ for (a), (b) and (c); double width for (d)). Time is scaled by the half-rate of the golden rule in the classical field (\ref{eq113}).}
\label{fig2}
\end{figure}
 In Figure \ref{fig1}, we plot the inverse of the decay time at $1/\eee$ of the signal as a function of the reduced wavenumber $\qb=\hbar cq/k_B T$ of the mode under consideration, for four values of the interaction strength $1/(\rho\xi^2)$ and for one or two values of $\epsilon=k_B T/mc^2$, see the disks. The unit of time is $2\Gamma_u^{-1}$, where the constant $\Gamma_u$ appears as a prefactor in the damping rate of Fermi's golden rule at the leading order in $\epsilon$, here for the classical field: \footnote{In expression (\ref{eq033}) for the quantum field, we replace $\bar{n}_k^{\rm lin}$ by $1/\kb$, neglect the term $1/2$ in front of $\bar{n}_k^{\rm lin}$ in the first integral and put the cutoff $\kb<\eta-\qb$ in the second.} 
\begin{equation}\label{eq113}
\boxed{\Gamma_q^{\mbox{\scriptsize clas}}=\Gamma_u \qb(\eta-\qb/2) \quad \mbox{with}\quad \Gamma_u = \frac{9k_B T\epsilon^2(1+\Lambda)^2}{4\pi\hbar(3\gamma)^{1/2}\rho\xi^2}}
\end{equation} As can be seen in panels (a) and (d) of Figure \ref{fig1}, the results normalized in this way do not depend significantly on $\epsilon$ at $\qb$ fixed and the $\epsilon\to 0$ limit appears to be reached; in the simulations, the inverse of the decay time at $1/\eee$ therefore varies indeed as $k_B T\epsilon^2$ at low temperature, as predicted by the golden rule and more generally by estimate (\ref{eq028}) at all orders in $H_3$. On the other hand, the $\qb$ dependence of the damping time is very poorly described by the golden rule, dashed line in Figure \ref{fig1}: as predicted by the general reasoning of section \ref{sec0}, the $\epsilon\to 0$ limit is therefore not sufficient in 2D to bring the phonon gas into a weak coupling regime. What section \ref{sec0} didn't say, however, is that the golden rule remains insufficient even if the underlying superfluid enters the weakly interacting regime $(\rho\xi^2)^{-1}\to 0$, see panel (d) where a perturbative calculation at the leading order in $H_3$ should a priori be legitimate; the behavior at weak $\qb$ predicted by (\ref{eq113}) is in fact particularly bad there. 

To make the inadequacy of the golden rule even more glaring, we plot in Figure \ref{fig2}, as a function of time, the deviation of the numerical signal from the pole approximation (\ref{eq014}) limited to second order at $H_3$ {\yc but taken to all orders in $\epsilon$} [see equation (\ref{eq124})] and transformed for the classical field (if we were at the leading order in $\epsilon$, {\yc it would simply be} the deviation from the exponential $\exp(-\Gamma_q^{\rm clas}t/2)$ of the golden rule), for the same interaction strengths as before but for a fixed value of the reduced wave number, $\qb=1/4$, large enough to make it easy to reach the thermodynamic limit in simulations. \footnote{The smaller $\qb$ is, the lower the damping rate of the mode and the greater the granularity of the energy levels of the phonon gas in the quantization box, and the more we need to increase the size of the system to reach the thermodynamic limit.} Here too, the $\epsilon\to 0$ limit is reached in simulations (when time $t$ is scaled as in (\ref{eq022}) with $\nu=2d-2=2$, which is done here by expressing it in units of $2/\Gamma_q^{\rm clas}$), the {\yc expected} deviation from the golden rule is indeed present and persists even in the weakly interacting limit $(\rho\xi^2)^{-1}\to 0$, at least on the real part.\footnote{If we move from panel (c) to panel (d) in Figure \ref{fig2}, the {\yct interaction} strength is divided by $\simeq 3$, the imaginary part is divided by $\simeq 2$ in maximum absolute value so may well tend towards zero, while the real part only decreases by a factor of $\simeq 1.3$.}

\section{Study at leading order in phonon-phonon coupling}
\label{sec2}

The classical field simulations of section \ref{sec1} established the inadequacy of Fermi's golden rule in describing phonon damping in dimension $d=2$; we have not shown this here, but in addition to disagreement on the decay time at $1/\eee$, they reveal, in logarithmic scale, a non-exponential decay of the {\yct signal} $s(t)$ \cite{TheseAlan}. In this section, we incriminate the approximation made in step (i) of section \ref{sec0}, and renounce it by retaining all the energy dependence of the self-energy in the integral expression (\ref{eq012}) of $s(t)$, continuing however to calculate $\Sigma_\qq(\zeta)$ at leading order in the phonon-phonon coupling and in $\epsilon$ (as in step (ii) of section \ref{sec0}). After a diagrammatic calculation of $\Sigma^{(2)}$, we obtain some analytical predictions put into a universal form and compare them with simulations.

\subsection{Calculation of the self-energy}
\label{sec2.1}

In the Hamiltonian $H$ of equation (\ref{eq001}), the cubic coupling {\yct $H_3$} seems to dominate over the quartic coupling {\yct $H_4$}, since it appears at order $\delta\hat{\rho}^3$ rather than $\delta\hat{\rho}^4$ in the expansion of the hydrodynamic Hamiltonian, see section \ref{sec1.1}. However, as $H_3$ does not conserve the phonon number, it must be treated  at least at second order of perturbation theory in the calculation of $\Sigma_\qq(\zeta)$, whereas $H_4$ gives a non-zero first order contribution, which is at first sight as large as that of $H_3$.\footnote{The contribution of $H_3$ {\yc to $\Sigma_\qq(\ii 0^+)$} involves the ratio $\langle\ |\mH_3|\ \rangle^2/\Delta E\approx \delta\hat{\rho}^{2\times 3-2}$, that of $H_4$ involves the matrix element $\langle\ |\mH_4|\ \rangle\approx \delta\hat{\rho}^4$, where we use the notations of equation (\ref{eq024}) and the fact that the denominator $\Delta E$, made up of energy differences of the quadratic Hamiltonian $H_2$, is $\approx\delta\hat{\rho}^2$.} The full calculation is required.

At first order in perturbation theory, there's no need to use Feynman {\yct diagrams}, as the contribution to $\Sigma_\qq(\zeta)$ is independent of the $\zeta$ variable and can be obtained by the following simple procedure, justifiable by the quantum master equation method of reference \cite{SCE}. We write the perturbation operator, here $H_4$, in representation $\hat{b},\hat{b}^\dagger$ and in normal order, with the help of bosonic commutation relations (all $\hat{b}^\dagger$ put on the left, all $\hat{b}$ put on the right). Terms that do not preserve phonon number do not contribute. The terms $\hat{b}^\dagger\hat{b}$, {\yc being quadratic}, are absorbed into a redefinition of the bare $\veps_\kk$ eigenenergies of the modes (as was expected from footnote \ref{note0}). Finally, we isolate in the {\yc remaining} \g{true} quartic coupling terms the contributions $\hat{b}^\dagger_\qq\hat{b}^\dagger_\kk\hat{b}_\kk\hat{b}_\qq$, whose thermal averaging over the modes $\kk\neq\qq$ gives the $\qq$-mode energy shift. All this is summarized by the compact expression 
\begin{equation}\label{eq120}
\Sigma_\qq^{(H_4)}(\zeta)=\int\frac{\dd^dk}{(2\pi)^d} \langle\qq,\kk|\mathcal{H}_4^{\rm true}
|\qq,\kk\rangle \bar{n}_k = \frac{\hbar^2q\Lambda_*}{4m\rho}\int\frac{\dd^dk}{(2\pi)^d} k\bar{n}_k \underset{\epsilon\to 0}{\approx} \frac{\bar{q}k_B T\epsilon^{d+1}}{\rho\xi^d}
\end{equation} where $\mH_4$ is the {\yct volumic} Hamiltonian such that $H_4=\mH_4/L^{d}$, the coupling constant $\Lambda_*$ is that of equation (\ref{eq032}), we have deduced the matrix element in the two-phonon state from expression (\ref{eq104}) of $H_4$ and expression (\ref{eq052}) of $\delta\hat{\rho}(\rr)$ and we have obtained the scaling behavior in the limit (\ref{eq020}) in the {\yct right-hand} side of equation (\ref{eq120}) as in section \ref{sec0}. As we see from the contribution (\ref{eq028}) of $H_3$ written for $n=1$, the contribution of $H_4$ in dimension $d=2$ (but not in dimension three) is subleading and can therefore be neglected when $\epsilon\to 0$. This is what we'll do from now on.
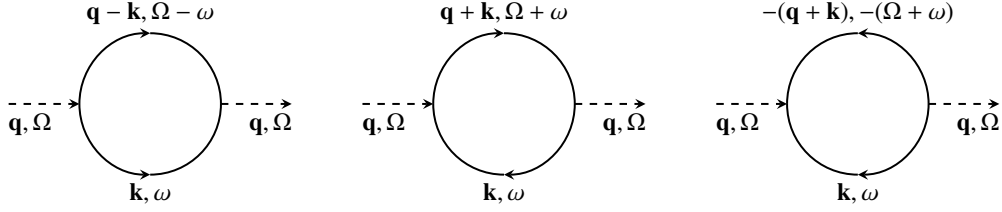
\begin{figure}[t]
\begin{center}
\begin{tikzpicture}
\draw[thick,dashed,->,>=stealth](0,0)--(1,0);
\draw[thick,->,>=stealth](1,0)arc(-180:-90:1);
\draw[thick,->,>=stealth](1,0)arc(180:90:1);
\draw[thick](2,-1)arc(-90:0:1);
\draw[thick](2,1)arc(90:0:1);
\draw[thick,dashed,->,>=stealth](3,0)--(4,0);
\node(a)at(0.3,0)[below]{$\vec{q},\Omega$};
\node(a)at(3.7,0)[below]{$\vec{q},\Omega$};
\node(a)at(2,1)[above]{$\vec{q}-\vec{k},\Omega-\omega$};
\node(a)at(2,-1)[below]{$\vec{k},\omega$};
\draw[thick,dashed,->,>=stealth](0+5,0)--(1+5,0);
\draw[thick](1+5,0)arc(-180:-90:1);
\draw[thick,->,>=stealth](1+5,0)arc(180:90:1);
\draw[thick,->,>=stealth](3+5,0)arc(0:-90:1);
\draw[thick](2+5,1)arc(90:0:1);
\draw[thick,dashed,->,>=stealth](3+5,0)--(4+5,0);
\node(a)at(0.3+5,0)[below]{$\vec{q},\Omega$};
\node(a)at(3.7+5,0)[below]{$\vec{q},\Omega$};
\node(a)at(2+5,1)[above]{$\vec{q}+\vec{k},\Omega+\omega$};
\node(a)at(2+5,-1)[below]{$\vec{k},\omega$};
\draw[thick,dashed,->,>=stealth](0+10,0)--(1+10,0);
\draw[thick](1+10,0)arc(-180:-90:1);
\draw[thick](1+10,0)arc(180:90:1);
\draw[thick,->,>=stealth](3+10,0)arc(0:-90:1);
\draw[thick,->,>=stealth](3+10,0)arc(0:90:1);
\draw[thick,dashed,->,>=stealth](3+10,0)--(4+10,0);
\node(a)at(0.3+10,0)[below]{$\vec{q},\Omega$};
\node(a)at(3.7+10,0)[below]{$\vec{q},\Omega$};
\node(a)at(2+10,1)[above]{$-(\vec{q}+\vec{k}),-(\Omega+\omega)$};
\node(a)at(2+10,-1)[below]{$\vec{k},\omega$};
\end{tikzpicture}
\end{center}
\caption{The self-energy $\Sigma(\qq,z)$ is given at order two in $H_3$ by one-loop diagrams, corresponding to Beliaev, Landau and non-resonant processes (from left to right in the figure). Variables $\Omega$ and $\omega$ are the Matsubara energies associated with the incoming $\qq$ and internal $\kk$ wave vectors. Inner lines (of any orientation and to be summed) are solid, outer lines (whose orientation is fixed by the choice of Green's function) are dashed.}
\label{fig6}
\end{figure}

At second order in $H_3$, it's best to use the method in reference \cite{FW}.\footnote{Since our phonon states already have the bosonic symmetry, we can dispense with the direct interaction and exchange lines introduced in \cite{FW}.} This method {\yc provides} a diagrammatic expansion of the self-energy $\Sigma(\qq,z)$ for particular values of the complex energy variable, $z=\Omega\in 2\ii\pi k_B T \mathbb{Z}$, where $\Omega$ is called Matsubara energy,\footnote{As a slight departure from the usual notation \cite{FW}, our Matsubara variables $\Omega$, $\omega$, etc., are not real frequencies but pure imaginary energies, i.e. we incorporate an additional factor $\ii\hbar$.} which is then completed by an analytic continuation to $\mathbb{C}\setminus\mathbb{R}$ \cite{FW}. Here, as in Figure \ref{fig6}, each diagram must have two vertices (we're at second order), where three phonon lines meet ($H_3$ is cubic in $\hat{b},\hat{b}^\dagger$). The only possible topology is one-loop and, since two loops deduced from each other by rotation of angle $\pi$ around their axis must be considered as identical, the four possible orientations of the two internal lines lead to only three distinct diagrams, represented in the order Beliaev, Landau and non-resonant in Figure \ref{fig6} (using the typology of equation (\ref{eq100})). Given the conservation of Matsubara energy and momentum at each vertex (the sum of incoming quantities must equal that of outgoing quantities), there remains a single independent Matsubara energy $\omega$ and wave vector $\kk$, on which to sum. Feynman's rules {\yc of} section 25 of reference \cite{FW} then lead for our phonon gas of zero chemical potential to 
\begin{equation}\label{eq121}
\Sigma^{(2)}(\qq,\Omega) = -k_B T \sum_{\omega\in 2\ii\pi k_B T\mathbb{Z}}\int\frac{\dd^dk}{(2\pi)^d} \Bigg\{\frac{(1/2)|\langle\kk,\qq-\kk|\mathcal{H}_3|\qq\rangle|^2}{(\omega-\veps_\kk)(\Omega-\omega-\veps_{\qq-\kk})}+\frac{|\langle\kk+\qq|\mathcal{H}_3|\kk,\qq\rangle|^2}{(\omega-\veps_\kk)(\Omega+\omega-\veps_{\kk+\qq})}+\frac{(1/2)|\langle 0|\mathcal{H}_3|\qq,\kk,-(\kk+\qq)\rangle|^2}{(\omega-\veps_\kk)(-\Omega-\omega-\veps_{-(\kk+\qq)})}
\Bigg\}
\end{equation} 
{\yc in the thermodynamic limit with $\mathcal{H}_3=H_3 L^{d/2}$, $|0\rangle$ the phonon vacuum and}, for each contribution, a $\kk$ integration domain ensuring that the wavevectors $\kk$, $\qq-\kk$, etc., satisfy the cutoff (\ref{eq103}). Also included are symmetry factors $1/2$ when the two branches of the loop have the same orientation and are therefore physically identical. Matsubara sums are calculated by means of the general expression \footnote{\label{notedemo} This is easily demonstrated by partial fraction decomposition of the summand with respect to variable $n$ and using the identity $\lim_{N\to +\infty} \sum_{n=-N}^{N} 1/(2\ii\pi n-\vepsb)=-(\bar{n}_\veps+1/2)\ \forall \veps\in{\yc\mathbb{C}\setminus 2\ii\pi\mathbb{Z}}$ resulting from the properties of the digamma function $\psi(z)$ {\yc such as the reflection formula}.} 
\begin{equation}\label{eq122}
\sum_{n\in\mathbb{Z}} \frac{1}{(2\ii\pi n-\vepsb_1)(2\ii\pi n-\vepsb_2)} = -\frac{\bar{n}_{\veps_1}-\bar{n}_{\veps_2}}{\vepsb_1-\vepsb_2}
\end{equation} where the real {\ycd or even complex} energies $\veps_j$ are two by two distinct, and we set $\vepsb=\veps/k_B T$ and $\bar{n}_\veps=1/(\exp\vepsb-1)$. We then use the fact that $\Omega/k_B T\in 2\ii\pi\mathbb{Z}$ to make this variable disappear from Bose's law, and the relation $\bar{n}_{-\veps}=-(1+\bar{n}_\veps)$ to bring us back to true occupation numbers (at positive energy). So, for example: 
\begin{equation}\label{eq123}
\bar{n}_{\Omega-\veps_{\qq-\kk}}=\bar{n}_{-\veps_{\qq-\kk}}=-(1+\bar{n}_{\qq-\kk})
\end{equation} The integrand becomes a rational function of the discrete variable $\Omega$, and its analytic continuation to $\mathbb{C}\setminus\mathbb{R}$ is achieved by the simple substitution $\Omega\to z$. We get 
\begin{multline}
\label{eq124}
\Sigma_\qq^{(2)}(\zeta)=\int\frac{\dd^dk}{(2\pi)^d}\Bigg[\frac{(1/2)|\langle\kk,\qq-\kk|\mathcal{H}_3|\qq\rangle|^2}{\zeta+\veps_\qq-(\veps_\kk+\veps_{\qq-\kk})}(\bar{n}_\kk+\bar{n}_{\qq-\kk}+1)+\frac{|\langle\kk+\qq|\mathcal{H}_3|\kk,\qq\rangle|^2}{\zeta+\veps_\qq+\veps_\kk-\veps_{\qq+\kk}}(\bar{n}_\kk-\bar{n}_{\qq+\kk})\\
-\frac{(1/2)|\langle 0|\mathcal{H}_3|\qq,\kk,-(\kk+\qq)\rangle|^2}{\zeta+\veps_\qq+\veps_{-(\kk+\qq)}+\veps_\kk}(\bar{n}_\kk+\bar{n}_{-(\qq+\kk)}+1)\Bigg\}
\end{multline}
 It remains to take the limit $\epsilon\to 0$ at $\qb$ and $\zetat$ fixed as in equations (\ref{eq020},\ref{eq021},\ref{eq022}) with, remember, exponents $\nu=\sigma=2$ in dimension $d=2$. Through the \g{small-denominator} effect described in section \ref{sec0}, each contribution to the integral (\ref{eq124}) is dominated by configurations where the three wave vectors in the energy denominator are quasi-colinear and of the same direction (within an angle $O(\epsilon)$). In the first contribution, it suffices to use expansion (\ref{eq025}) with $\kk'=\qq-\kk$ (having $\kk$ and $\kk'$ quasicollinear and of the same direction then implies $k<q$); in the second, we use (\ref{eq025}) with $\kk'=\qq$ which imposes nothing on $k$; in the third, we are dealing with a sum with positive coefficients of three phonon energies, which can never be $O(k_B T\epsilon^2$) (it is always $\approx k_B T\epsilon^0$), so that the corresponding contribution is subleading and negligible at low temperatures. \footnote{After these expansions, the angular integral is Lorentzian and can therefore be performed very well using Cauchy's formula, by closing the integration path with an infinite semicircle in the complex plane, after performing the change of variable $\theta=\epsilon\tilde{\theta}$ {\yc on the angle $\theta$ between $\qq$ and $\kk$} and sending the integration bounds $\pm\pi/\epsilon$ over $\tilde{\theta}$ to infinity. {\ycd In the first contribution, the limiting behavior $\theta'\sim-k\theta/(q-k)$ of the angle $\theta'$ between $\qq$ and $\kk'$ was used beforehand.}} Respecting the order of the contributions in (\ref{eq124}), and restricting ourselves to the strict upper half {\yc of the complex} plane as in expression (\ref{eq012}) of the signal, we get:\footnote{\label{notegen} The value of $\tilde{\Sigma}_{\qb}^{(2,2)}(\zetat)$ in the strict lower half {\yc of the complex} plane can be deduced from the general relation valid for all orders in phonon coupling, ${\yc\Sigma_\qq(\zeta)^*}=\Sigma_\qq(\zeta^*)$.} 
\begin{equation}\label{eq125}
\boxed{\tilde{\Sigma}_{\qb}^{(2,2)}(\zetat)\stackrel{\im\zetat>0}{=}\frac{9(1+\Lambda)^2(2\qb)^{1/2}}{64\ii\pi\rho\xi^2} \Bigg\{\int_0^{\qb} \dd\kb\, \frac{[\kb(\qb-\kb)]^{3/2}(1+\bar{n}_k^{\rm lin}+\bar{n}_{q-k}^{\rm lin})}{\left[(3\gamma/8)\qb\kb(\qb-\kb)+\tilde{\zeta}\right]^{1/2}} +2 \int_0^{\eta-\qb} \dd\bar{k}\frac{[\kb(\qb+\kb)]^{3/2}(\bar{n}^{\rm lin}_k-\bar{n}^{\rm lin}_{k+q})}{\left[(3\gamma/8)\qb\kb(\kb+\qb)-\tilde{\zeta}\right]^{1/2}}\Bigg\}}
\end{equation} where the occupation numbers marked by the superscript \g{lin} are those (\ref{eq029}) of the phonon modes for the linearized dispersion relation. In the left-hand side of (\ref{eq125}), the second integer $2$ in the exponent reminds us that $\Sigma^{(2)}$ is calculated to second order in $\epsilon$; the tildes on $\Sigma$ and $\zeta$ remind us that a factor $k_BT\epsilon^2$ has been taken out of these quantities as in (\ref{eq021},\ref{eq022}). Finally, despite the ever-present threat of quantum field fluctuations, we can take the limit of an infinite $\eta$ cutoff without triggering ultraviolet divergence, so without having to renormalize any quantities whatsoever. Then, if we make $\zetat$ tend towards zero by positive imaginary parts in (\ref{eq125}), we find equation (\ref{eq033}).

\subsection{Analytical studies: one-parameter universality}
\label{sec2.2}

Remarkably, it is possible, by a simple change of variable, to ensure that in the leading-order theory {\yc in $\epsilon$ and in the coupling} treated in this section \ref{sec2}, there remains only one parameter $u$ containing all physical parameters such as the interaction strength $1/\rho\xi^2$, the curvature parameter $\gamma$ and the coupling constant $\Lambda$, i.e.\footnote{For the same to apply in the classical field model, we would also have to incorporate the cut-off parameter $\eta$ into $u$, i.e. introduce the quantity $u_{\rm clas}=\eta u$. This is unnecessary here, as we have taken $\eta=1$ in the simulations. As a curiosity, let's point out, for the classical field and in the limit $\qb\ll 1$, that the signal $s^{(2,2)}(t)$ defined below ceases to start at value $1$ in $t=0$ (as it should physically) if $u_{\rm clas}<1$, which we link mathematically to the appearance of a pole of {\yct the} Green's function on the real axis. Consequently, it would be interesting to know whether all two-dimensional physical systems (even in very strong interaction, provided it's short-range) lead to $u>1$.} 
\begin{equation}\label{eq112}
u=\frac{\pi\rho\xi^2\gamma^{3/2}}{\sqrt{3}(1+\Lambda)^2}
\end{equation} In fact, the new definitions marked with a Czech accent, 
\begin{equation}\label{eq130}
\check{t}=\frac{3\gamma\qb}{8u}\tilde{t}=\frac{3\gamma\epsilon^2 c q t}{8u}\quad ; \quad \zetac=\frac{8u}{3\gamma\qb}\zetat=\frac{8u\zeta}{3\gamma\epsilon^2\hbar c q} 
\end{equation} lead to a new self-energy and a new signal expression, clearly involving only $u$ and the reduced variables $\qb$ and $\check{t}$: 
\begin{equation}\label{eq131}
\check{\Sigma}^{(2,2)}_{\qb}(\check{\zeta}) = \frac{8u}{3\gamma\qb} \tilde{\Sigma}^{(2,2)}_{\qb}(\zetat) \stackrel{\im\zetac>0}{=} \frac{1}{\ii} \int_0^{\qb} \frac{\dd\kb}{2\qb} \frac{[\kb(\qb-\kb)]^{3/2} (1+\bar{n}^{\rm lin}_k+\bar{n}^{\rm lin}_{q-k})}{[\kb(\qb-\kb)+\check{\zeta}/u]^{1/2}}+\frac{1}{\ii} \int_0^{+\infty} \frac{\dd\kb}{\qb}\frac{[\kb(\qb+\kb)]^{3/2} (\bar{n}^{\rm lin}_k-\bar{n}^{\rm lin}_{q+k})}{[\kb(\qb+\kb)-\check{\zeta}/u]^{1/2}}
\end{equation} and 
\begin{equation}\label{eq132}
s^{(2,2)}(t) = \int_{C_+} \frac{\dd\check{\zeta}}{2\ii\pi} \eee^{-\ii\check{\zeta}\check{t}} \check{\mathcal{G}}_{\qb}^{(2,2)}(\zetac)  \quad\mbox{with}\quad \check{\mathcal{G}}_{\qb}^{(2,2)}(\zetac)=\frac{1}{\check{\zeta}-\check{\Sigma}^{(2,2)}_{\qb}(\check{\zeta})}
\end{equation} In this case, we speak of one-parameter universality.
\begin{figure}[t]
\begin{center}
\begin{tikzpicture}[x=8mm,y=8mm]
\draw[thick,->,>=stealth](0,0)--(7,0);
\draw[thick,->,>=stealth,color=black,->,>=stealth](7,0.3)--(4,0.3);
\draw[thick,color=black](4,0.3)--(0,0.3);
\node(a)at(5,0.3)[above,color=black]{$C_+$};
\node(a)at(7,0)[below]{${\rm Re\,}\check{\zeta}$};
\draw[thick,->,>=stealth](3,-1)--(3,3);
\node(a)at(3,3)[above]{${\rm Im\,}\check{\zeta}$};
\node(a)at(0.2,3)[above]{$(a)$};
\node(a)at(0.8,0.3)[above]{$-u\bar{q}^2/4$};
\draw[dashed,ultra thick](1,0)--(7,0);
\node(a)at(1,0)[]{$\bullet$};
\draw[thick,color=white](3,-4.25)--(3,-1);
\end{tikzpicture}
\quad
\begin{tikzpicture}[x=8mm,y=8mm]
\draw[thick,->,>=stealth](-1,0)--(7,0);
\draw[thick,->,>=stealth,color=black,->,>=stealth](7,0.3)--(4,0.3);
\draw[thick,color=black](4,0.3)--(-1,0.3);
\node(a)at(5,0.3)[above,color=black]{$C_+$};
\node(a)at(7,0)[below]{$\re\check{\zeta}$};
\draw[thick,->,>=stealth](3,-1)--(3,3);
\node(a)at(3,3)[above]{$\im\check{\zeta}$};
\node(a)at(-0.8,3)[above]{$(b)$};
\node(a)at(0.8,0.3)[above]{$-u\bar{q}^2/4$};
\draw[dashed,ultra thick](1,0)--(-1,-2);
\node(a)at(1,0)[]{$\bullet$};
\draw[thick,color=blue,->,>=stealth](3,0.3)+(180:4)arc(-180:-155:4);
\draw[thick,color=blue,->,>=stealth](3,0.3)+(-150:4)arc(-150:-60:4);
\draw[thick,color=blue](3,0.3)+(-60:4)arc(-60:0:4);
\draw[thick,color=red,->,>=stealth](-0.5-0.08,-1.5+0.16)--(1-0.08,0+0.16);
\draw[thick,color=red,->,>=stealth](1+0.05,0-0.2)--(-0.5+0.05,-1.5-0.2);
\draw[thick,color=red](1,0)+(120:0.18)arc(120:-48:0.18);
\draw[thick,->,>=stealth](1,0)+(-0:0.5)arc(0:-135:0.5);
\node(a)at(1.3,-0.7)[]{$\phi$};
\node(a)at(4,-1)[ultra thick,color=red]{$\times$};
\node(a)at(4,-1.2)[ultra thick,color=black,below]{$\check{\zeta}_{\bar{q}}$};
\end{tikzpicture}
\end{center}
\caption{(a) Singularities of Green's function $\check{\mG}^{(2,2)}_{\qb}(\zetac)$ in the complex plane and (b) integration path followed in calculating signal (\ref{eq132}) to obtain the first deviation (\ref{eq151}) from the pole approximation. The change of variable (\ref{eq130}) $\zetac=8(z-\veps_\qq)u/(3\gamma\hbar c q \epsilon^2)$ was performed to reveal a one-parameter universality $u$ (\ref{eq112}). Black horizontal line: original integration path in (\ref{eq132}). Black disk: $\check{\zeta}\mapsto\check{\mG}^{(2,2)}_{\qb}(\zetac)$ branch point. Dashed half-line: original branch cut of $\check{\zeta}\mapsto\check{\mG}^{(2,2)}_{\qb}(\zetac)$ in (a), and rotated by an angle $\phi\in]-\pi,-\pi/2[$ in (b) by analytic continuation of $\check{\zeta}\mapsto\check{\mG}^{(2,2)}_{\qb}(\zetac)$ from the upper to the lower half-plane, resulting in the $\check{\zeta}_{\qb}$ pole (red cross). In (b), to apply Cauchy's theorem, we close $C_+$ with two portions of a circle at infinity (in blue) and a contour (in red) which bypasses the displaced branch cut.}
\label{fig5}
\end{figure}
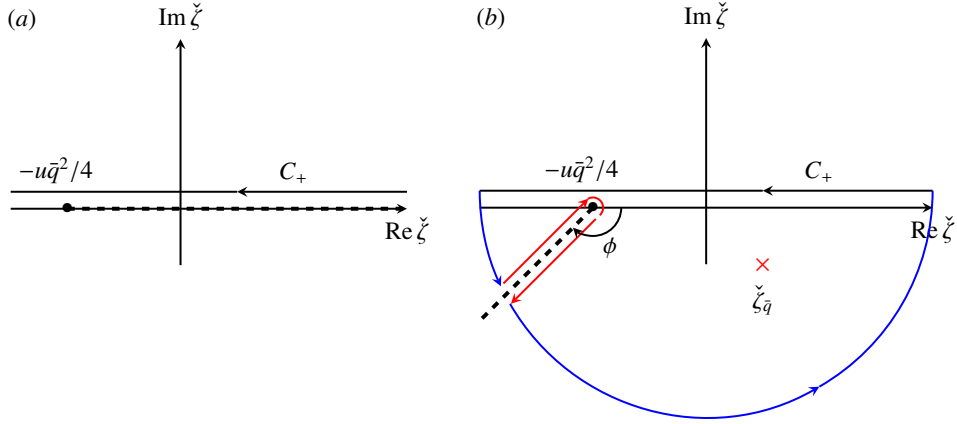

Let's carry out the analytic continuation of function (\ref{eq131}) from the upper to the lower half-plane through its branch cut on the real axis.\footnote{There's a mathematical subtlety here that the reader will forgive us for. Implicitly, from expression (\ref{eq131}) for $\check{\Sigma}_{\qb}^{(2,2)}(\zetac)$, valid only for $\im\zetac>0$, we define a function in all $\mathbb{C}\setminus\mathbb{R}$ by applying (\ref{eq131}) as it is for $\im\zetac<0$; the branch cut $[-u\qb^2/4,+\infty[$ mentioned in the text and shown in Figure \ref{fig5}a corresponds to this function rather than to the true self-energy $\check{\Sigma}_{\qb}^{(2,2)}(\check{\zeta})$, whose branch cut is $\mathbb{R}$ in its entirety and which would therefore be less convenient to analytically continuate from the upper half-plane to the lower half-plane, for the same final result (we see that it's $\mathbb{R}$ in its entirety (i) by considering the energy denominator $\zeta+\Delta E$ in the Landau contribution of equation (\ref{eq124}) and checking that $\Delta E$ spans $\mathbb{R}$ when $\theta/\epsilon$ and $\kb$ span $\mathbb{R}^+$ in (\ref{eq025}) with $\kk'=\qq$, or (ii) by using the relation $\textrm{i}z^{1/2}=-\mathrm{signe}(\im z)(-z)^{1/2}$, $\forall z\in\mathbb{C}\setminus\mathbb{R}$, to make the factor $\ii$ disappear in (\ref{eq131}) and obtain a physically correct expression over all $\mathbb{C}\setminus\mathbb{R}$ satisfying $\Sigma_{\qq}(\zeta)^*=\Sigma_{\qq}(\zeta^*)$, whose branch cut is then shown to be $\mathbb{R}$).} This will enable us, as in reference \cite{CCTbordeaux}, to separate the signal into the contribution of a contour around a displaced branch cut (power-law decreasing contribution at long times) and the contribution of a pole in the analytically continuated Green's function (exponentially decreasing contribution in time), see our Figure \ref{fig5}b, which improves physical understanding. First, we need to identify the branch points (immutable singularities) of $\check{\Sigma}_{\qb}^{(2,2)}(\check{\zeta})$ on the real axis. Since the values of $\check{\zeta}$ that make the square root argument in the denominator vanish, span the interval $[-u\qb^2/4,0]$ when $\kb$ spans $[0,\qb]$ in the first contribution to (\ref{eq131}), and span $[0,+\infty[$ when $\kb$ spans $\mathbb{R}^+$ in the second, we might expect to have two branch points, $\check{\zeta}=-u\qb^2/4$ and $\check{\zeta}=0$. \footnote{For example, $\zetac\mapsto\int_a^b\mathrm{d}x/(\zetac-x)^{1/2}=2(\zetac-b)^{1/2}-2(\zetac-a)^{1/2}$ has the real numbers $a$ and $b$ as branch points.} In reality, by subtle compensation between the two contributions, only $\zetac=-u\qb^2/4$ remains; the branch cut $[-u\qb^2/4,+\infty[$ starts from this point, as shown in Figure \ref{fig5}a.\footnote{\label{note30} If we return to variable $\zetat$, the branch point is at $-3\gamma\qb^3/32$.}. The branch cut can then be rotated by an angle $\phi \in ]-\pi,-\pi/2[$ to bring it into the third quadrant, by analytic continuation of $\check{\Sigma}^{(2,2)}_{\qb}(\check{\zeta})$ from the upper half-plane to the lower half-plane, the new branch cut being the half-line $-u\qb^2/4+\exp(\ii\phi)\mathbb{R}^+$, see Figure \ref{fig5}b. As shown in \ref{app1} and agreeing that $z^{3/2}=z\sqrt{z}$, the function continuated in this way is written: 
\begin{equation}\label{eq140}
\check{\Sigma}^{(2,2)}_{\qb\downarrow}(\check{\zeta})= \frac{\eee^{3\ii\phi/2}}{\ii} \int_0^{+\infty} \frac{\dd\kb}{\qb} 
\frac{(\kb^2-\eee^{-\ii\phi}\qb^2/4)^{3/2}}{[\kb^2-\eee^{-\ii\phi}(\qb^2/4+\zetac/u)]^{1/2}} \Bigg[
\frac{1}{\exp(\eee^{\ii\phi/2}\kb-\qb/2)-1}-\frac{1}{\exp(\eee^{\ii\phi/2}\kb+\qb/2)-1} \Bigg]
\end{equation} In particular, it is analytic on a neighborhood of $\zetac=0$ (see \ref{app1}, which also gives its value and that of its derivative at this point). On the other hand, the corresponding Green's function $\check{\mG}_{\qb\downarrow}^{(2,2)}(\zetac)$ has a pole $\check{\zeta}_{\qb}$ in the fourth quadrant, as we have verified numerically.\footnote{If angle $\phi$ is too close to $-\pi/2$ (which we'll avoid in the following) and if $\qb$ is quite small, a second pole of imaginary part $<0$, of uncertain physical interpretation, may appear to the left of the displaced branch cut.} 

\begin{SCfigure}
\includegraphics[width=10cm,clip=]{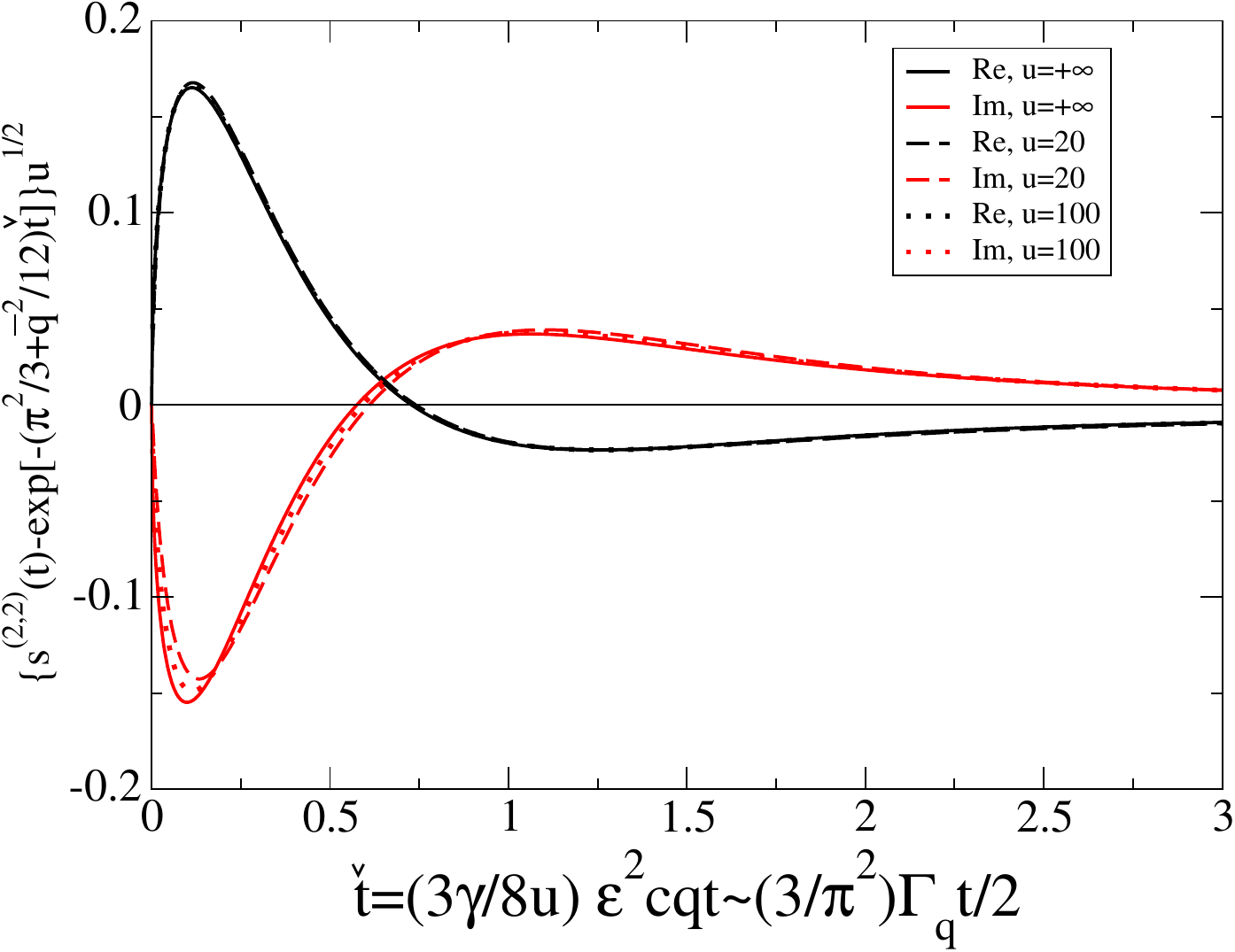}
\caption{In quantum theory and for a self-energy calculated at order two in $H_3$ and at order two in $\epsilon=k_B T/mc^2$, deviation of signal (\ref{eq012}) from Fermi's golden rule {\yc exponential $\exp(-\Gamma_qt/2)$} multiplied by $u^{1/2}$ (the universal parameter $u$ is given by (\ref{eq112}), {\yc the rate $\Gamma_q$ by (\ref{eq033})}) and plotted as a function of reduced time (\ref{eq130}) for real and imaginary parts, for increasing values of $u$ (the underlying two-dimensional superfluid is more and more weakly interacting) but for a scaled wave number as in (\ref{eq150}) {\yc kept constant,} $Q=1$. Curves for finite values of $u$: obtained by numerical integration of expression (\ref{eq132}). Curves for $u=+\infty$: taken from analytical prediction (\ref{eq151}) (they simply correspond to $\re S(\check{t})$ and $\im S(\check{t})$).}
\label{fig4}
\end{SCfigure}

Rather than numerically evaluating the contour and pole contributions to the signal, let's now extract analytically from form (\ref{eq140}) the \g{very substance} of the theory at order two in $H_3$, noting for example that it is illusory to calculate the position of the $\zetac_{\qb}$ pole at all orders in $1/u$ since the self-energy is calculated here only at order one in $1/u$. On the other hand, limiting ourselves to order zero in $1/u$, which would mean taking the $u\to+\infty$ limit at $\qb$ fixed, would be too poor, as the signal would then be reduced to the exponential of Fermi's golden rule. As can be seen from (\ref{eq131}), the new self-energy depends on $u$ only through the ratio $\zetac/u$; so it tends towards its value at $\textrm{i}0^+$ when $u$ tends to $+\infty$ under the integral sign ($\zetac$ and $\check{t}$ being fixed in (\ref{eq132})). A more interesting result is obtained by taking the limit $u\to+\infty$ for 
\begin{equation}\label{eq150}
Q=u^{1/2}\bar{q}\quad\mbox{and}\quad \check{t}\quad\mbox{fixed}
\end{equation} so that the {\yct branch} point in Figure \ref{fig5} remains fixed instead of being rejected at infinity (which would effectively make the contour contribution disappear). We then find a correction of order $u^{-1/2}$ to Fermi's golden rule {\yc signal}; as $u^{-1}\ll u^{-1/2}$ {\yct one would expect} that we can keep this correction without having to calculate the self-energy at the next order (order four) in $H_3$, {\yc even if we'll see} in section \ref{sec4} that this is not the case. Let's report the technical details in \ref{app2} and give the result directly {\yc in quantum theory}: 
\begin{equation}\label{eq151}
s^{(2,2)}(t) \stackrel{Q\,\mbox{\scriptsize {\yct and}}\,\check{t}\,\mbox{\scriptsize fixed}}{\underset{u\to+\infty}{=}}\eee^{-\pi^2\check{t}/3} + \frac{S(\check{t})}{u^{1/2}}+O(1/u) \quad\mbox{with}\quad S(\check{t})=S_{\rm contour}(\check{t})+S_{\mbox{\scriptsize pole}}(\check{t})
\end{equation} The first contribution is the real-valued exponential of the golden rule in limit $\qb\to 0$. The second is divided into the contour and pole contributions established in \ref{app2}: 
\begin{equation}\label{eq152}
S_{\rm contour}(\check{t})= -\frac{\eee^{\textrm{i}Q^2\check{t}/4}}{\pi}\int_0^{\eee^{\ii(\phi+\pi)}(+\infty)} \mathrm{d}z \frac{\eee^{\textrm{i}z \check{t}}\int_0^{\pi/2} \dd\alpha \left(z\sin^2\alpha+\frac{Q^2}{4}\right)^{1/2}}{(\frac{Q^2}{4}+z-\ii\frac{\pi^2}{3})^2} \quad ; \quad S_{\mbox{\scriptsize pole}}(\check{t})=\eee^{-\pi^2\check{t}/3} (\delta \check{Z}-\textrm{i}\delta \check{\zeta}_{\qb}\check{t})
\end{equation} where angle $\phi$ is freely chosen in $]-\pi,-\pi/2[$ (result (\ref{eq152}) does not depend on it). In a power expansion of $u^{-1/2}$, $\delta\check{\zeta}_{\qb}$ is the first deviation of the pole from the pole approximation $-\ii\pi^2/3$ (multiplied by $u^{1/2}$) and $\delta\check{Z}$ is the first deviation from one of the residue of $\check{\mG}_{\qb\downarrow}^{(2,2)}(\zetac)$ (also multiplied by $u^{1/2}$): \footnote{As signal (\ref{eq012}) starts exactly at one, $S(0)=0$ so the contour contribution for $\check{t}=0$ must exactly cancel out that of the pole residue's deviation from one. Verification is non-trivial. The main idea is to reduce $S_{\rm contour}(0)$ to a simple integral by inverting the integration on $z$ and $\alpha$ in (\ref{eq152}) (a primitive function of $(z+a)^{1/2}/(z+b)^2$ is known, where $a$ and $b$ are two constants). {\yc  Moreover, we note that $\re\delta\check{Z}>0$ for all $Q$: the first correction in $u^{-1/2}$ makes the residue $\check{Z}$ of modulus $>1$ and therefore the pole $\check{\zeta}_{\qb}$ of spectral weight $>1$.}}
\begin{equation}\label{eq154}
\delta\check{\zeta}_{\qb} = \int_0^{+\infty} \mathrm{d}K \left[1-\frac{(K^2+Q^2/4)^{1/2}}{(K^2-\ii\frac{\pi^2}{3}+\frac{Q^2}{4})^{1/2}}\right]\quad;\quad 
\delta \check{Z}=\frac{1}{2} \int_0^{+\infty} \mathrm{d}K \frac{(K^2+Q^2/4)^{1/2}}{(K^2-\ii\frac{\pi^2}{3}+\frac{Q^2}{4})^{3/2}} 
\end{equation} In Figure \ref{fig4}, we show function $S(\check{t})$ of equation (\ref{eq151}) for $Q=1$, together with predictions drawn numerically from $\check{\Sigma}_{\qb}^{(2,2)}{\yc(\check{\zeta})}$ for finite values of $u$. It can be seen that $u=20$ has already reached the asymptotic regime, so the same is expected for the parameters of the classical field simulations in Fig.~\ref{fig2}d, which correspond to $u\simeq 18.14$ and $Q\simeq 1.06$. So function $S(\check{t})$ is not just a mathematical curiosity, and its predictions are not without practical interest. Let's keep two of them, due to the contour: the behavior at short times is square-root, which accounts for the rapid rise of the signal in Figure \ref{fig4}, and the behavior at long times is sinusoidal damped as the inverse of time, its continuous connection to that at short times explaining the sign changes on this figure,\footnote{If $\check{t}\to +\infty$, the integral giving $S_{\rm contour}(\check{t})$ is dominated by values of $z$ close to zero and $z$ can be replaced by $0$ in the integrand, everywhere except in $\exp(\textrm{i}z\check{t})$. If $\check{t}\to 0^+$, the integral giving $\exp(\textrm{i}Q^2\check{t}/4)S_{\rm contour}(\check{t})-S_{\rm contour}(0)$ is dominated by values of $z$ close to infinity, so reduces to the Fresnel integral $(-1/\pi)\int_0^{+\infty} \mathrm{d}z [\exp(\textrm{i}z\check{t})-1]/z^{3/2}$ whose value is known (we've taken $\phi+\pi=0^+$ for simplicity).}\footnote{\label{note2} A more explicit expression of the signal can be obtained in the limit $Q\to 0$:
   $S(\check{t})=\left(\frac{\check{t}}{\pi}\right)^{1/2}\eee^{-\ii\pi/4}-\frac{\sqrt{3}}{2\pi} \eee^{-\ii\pi/4}\left(\frac{2\pi^2\check{t}}{3}-1\right)\erfi(\sqrt{\pi^2\check{t}/3})\eee^{-\pi^2\check{t}/3}$, where $\erfi$ is the imaginary error function, which has the same behavior at short times as in (\ref{eq155}) but the new behavior at long times $S(\check{t}) \underset{\check{t}\to +\infty}{\sim} - \frac{\eee^{-\ii\pi/4}}{2\pi^{1/2}} \left(\frac{3}{\pi^2}\right)^2 \frac{1}{\check{t}^{3/2}}$. Since function $\erfi$ is, despite its name, real-valued on the real axis, we find that $\re S(\check{t})=-\im S(\check{t})$ in this limit, a property of which a trace remains in Figure \ref{fig4}.} 
\begin{equation}\label{eq155}
S(\check{t}) \underset{\check{t}\to 0^+}{\sim} 2\eee^{-\ii\pi/4} (\check{t}/\pi)^{1/2} \quad\quad ; \quad\quad
S(\check{t})\underset{\check{t}\to +\infty}{\sim} \frac{-\textrm{i}Q/4}{(\frac{Q^2}{4}-\ii\frac{\pi^2}{3})^2} \frac{\eee^{\textrm{i}Q^2\check{t}/4}}{\check{t}}
\end{equation}
 
\subsection{Comparison with classical field simulations}
\label{sec2.3}

In order to compare with the simulations of section \ref{sec1}, we transpose the quantum treatments of sections \ref{sec2.1} and \ref{sec2.2} to the classical field: in equation (\ref{eq125}), we must keep the cutoff parameter $\eta$ finite (it is one in the simulations), neglect the term $1$ in front of the occupation numbers $\bar{n}^{\rm lin}_k$ and replace these by the classical equipartition law $1/\kb$. As a test, note that the corresponding value of $\Sigma_\qq^{(2,2)}(\textrm{i}0^+)$ is then written as $-\ii\hbar\Gamma_q^{\mbox{\scriptsize clas}}/2$, where $\Gamma_q^{\mbox{\scriptsize clas}}$ is the damping rate (\ref{eq113}), as it should be. In what follows, we of course retain the full dependence of $\tilde{\Sigma}_\qq^{(2,2)}(\zetat)$ on $\zetat$ when calculating signal $s^{(2,2)}(t)$ by Green's function.

The corresponding predictions for the decay time at $1/\eee$ are shown in thick black solid lines on each panel of Figure \ref{fig1}. They are in all cases closer to the simulations than the golden rule, but paradoxically the improvement is clearest in the strong interaction regime (a) and becomes very modest in the weakly interacting regime (d). This deterioration was, however, to be expected from the reasoning preceding equation (\ref{eq150}): if $\rho\xi^2$ increases at a fixed $\qb$, the second-order Green's function calculation in $H_3$ must reduce to the golden rule, thus deviating from the simulations.

Analysis of Figure \ref{fig2} showing the time dependence of the signal gives us a further clue as to what happens when we reduce the {\yct interaction} strength, $(\rho \xi^2)^{-1}\to 0$. The predictions derived from $\tilde{\Sigma}_\qq^{(2,2)}(\zetat)$ are this time shown as dashed lines (those from the golden rule would be identically zero). They come remarkably close to the simulations for the imaginary part of the {\yct deviation} $s(t)-s_{\mbox{\scriptsize pole}}(t)$ (the red dashed line almost merges with the numerical results in panel (d)), but they show a persistent, and even increasing in relative values, disagreement on the real part, which explains the previous failure on the decay time at $1/\eee$.

The anomaly exhibited by the real part in Figure \ref{fig2}d is striking if we recall the analytical predictions of Figure \ref{fig4} (even if they are for the quantum field): the real part and the imaginary part, for the corresponding values of $u$ and $Q$ ($u\simeq 18.14$ and $Q\simeq 1.06$, remember), should be more or less symmetrical to each other with respect to the abscissa axis (see also footnote \ref{note2}), which is far from being the case in the simulations. In conclusion, the calculation of the signal with the self-energy $\tilde{\Sigma}^{(2,2)}_{\yc\qb}(\zetat)$, by fault of the real part, fails to account for the first deviation from the golden rule in the limit $u\to +\infty$ where it had every chance of excelling.

\section{Study at fourth order in $H_3$}
\label{sec3}

To understand the persistent discrepancy in the real part of the signal between the numerical simulations in section \ref{sec1} and the theoretical study at second-order in $H_3$ in section \ref{sec2}, even when the two-dimensional superfluid enters the weakly interacting limit $(\rho\xi^2)^{-1}\to 0$, we calculate the self-energy at fourth order in $H_3$, either {\yct with} the usual diagrammatic formalism in section \ref{sec3.1}, or by a Landau-style reasoning in section \ref{sec3.2}, which provides an interesting physical insight and leads to a more compact expression. In section \ref{sec3.3}, however, we show that the fourth-order result is useless in the low-temperature limit $\epsilon=k_B T/mc^2\to 0$ of interest here.

\subsection{Calculation with the diagrammatic formalism of reference \cite{FW}}
\label{sec3.1}

\paragraph{The two topologies}
 We directly omit the quartic Hamiltonian $H_4$ in equation (\ref{eq001}) and the non-resonant terms $H_3^{\mbox{\yct\scriptsize(non\, res.)}}$ in the cubic Hamiltonian (\ref{eq100}), as their contributions to $\Sigma^{(4)}$ are subleading in the limit $\epsilon\to 0$ (this is obvious for $H_3^{\mbox{\yct\scriptsize(non\, res.)}}$, which does not benefit from the \g{small-denominator} effect, and this will be justified for $H_4$ in section \ref{sec3.2}). The coupling Hamiltonian therefore reduces to the terms $H_3^{(+/-)}$ given in equation (\ref{eq101}). In the corresponding fourth-order Feynman diagrams, each of the four vertices is the meeting point of three phonon lines, leading to two possible topologies, (i) inner-loop and (ii) bridge, as shown in Figure \ref{fig8}. In the first topology, we lift the degeneracy by arbitrarily putting the inner loop in the upper branch, and remember (as in section \ref{sec2}) that swapping the two inner-loop lines leads to the same diagram; {\yc after orientation of internal lines,} this leaves eight distinct diagrams. In the second topology, the degeneracy is lifted by orienting the bridge line downwards; six distinct diagrams remain. In reality, in the terminology of Figure \ref{fig8}, the contributions of diagrams $B_3$ and $B_4$ to $\Sigma^{(4)}$ double, as do those of $P_2$ and $P_3$ and those of $P_4$ and $P_6$.\footnote{There is indeed a non-trivial symmetry connecting certain diagrams. To see this, let's apply a vertical axis rotation of angle $\pi$ to each diagram, then change the orientation of all lines (internal and external); this does not change the value of the diagram's contribution to $\Sigma^{(4)}$. Most diagrams are {\yct mapped} on themselves in this way, with the exception of pairs $\{B_3,B_4\}$, $\{P_2,P_3\}$ and $\{P_4,P_6\}$, whose two partners are simply exchanged (remember, in the case of bridge diagrams, that a horizontal axis rotation of angle $\pi$, which restores the original orientation of the vertical line, does not change the value of the diagram). This means that $\Sigma_{B_3}^{{\yc(4)}}(\qq,z)=\Sigma_{B_4}^{{\yc(4)}}(\qq,z)$, $\Sigma_{P_2}^{{\yc(4)}}(\qq,z)=\Sigma_{P_3}^{{\yc(4)}}(\qq,z)$ and $\Sigma_{P_4}^{{\yc(4)}}(\qq,z)=\Sigma_{P_6}^{{\yc(4)}}(\qq,z)$ as announced.} Only eleven truly independent diagrams remain.
\begin{figure}[p]
\begin{center}
\begin{tikzpicture}[x=10mm,y=5mm]
\draw[thick,->,>=stealth](1,0)arc(180:135:1);
\draw[thick,->,>=stealth](2,0)+(135:1)arc(135:25:1);
\draw[thick](2,0)+(25:1)arc(25:0:1);
\draw[thick,dashed,->,>=stealth](3,0)--(4,0);
\node(a)at(0,0.7)[above]{$B_1$};
\draw[thick,dashed,->,>=stealth](0,0)--(1,0);
\draw[thick,->,>=stealth](1,0)arc(-180:-90:1);
\draw[thick](2,0)+(-90:1)arc(-90:0:1);
\draw[thick,fill=white] (2,1) circle (0.5);
\draw[thick,->,>=stealth](2,1)+(180:0.5)arc(180:90:0.5);
\draw[thick,->,>=stealth](2,1)+(-180:0.5)arc(-180:-90:0.5);
\node(a)at(0,0)[below]{\small $\vec{q}$};
\node[below](b)at(2,-1) {\small $\vec{q}_1^\prime$};
\node[](b)at(0.8,0.7) {};
\node[above](b)at(2,1.5) {\small $\vec{q}_2^\prime$};
\node[below](b)at(2,0.5) {\small $\vec{q}_3^\prime$};
\node[](b)at(3.2,0.7) {};
\node(a)at(4,0)[below]{\color{black} {\small $\vec{q}$}};
\end{tikzpicture}
\quad
\begin{tikzpicture}[x=10mm,y=5mm]
\draw[thick,->,>=stealth](1,0)arc(180:135:1);
\draw[thick,->,>=stealth](2,0)+(135:1)arc(135:25:1);
\draw[thick](2,0)+(25:1)arc(25:0:1);
\draw[thick,dashed,->,>=stealth](3,0)--(4,0);
\node(a)at(0,0.7)[above]{$B_2$};
\draw[thick,dashed,->,>=stealth](0,0)--(1,0);
\draw[thick,->,>=stealth](1,0)arc(-180:-90:1);
\draw[thick](2,0)+(-90:1)arc(-90:0:1);
\draw[thick,fill=white] (2,1) circle (0.5);
\draw[thick,->,>=stealth](2,1)+(180:0.5)arc(180:90:0.5);
\draw[thick,->,>=stealth](2,1)+(0:0.5)arc(0:-90:0.5);
\node(a)at(0,0)[below]{\small $\vec{q}$};
\node[below](b)at(2,-1) {\small $\vec{q}_1^\prime$};
\node[](b)at(0.8,0.7) {};
\node[below](b)at(2,0.5) {\small $\vec{q}_2$};
\node[above](b)at(2,1.5) {\small $\vec{q}_2^\prime$};
\node[](b)at(3.2,0.7) {};
\node(a)at(4,0)[below]{\color{black} {\small $\vec{q}$}};
\end{tikzpicture}
\quad
\begin{tikzpicture}[x=10mm,y=5mm]
\draw[thick,->,>=stealth](1,0)arc(180:135:1);
\draw[thick](2,0)+(135:1)arc(135:45:1);
\draw[thick,->,>=stealth](2,0)+(0:1)arc(0:45:1);
\draw[thick,dashed,->,>=stealth](3,0)--(4,0);
\node(a)at(0,0.7)[above]{$B_3$};
\draw[thick,dashed,->,>=stealth](0,0)--(1,0);
\draw[thick,->,>=stealth](1,0)arc(-180:-90:1);
\draw[thick](2,0)+(-90:1)arc(-90:0:1);
\draw[thick,fill=white] (2,1) circle (0.5);
\draw[thick,->,>=stealth](2,1)+(180:0.5)arc(180:90:0.5);
\draw[thick,->,>=stealth](2,1)+(0:0.5)arc(0:-90:0.5);
\node(a)at(0,0)[below]{\small $\vec{q}$};
\node[below](b)at(2,-1) {\small $\vec{q}_1^\prime$};
\node[](b)at(0.8,0.7) {};
\node[below](b)at(2,0.5) {\small $\vec{q}_2$};
\node[above](b)at(2,1.5) {\small $\vec{q}_2^\prime$};
\node[](b)at(3.2,0.7) {};
\node(a)at(4,0)[below]{\color{black} {\small $\vec{q}$}};
\end{tikzpicture}

\bigskip

\begin{tikzpicture}[x=10mm,y=5mm]
\draw[thick](1,0)arc(180:155:1);
\draw[thick,->,>=stealth](2,0)+(90:1)arc(90:155:1);
\draw[thick,->,>=stealth](2,0)+(90:1)arc(90:25:1);
\draw[thick](2,0)+(25:1)arc(25:0:1);
\draw[thick,dashed,->,>=stealth](3,0)--(4,0);
\node(a)at(0,0.7)[above]{$B_4$};
\draw[thick,dashed,->,>=stealth](0,0)--(1,0);
\draw[thick,->,>=stealth](1,0)arc(-180:-90:1);
\draw[thick](2,0)+(-90:1)arc(-90:0:1);
\draw[thick,fill=white] (2,1) circle (0.5);
\draw[thick,->,>=stealth](2,1)+(180:0.5)arc(180:90:0.5);
\draw[thick,->,>=stealth](2,1)+(0:0.5)arc(0:-90:0.5);
\node(a)at(0,0)[below]{\small $\vec{q}$};
\node[below](b)at(2,-1) {\small $\vec{q}_2^\prime$};
\node[](b)at(0.8,0.7) {};
\node[below](b)at(2,0.5) {\small $\vec{q}_2$};
\node[above](b)at(2,1.5) {\small $\vec{q}_1^\prime$};
\node[](b)at(3.2,0.7) {};
\node(a)at(4,0)[below]{\color{black} {\small $\vec{q}$}};
\end{tikzpicture}
\quad
\begin{tikzpicture}[x=10mm,y=5mm]
\draw[thick](1,0)arc(180:155:1);
\draw[thick,->,>=stealth](2,0)+(90:1)arc(90:155:1);
\draw[thick](2,0)+(35:1)arc(35:90:1);
\draw[thick,->,>=stealth](2,0)+(0:1)arc(0:35:1);
\draw[thick,dashed,->,>=stealth](3,0)--(4,0);
\node(a)at(0,0.7)[above]{$B_5$};
\draw[thick,dashed,->,>=stealth](0,0)--(1,0);
\draw[thick,->,>=stealth](1,0)arc(-180:-90:1);
\draw[thick](2,0)+(-90:1)arc(-90:0:1);
\draw[thick,fill=white] (2,1) circle (0.5);
\draw[thick,->,>=stealth](2,1)+(180:0.5)arc(180:90:0.5);
\draw[thick,->,>=stealth](2,1)+(0:0.5)arc(0:-90:0.5);
\node(a)at(0,0)[below]{\small $\vec{q}$};
\node[below](b)at(2,-1) {\small $\vec{q}_1^\prime$};
\node[](b)at(0.8,0.7) {};
\node[below](b)at(2,0.5) {\small $\vec{q}_2$};
\node[above](b)at(2,1.5) {\small $\vec{q}_2^\prime$};
\node[](b)at(3.2,0.7) {};
\node(a)at(4,0)[below]{\color{black} {\small $\vec{q}$}};
\end{tikzpicture}
\quad
\begin{tikzpicture}[x=10mm,y=5mm]
\draw[thick](1,0)arc(180:155:1);
\draw[thick,->,>=stealth](2,0)+(90:1)arc(90:155:1);
\draw[thick](2,0)+(35:1)arc(35:90:1);
\draw[thick,->,>=stealth](2,0)+(0:1)arc(0:35:1);
\draw[thick,dashed,->,>=stealth](3,0)--(4,0);
\node(a)at(0,0.7)[above]{$B_6$};
\node[below](b)at(2,-1) {\small $\vec{q}_1^\prime$};
\draw[thick,dashed,->,>=stealth](0,0)--(1,0);
\draw[thick,->,>=stealth](1,0)arc(-180:-90:1);
\draw[thick](2,0)+(-90:1)arc(-90:0:1);
\draw[thick,fill=white] (2,1) circle (0.5);
\draw[thick,->,>=stealth](2,1)+(0:0.5)arc(0:90:0.5);
\draw[thick,->,>=stealth](2,1)+(0:0.5)arc(0:-90:0.5);
\node(a)at(0,0)[below]{\small $\vec{q}$};
\node[below](b)at(2,-1) {};
\node[](b)at(0.8,0.7) {};
\node[below](b)at(2,0.5) {\small $\vec{q}_3$};
\node[above](b)at(2,1.5) {\small $\vec{q}_2$};
\node[](b)at(3.2,0.7) {};
\node(a)at(4,0)[below]{\color{black} {\small $\vec{q}$}};
\end{tikzpicture}

\bigskip
\begin{tikzpicture}[x=10mm,y=5mm]
\draw[thick,->,>=stealth](1,0)arc(180:135:1);
\draw[thick,->,>=stealth](2,0)+(135:1)arc(135:25:1);
\draw[thick](2,0)+(25:1)arc(25:0:1);
\draw[thick,dashed,->,>=stealth](3,0)--(4,0);
\node(a)at(0,0.7)[above]{$B_7$};
\draw[thick,dashed,->,>=stealth](0,0)--(1,0);
\draw[thick,->,>=stealth](2,0)+(0:1)arc(0:-90:1);
\draw[thick](2,0)+(-180:1)arc(-180:-90:1);
\draw[thick,fill=white] (2,1) circle (0.5);
\draw[thick,->,>=stealth](2,1)+(180:0.5)arc(180:90:0.5);
\draw[thick,->,>=stealth](2,1)+(-180:0.5)arc(-180:-90:0.5);
\node(a)at(0,0)[below]{\small $\vec{q}$};
\node[below](b)at(2,-1) {\small $\vec{q}_2$};
\node[](b)at(0.8,0.7) {};
\node[below](b)at(2,0.5) {\small $\vec{q}_2^\prime$};
\node[above](b)at(2,1.5) {\small $\vec{q}_1^\prime$};
\node[](b)at(3.2,0.7) {};
\node(a)at(4,0)[below]{\color{black} {\small $\vec{q}$}};
\end{tikzpicture}
\quad
\begin{tikzpicture}[x=10mm,y=5mm]
\draw[thick,->,>=stealth](1,0)arc(180:135:1);
\draw[thick,->,>=stealth](2,0)+(135:1)arc(135:25:1);
\draw[thick](2,0)+(25:1)arc(25:0:1);
\draw[thick,dashed,->,>=stealth](3,0)--(4,0);
\node(a)at(0,0.7)[above]{$B_8$};
\node[above](b)at(2,1.5) {\small $\vec{q}_1^\prime$};
\draw[thick,dashed,->,>=stealth](0,0)--(1,0);
\draw[thick,->,>=stealth](2,0)+(0:1)arc(0:-90:1);
\draw[thick](2,0)+(-180:1)arc(-180:-90:1);
\draw[thick,fill=white] (2,1) circle (0.5);
\draw[thick,->,>=stealth](2,1)+(180:0.5)arc(180:90:0.5);
\draw[thick,->,>=stealth](2,1)+(0:0.5)arc(0:-90:0.5);
\node(a)at(0,0)[below]{\small $\vec{q}$};
\node[below](b)at(2,-1) {\small $\vec{q}_2$};
\node[](b)at(0.8,0.7) {};
\node[below](b)at(2,0.5) {\small $\vec{q}_3$};
\node[above](b)at(2,1.5) {};
\node[](b)at(3.2,0.7) {};
\node(a)at(4,0)[below]{\color{black} {\small $\vec{q}$}};
\end{tikzpicture}

\bigskip
\bigskip
\begin{tikzpicture}
\draw[thick,->,>=stealth](1,0)arc(180:135:1);
\draw[thick,->,>=stealth](2,0)+(135:1)arc(135:45:1);
\draw[thick](2,0)+(45:1)arc(45:0:1);
\draw[thick,dashed,->,>=stealth](3,0)--(4,0);
\node(a)at(0,0.7)[above]{$P_1$};
\draw[thick,dashed,->,>=stealth](0,0)--(1,0);
\node(a)at(0,0)[below]{\color{black} {\small $\vec{q}$}};
\node(a)at(0.8,0.7)[]{\small  {\color{red} $\vec{q}_1^\prime$}};
\node(a)at(3.2,0.7)[]{\small {\color{black} $\vec{q}_2^\prime$} };
\node(a)at(0.8,-0.7)[]{\small {\color{black} $\vec{q}_1^\prime$} };
\node(a)at(3.2,-0.7)[]{\small  {\color{red} $\vec{q}_2^\prime$}};
\node(a)at(2,0)[]{\small {\color{black} $\vec{q}_3^\prime$}\:\: {\color{red} $\vec{q}_2$}};
\node(a)at(4,0)[below]{\color{black} {\small $\vec{q}$}};
\draw[thick,->,>=stealth](1,0)arc(-180:-135:1);
\draw[thick,->,>=stealth](2,0)+(-135:1)arc(-135:-45:1);
\draw[thick](2,0)+(-45:1)arc(-45:0:1);
\draw[thick,->,>=stealth](2,1)--(2,0);
\draw[thick](2,0)--(2,-1);
\end{tikzpicture}
\quad
\begin{tikzpicture}
\draw[thick,->,>=stealth](1,0)arc(180:135:1);
\draw[thick](2,0)+(135:1)arc(135:45:1);
\draw[thick,->,>=stealth](2,0)+(0:1)arc(0:45:1);
\draw[thick,dashed,->,>=stealth](3,0)--(4,0);
\node(a)at(0,0.7)[above]{$P_2$};
\draw[thick,dashed,->,>=stealth](0,0)--(1,0);
\node(a)at(0,0)[below]{\color{black} {\small $\vec{q}$}};
\node(a)at(0.8,0.7)[]{\small  {\color{red} $\vec{q}_1^\prime$}};
\node(a)at(3.2,0.7)[]{\small {\color{black} $\vec{q}_2$}};
\node(a)at(0.8,-0.7)[]{\small {\color{black} $\vec{q}_1^\prime$}};
\node(a)at(3.2,-0.7)[]{\small  {\color{red} $\vec{q}_2^\prime$}};
\node(a)at(2,0)[]{\small {\color{black} $\vec{q}_2^\prime$} \:\: {\color{red} $\vec{q}_2$}};
\node(a)at(4,0)[below]{\color{black} {\small $\vec{q}$}};
\draw[thick,->,>=stealth](1,0)arc(-180:-135:1);
\draw[thick,->,>=stealth](2,0)+(-135:1)arc(-135:-45:1);
\draw[thick](2,0)+(-45:1)arc(-45:0:1);
\draw[thick,->,>=stealth](2,1)--(2,0);
\draw[thick](2,0)--(2,-1);
\end{tikzpicture}
\quad
\begin{tikzpicture}
\draw[thick,->,>=stealth](1,0)arc(180:135:1);
\draw[thick,->,>=stealth](2,0)+(135:1)arc(135:45:1);
\draw[thick](2,0)+(45:1)arc(45:0:1);
\draw[thick,dashed,->,>=stealth](3,0)--(4,0);
\node(a)at(0,0.7)[above]{$P_3$};
\draw[thick,dashed,->,>=stealth](0,0)--(1,0);
\node(a)at(0,0)[below]{\color{black} {\small $\vec{q}$}};
\node(a)at(0.8,0.7)[]{\small {\color{red} $\vec{q}_1^\prime$}};
\node(a)at(3.2,0.7)[]{\small {\color{black} $\vec{q}_2^\prime$}};
\node(a)at(0.8,-0.7)[]{\small {\color{black} $\vec{q}_2$}};
\node(a)at(3.2,-0.7)[]{\small  {\color{red} $\vec{q}_2^\prime$}};
\node(a)at(2,0)[]{\small {\color{black} $\vec{q}_1^\prime$} \:\: {\color{red} $\vec{q}_2$}};
\node(a)at(4,0)[below]{\color{black} {\small $\vec{q}$}};
\draw[thick](1,0)arc(-180:-135:1);
\draw[thick,->,>=stealth](2,0)+(-45:1)arc(-45:-135:1);
\draw[thick](2,0)+(0:1)arc(0:-45:1);
\draw[thick,->,>=stealth](2,0)+(-90:1)arc(-90:-45:1);
\draw[thick,->,>=stealth](2,1)--(2,0);
\draw[thick](2,0)--(2,-1);
\end{tikzpicture}

\bigskip
\begin{tikzpicture}
\draw[thick,->,>=stealth](1,0)arc(180:135:1);
\draw[thick,->,>=stealth](2,0)+(135:1)arc(135:45:1);
\draw[thick](2,0)+(45:1)arc(45:0:1);
\draw[thick,dashed,->,>=stealth](3,0)--(4,0);
\node(a)at(0,0.7)[above]{$P_4$};
\node(a)at(0.8,0.7)[]{\small  {\color{red} $\vec{q}_1^\prime$}};
\draw[thick,dashed,->,>=stealth](0,0)--(1,0);
\node(a)at(0,0)[below]{\color{black} {\small $\vec{q}$}};
\node(a)at(0.8,0.7)[]{\small };
\node(a)at(3.2,0.7)[]{\small {\color{black} $\vec{q}_1^\prime$}};
\node(a)at(0.8,-0.7)[]{\small {\color{black} $\vec{q}_2$}};
\node(a)at(3.2,-0.7)[]{\small {\color{red} $\vec{q}_3$}};
\node(a)at(2,0)[]{\small {\color{black} $\vec{q}_2^\prime$} \:\: {\color{red} $\vec{q}_2$}};
\node(a)at(4,0)[below]{\color{black} {\small $\vec{q}$}};
\draw[thick](1,0)arc(-180:-135:1);
\draw[thick,->,>=stealth](2,0)+(-45:1)arc(-45:-135:1);
\draw[thick,->,>=stealth](2,0)+(0:1)arc(0:-45:1);
\draw[thick,->,>=stealth](2,1)--(2,0);
\draw[thick](2,0)--(2,-1);
\end{tikzpicture}
\quad
{\color{red}
\begin{tikzpicture}
\draw[thick,->,>=stealth](1,0)arc(180:135:1);
\draw[thick](2,0)+(135:1)arc(135:45:1);
\draw[thick,->,>=stealth](2,0)+(0:1)arc(0:45:1);
\draw[thick,dashed,->,>=stealth](3,0)--(4,0);
\node(a)at(0,0.7)[above]{$P_5$};
\draw[thick,dashed,->,>=stealth](0,0)--(1,0);
\node(a)at(0,0)[below]{\color{black} {\small $\vec{q}$}};
\node(a)at(0.8,0.7)[]{\small  {\color{red} $\vec{q}_1^\prime$}};
\node(a)at(3.2,0.7)[]{\small {\color{black} $\vec{q}_3$}};
\node(a)at(0.8,-0.7)[]{\small {\color{black} $\vec{q}_2$}};
\node(a)at(3.2,-0.7)[]{\small  {\color{red} $\vec{q}_2^\prime$}};
\node(a)at(2,0)[]{\small {\color{black} $\vec{q}_1^\prime$} \:\: {\color{red} $\vec{q}_2$}};
\node(a)at(4,0)[below]{\color{black} {\small $\vec{q}$}};
\draw[thick](1,0)arc(-180:-135:1);
\draw[thick,->,>=stealth](2,0)+(-90:1)arc(-90:-135:1);
\draw[thick,->,>=stealth](2,0)+(-135:1)arc(-135:-45:1);
\draw[thick](2,0)+(-45:1)arc(-45:0:1);
\draw[thick,->,>=stealth](2,1)--(2,0);
\draw[thick](2,0)--(2,-1);
\end{tikzpicture}
}
\quad
\begin{tikzpicture}
\draw[thick](1,0)arc(180:135:1);
\draw[thick,->,>=stealth](2,0)+(90:1)arc(90:135:1);
\draw[thick](2,0)+(135:1)arc(135:45:1);
\draw[thick,->,>=stealth](2,0)+(0:1)arc(0:45:1);
\draw[thick](2,1)arc(90:0:1);
\draw[thick,dashed,->,>=stealth](3,0)--(4,0);
\node(a)at(0,0.7)[above]{$P_6$};
\node(a)at(3.2,-0.7)[]{\small  {\color{red} $\vec{q}_1^\prime$}};
\draw[thick,dashed,->,>=stealth](0,0)--(1,0);
\node(a)at(0,0)[below]{\color{black} {\small $\vec{q}$}};
\node(a)at(0.8,0.7)[]{\small {\color{red} $\vec{q}_2$}};
\node(a)at(3.2,0.7)[]{\small {\color{black} $\vec{q}_2$}};
\node(a)at(0.8,-0.7)[]{\small {\color{black} $\vec{q}_2^\prime$}};
\node(a)at(3.2,-0.7)[]{\small };
\node(a)at(2,0)[]{\small {\color{black} $\vec{q}_1^\prime$} \:\: {\color{red} $\vec{q}_3$}};
\node(a)at(4,0)[below]{\color{black} {\small $\vec{q}$} };
\draw[thick,->,>=stealth](1,0)arc(-180:-135:1);
\draw[thick,->,>=stealth](2,0)+(-135:1)arc(-135:-45:1);
\draw[thick](2,0)+(-45:1)arc(-45:0:1);
\draw[thick,->,>=stealth](2,1)--(2,0);
\draw[thick](2,0)--(2,-1);
\end{tikzpicture}
\end{center}
\caption{Contributions to self-energy $\Sigma(\qq,z)$ at fourth order in $H_3^{(+)}+H_3^{(-)}$ are given by eight inner-loop diagrams (numbered $B_1$ to $B_8$) and six bridge diagrams (numbered $P_1$ to $P_6$). The wave vectors $\qq_i$ $(2\leq i\leq n$) and $\qq'_j$ ($1\leq j\leq n'$) near the inner lines are those of the incoming and outgoing phonons of the four-phonon $n\to n'$ collisional processes in the Landau-style physical interpretation of section \ref{sec3.2} (for bridge diagrams: wave vectors of type I contributions in black, type II in red, in the sense of equation {\yct(\ref{eq233})}); for the sake of simplicity, wave vectors of virtual phonons are not shown (they are deduced by conservation of momentum at each vertex). {\yc The contributions of the non-resonant processes of $H_3$, subleading in the $\epsilon\to 0$ limit, are not shown. On the other hand, once this limit is taken, the {\yc diagram} $P_5$ (more precisely, its type II contribution, hence the red color) diverges linearly for $\zetat\equiv (z-\veps_\qq)/k_B T\epsilon^2=0$, revealing a non-physical pole in Green's function, see section \ref{sec3.3}.}}
\label{fig8}
\end{figure}
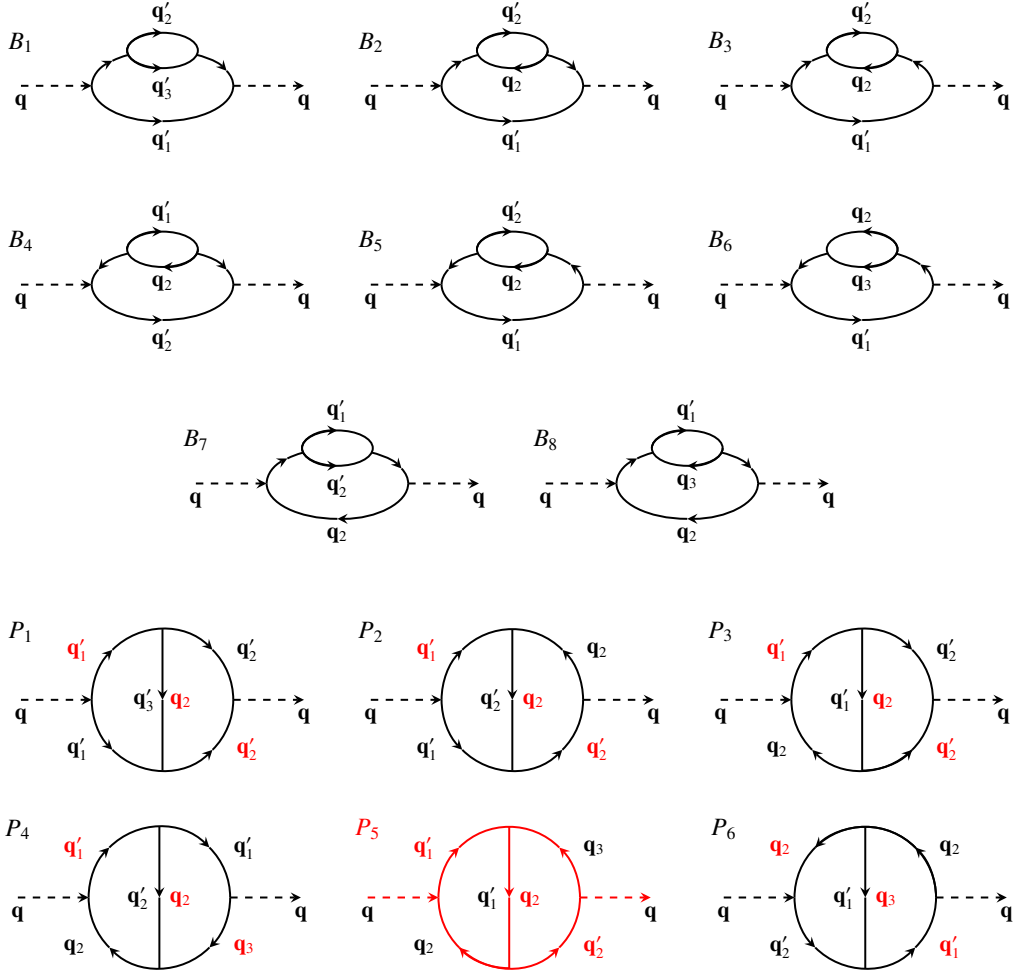
 \begin{figure}[t]
\begin{center}
\begin{tikzpicture} 
\draw[thick,dashed,->,>=stealth](-0.5,0)--(0.5,0);
\draw[thick] (2,0) ellipse (1.5 and 1);
\draw[thick,fill=white] (2,0.95) ellipse (0.6 and 0.5);
\draw[thick,red](2,-2)--(2,2.5);
\draw[thick,dashed,->,>=stealth](3.5,0)--(4.5,0);
\node(a)at(-0.2,0)[below]{$\vec{q},\Omega$};
\node(a)at(4.2,0)[below]{$\vec{q},\Omega$};
\node(a)at(2.1,1.8)[above]{$\vec{k\,}^\prime,\omega^\prime$};
\node[draw,rectangle,fill=white,text=blue](b)at(2,-1) {$s_1$};
\node[draw,rectangle,fill=white,text=blue](b)at(2,0.5) {$s_4$};
\node[draw,rectangle,fill=white,text=blue](b)at(2,1.5) {$s_3$};
\node[draw,rectangle,fill=white,text=blue](b)at(0.8,0.7) {$s_2$};\node(a)at(0.8,1.0)[above]{$\vec{k},\omega$};
\node[draw,rectangle,fill=white,text=blue](b)at(3.2,0.7) {$s_5$};\draw[thick,red](1.25,-2)--(1.25,2.5); 
\draw[thick,red](2.75,-2)--(2.75,2.5);
\draw[thick,dashed,->,>=stealth](-0.5+7,0)--(0.5+7,0);
\draw[thick] (2+7,0) circle (1.5 );
\draw[thick,dashed,->,>=stealth](3.5+7,0)--(4.5+7,0);
\node(a)at(-0.2+7,0)[below]{$\vec{q},\Omega$};
\node(a)at(4.2+7,0)[below]{$\vec{q},\Omega$};
\node[draw,rectangle,fill=white,text=blue](b)at(0.8+7,0.7) {$s_1$};
\node[draw,rectangle,fill=white,text=blue](b)at(3.2+7,0.7) {$s_3$};
\node[draw,rectangle,fill=white,text=blue](b)at(0.8+7,-0.7) {$s_2$};
\node[draw,rectangle,fill=white,text=blue](b)at(3.2+7,-0.7) {$s_4$};
\draw[thick,red](1.25+7,-1.7)--(1.25+7,2.2);
\draw[thick,red](2.75+7,-1.7)--(2.75+7,2.2);
\draw[thick,red](1.25+7,-2)--(2.75+7,2.5);
\draw[thick,red](2.75+7,-2)--(1.25+7,2.5);
\node[text=red](b)at(2.5+7,2.5) {${\rm I}$};
\node[text=red](b)at(1.5+7,2.5) {${\rm II}$};
\node(a)at(0.6+7,1)[above]{$\vec{k},\omega$};
\node(a)at(3.6+7,1)[above]{$\vec{k}^\prime,\omega'$};
\draw[thick](2+7,-1.5)--(2+7,1.5); 
\node[draw,rectangle,fill=white,text=blue](b)at(2+7,0) {$s_5$};
\end{tikzpicture}
\end{center}
\caption{In the expansion of the self-energy $\Sigma(\qq,z)$, generic forms of fourth-order diagrams in $H_3$ for inner-loop (left-hand side) or bridge diagrams (right-hand side). The boxes contain the possible $s_j=\pm 1$ orientations of the internal lines and indicate their numbering $j$. Given the conservation of the Matsubara energy-momentum at each vertex, there are only two independent internal wave vectors $\kk$, $\kk'$ and Matsubara energies $\omega$, $\omega'$; the others are given by equations (\ref{eq211}) (inner loop) and (\ref{eq231}) (bridge). The red lines are a graphical construction for finding the energy denominators in the simplified forms (\ref{eq213}) and (\ref{eq233}) of the double Matsubara sums, which apply to leading order in $\epsilon$ when all internal wave vectors make a small angle $O(\epsilon)$ with $\qq$. In the simplified form (\ref{eq213}), the quantity $\zeta_1$ corresponds to the two outer red lines of the inner-loop diagram (indeed, $s_2=s_5$ at the leading order in $\epsilon$) and $\zeta_2$ corresponds to the central red line. In the simplified form (\ref{eq233}), the quantity $\zeta_1$ corresponds to the left vertical red line of the bridge diagram, $\zeta_2$ to the other vertical red line, $\zeta_3$ to the {\yc oblique} line numbered II and $\zeta_4$ to the {\yc oblique} line numbered I.}
\label{fig7}
\end{figure}
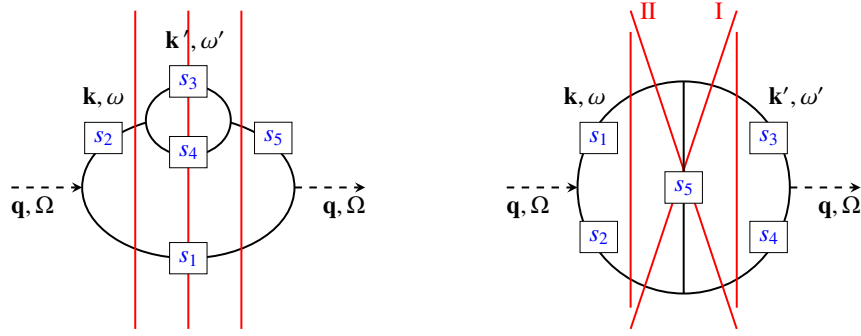
 
\paragraph{Calculating a generic inner-loop diagram}
 It's convenient to number the internal lines from $j=1$ to $5$ as on the left-hand side of Figure \ref{fig7}. Each line has an orientation $s_j\in\{-1;1\}$ (left to right if $s_j=1$, right to left if $s_j=-1$), a wave vector $\kk_j$ and a Matsubara energy $\omega_j\in 2\ii\pi k_B T \mathbb{Z}$. In Figure \ref{fig8}, we have chosen to always have $s_3\geq s_4$ in order to avoid the double counting of diagrams mentioned above. As independent quadrivectors, we've chosen those in lines 2 and 3, which we've renamed as follows:  
\begin{equation}\label{eq210}
(\kk_2,\omega_2)=(\kk,\omega) \quad ; \quad (\kk_3,\omega_3)=(\kk',\omega')
\end{equation} The others can be deduced by energy-momentum conservation at each vertex. For example, the sum of the wave vectors arriving at a given vertex is equal to the sum of the wave vectors departing from it, giving: 
\begin{equation}\label{eq211}
\kk_1=s_1(\qq-s_2\kk) \quad ; \quad \kk_4=s_4(s_2\kk-s_3\kk') \quad ; \quad \kk_5=s_2 s_5 \mathbf{k}\end{equation} The same relations apply to Matsubara energies. Given the rules set out in section 25 of reference \cite{FW}, the contribution to the self-energy of the generic inner-loop diagram $B_i$ is: 
\begin{equation}\label{eq212}
\Sigma_{B_i}^{(4)}(\qq,z) = \frac{(-k_B T)^2}{1+\delta_{s_3,s_4}} \int\frac{\dd^dk}{(2\pi)^d}\int\frac{\dd^dk'}{(2\pi)^d}
\mA(\kk_1,\kk_2;\qq)\mA(\kk_3,\kk_4;\kk_2)\mA(\kk_3,\kk_4;\kk_5)\mA(\kk_1,\kk_5;\qq)
\sum_{\omega,\omega'\in 2\ii\pi k_B T\mathbb{Z}}\frac{1}{\prod_{j=1}^{5} (\omega_j-\veps_{\kk_j})}
\end{equation} where we temporarily have $z=\Omega\in 2\ii\pi k_B T\mathbb{Z}$ (see section \ref{sec2.1}). To write the transition amplitudes at each vertex, we used the invariance of function $\mA$ of equation (\ref{eq102}) by permutation of its three arguments and hence the fact that matrix elements $\langle \kk_2,\kk_3|\mathcal{H}_3|\kk_1\rangle$, $\langle \kk_1,\kk_3|\mathcal{H}_3|\kk_2\rangle$ and $\langle \kk_1,\kk_2|\mathcal{H}_3|\kk_3\rangle$ are all equal to $\mA(\kk_2,\kk_3;\kk_1)$. The denominator of the symmetry factor {\yc in front of the integrals} is $2$ if the two lines of the inner loop have the same orientation (to compensate for the double counting of configurations differing only by the exchange of quadrivectors $(\kk_3,\omega_3)$ and $(\kk_4,\omega_4)$ and in this case physically identical); otherwise, these two lines are distinguishable and the symmetry factor is $1$. The Matsubara double sum is calculated explicitly in \ref{app3}, which also shows that its expression simplifies in the limit of small angles $O(\epsilon)\to 0$ between the wave vectors, where we can restrict to $s_2=s_5$ (otherwise, the diagram's contribution is subleading in $\epsilon$, see equation (\ref{eq215}) below): 
\begin{equation}\label{eq213}
\sum_{\omega,\omega'\in 2\ii\pi k_B T\mathbb{Z}}\frac{(k_B T)^2}{\prod_{j=1}^{5}(\omega_j-\veps_{\kk_j})} \stackrel{\mbox{\scriptsize diag.}\, B}{\underset{\epsilon\to 0}{\sim}} -s_1 s_4 \frac{(\bar{n}_{\veps_3}-\bar{n}_{-s_3s_4\veps_4})(\bar{n}_{-s_1\veps_1}-\bar{n}_{s_2\veps_2})}{(\zeta-\zeta_1)^2(\zeta-\zeta_2)}
\end{equation} with variable $\zeta=z-\veps_\qq$ and $\zeta_j$ all of order $k_B T\epsilon^2$, 
\begin{equation}\label{eq214}
\zeta_1\equiv s_1\veps_{\kk_1}+s_2\veps_{\kk_2}-\veps_\mathbf{q}\quad ; \quad \zeta_2\equiv s_1\veps_{\kk_1}+s_3\veps_{\kk_3}+s_4\veps_{\kk_4}-\veps_{\qq}
\end{equation} The dependence in $\zeta$ of form (\ref{eq213}) on the right-hand  side is easy to interpret physically since $\zeta_1+\veps_\qq$ and $\zeta_2+\veps_\qq$ are the energies of the intermediate states marked by the vertical lines on the left-hand side of Figure \ref{fig7}: in fact, they are the sums of the energies of the internal lines $j$ intersected by each vertical line counted algebraically, i.e. weighted by the orientations $s_j$ (if $s_j=-1$, the phonon on line $j$ is going backward in time and its energy must be counted negatively). This validates estimate (\ref{eq024}) by reproducing its energy denominators. 

To go to the $\epsilon\to 0$ limit in (\ref{eq212}), we directly set the angles to zero in each coupling amplitude $\mA$ (which brings out a factor $3(1+\Lambda)$), then expand the energy denominators of (\ref{eq213}) on the model of (\ref{eq025}) ; for the \g{small-denominator} effect to take full action, the internal vectors $\kk_j$ must all make a small angle $O(\epsilon)$ with the incoming wave vector $\qq$, which imposes\footnote{When $\kk_2=\kk$ and $\kk_3=\kk'$ form a zero angle with $\qq$, the other wave vectors $\kk_j$ given by (\ref{eq211}) must satisfy the condition $\kk_j\cdot\qq=k_j q$.} 
\begin{equation}\label{eq215}
\sig(q-s_2k)=s_1 \quad ; \quad \sig(s_2k-s_3k')=s_4 \quad ; \quad s_2s_5=1
\end{equation} The first two conditions in (\ref{eq215}) restrict the range of variation of wave numbers $k$ and $k'$. The last, which is independent of them, eliminates diagrams $B_3$ and $B_4$, of subleading contributions. Remembering the definitions (\ref{eq022},\ref{eq004}) of $\zetat$ and (\ref{eq020}) of $\qb$, we arrive in dimension $d=2$ at the result (for $\im\zetat\neq 0$ but of arbitrary sign): 
\begin{equation}\label{eq220}
\tilde\Sigma^{(4,2)}_{B_i}(\qb,\tilde{\zeta})=\frac{\qb}{1+\delta_{s_3,s_4}}\left(\frac{9(1+\Lambda)^2}{8\rho\xi^2}\right)^2\int_{\mathcal{D}_1^B}\frac{\kb\mathrm{d}\kb}{(2\pi)^2} \int_{\mathcal{D}_2^B(\kb)} \frac{\kb'\mathrm{d}\kb'}{(2\pi)^2} \kb^2\kb'(\qb-s_2\kb)(s_2\kb-s_3\kb')(\bar{n}^{\rm lin}_{k'}-\bar{n}^{\rm lin}_{k'-s_2 s_3 k})(\bar{n}^{\rm lin}_{s_2k}-\bar{n}^{\rm lin}_{s_2k-q}) I(\kb,\kb')
\end{equation} 
with {\yc the occupation numbers (\ref{eq029}) extended to negative wavenumbers and} the angular integral\footnote{In the integral, before taking the limit $\epsilon\to 0$, we set $\tilde{\theta}=\theta/\epsilon$ and $\tilde{\phi}=(\theta-\theta')/\epsilon$, where $\theta$ is the angle between $\qq$ and $\kk$, $\theta'$ the angle between $\qq$ and $\kk'$.} 
\begin{equation}\label{eq221}
I(\kb,\kb')=\int_{\mathbb{R}} \dd\tilde\theta \int_{\mathbb{R}} \dd\tilde\phi \frac{1}{[\tilde{\zeta}-(A_1^B+B_1^B\tilde\theta^2)]^{2}[\tilde{\zeta}-(A^B_2+B^B_1\tilde\theta^2+B_2^B\tilde\phi^2)]}
\end{equation} involving coefficients 
\begin{equation}\label{eq222}
A_1^B=\frac{3\gamma}{8}(\kb-s_2 \qb)\bar{k}\bar{q}\quad ; \quad A_2^B=\frac{3\gamma}{8}\bar{k}\kb' (s_2\kb'-s_3 \kb) +A_1^{B} \quad ; \quad B_1^B=\frac{-\kb\qb/2}{\kb-s_2 \qb} \quad ; \quad B_2^B=\frac{-\bar{k}\kb'/2}{s_2\kb'-s_3 \kb}
\end{equation} and integration domains {\yc$\mathcal{D}_1^B$} over $\kb$, then {\yc$\mathcal{D}_2^{B}(\kb)$} over $\kb'$ at fixed $\kb$, satisfying conditions (\ref{eq215}) and cut-off (\ref{eq103}) $\kb_j<\eta$, $1\leq j\leq 5$. In the left-hand side of (\ref{eq220}), the second integer in the exponent reminds us {\yc to which order in $\epsilon$ the function} $\Sigma^{(4)}$ is calculated {\yc(here to second order)}, and the tilde over $\Sigma$ means, as in (\ref{eq021}), that {\yc the result} has been divided by $k_BT\epsilon^2$. Finally, we check on a case-by-case basis that expression (\ref{eq220}) has a finite limit when the cutoff $\eta\to+\infty$ (for the quantum field).\footnote{For example, in the case of {\yc the diagram} $B_8$ ($s_1=s_4=-1$, $s_2=s_3=s_5=1$), the domains {\yc$\mathcal{D}_1^B=[\qb,\eta]$ and $\mathcal{D}_2^B(\kb)=[\kb,\eta]$} are unbounded when $\eta\to +\infty$ but the bosonic amplification factors $(\bar{n}_{k'}^{\rm lin}-\bar{n}_{k'-k}^{\rm lin})(\bar{n}_k^{\rm lin}-\bar{n}_{k-q}^{\rm lin})=O[\exp(-\kb-\kb'')]$ are {\yct rapidly decreasing} in the two decoupled integration variables $(\kb,\kb''=\kb'-\kb)$ and {\yct suppress} any power-law divergence that might arise from the other factors ; in the case of $B_7$ ($s_1=-1$, $s_2=s_3=s_4=s_5=1$), $\mathcal{D}_1^B=[\qb,\eta]$ and $\mathcal{D}_2^B(\kb)=[0,\kb]$ remain unbounded and the bosonic factors $(1+\bar{n}_{k'}^{\rm lin}+\bar{n}_{k-k'}^{\rm lin})(\bar{n}_k^{\rm lin}-\bar{n}_{k-q}^{\rm lin})=O(\exp(-\kb))$ are now {\yct rapidly decreasing} only in variable $\kb$ but, since $\kb'<\kb$, the integral over $\kb'$ at fixed $\kb$ is at most $O(\kb^\alpha)$, $\alpha>0$, and the integral over $\kb$ converges if $\eta\to +\infty$.}

\paragraph{Calculating a generic bridge diagram}
 The beginning is very similar to the previous case. We number the internal lines as on the right-hand side of Figure \ref{fig7}, with the same sign convention, except for the vertical line (the \g{bridge}), with sign $s_5=1$ when oriented downwards (which we have limited ourselves to in Figure \ref{fig8} to avoid double counting). We keep as independent quadrivectors 
\begin{equation}\label{eq230}
(\kk_1,\omega_1)=(\kk,\omega) \quad ;\quad (\kk_3,\omega_3)=(\kk',\omega')
\end{equation} the others deduced by energy-momentum conservation, e.g. for wave vectors: 
\begin{equation}\label{eq231}
\kk_2=s_2(\qq-s_1\kk) \quad ;\quad \kk_4=s_4(\qq-s_3\kk') \quad ;\quad \kk_5=s_5(s_1\kk-s_3\kk')
\end{equation} The contribution of diagram $P_i$ to the self-energy is then written \cite{FW}: 
\begin{equation}\label{eq232}
\Sigma_{P_i}^{(4)}(\qq,z) = (-k_B T)^2\int\frac{\dd^dk}{(2\pi)^d}\int\frac{\dd^dk'}{(2\pi)^d}
\mA(\kk_1,\kk_2;\qq)\mA(\kk_3,\kk_5;\kk_1)\mA(\kk_2,\kk_5;\kk_4)\mA(\kk_3,\kk_4;\qq)
\sum_{\omega,\omega'\in 2\ii\pi k_B T\mathbb{Z}}\frac{1}{\prod_{j=1}^{5} (\omega_j-\veps_{\kk_j})}
\end{equation} No need here to divide by a symmetry factor. The explicit calculation of the Matsubara double sum is given in \ref{app3}, which also establishes its simplified expression in the limit of small angles $O(\epsilon)$ between the wave vectors: 
\begin{equation}\label{eq233}
\sum_{\omega,\omega'\in 2\ii\pi k_B T\mathbb{Z}}\frac{(k_B T)^2}{\prod_{j=1}^{5}(\omega_j-\veps_{\kk_j})} \stackrel{\mbox{\scriptsize diag.}\, P}{\underset{\epsilon\to 0}{\sim}} 
-s_1 s_2 s_4 s_5 \frac{(\bar{n}_{-s_2\veps_2}-\bar{n}_{s_1\veps_1})(\bar{n}_{\veps_3}-\bar{n}_{-s_3s_5\veps_5})}
{(\zeta-\zeta_1)(\zeta-\zeta_2)(\zeta-\zeta_4)}
-s_1 s_2 s_4 s_5 \frac{(\bar{n}_{-s_3s_5\veps_5}-\bar{n}_{-s_3s_4\veps_4})(\bar{n}_{s_1\veps_1}-\bar{n}_{-s_2\veps_2})}
{(\zeta-\zeta_1)(\zeta-\zeta_2)(\zeta-\zeta_3)}
\end{equation} with variable $\zeta$ and $\zeta_j$ all of order $k_BT\epsilon^2$, 
\begin{equation}\label{eq238}
\zeta_1\equiv s_1\veps_{\kk_1}+s_2\veps_{\kk_2}-\veps_\qq\quad ; \quad \zeta_2=s_3\veps_{\kk_3}+s_4\veps_{\kk_4}-\veps_\mathbf{q}\quad ; \quad \zeta_3\equiv s_1\veps_{\kk_1}+s_4\veps_{\kk_4}-s_5\veps_{\kk_5}-\veps_{\qq} \quad ;\quad \zeta_4\equiv s_2\veps_{\kk_2}+s_3\veps_{\kk_3}+s_5\veps_{\kk_5}-\veps_{\qq}
\end{equation} There are this time two contributions to the right-hand side, called type I and type II in this order. Quantities $\zeta_j$ are energy differences between intermediate and initial states, as in the inner-loop case, except that four straight lines must be introduced to identify them, see right-hand side of Figure \ref{fig7}: two vertical lines for $\zeta_1$ and $\zeta_2$, two oblique lines for $\zeta_3$ and $\zeta_4$. Indeed, from the point of view of the oblique line marked I on the figure, the internal line of the bridge is oriented from left to right if $s_5=1$, and $\zeta_4$ must have a plus sign in front of $s_5\veps_{\kk_5}$ in equation (\ref{eq238}); for oblique line II, the opposite is true, hence the minus sign in front of $s_5\veps_{\kk_5}$ in $\zeta_3$.

When $\epsilon\to 0$, the limit of small angles in the energy denominators and in the coupling amplitudes $\mA$ is taken as before, except that a distinction must be made between the two types. For type I contribution, we obtain (for $\im\zetat\neq 0$ but of arbitrary sign) 
\begin{equation}\label{eq240}
\tilde\Sigma^{(4,2)}_{P_{i}^{\rm I}}(\qb,\tilde{\zeta})=\qb\left(\frac{9(1+\Lambda)^2}{8\rho\xi^2}\right)^2\int_{\mathcal{D}_1^P} \frac{\kb\mathrm{d}\kb}{(2\pi)^2} \int_{\mathcal{D}_2^P(\kb)} \frac{\kb'\mathrm{d}\kb'}{(2\pi)^2} \bar{k}\kb'(\qb-s_1\kb)(\qb-s_3\kb')(\kb-s_1s_3\kb')(\bar{n}^{\rm lin}_{k'}-\bar{n}^{\rm lin}_{k'-s_1 s_3 k})(\bar{n}^{\rm lin}_{s_1k}-\bar{n}^{\rm lin}_{s_1k-q}) J(\kb,\kb')
\end{equation} with the angular integral 
\begin{equation}\label{eq241}
J(\kb,\kb')=\int_{\mathbb{R}} \dd\tilde\theta \int_{\mathbb{R}} \dd\tilde\theta' \frac{1}{[\tilde{\zeta}-(A_1^P+B_1^P\thetat^2)][\tilde{\zeta}-(A_2^P+B_2^P\thetat'^2)][\tilde{\zeta}-(A_4^P+B_1^P\thetat^2+B_4^P(\thetat-\thetat')^2)]}
\end{equation} For type II contribution, we obtain (again for $\im\zetat\neq 0$ but of arbitrary sign) 
\begin{equation}\label{eq242}
\tilde\Sigma^{(4,2)}_{P_{i}^{\rm II}}(\qb,\tilde{\zeta})=\qb\left(\frac{9(1+\Lambda)^2}{8\rho\xi^2}\right)^2\int_{\mathcal{D}_1^P} \frac{\kb\mathrm{d}\kb}{(2\pi)^2} \int_{\mathcal{D}_2^P(\kb)} \frac{\kb'\mathrm{d}\kb'}{(2\pi)^2} \bar{k}\kb'(\qb-s_1\kb)(\qb-s_3\kb')(\kb-s_1s_3\kb')(\bar{n}^{\rm lin}_{k'-s_3q}-\bar{n}^{\rm lin}_{k'-s_1 s_3 k})(\bar{n}^{\rm lin}_{s_1k}-\bar{n}^{\rm lin}_{s_1k-q}) K(\kb,\kb')
\end{equation} with the angular integral 
\begin{equation}\label{eq243}
K(\kb,\kb')=\int_{\mathbb{R}} \dd\tilde\theta \int_{\mathbb{R}} \dd\tilde\theta' \frac{1}{[\tilde{\zeta}-(A_1^P+B_1^P\thetat^2)][\tilde{\zeta}-(A_2^P+B_2^P\thetat'^2)][\tilde{\zeta}-(A_3^P+B_2^P\thetat'^2+B_3^P(\thetat-\thetat')^2)]}
\end{equation} Irrespective of type, the coefficients are 
\begin{equation}\label{eq244}
A_1^P=\frac{3\gamma}{8}(\kb-s_1\qb)\kb\bar{q}\quad ; \quad A_2^P=\frac{3\gamma}{8}(\kb'-s_3\qb)\kb'\bar{q}\quad ; \quad A_4^P=A_1^P+\frac{3\gamma}{8}\bar{k}\kb' (s_1\kb'-s_3\kb) \quad ; \quad A_3^P=A_1^P+A_2^P-A_4^P
\end{equation} \begin{equation}\label{eq245}
B_1^P=\frac{-\kb\qb/2}{\kb-s_1\qb} \quad ; \quad B_2^P=\frac{-\kb'\qb/2}{\kb'-s_3\qb} \quad ; \quad B_4^P=\frac{-\kb\kb'/2}{s_1\kb'-s_3\kb} \quad ; \quad B_3^P=-B_4^P
\end{equation} The integration domains $\mathcal{D}_1^P$ and $\mathcal{D}_2^P(\kb)$ are also independent of type; they take into account the cut-off (\ref{eq103}) on all internal wave vectors $\kk_j$, $1\leq j\leq 5$, as well as conditions 
\begin{equation}\label{eq239}
\sig(q-s_1k)=s_2 \quad ;\quad \sig(q-s_3k')=s_4 \quad;\quad \sig(s_1k-s_3k')=s_5
\end{equation} ensuring that $\kk_2$, $\kk_4$ and $\kk_5$ are collinear with and in the same direction as $\qq$ as soon as $\kk$ and $\kk'$ are. For all $s_j$ orientations taken from Figure \ref{fig8}, conditions (\ref{eq239}) are compatible, so none of the $P_i$ diagrams is subleading in $\epsilon$. Finally, we verify the existence of a finite limit in (\ref{eq240},\ref{eq242}) when cutoff $\eta\to +\infty$ (for the quantum field).

\subsection{Calculation using Landau approach}
\label{sec3.2}

The calculation of $\Sigma^{(4)}$ as in reference \cite{FW}, although systematic and rigorous, does not explicitly say which collisional processes between phonons lie behind each diagram in Figure \ref{fig8}. Here we develop a more heuristic method, inspired by reference \cite{LK} and valid to the leading order in $\epsilon$, which answers the question, gives the result in compact form (rather than spread into a dozen or so integral contributions) and makes it easy to show that the contribution of $H_4$ is subleading in the $\epsilon\to 0$ limit.

Reference \cite{LK} explains, for Hamiltonian (\ref{eq001}), how to calculate by an extended Fermi golden rule the damping rate of  phonon mode $\qq$ due to $n\to n'$ collisions, $n$ being the number of incoming $\qq_i$ phonons (they include one and only one phonon of wave-vector $\qq$) and $n'$ the number of outgoing $\qq'_j$ phonons (they do not include any phonon of wave-vector $\qq$). The ordinary cases of Beliaev and Landau processes correspond to $(n,n')=(1,2)$ and $(n,n')=(2,1)$, they have $n+n'=3$ phonons; here, it will be about four-phonon processes, like $(n,n')=(2,2)$. We start by writing kinetic equations for the phonon occupation numbers, provisionally assumed to be out of equilibrium, counting positively the direct process $n\to n'$ (which tends to empty mode $\qq$) and negatively the inverse process $n'\to n$ (which tends to fill it):  
\begin{multline}
\label{eq250}
\frac{\dd}{\mathrm{d}t} {n}_\mathbf{q}= -\frac{1}{(n-1)!}\int\left(\prod_{i=2}^{n}\frac{\dd^dq_i}{(2\pi)^d}\right)\frac{1}{n'!}\int\left(\prod_{j=1}^{n'}\frac{\dd^dq'_j}{(2\pi)^d}\right)\frac{2\pi}{\hbar} \mA^2_{\ii\to \ff} \delta(E_\ii-E_\ff) (2\pi)^d \delta\left(\sum_{i=1}^{n}\qq_i-\sum_{j=1}^{n'}\qq'_j\right)\\
\times\left[\prod_{i=1}^{n}n_{\qq_i}\prod_{j=1}^{n'}(1+n_{\qq'_j})-\prod_{i=1}^{n} (1+n_{\qq_i}) \prod_{j=1}^{n'}n_{\qq'_j}\right]
\end{multline}
 We have taken the thermodynamic limit, and divided by symmetry numbers taking into account the invariance of the process by permutation of the $n-1$ last incoming wave vectors (we agree that $\qq_1=\qq$) and by permutation of the $n'$ outgoing wave vectors. We recognize in (\ref{eq250}) the Dirac distribution of energy conservation, the factor $2\pi/\hbar$ and the square of a transition amplitude $\mA_{\ii\to\ff}$ between initial state $|\ii\rangle=|(\qq_i)_{1\leq i\leq n}\rangle$ and final state $|\ff\rangle=|(\qq'_j)_{1\leq j\leq n'}\rangle$ typical of the golden rule, as well as a Dirac distribution of momentum conservation and the bosonic amplification factors. The principle of microreversibility was used to factorize the squared amplitude of the direct process. 

We then take the excitation procedure (\ref{eq010}), where all phonon modes are at thermal equilibrium except $\qq$, hence the substitution $n_\kk\to \bar{n}_\kk$, $\kk\neq\qq$, and linearize (\ref{eq250}) around $n_\qq=\bar{n}_\qq$ for weak excitation, which gives the back-to-equilibrium equation $(\dd/\mathrm{d}t)(n_\qq-\bar{n}_\qq)|_{t=0^+}=-\Gamma_\qq^{n\to n'}(n_\qq-\bar{n}_\qq)$ with rate 
\begin{multline}
\label{eq251}
\Gamma_\qq^{n\to n'}=\frac{1}{(n-1)!}\int\left(\prod_{i=2}^{n}\frac{\dd^dq_i}{(2\pi)^d}\right)\frac{1}{n'!}\left(\int\prod_{j=1}^{n'}\frac{\dd^dq'_j}{(2\pi)^d}\right)\frac{2\pi}{\hbar} \mA^2_{\ii\to \ff} \delta(E_\ii-E_\ff) (2\pi)^d \delta\left(\sum_{i=1}^{n}\qq_i-\sum_{j=1}^{n'}\qq'_j\right) \\ {\yct\times} \left[\prod_{i=2}^{n}\bar{n}_{\qq_i}\prod_{j=1}^{n'}(1+\bar{n}_{\qq'_j})-\prod_{i=2}^{n} (1+\bar{n}_{\qq_i}) \prod_{j=1}^{n'}\bar{n}_{\qq'_j}\right]
\end{multline}
 where, unlike equation (\ref{eq250}), all products on the incoming wave vectors exclude mode $\qq$ and therefore start at $i=2$. The difference with the golden rule lies in the calculation of amplitude $\mA_{\ii\to\ff}$. Usually, we simply have $\mA_{\ii\to\ff}=\langle\ff|\mH_3|\ii\rangle$. Reference \cite{LK} deals precisely with the useful case of four-phonon collisions, where $\mA_{\ii\to\ff}$ is obtained by treating {\yc$H_3=\mathcal{H}_3/L^{d/2}$} at second order and {\yc$H_4=\mathcal{H}_4/L^d$} at first order of perturbation theory: 
\begin{equation}\label{eq252}
\mA_{\ii\to\ff}=\langle\ff|\mH_4|\ii\rangle + \sum_{|\lambda \rangle} \frac{\langle\ff|\mH_3|\lambda\rangle\langle\lambda|\mH_3|\ii\rangle}{E_\ii-E_\lambda}
\end{equation} and the sum is taken over all possible intermediate phonon Fock states $|\lambda\rangle$.  A simple power counting as in section \ref{sec0} shows that $\mH_4$'s contribution to (\ref{eq252}) is subleading in the small-angle limit $\epsilon\to 0$;\footnote{In view of equations (\ref{eq052},\ref{eq104}), the contribution of $\mH_4$ to (\ref{eq252}) is of order $(mc^2/\rho)\epsilon^2$; in view of equations (\ref{eq025},\ref{eq027}), that of $\mH_3$ is of order $(mc^2/\rho)\epsilon^0$.} so we omit it in the following. In the energy denominator, we have chosen the reference energy $E_\ii$ of the initial state rather than $E_\ff$ of the final state, because we are talking about a $\ii\to\ff$ transition, as recalled by the choice of $\ff$ in bra and $\ii$ in ket in the matrix element in the numerator, which is not a serious physical reason. Fortunately, this doesn't matter, since the amplitude is used in (\ref{eq250}) on the energy shell ($E_\ii=E_\ff$): the result is invariant by exchanging the roles of $\ii$ and $\ff$ in any term of the sum in the right-hand side of (\ref{eq252}): \begin{equation}\label{eq253}
\frac{\langle\ff|\mH_3|\lambda\rangle\langle\lambda|\mH_3|\ii\rangle}{E_\ii-E_\lambda} \mbox{(direct direction)} \longleftrightarrow \frac{\langle\ii|\mH_3|\lambda\rangle\langle\lambda|\mH_3|\ff\rangle}{E_\ff-E_\lambda} \mbox{(inverse direction)}
\end{equation} The total damping rate is obtained by summing $\Gamma_\qq^{n\to n'}$ over the collisional processes $n\to n'$ relevant to the order under consideration. From the following obvious rules, (i) the diagram giving the transition amplitude $\mA_{\ii\to\ff}$ is connected, i.e. there is no possible decoupling of the $n\to n'$ process into two independent collisional processes, (ii) any intermediate state (between successive actions of $\mH_3$) must differ from the initial and final states, we find at order four in $H_3^{(+)}+H_3^{(-)}$ as the only possibilities the processes $1\to 3$, $2\to 2$ and $3\to 1$: \footnote{In reference \cite{LK}, {\yc unlike in our case,} the acoustic branch has a concave start {\yc($\gamma<0$ in equation (\ref{eq003}))}: only process $2\to 2$ preserves momentum-energy and gives a non-zero contribution to (\ref{eq254}); however, it was miscalculated, and the error has been corrected in reference \cite{Annalen}.} 
\begin{equation}\label{eq254}
\Gamma^{(4)}_\qq=\Gamma_\qq^{1\to 3}+\Gamma_\qq^{2\to 2}+\Gamma_\qq^{3\to 1}
\end{equation}
 Heuristically, we generalize this formalism to the calculation of the fourth-order contribution in $H_3^{(+)}+H_3^{(-)}$ to the self-energy of mode $\qq$. As with the damping rate, we assume that it is the sum of the contributions of the above-mentioned $n\to n'$ collisional processes: 
\begin{equation}\label{eq255}
\Sigma^{(4)}(\qq,z)=\Sigma_{1\to 3}(\qq,z)+\Sigma_{2\to 2}(\qq,z)+\Sigma_{3\to 1}(\qq,z)
\end{equation} Knowing that the energy-conserving Dirac distribution in (\ref{eq251}) ultimately comes from the identity
\begin{equation}\label{eq256}
\lim_{\delta\to 0^+} \im \frac{1}{E_\ii+\ii\delta-E_\ff}=-\pi\delta(E_\ii-E_\ff)\quad,
\end{equation} 
that the displaced variable $\zeta=z-\veps_\qq$ plays the role of $\ii\delta$ in the pole approximation (\ref{eq014}) {\yc and a self-energy differs from a rate by a factor of $-\ii\hbar/2$}, we arrive without hesitation at the form 
\begin{multline}
\label{eq257}
\Sigma_{n\to n'}^{(4)}(\qq,z)=\frac{1}{(n-1)!} \int\left(\prod_{i=2}^{n}\frac{\dd^dq_i}{(2\pi)^d}\right)\frac{1}{n'!} \int\left(\prod_{j=1}^{n'} \frac{\dd^dq'_j}{(2\pi)^d}\right) \frac{\mathcal{A}_{\ii\to\ff}^2(\qq,z)}{\zeta+E_\ii-E_\ff}(2\pi)^d \delta\left(\sum_{i=1}^{n}\qq_i-\sum_{j=1}^{n'}\qq'_j\right) \\
\times\left[\prod_{i=2}^{n}\bar{n}_{\qq_i}\prod_{j=1}^{n'}(1+\bar{n}_{\qq'_j})-\prod_{i=2}^{n} (1+\bar{n}_{\qq_i}) \prod_{j=1}^{n'}\bar{n}_{\qq'_j}\right]
\end{multline}
 The only pitfall lies in the construction of the $\zeta$ dependence of the {\yc$\mA_{\ii\to\ff}$} transition amplitude. The choice of $E_\ii$ or $E_\ff$ as reference energy is no longer irrelevant. Choosing $E_\ii$ as in the left-hand side of (\ref{eq253}) means taking the point of view of a $\ii\to\ff$ transition, i.e. travelling through time in the ordinary direction; the correct integration path is precisely $C_+$, the one spanned by variable $\zeta$ in (\ref{eq012}); $E_\ii$ must therefore be replaced by $\zeta+E_\ii$ in the energy denominator internal to {\yc$\mA_{\ii\to\ff}$}. On the other hand, choosing $E_\ff$ as in the right-hand side of (\ref{eq253}) means taking the opposite point of view of a $\ff\to\ii$ transition, i.e. going backward in time; the correct integration path for {\yct the} Green's function is then $C_-$ \cite{CCTbordeaux}, which runs in the lower half-plane parallel to the real axis from $-\infty$ to $+\infty$ and is spanned by variable $-\zeta$ in (\ref{eq012}); $E_\ff$ must therefore be replaced by $-\zeta+E_\ff$ in the energy denominator internal to {\yc$\mA_{\ii\to\ff}$}. 

To decide according to the intermediate state $|\lambda\rangle$, we need to rely on the following property. By construction, function $\Sigma^{(4)}(\qq,z)$ depends on the mode through its wave vector $\qq$ but not its energy $\veps_\qq$ (when using the original complex energy $z$ as a variable), and therefore cannot have an energy denominator containing $\veps_\qq$. The opposite is true when using the displaced variable $\zeta$. This means that, if in the direct direction of the transition amplitude, the action on $|\ii\rangle$ of operator $\mH_3^{(\pm)}$ does not cause the phonon of wave vector $\qq$ to disappear, {\ycd its energy $\veps_\qq$ appears in $E_\lambda$ and therefore disappears in $\zeta+E_\ii-E_\lambda$}, and reverse direction of the transition amplitude must be used, this rule being applied intermediate state by intermediate state. Hence our final prescription on the transition amplitude to be included in (\ref{eq257}):\footnote{{\ycd In reality, the collisional processes $n\to n'$ with $n$ incoming phonons $\qq_i$ and $n'$ outgoing phonons $\qq'_j$ occur in a thermal bath, and the modes of the virtual phonons $\kk$ (neither incoming nor outgoing) are not initially empty. The corresponding bosonic amplification factors $(n_\kk+1)^{1/2}$ and $n_\kk^{1/2}$ would therefore have to be taken into account in the expression (\ref{eq258}) of the transition amplitude, and the contribution of the new allowed processes, in which a virtual phonon $\kk$ is annihilated before being recreated (which is forbidden in vacuum), would have to be included. However, these two modifications exactly cancel each other and the expression (\ref{eq258}) remains unchanged. Let's check this on a single term of the first sum in (\ref{eq258}). To this aim, let's describe in more detail the corresponding transition $|\ii\rangle\to|\lambda\rangle\to|\ff\rangle$ by noting $|\ii\rangle=|\ii_1\rangle\otimes|\ii_2\rangle\otimes|n_\kk:\kk\rangle$ and $|\ff\rangle=|\ff_1\rangle\otimes|\ff_2\rangle\otimes|n_\kk: \kk\rangle$, where $|\ii_1\rangle$ ($|\ii_2\rangle$) are the incoming photons annihilated by the first (second) action of $\mathcal{H}_3$, $|\ff_1\rangle$ ($|\ff_2\rangle$) are the outgoing phonons created by the first (second) action of $\mathcal{H}_3$ and $|n_\kk: \kk\rangle$ the Fock state with $n_\kk$ quanta in the virtual phonon mode, so the intermediate state is written $|\lambda\rangle=|\ff_1\rangle\otimes|\ii_2\rangle\otimes|n_\kk+1:\kk\rangle$
(we don't take into account the other modes, spectators, nor the thermal occupation of incoming and outgoing modes, since their bosonic factors are included in (\ref{eq250},\ref{eq257})). In this example, $\qq\notin |\lambda\rangle$ hence $\qq\in|\ii_1\rangle$.
The components of $\mathcal{H}_3$ ensuring the transitions $|\ii\rangle\to|\lambda\rangle$ and $|\lambda\rangle\to|\ff\rangle$ are written respectively $W_1=\mathcal{A}_1\hat{b}^\dagger_\kk(\prod_{\qq'_j\in|\ff_1\rangle} \hat{b}^\dagger_{\qq'_j})(\prod_{\qq_i\in|\ii_1\rangle} \hat{b}_{\qq_i})$ and $W_2=\mathcal{A}_2(\prod_{\qq'_j\in|\ff_2\rangle} \hat{b}^\dagger_{\qq'_j})(\prod_{\qq_i\in|\ii_2\rangle} \hat{b}_{\qq_i})\hat{b}_\kk$.
In the presence of the thermal bath, (i) the numerator $\langle\ff|\mH_3^{(\pm)}|\lambda\rangle\langle\lambda|\mH_3^{(\pm)}|\ii\rangle=\langle\ff|W_2|\lambda\rangle\langle\lambda|W_1|\ii\rangle$ becomes $\mathcal{A}_1\mathcal{A}_2(1+n_\kk)$, (ii) the new associated process, corresponding to the inverse order of action of $W_1$ and $W_2$, has numerator $\langle\ff|W_1|\lambda'\rangle\langle\lambda'|W_2|\ii \rangle=\mathcal{A}_1\mathcal{A}_2n _\kk$, (iii) the new intermediate state $|\lambda'\rangle=|\ii_1\rangle\otimes|\ff_2\rangle\otimes|n_\kk-1: \kk\rangle$ contains the phonon $\qq$ so its energy denominator must be taken from the second sum of (\ref{eq258}), and happens to be the exact opposite of the denominator of the former process, the calculation giving $-\zeta+E_\ff-E_{\lambda'}=-(\zeta+E_\ii-E_\lambda)$. Ultimately, the linear contributions in $n_\kk$ of the two processes exactly cancel each other in the transition amplitude $\mathcal{A}_{\ii\to\ff}(\qq,z)$, which allows us to forget the thermal background as is done in (\ref{eq258}). Similar reasoning, but limited to $2\to 2$ processes on the energy shell, can already be found in reference \cite{Annalen}.
}}
\begin{equation}\label{eq258}
\boxed{\mathcal{A}_{\ii\to\ff}(\qq,z) = \sum_{|\lambda\rangle\, / \, \qq\notin|\lambda\rangle} \frac{\langle \ff| \mH_3^{(\pm)} |\lambda\rangle \langle\lambda| \mH_3^{(\pm)}|\ii\rangle}{\zeta+E_\ii-E_\lambda} 
+
\sum_{|\lambda\rangle\, / \, \qq\in|\lambda\rangle} \frac{\langle \ii| \mH_3^{(\pm)} |\lambda\rangle \langle\lambda| \mH_3^{(\pm)}|\ff\rangle}{-\zeta+E_\ff-E_\lambda}} 
\end{equation} From expression (\ref{eq258}) we derive the transition amplitude for process $1\to 3$ (knowing that $\qq'_1+\qq'_2+\qq'_3=\qq$): 
\begin{multline}
\label{eq260}
\mathcal{A}_{\qq\to\qq'_1,\qq'_2,\qq'_3}(\qq,\zeta)= \frac{\mA(\qq'_1,\qq'_2;\qq'_1+\qq'_2)\mA(\qq'_1+\qq'_2,\qq'_3;\qq)}{\zeta+\epsilon_\qq-(\epsilon_{\qq'_1+\qq'_2}+\epsilon_{\qq'_3})} +\frac{\mA(\qq'_2,\qq'_3;\qq'_2+\qq'_3)\mA(\qq'_2+\qq'_3,\qq'_1;\qq)}{\zeta+\epsilon_\qq-(\epsilon_{\qq'_2+\qq'_3}+\epsilon_{\qq'_1})} \\ +\frac{\mA(\qq'_1,\qq'_3;\qq'_1+\qq'_3)\mA(\qq'_1+\qq'_3,\qq'_2;\qq)}{\zeta+\epsilon_\qq-(\epsilon_{\qq'_1+\qq'_3}+\epsilon_{\qq'_2})}
\end{multline}
 then for process $2\to 2$ (knowing that $\qq'_1+\qq'_2=\qq+\qq_2$): 
\begin{multline}
\label{eq262}
\mathcal{A}_{\qq,\qq_2\to\qq'_1,\qq'_2}(\qq,\zeta)= \frac{\mA(\qq'_1,\qq'_2;\qq'_1+\qq'_2)\mA(\qq,\qq_2;\qq+\qq_2)}{\zeta+\epsilon_{\qq}+\epsilon_{\qq_2}-\epsilon_{\qq+\qq_2}} -\frac{\mA(\qq'_1-\qq,\qq'_2;\qq_2)\mA(\qq'_1-\qq,\qq;\qq'_1)}{\zeta+\epsilon_{\qq}+\epsilon_{\qq'_1-\qq}-\epsilon_{\qq'_1}} \\ +\frac{\mA(\qq'_2-\qq_2,\qq_2;\qq'_2)\mA(\qq'_2-\qq_2,\qq'_1;\qq)}{\zeta+\epsilon_{\qq}-(\epsilon_{\qq'_1}+\epsilon_{\qq'_2-\qq_2})} +\frac{\mA(\qq-\qq'_2,\qq_2;\qq'_1)\mA(\qq-\qq'_2,\qq'_2;\qq)}{\zeta+\epsilon_{\qq}-(\epsilon_{\qq'_2}+\epsilon_{\qq-\qq'_2})}-\frac{\mA(\qq'_2-\qq,\qq'_1;\qq_2)\mA(\qq'_2-\qq,\qq;\qq'_2)}{\zeta+\epsilon_{\qq}+\epsilon_{\qq'_2-\qq}-\epsilon_{\qq'_2}}
\end{multline}
 and finally for process $3\to 1$ (knowing that $\qq_1'=\qq+\qq_2+\qq_3$): 
\begin{multline}
\label{eq264}
\mathcal{A}_{\qq,\qq_2,\qq_3\to \qq_1'}(\qq,\zeta)= -\frac{\mA(\qq_2,\qq_3;\qq_2+\qq_3)\mA(\qq_2+\qq_3,\qq;\qq_1')}{\zeta+\epsilon_{\qq}+\epsilon_{\qq_2+\qq_3}-\epsilon_{\qq_1'}}+\frac{\mA(\qq+\qq_2,\qq_3;\qq_1')\mA(\qq,\qq_2;\qq+\qq_2)}{\zeta+\epsilon_{\qq}+\epsilon_{\qq_2}-\epsilon_{\qq+\qq_2}}\\ +\frac{\mA(\qq+\qq_3,\qq_2;\qq_1')\mA(\qq,\qq_3;\qq+\qq_3)}{\zeta+\epsilon_{\qq}+\epsilon_{\qq_3}-\epsilon_{\qq+\qq_3}}
\end{multline}
 In these equations (\ref{eq260},\ref{eq262},\ref{eq264}), the three contributions where we had to choose the opposite direction in (\ref{eq253}), i.e. take $E_\ff$ rather than $E_\ii$ as the reference energy, are easy to spot as they start with a minus sign. 

Our prescription (\ref{eq257},{\yc\ref{eq258}}) gives simple but generally different results from the exact fourth-order diagrammatic theory, given the expressions for Matsubara double sums in \ref{app3}. Remarkably, however, once these sums have been replaced by their simplified expressions (\ref{eq213},\ref{eq233}), which is legitimate at small angles, it is in perfect agreement with the value of $\Sigma^{(4)}$ predicted by the diagrams in Figure \ref{fig8}; this can be verified with the aid of the correspondence table \ref{table1}, obtained from the coupling amplitudes in (\ref{eq212},\ref{eq232}) and the energy denominators in (\ref{eq213},\ref{eq233}), and which makes it possible to associate with each internal line on this figure an incoming $\qq_i$, outgoing $\qq'_j$ or virtual (linear combination of the previous ones) phonon of a $n\to n'$ process, thus giving a clear physical meaning to the corresponding diagram.\footnote{{\ycd The verification is facilitated by the following result: according to our Landau prescription, each diagrammatic contribution to the Matsubara sums (\ref{eq213},\ref{eq233}) is given by $\sigma/\prod_j (\zeta-\zeta_j)$ with $\sigma=s(f_1 f_2-f_3f_4)$, where (i) the global sign $s$ is the product of the orientations $s_i$ of the virtual phonon lines (indeed, if a virtual phonon goes back in time, its energy appears with a plus sign in the denominator in the corresponding $(\zeta-\zeta_j)$ factor, so it contributes through the second sum of (\ref{eq258}) and a global minus sign must be included), (ii) each $f_k$ is the product of the bosonic amplification factors of the lines connected to the vertex $k$ (the line $i$ contributes to $f_k$ by a factor $\bar{n}_{\veps_i}$ if it goes towards the vertex, or by a factor $1+\bar{n}_{\veps_i}$ if it goes away from it, which is written $s_i(1+\bar{n}_{s_i\veps_i})$ if the line is to the right of the vertex and $s_i\bar{n}_{s_i\veps_i}$ if it is to the left), excluding, however, external lines and virtual phonon lines, whose occupation numbers do not appear in the prescription (\ref{eq257},\ref{eq258}). Diagram vertices are numbered from left to right; in the case of bridge diagrams, the degeneracy of internal vertex numbering is lifted by rotating the vertical phonon line by a small angle to the left (type I, effective orientation $s_5$) or to the right (type II, effective orientation $-s_5$). One thus finds, all signs factorized, $\sigma_B=s_1s_2s_3s_4s_5[(1+\bar{n}_{s_1\veps_1})(1+\bar{n}_{s_3\veps_3})(1+\bar{n}_{s_4\veps_4})- \bar{n}_{s_1\veps_1}\bar{n}_{s_3\veps_3}\bar{n}_{s_4\veps_4}]$ for inner-loop diagrams ($s=s_2s_5$), $\sigma_{P^{\rm I}}=s_1s_2s_3s_4s_5[(1+\bar{n}_{s_2\veps_2})(1+\bar{n}_{s_3\veps_3})(1+\bar{n}_{s_5\veps_5})-\bar{n}_{s_2\veps_2} \bar{n}_{s_3\veps_3} \bar{n}_{s_5\veps_5}]$ and $\sigma_{P^{\rm II}}=s_1s_2s_3s_4(-s_5)[(1+\bar{n}_{s_1\veps_1})(1+\bar{n}_{s_4\veps_4})(1+\bar{n}_{-s_5\veps_5})-\bar{n}_{s_1\veps_1} \bar{n}_{s_4\veps_4} \bar{n}_{-s_5\veps_5}]$ for bridge diagrams ($s=s_1s_4$ for type I, $s=s_2s_3$ for type II). We then find exactly the simplified forms (\ref{eq213},\ref{eq233}) if we use energy conservation at the second vertex, respectively $s_2\veps_2=s_3\veps_3+s_4\veps_4$, $s_1\veps_1=s_3\veps_3+s_5\veps_5$ and $s_2\veps_2=s_4\veps_4+(-s_5)\veps_5$, through the identity $\bar{n}_{a+b}=\bar{n}_a\bar{n}_b/(1+\bar{n}_a+\bar{n}_b)$, and the equality $s_2=s_5$ in the inner-loop diagrams.}} Ultimately, the prescription (\ref{eq257},{\yc\ref{eq258}}) leads in $d=2$ to a rigorously equivalent but much more compact expression of $\Sigma^{(4)}$ at leading order in $\epsilon$: 
\begin{equation}\label{eq269}
\boxed{
\tilde{\Sigma}_{\qb}^{(4,2)}(\tilde{\zeta})=\frac{1}{3}\int\!\!\!\!\int_{\mathbb{T}_{\qb}} \frac{\qb'_1\mathrm{d}\qb'_1}{(2\pi)^2} \frac{\qb'_2\mathrm{d}\qb'_2}{(2\pi)^2} \Phi(\qb'_1,\qb'_2) + \int\!\!\!\!\int_{\mathbb{R}^{+2}\setminus\mathbb{T}_{\qb}} \frac{\qb'_1\mathrm{d}\qb'_1}{(2\pi)^2} \frac{\qb'_2\mathrm{d}\qb'_2}{(2\pi)^2} \Phi(\qb'_1,\qb'_2) + \int\!\!\!\!\int_{\mathbb{R}^{+2}}\frac{\qb_2\mathrm{d}\qb_2}{(2\pi)^2}\frac{\qb_3\mathrm{d}\qb_3}{(2\pi)^2} \Phi(-\qb_2,-\qb_3)}
\end{equation} where we have kept the order of processes $1\to 3$, $2\to 2$ and $3\to 1$, $\mathbb{T}_{\qb}$ is the triangle $\{(\qb'_1,\qb'_2)\in\mathbb{R}^{+2}\;|\; \qb'_1+\qb'_2\leq \qb\}$ and, as before, the tilde on $\Sigma_{\qb}$ indicates a division by $k_B T\epsilon^2$. One and the same function is involved in the integrand of the $n\to n'$ contributions, only the integration domain changes with the process under consideration: \footnote{As far as the bosonic factors are concerned, this is shown by making use of the relation $\bar{n}_{-k}^{\rm lin}=-(1+\bar{n}_k^{\rm lin})$, $\forall k\in\mathbb{R}^*$.} 
\begin{multline}
\label{eq270}
\Phi(\qb'_1,\qb'_2)=\frac{1}{2} \bar{q}\left(\frac{9(1+\Lambda)^2}{8\rho\xi^2}\right)^2 \qb'_1 \qb'_2 (\qb'_1+\qb'_2-\qb) [(1+\bar{n}^{\rm lin}_{q'_1}+\bar{n}^{\rm lin}_{q'_2})\bar{n}^{\rm lin}_{q'_1+q'_2-q}-\bar{n}^{\rm lin}_{q'_1}\bar{n}^{\rm lin}_{q'_2}] \\ \times \int_{\mathbb{R}}\dd\thetat_1\int_{\mathbb{R}}\dd\thetat_2\frac{\left\{\frac{\qb'_1+\qb'_2}{\zetat+\qb(\qb'_1+\qb'_2)(\qb'_1+\qb'_2-\qb)\left[\frac{(\qb'_1\thetat_1+\qb'_2\thetat_2)^2}{2(\qb'_1+\qb'_2-\qb)^2(\qb'_1+\qb'_2)^2}-\frac{3\gamma}{8}\right]}+\frac{\qb-\qb'_1}{\zetat+\bar{q}\qb'_1(\qb-\qb'_1)\left[-\frac{\thetat_1^2}{2(\qb'_1-\qb)^2}+\frac{3\gamma}{8}\right]}+\frac{\qb-\qb'_2}{\zetat+\bar{q}\qb'_2(\qb-\qb'_2)\left[-\frac{\thetat_2^2}{2(\qb'_2-\qb)^2}+\frac{3\gamma}{8}\right]}\right\}^2}{\tilde{\zeta}- \frac{\qb'_1(\qb'_2-\qb)\thetat_1^2+\qb'_2(\qb'_1-\qb)\thetat_2^2-2\qb'_1 \qb'_2\thetat_1 \thetat_2}{2(\qb'_1+\qb'_2-\qb)}+\frac{3\gamma}{8} (\qb'_1+\qb'_2)(\qb'_1-\qb)(\qb'_2-\qb)}
\end{multline}

 \begin{table}[t]
\centerline{\small
\begin{tabular}{|c|c|c|c|c|c|}
$B_1$ & all the squares of $1\to 3$ & $P_{1}^{\rm I}$  &  all crossed terms of $1\to 3$ & $P_{1}^{\rm II}$ & $3^{\rm rd}$ term $\times$ $4^{\rm th}$ term of $2\to 2$\\
$B_2$ & squares of $3^{\rm rd}$ and $4^{\rm th}$ terms of $2\to 2$ & $P_{2}^{\rm I}$ & $1^{\rm st}$ term $\times$ $3^{\rm rd}$ term of $2\to 2$ & $P_{2}^{\rm II}$ & $3^{\rm rd}$ term $\times$ $5^{\rm th}$ term of $2\to 2$\\
$B_3$ & $2^{\rm nd}$ term $\times$ $3^{\rm rd}$ term of $2\to 2$ & $P_{3}^{\rm I}$ & $1^{\rm st}$ term $\times$ $4^{\rm th}$ term of $2\to 2$ & $P_{3}^{\rm II}$ & $2^{\rm nd}$ term $\times$ $4^{\rm th}$ term of $2\to 2$\\
$B_4$ & $4^{\rm th}$ term $\times$ $5^{\rm th}$ term of $2\to 2$ & $P_{4}^{\rm I}$ & $1^{\rm st}$ term $\times$ $2^{\rm nd}$ term of $2\to 2$ & $P_{4}^{\rm II}$ & $1^{\rm st}$ term $\times$ $3^{\rm rd}$ term of $3\to 1$\\
$B_5$ & squares of $2^{\rm nd}$ and $5^{\rm th}$ terms of $2\to 2$ & $P_{5}^{\rm I}$ & $2^{\rm nd}$ term $\times$ $3^{\rm rd}$ term of $3\to 1$ & $P_{5}^{\rm II}$ & $2^{\rm nd}$ term $\times$ $5^{\rm th}$ term of $2\to 2$\\
$B_6$ & square of $1^{\rm st}$ term of $3\to 1$ & $P_{6}^{\rm I}$ & $1^{\rm st}$ term $\times$ $5^{\rm th}$ term of $2\to 2$ & $P_{6}^{\rm II}$ & $1^{\rm st}$ term $\times$ $2^{\rm nd}$ term of $3\to 1$\\
$B_7$ & square of $1^{\rm st}$ term of $2\to 2$ & & & & \\
$B_8$ & squares of $2^{\rm nd}$ and $3^{\rm rd}$ terms of $3\to 1$ & & & & 
\end{tabular}
}
\caption{In the calculation of the self-energy at fourth-order in $H_3$, correspondence between the Feynman diagrams in Figure \ref{fig8} and the square or crossed terms (double products) of the quantities $\mA^2_{\ii\to\ff}(\qq,z)$ involved in Landau prescription (\ref{eq257}); the various terms of the $\mA_{\ii\to\ff}(\qq,z)$ transition amplitudes referred to are those of equations (\ref{eq260},\ref{eq262},\ref{eq264}) written in that order. For bridge diagrams, it was necessary to distinguish between type I and type II, as in (\ref{eq233}). The correspondence holds after integration on the wave vectors as in (\ref{eq257}), which makes, for example, the squares of each of the terms of (\ref{eq260}), which differ by a permutation of $\qq_j'$, or the double products of $2^{\rm nd}$ and $4^{\rm th}$ terms and $5^{\rm th}$ and $3^{\rm rd}$ terms of (\ref{eq262}), which are mapped one onto the other by exchange of $\qq_1'$ and $\qq_2'$, give the same contribution.}
\label{table1}
\end{table}
 
\subsection{Failure of order four in the limit $\epsilon=k_B T/mc^2\to 0$}
\label{sec3.3}

\paragraph{An unexpected divergence}
 As a first simple application of the expansion of the self-energy to order four in $H_3$, we had decided to calculate {\yct $\tilde{\Sigma}_{\qb}^{(4,2)}(\textrm{i}0^+)$} numerically\footnote{The angular integrals are calculated analytically, only the integration over the wavenumbers is done by computer.} to see if, through a better approximation of $\Sigma_{\qq}(\textrm{i}0^+)$ in equation (\ref{eq014}), this would bring the pole approximation closer to the numerical results by reducing the deviation in Figure \ref{fig2}d. To our surprise, we discovered that $\tilde{\Sigma}_{\yc\qb}^{(4,2)}(\ii\delta)$ diverges as $1/\ii\delta$ when $\delta\to 0^+$! We then confirmed this result analytically, showing that\footnote{To facilitate the calculation of angular integrals, we have established (\ref{eq280}) first in the case where $\zetat\to 0$ by positive imaginary parts, then extended to the case $\im\zetat<0$ as in footnote \ref{notegen}.} 
\begin{equation}\label{eq280}
\boxed{\tilde{\Sigma}_{\qb}^{(4,2)}(\zetat)\underset{\zetat\to 0}{\sim} \frac{-2}{3\pi^2\gamma\zetat} \left(\frac{9(1+\Lambda)^2}{8\rho\xi^2}\right)^2 \int_{\qb}^{\eta} \kb\dd\bar{k}\int_0^{\eta-\kb} \kb'\dd\kb' (\kb-\qb)(\qb+\kb')(\bar{n}_{k-q}^{\rm lin}-\bar{n}_{k}^{\rm lin})(\bar{n}_{k'+q}^{\rm lin}-\bar{n}_{k'+k}^{\rm lin})}
\end{equation} This divergence comes from the angular integral $K(\kb,\kb')$ (\ref{eq243}) of bridge diagram $P_5$. Let's analyze the behavior of (\ref{eq243}) in the $\zetat\to 0$ limit for any orientation $s_j$ of the internal lines. If it is divergent, it is necessarily dominated by a neighborhood of the cancellation locus of the integrand denominator in the $(\thetat,\thetat')$ plane for $\zetat=0$. As the curvature parameter $\gamma$ is $>0$, this locus is the union of the two vertical lines $\thetat=\veps(-A_1^P/B_1^P)^{1/2}$ ($\veps=\pm 1$) where the first factor in the denominator vanishes, the two horizontal lines $\thetat'=\veps'(-A_2^P/B_2^P)^{1/2}$ ($\veps'=\pm 1$) where the second factor vanishes, and the conic making the third factor vanish. The unfortunate thing is that the conic always passes through the intersection points $P_{\veps,\veps'}\equiv(\veps(-A_1^P/B_1^P)^{1/2},\veps'(-A_2^P/B_2^P)^{1/2})$ of the horizontal and vertical lines, whatever the wave numbers $\kb$, $\kb'$ and $\qb$, on the sole condition that $\veps\veps'=s_2 s_4$. This follows from the identity on the coefficients (\ref{eq244},\ref{eq245}): 
\begin{equation}\label{eq281}
4A_1^P A_2^P B_1^P B_2^P (B_3^P)^2 = [A_2^P B_1^P(B_2^P+B_3^P)+A_1^P B_2^P B_3^P-A_3^P B_1^P B_2^P]^2
\end{equation} Since the angular integral is two-dimensional, a possible divergent part $K_{\rm div}(\kb,\kb')$ of $K(\kb,\kb')$ can only come from a small neighborhood of the triple intersection points $P_{s_2,s_4}$ and $P_{-s_2,-s_4}$, where we replace each factor in the denominator by its linear approximation in $\tau=\thetat-\veps(-A_1^P/B_1^P)^{1/2}$ and $\tau'=\thetat'-\veps'(-A_2^P/B_2^P)^{1/2}$. This leaves 
\begin{equation}\label{eq282}
K_{\rm div}(\kb,\kb')=\int_{\mathbb{R}^2} \frac{2\mathrm{d}\tau\mathrm{d}\tau'}{[\zetat-(\frac{3\gamma}{4})^{1/2}s_1\qb\bar{k}\tau][\zetat-(\frac{3\gamma}{4})^{1/2}s_3\qb\kb' \tau']\{\zetat-(\frac{3\gamma}{4})^{1/2}s_3\kb'[s_1\bar{k}\tau +(\qb-s_1\kb) \tau']\}}=- \frac{8\pi^2(1-s_2)(1-s_1s_3)}{3\gamma\tilde{\zeta}\qb\kb\kb' (s_1\kb-s_3\kb')}
\end{equation} where we have been able to extend the integral to the whole $\mathbb{R}^2$ since it converges to infinity, and have been able to restrict ourselves to the point $P_{s_2,s_4}$ (at the cost of adding a global factor $2$) by parity invariance $(\thetat,\thetat')\to (-\thetat,-\thetat')$. In the right-hand side of (\ref{eq282}), we have explicitly calculated the integral over $\tau'$ and then over $\tau$ by the residue method (the integration path must be closed by a semicircle at infinity in the upper or lower half-plane containing the minimum number of poles in view of constraints (\ref{eq239})). The divergent part is therefore non-zero only if $s_2=s_1s_3=-1$, i.e. if the lower left branch of the bridge loop is oriented to the left and the two upper branches have opposite orientations, which selects as announced diagram $P_5$ in Figure \ref{fig8}. {\yct Inserting} the corresponding value of $K_{\rm div}(\kb,\kb')$ {\yct into} (\ref{eq242}) yields equation (\ref{eq280}). The same technique applied to type I contribution of bridge diagrams leads in (\ref{eq241}) to {\yc a divergent part of the form}\footnote{In fact, the coefficients also satisfy the relationship obtained by exchanging indices $1$ and $2$ and replacing index $3$ by index $4$ in (\ref{eq281}).}\footnote{In the case of inner-loop diagrams, the existence of a triple nodal point is less surprising since the denominator of (\ref{eq221}) comprises a squared trinomial. In fact, there are four of them, playing equivalent roles. For the sign choices $\veps=s_1 s_2$ and $\veps'=s_2 s_3 s_4$, the coefficients of $\tau$ and $\tau'$ in the linearized forms are all negative; setting them to $-1$ and taking $\zetat=\ii\delta$ to simplify, we get $\int_{[-1,1]^2} 4\mathrm{d}\tau\mathrm{d}\tau'/[(\ii\delta-\tau)^2 (\ii\delta -\tau-\tau')]$, which is $O(1)$ when $\delta\to 0^+$.} 
\begin{equation}\label{eq283}
J_{\rm div}(\kb,\kb')=\int_{\mathbb{R}^2} \frac{2\mathrm{d}\tau\mathrm{d}\tau'}{[\zetat-(\frac{3\gamma}{4})^{1/2}s_1\qb\bar{k}\tau][\zetat-(\frac{3\gamma}{4})^{1/2}s_3\qb\kb' \tau']\{\zetat-(\frac{3\gamma}{4})^{1/2}s_1\kb[(\qb-s_3\kb')\tau +s_3\kb'\tau']\}}=\frac{8\pi^2(1-s_1)(1-s_2s_4)}{3\gamma\tilde{\zeta}\qb\kb\kb'(s_1\kb-s_3\kb')}
\end{equation} and concludes that there is no divergence ({\yc to have $J_{\rm div}\neq 0$,} the upper left branch of the bridge loop would have to be oriented to the left, and the two lower branches would have to be of opposite orientations, which does not occur in Figure \ref{fig8}).

\paragraph{Physical implications}
 From the expansion of the self-energy to order four in $H_3$, and still to leading order (two) in $\epsilon$, we form a new approximation to {\yct the} Green's function, 
\begin{equation}\label{eq289}
\tilde{\mathcal{G}}_{\qb}^{(\leq 4,2)}(\zetat)=\frac{1}{\zetat-\tilde{\Sigma}_{\qb}^{(2,2)}(\zetat)-\tilde{\Sigma}_{\qb}^{(4,2)}(\zetat)}
\end{equation} One might think that it is more accurate and performs better than the one in section \ref{sec2}. Unfortunately, because of the divergence (\ref{eq280}), this is not the case: in the weakly interacting regime, which is the most favorable, it has a pole $\zetat^{n\phi}_{\qb}$ in the upper half-plane (and its complex conjugate in the lower half-plane), in the absence, it should be pointed out, of any analytic continuation, hence the exponent $n\phi$ (\g{non physical}). {\yc This} violates the fundamental analyticity property of {\yct the} Green's function on $\mathbb{C}\setminus\mathbb{R}$ \cite{FW}, and leads, for example, to a signal diverging exponentially in time in (\ref{eq012}) if path $C_+$ passes above $\zetat^{n\phi}_{\qb}$, or not starting at one if it passes below. To show the existence of this pole, let's reduce the implicit equation 
\begin{equation}\label{eq290}
\zetat_{\qb}^{n\phi} = \tilde{\Sigma}_{\qb}^{(2,2)}(\zetat_{\qb}^{n\phi})+\tilde{\Sigma}_{\qb}^{(4,2)}(\zetat_{\qb}^{n\phi})
\end{equation} to its leading order, assuming that $\zetat_{\qb}^{n\phi}$ tends to zero as $(\rho\xi^2)^{-1}$ (by positive imaginary parts). Since $\tilde{\Sigma}_{\qb}^{(2,2)}(\zetat)$ is already of order one in $(\rho\xi^2)^{-1}$, and is a derivable function of $\zetat$ at $\textrm{i}0^+$ (see \ref{app1}), we can replace it by its value at this point. In the case of $\tilde{\Sigma}_{\qb}^{(4,2)}(\zetat)$, we keep its divergent part (\ref{eq280}), as the second-order prefactor $(\rho\xi^2)^{-2}$ it contains is promoted to first order by division by $\zetat$, but we neglect its regular part, of second-order near $\textrm{i}0^+$. By making the universalizing change of variables (\ref{eq112},\ref{eq130},\ref{eq131}) {\yct as in} section \ref{sec2}, we are left with 
\begin{equation}\label{eq291}
\zetac^{n\phi}_{\qb} = \check{\Sigma}_{\qb}^{(2,2)}(\textrm{i}0^+)-\frac{(2/\qb^{2})\int_{\qb}^{\eta}\kb\,\dd\bar{k}\int_0^{\eta-\kb}\kb'\dd\kb'(\kb-\qb)(\kb'+\qb){\yc(\bar{n}_{k-q}^{\rm lin}-\bar{n}_{k}^{\rm lin})(\bar{n}_{k'+q}^{\rm lin}-\bar{n}_{k'+k}^{\rm lin})}}{\zetac^{n\phi}_{\qb}}
\end{equation} where the first term in the right-hand side is given in (\ref{eq141}). Since the integral in the numerator is $>0$, equation (\ref{eq291}) on $\zetac^{n\phi}_{\qb}$ has a pure imaginary solution in the upper half-plane.\footnote{Let us set $\zetac^{n\phi}_{\qb}=\textrm{i}y$. Then {\yc(\ref{eq291}) is written as} $y=-2B+A/y$ where the coefficients $A$ and $B$ are real $>0$. The resulting second-degree equation on $y$ has one (and only one) positive root, $y_0=A/[(A+B^2)^{1/2}+B]$. When $\qb\to 0$, $B$ has a finite, non-zero limit in view of (\ref{eq141}), then $A$ diverges as $1/\qb$, so that $\zetac^{n\phi}_{\qb}$ diverges as $\qb^{-1/2}$ and $\zetat_{\qb}^{n\phi}$ tends to zero as $\qb^{1/2}$. Thus, condition $\im\zetat^{n\phi}_{\qb}\lll|\tilde{\Sigma}_{\qb}^{(2,2)}(\textrm{i}0^+)|=O(\qb)$, which makes the non-physical pole without effect on the signal in its time half-width, cannot be satisfied at wavenumber $\qb\lesssim 1$. It can be satisfied at large wavenumber $\qb\gg 1$, where $B\approx \qb^2$ in view of (\ref{eq141}) and $A\sim 2\exp(-\qb)$.} For the sake of completeness, let's point out that $\tilde{\Sigma}_{\qb}^{(4,2)}(\zetat)$ has another linear divergence on the real axis, at $\zetat_0=-3\gamma\qb^3/32$,\footnote{Note that $\zetat_0$ is also the branch point of $\tilde{\Sigma}_{\qb}^{(2,2)}(\zetat)$, see section \ref{sec2.1} and footnote \ref{note30}.} as shown by \ref{app4}, which is less damaging as it does not imply the existence of a non-physical pole of {\yct the} Green's function in the weakly interacting limit (it suffices to have $|\tilde{\Sigma}_{\qb}^{(2,2)}(\textrm{i}0^+)|\lll|\re\zetat_0|$ for this divergence to have little effect on signal damping in its half-width). 

\paragraph{Applicability of order four}
 Does the appearance of a non-physical pole force us to question the validity of the diagrammatic expansion and, by extension, that of the many-body Green's function method? No. The theory was intended to be used in the $(\rho\xi^2)^{-1}\to 0$ weakly interacting limit at fixed reduced temperature $\epsilon=k_B T/mc^2$. However, we took the limit $\epsilon\to 0$ with a fixed order of expansion in $(\rho\xi^2)^{-1}$, which fails in dimension $d=2$. Let's show that everything is fine, i.e. the non-physical pole disappears, when $\rho\xi^2\epsilon^2$ is large enough (at fixed $\qb$). 

To this end, let's see how the local analysis leading to (\ref{eq282}) is modified when $\epsilon$ is small but not infinitesimal, by restricting ourselves directly to diagram $P_5$, i.e. taking $s_1=s_4=s_5=1$, $s_2=s_3=-1$. If $\zetat=0$, type II energy denominator in (\ref{eq233}), which is also that of (\ref{eq282}), is written $(-\zeta_1)(-\zeta_2)(-\zeta_3)$, where the energy defects $\zeta_i$ are those in equation (\ref{eq238}). The first factor $-\zeta_1=-(\veps_{\kk}-\veps_{\kk-\qq}-\veps_{\qq})$ is a function of the single angle $\theta=\epsilon\thetat$ between $\qq$ and $\kk$, and continues to vanish on two vertical lines in the plane $(\thetat,\thetat')$, at slightly modified abscissas, which negligibly changes (by a factor $1+O(\epsilon)^2$) the coefficient of $\tau$ in (\ref{eq282}); the second factor $-\zeta_2=-(-\veps_{\kk'}+\veps_{\qq+\kk'}-\veps_{\qq})$ is a function of the single angle $\theta'=\epsilon\thetat'$ between $\qq$ and $\kk'$, and continues to vanish on two horizontal lines, slightly displaced, with the same negligible consequences on the coefficient of $\tau'$ in (\ref{eq282}). On the other hand, the cancellation locus of the third factor, namely $-\zeta_3=-(\veps_{\kk}+\veps_{\qq+\kk'}-\veps_{\kk+\kk'}-\veps_{\qq})$, is no longer exactly a conic in the $(\thetat,\thetat')$ plane and now passes slightly to the side of the intersection points $P_{-1,1}$ and $P_{1,-1}$ of the horizontal and vertical lines: this changes very slightly the coefficients of the linear combination of $\tau$ and $\tau'$ in (\ref{eq282}), but above all adds, next to $\zetat$, the now non-zero value of $-\zetat_3$ in $P_{-1,1}$: 
\begin{equation}\label{eq292}
(-\zetat_3)_{P_{-1,1}} \underset{\epsilon\to 0}{\sim} -\frac{27}{64}\gamma^2\epsilon^2\kb\kb'(\kb-\qb)(\kb+\kb')(\kb'+\qb)
\end{equation} We recalculate the right-hand side of (\ref{eq282}) accordingly: 
\begin{equation}\label{eq293}
K_{\rm div}^{P_5}(\kb,\kb')=-\frac{32\pi^2}{3\gamma\qb\kb\kb'(\kb+\kb')} \frac{1}{\zetat-\frac{27}{64}\gamma^2\epsilon^2\qb\kb\kb'(\kb-\qb)(\kb'+\qb)}
\end{equation} The new equation on the non-physical pole 
\begin{equation}\label{eq294}
\zetac^{n\phi}_{\qb} = \check{\Sigma}_{\qb}^{(2,2)}(\textrm{i}0^+)-\frac{2}{\qb^{2}}\int_{\qb}^{\eta}\kb\,\dd\bar{k}\int_0^{\eta-\kb}\kb'\dd\kb' \frac{(\kb-\qb)(\kb'+\qb){\yc(\bar{n}_{k-q}^{\rm lin}-\bar{n}_{k}^{\rm lin})(\bar{n}_{k'+q}^{\rm lin}-\bar{n}_{k'+k}^{\rm lin})}}{\zetac^{n\phi}_{\qb}-\frac{9}{8}\gamma u \epsilon^2\kb\kb'(\kb-\qb)(\kb'+\qb)}
\end{equation} ceases to have a solution in the upper half-plane when $\gamma u\epsilon^2>h(\qb)$, where $u$ is the universal parameter (\ref{eq112}) and $h(\qb)$ is a function exclusively of $\qb$ (and of the cut-off $\eta$ if it is not sent to infinity in quantum theory), as can be seen by formally taking the limit $\gamma u \epsilon^2\to +\infty$ in (\ref{eq294}). For the classical field, we find numerically that $h(\qb)$ is a decreasing function of $\qb$; the problem of applicability of fourth-order theory for small values of $\epsilon$ therefore remains in the limit $\qb \to 0$.\footnote{{\ycd For the classical field ($\eta=1$), we have the equivalent $h(\qb)\sim \frac{8}{9}(\pi^2/2\qb)^{4/3}$ when $\qb \to 0$ (in practice $\qb^{1/3}\ll 1$). This is obtained by writing the right-hand side of (\ref{eq294}) to leading order in $\qb$ (in particular, we replace $\qb$ by zero in the first integration bound and in the denominator of the integrand, and replace the numerator by its linear approximation in $\qb$), then showing that the double integral is dominated by a neighborhood of $(\kb,\kb')=(0,0)$ of radius $h(\qb)^{-1/4}\ll 1$ when $\qb \to 0$, which allows us to replace the upper integration bounds by $+\infty$. The result is an analytically calculable double integral and the equation $\zetac^{n\phi}_{\qb}=-\ii+(1+\ii)\pi^2/[2\qb(\frac{9}{8}\gamma u\epsilon^2)^{3/4}(\zetac^{n\phi}_{\qb})^{1/4}]$. It remains to express the fact that the non-physical pole has a zero imaginary part when it disappears.}} {\ycd To see this more simply, we can evaluate $\check{\Sigma}_{\qb}^{(4)}(\ii 0^+)$ (fourth-order in $H_3$ contribution considered at all orders in $\epsilon$) in the limit $\epsilon\to 0$. All we need to do is replace $\check{\zeta}_{\qb}^{n\phi}$ by $\ii 0^+$ in the double integral in the right-hand side of (\ref{eq294}), which we can then calculate explicitly for the quantum field (cutoff $\eta\to+\infty$):
\be
\label{eq296}
\check{\Sigma}_{\qb}^{(4)}(\ii 0^+)\underset{\epsilon\to 0}{\sim} \frac{16 g_2(\eee^{-\qb})}{9\gamma u \epsilon^2 \qb^2} \quad \mbox{hence} \quad \Sigma_{\qb}^{(4)}(\ii 0^+)\underset{\epsilon\to 0}{\sim} k_B T \left[\frac{(1+\Lambda)^2}{\gamma^{3/2}\rho\xi^2}\right]^2 \frac{2 g_2(\eee^{-\qb})}{\pi^2\qb}
\ee
At low temperature $\epsilon \ll 1$, the fourth order in $H_3$ (\ref{eq296}) is therefore a small correction to the second order (\ref{eq141}) only if $\gamma u\epsilon^2$ is much larger than a certain function of $\qb$, namely $1/\qb^2$ at low $\qb$. \footnote{{\yc The result (\ref{eq296}) does not contradict equation (\ref{eq280}) because the order of the limits $\zetat\to 0$ and $\epsilon\to 0$ is not the same.}}}

\paragraph{Conclusion}
 At a fixed reduced wave number $\qb$ of the phonon mode {\yc whose damping is being studied}, the condition for the applicability of diagrammatic theory at order four in $H_3$ is not only that the universal parameter $u$ is large enough (the expected weakly interacting regime) but also that the product ${\yc\gamma}u\epsilon^2$ is large enough. The perturbative approach therefore presents a fundamental incompatibility with the limit studied in this article, the low-temperature $\epsilon\to 0$ limit with fixed interaction strength. In particular, for the numerical simulations in Figure \ref{fig2}, where $\qb=1/4$ and $\gamma=1$, we find, by following with Newton's method the root of equation (\ref{eq294}), written for the classical field, until it disappears, that we should have $u\epsilon^2>10.6$ or $\rho\xi^2>23.4$ for $\epsilon=1/2$ and $\rho\xi^2>52.6$ for $\epsilon=1/3$. However, $\rho\xi^2$ is at most $10$ in Figure \ref{fig2}; the fourth-order theory developed in this section is therefore inapplicable.

\section{Nonperturbative theories in $H_3$}
\label{sec4}

As we saw in section \ref{sec3}, the theory at fourth order in $H_3$ is not usable in dimension $d=2$ in the low-temperature limit $\epsilon=k_B T/mc^2\to 0$. To account for the discrepancy between the theory at second order in $H_3$ and simulations on the decay time at $1/\eee$ of the phonon mode $\qq$ and on the real part of the associated signal (\ref{eq012}) in the weakly interacting regime $(\rho\xi^2)^{-1}\ll 1$, see figures \ref{fig1}d and \ref{fig2}d, we propose here a non-perturbative theory guided by physical considerations. In brief, {\yc we omit the finite part and} we regularize the divergent {\yc part of the fourth-order} self-energy by giving a finite lifetime to the so-called virtual (or intermediate) phonons of four-phonon collisional processes; this brings out a contribution that turns out to be second-order in $H_3$ and thus corrects the eigenenergy expression to the leading order in the {\yct interaction} strength. We present an atomic physics-type regularization in section \ref{sec4.1} (virtual phonons $\KK$ acquire a complex energy $\veps_\KK\to \veps_\mathbf{K}+\Sigma_\KK^{(2,2)}(\textrm{i}0^+)=\veps_\KK-(\ii\hbar/2)\Gamma_\KK$) and a condensed matter-type regularization in section \ref{sec4.2} (the frequency dependence of the self-energy $\Sigma_\KK^{(2,2)}(\zeta)$ is retained in the complex energy of virtual phonons). Finally, in section \ref{sec4.3}, we give the new form of the theory in the $(\rho\xi^2)^{-1}\to 0$ limit, common to both regularizations and successfully replacing Fermi's golden rule.

\subsection{Atomic physics-type regularization}
\label{sec4.1}

We learn from section \ref{sec3.3} that the addition of an infinitesimal complex part $\textrm{i}0^+$ to $\zeta=z-\veps_\qq$, very common in the Green's function method, is not sufficient to give a finite value to $\tilde{\Sigma}_\qq^{(4,2)}(\zeta=0)$. Collisions between phonons, which render them unstable, naturally provide a non-infinitesimal imaginary contribution next to $\zeta$ in the energy denominators. Following the prescription of atomic physics, we therefore give virtual phonons a complex energy 
\begin{equation}\label{eq300}
\veps_\KK\to\veps_\KK+\Sigma_\KK^{(2,2)}(\textrm{i}0^{\pm})=\veps_\KK\mp \frac{\ii\hbar}{2}\Gamma_\mathbf{K}\end{equation} where $\Gamma_\KK$ is the phonon-$\KK$ damping rate of Fermi's golden rule to order two in $\epsilon$, see equation (\ref{eq033}). Virtual phonons, by definition neither incoming nor outgoing in Landau's formulation of section \ref{sec3.2}, are those whose wave vectors are not shown in {\yc the diagrams of} Figure \ref{fig8}.  The sign to be taken in (\ref{eq300}) in front of $\Gamma_\KK$ or in the argument of $\Sigma_\KK$ is the upper one if the phonon goes forward in time (internal line oriented to the right) or the lower one if the phonon goes backward in time (internal line oriented to the left).\footnote{In Landau's formulation, we choose $-\ii(\hbar/2)\Gamma_\KK$ in the first sum of equation (\ref{eq258}) and its opposite in the second sum, for the reasons explained below (\ref{eq257}) {\yc (in the second sum, the transition amplitudes are reversed in time),} which leads to the same result {\yc as in the diagrammatic formulation}.} On the other hand, we don't give a complex energy to incoming or outgoing phonons as this would question energy conservation in the gas during collisional processes. Consequently, there is no need to modify $\Sigma_{\qq}(\zeta)$ to order two in $H_3$, which has no virtual phonons. 

\begin{figure}[t]
\includegraphics[width=8cm,clip=]{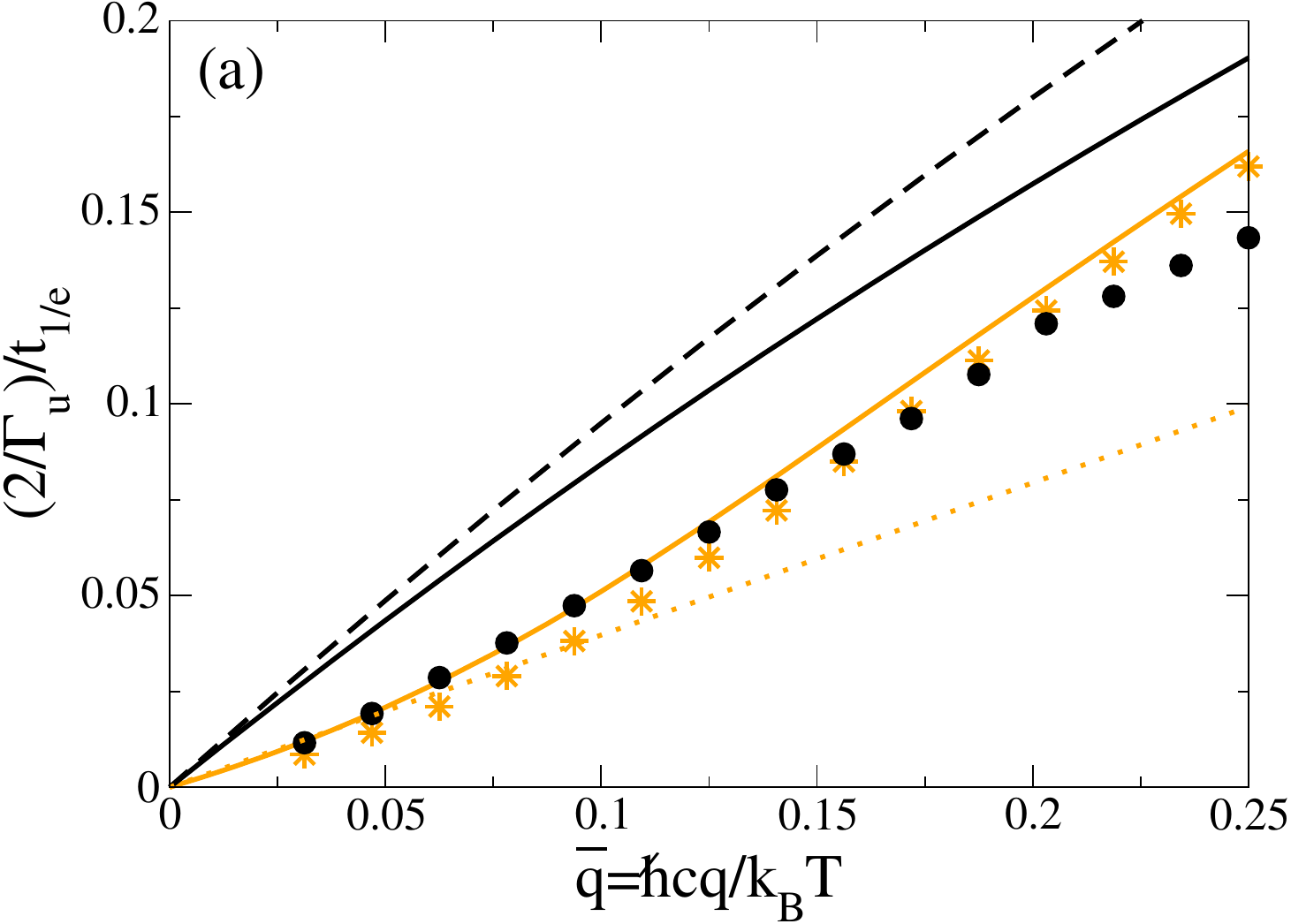} \includegraphics[width=8cm,clip=]{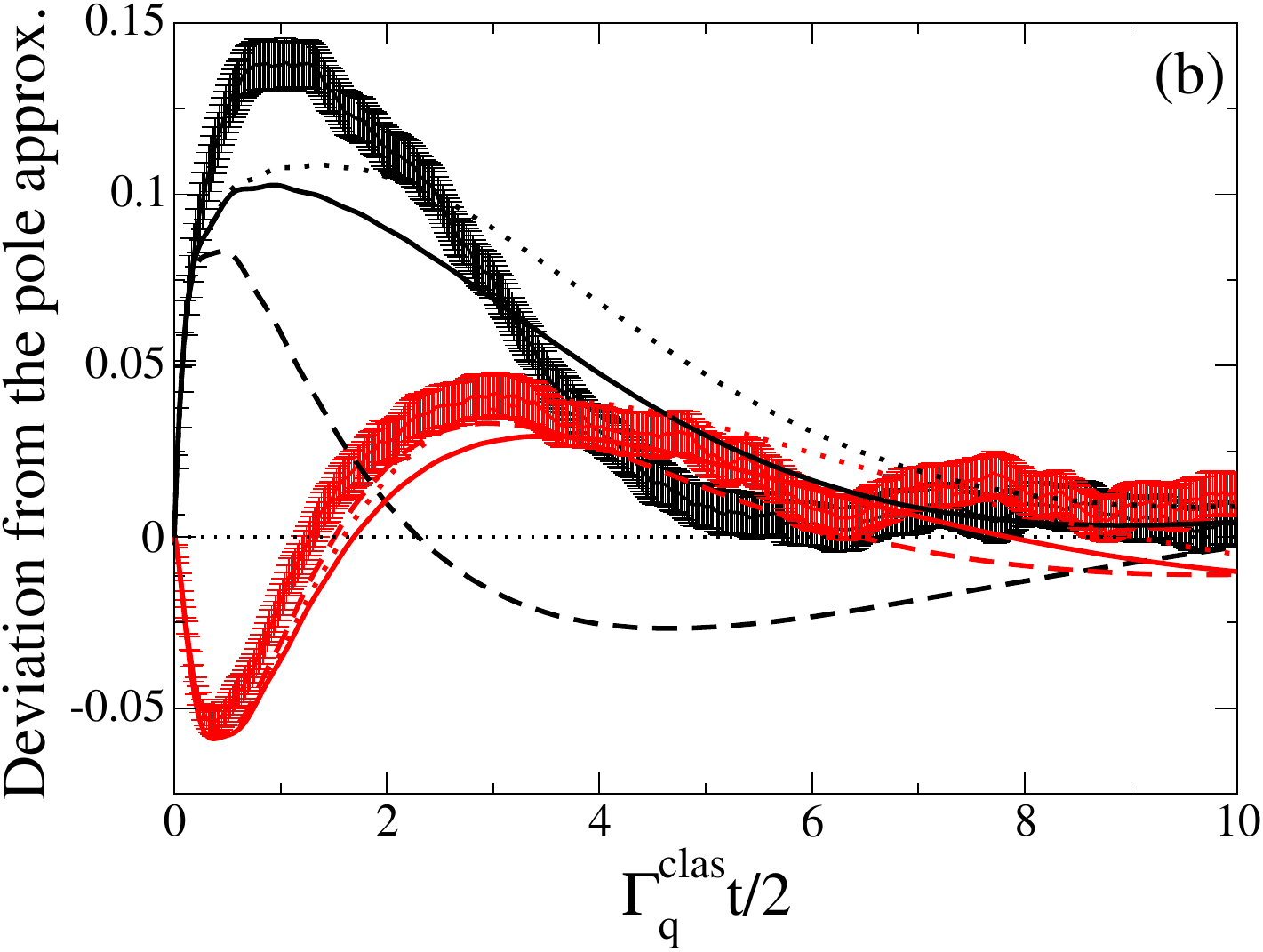}
\caption{(a) Inverse of the decay time at $1/\eee$ of the modulus of signal $s(t)$ (\ref{eq012}) as a function of the wavenumber $q$ of the modes along $Ox$, for the parameters of Figure \ref{fig1}d, in particular $(\rho\xi^2)^{-1}=0.1$ (this is the weakest interaction strength in Figure \ref{fig1}). Black discs: numerical experiment with $\epsilon=1/2$. Black dashed line: Fermi's golden rule (\ref{eq113}) for the classical phonon-field and in the $\epsilon\to 0$ limit. Black solid line: many-body Green’s function method with the $\tilde{\Sigma}^{(2,2)}_{\yc\qb}(\zetat)$ (\ref{eq125}) self-energy (leading order in $H_3$ and $\epsilon$). Orange solid line: idem with the $\tilde{\Sigma}^{(2,2)}_{\mbox{\tiny corr}}(\qb,\zetat)=\tilde{\Sigma}^{(2,2)}_{\yc\qb}(\zetat)+\tilde{\Sigma}^{(4,2)}_{\mbox{\tiny div. {\yct reg.}}}(\qb,\zetat)$ self-energy of equation (\ref{eq303}) (we keep only the fourth-order divergent part in $H_3$ and regularize it according to the atomic physics prescription by giving by hand Fermi's golden rule complex energy to the virtual phonons). Orange stars: idem with the $\tilde{\Sigma}^{(4,2)}_{\mbox{\tiny div. {\yct reg.}}}(\qb,\zetat)$ component of the self-energy (\ref{eq310}) obtained by a condensed matter physics-type regularization (virtual phonons $\KK$ are given a complex energy $\Sigma^{(2,2)}_{\KK}(\zeta_{\KK})$ depending on frequency i.e. on variable $\zeta$ in (\ref{eq012})). Orange dotted line: tangent at the origin (\ref{eq313}) for the $(\rho\xi^2)^{-1}\to 0$ limit of regularized theories. All self-energies are written in their classical field versions to allow comparison with numerics. The $\Gamma_u$ rate used as a unit is that of equation (\ref{eq113}) giving $\Gamma_q^{{\yct\rm clas}}$. (b) Signal deviation from the pole approximation {\yc $s(t)-s_{\mbox{\scriptsize pole}}(t)$} for parameters in Figure \ref{fig2}d, in particular {\yc$\qb=1/4$ and} $(\rho\xi^2)^{-1}=0.1$. Curves with error bars: numerical experiment. Dashed line: Green's function method with self-energy $\tilde{\Sigma}^{(2,2)}_{\yc\qb}(\zetat)$. Solid line: idem taking the self-energy $\tilde{\Sigma}^{(2,2)}_{\mbox{\tiny corr}}(\qb,\zetat)$ from equation (\ref{eq303}). Dotted line: idem using expression (\ref{eq310}) for the $\tilde{\Sigma}^{(4,2)}_{\mbox{\tiny div. {\yct reg.}}}(\qb,\zetat)$ component of self-energy $\tilde{\Sigma}^{(2,2)}_{\yc\qb}(\zetat)$. The real part of the signal is in black, the imaginary part in red.}
\label{fig3}
\end{figure}

We apply procedure (\ref{eq300}) to the divergent part of $\tilde{\Sigma}^{(4,2)}_{\qb}$, which comes from the type II contribution of diagram $P_5$, and omit all other contributions to $\tilde{\Sigma}^{(4,2)}_{\qb}$, which turn out to be $O((\rho\xi^2)^{-2})$ and subleading when $(\rho\xi^2)^{-1}\to 0$.\footnote{In the absence of divergence at $\zeta=0$, adding an imaginary part to $\veps_\KK$ does not change the leading order of the diagram, which remains fourth-order in $H_3$.} As shown in Figure \ref{fig8}, the virtual phonons are then {\ycd $\kk_2=\kk-\qq$} and $\kk_3=\kk'$ in the notations of Figure \ref{fig7}. Both go backward in time ($s_2=s_3=-1$), so the quantities $\zeta_1$ and $\zeta_2$ in (\ref{eq233}), {\ycd that are written $\zeta_1=\veps_\kk-\veps_{\kk-\qq}-\veps_\qq$ and $\zeta_2=-\veps_{\kk'}+\veps_{\kk'+\qq}-\veps_{\qq}$,} and the coefficients $A_1^P$ and $A_2^P$ in (\ref{eq243}) are changed as follows, 
\begin{equation}\label{eq301}
\zeta_1\to\zeta_1-\ii\frac{\hbar}{2}\Gamma_{\yc\kk-\qq} \quad ;\quad  \zeta_2\to\zeta_2-\ii\frac{\hbar}{2}\Gamma_{\kk'} \quad ; \quad A_1^P\to A_1^P-\frac{\ii}{2}\tilde{\Gamma}_{\kb-\qb} \quad ; \quad A_2^P\to A_2^P-\frac{\ii}{2}\tilde{\Gamma}_{\kb'}
\end{equation} the order of the factors in the denominator of (\ref{eq243}) respecting that of (\ref{eq233}); here $\tilde{\Gamma}_{\bar{K}}=\hbar\Gamma_{\KK}/k_BT\epsilon^2$ in the spirit of equation (\ref{eq022}). {\yc On the other hand, the quantity $\zeta_3$ does not involve any virtual phonon and remains unchanged.} The previously diverging part of the angular integral (\ref{eq282}) becomes for $\im\zetat>0$:\footnote{We continue to linearize around the same triple intersection point of the cancellation loci, grouping the complex corrections with variable $\zetat$, so that in the {\yct integral in} (\ref{eq282}), we replace $\zetat$ by $\zetat+(\ii/2)\tilde{\Gamma}_{\kb-\qb}$ in the first factor and by $\zetat+(\ii/2)\tilde{\Gamma}_{\kb'}$ in the second factor in the denominator.} 
\begin{equation}\label{eq302}
K_{\mbox{\scriptsize div. {\yct reg.}}}^{P_5^{\rm II}}(\kb,\kb')=-\frac{32\pi^2}{3\gamma\qb\kb\kb'} \frac{1}{\zetat(\kb+\kb')+\frac{\ii}{2}\kb'\tilde{\Gamma}_{\kb-\qb}+\frac{\ii}{2}(\kb-\qb)\tilde{\Gamma}_{\kb'}}
\end{equation} After replacing $K(\kb,\kb')$ by (\ref{eq302}) in (\ref{eq242}) and adding the contribution of second order in $H_3$, we end up with the non-perturbative prescription for the self-energy correcting the result of section \ref{sec2}: 
\begin{equation}\label{eq303}
\boxed{\tilde{\Sigma}_{\rm corr}^{(2,2)}(\qb,\zetat) = \tilde{\Sigma}_{\qb}^{(2,2)}(\zetat) + \tilde{\Sigma}^{(4,2)}_{\mbox{\scriptsize div. {\yct reg.}}}(\qb,\zetat)}
\end{equation} where the first contribution is that of equation (\ref{eq125}) and the second, a regularization of the divergent part of $\tilde{\Sigma}_{\qb}^{(4,2)}(\zetat)$, is written, again for $\im\zetac>0$: 
\begin{equation}\label{eq304}
\boxed{\tilde{\Sigma}^{(4,2)}_{\mbox{\scriptsize div. {\yct reg.}}}(\qb,\zetat)=-\frac{2}{3\pi^2\gamma} \left(\frac{9(1+\Lambda)^2}{8\rho \xi^2}\right)^2
\int_{\qb}^\eta \dd\bar{k}\int_0^{\eta-\kb} \dd\kb' \frac{\bar{k}\kb' (\kb-\qb)(\kb'+\qb)(\kb+\kb')}
{\tilde{\zeta}(\kb+\kb')+\frac{\ii}{2}\kb'\tilde{\Gamma}_{\kb-\qb}+\frac{\ii}{2}(\kb-\qb)\tilde{\Gamma}_{\kb'}} 
(\bar{n}^{\rm lin}_{k'+q}-\bar{n}^{\rm lin}_{k'+k})(\bar{n}^{\rm lin}_{k-q}-\bar{n}^{\rm lin}_{k})}
\end{equation} By inserting function (\ref{eq303}) into equation (\ref{eq023}) (with $\nu=\sigma$ in dimension $d=2$) in its classical field version, we obtain a much better agreement with numerical simulations for the weakest interaction strength, see the orange solid line in Figure \ref{fig3}a and the black solid line in Figure \ref{fig3}b, to be compared respectively with the black solid line in Figure \ref{fig3}a and the black dashed line in Figure \ref{fig3}b corresponding to the uncorrected second order in $H_3$. 

\subsection{Condensed matter regularization}
\label{sec4.2}

It is well known in condensed matter physics that the complex energy of a quasiparticle in an interacting system depends on the frequency at which it is observed, i.e. the frequency at which the response functions are measured. This is the essence of $\Sigma$ in (\ref{eq012}). In (\ref{eq300}), therefore, we need to replace the fixed complex energy $\mp(\ii\hbar/2)\Gamma_\KK$ of the virtual phonon {\yc$\KK$} with a function of {\yc the integration variable} $\zeta$: 
\begin{equation}\label{eq305} 
\veps_\kk\to \veps_\KK+\Sigma_\KK^{(2)}(\pm(\zeta+\Delta E)-\veps_\KK)
\end{equation} where we need to take the sign $+$ if the phonon $\KK$ goes forward in time, and the sign $-$ if it goes backward. The quantity $\Delta E$ is obtained by reinterpreting the energy denominator in which phonon $\KK$ participates in $\Sigma^{(4)}$ as that of the free propagator of this phonon with complex energy $z_{\KK}=\zeta+\Delta E$. So, if phonon $\KK$ goes forward in time (sign $s_\KK=+1$ before $\veps_\KK$ in (\ref{eq238})), it will contribute to $\Sigma^{(4)}$ in simplified forms (\ref{eq213},\ref{eq233}) through factor $1/(\zeta+\Delta E-\veps_\KK)$, which is precisely its unperturbed propagator at complex energy $z_\KK$ and therefore at complex energy redefined as in (\ref{eq013}) $\zeta_\KK=z_\KK-\veps_\KK$. We therefore propose to replace it by the Green's function (\ref{eq012}) of phonon $\KK$ {\yc written as in (\ref{eq013})}, with $\Sigma_{\KK}$ calculated to order two in $H_3$: \begin{equation}\label{eq306}
\frac{1}{\zeta+\Delta E-\veps_\KK}\to \frac{1}{\zeta+\Delta E-\veps_\KK-\Sigma_\KK^{(2)}(\zeta+\Delta E-\veps_\KK)}
\end{equation} which explains the substitution (\ref{eq305}). If the virtual phonon goes backward in time (sign $s_\KK=-1$), we reduce to the previous case through the chain of transformations 
\begin{equation}\label{eq307}
\frac{1}{\zeta+\Delta E+\veps_\KK} = -\frac{1}{-(\zeta+\Delta{\ycd E})-\veps_\KK} \longrightarrow -\frac{1}{-(\zeta+\Delta E)-\veps_\KK-\Sigma_\KK^{(2)}(-(\zeta+\Delta E{\yc)-}\veps_\KK)} = \frac{1}{\zeta+\Delta E+\veps_\KK+\Sigma_\KK^{(2)}(-(\zeta+\Delta E{\yc)-}\veps_\KK)}
\end{equation} again in accordance with (\ref{eq305}). In practice, as in section \ref{sec4.1}, we keep only the divergent part of $\tilde{\Sigma}^{(4,2)}_{\qb}$ (coming from the type II contribution of diagram $P_5$), in which the virtual phonons {\yc both go backward in time and} are of wave vector $\KK=\kk-\qq$ (with $\Delta E=\veps_{\qq}-\veps_\kk$ {\ycd so that $\zeta-\zeta_1=\zeta+\Delta E+\veps_{\kk-\qq}$}) or $\KK=\kk'$ (with $\Delta E=\veps_{\qq}-\veps_{\kk'+\qq}$ {\ycd so that $\zeta-\zeta_2=\zeta+\Delta E+\veps_{\kk'}$}). After regularization, the divergent part (\ref{eq282}) of the angular integral becomes 
\begin{equation}\label{eq308}
K_{\mbox{\scriptsize div. {\yct reg.}}}^{P_5^{\rm II}}(\kb,{\yct\kb'}) = \int_{\mathbb{R}^2} \frac{2\mathrm{d}\tau \mathrm{d}\tau'}{[\zetat-\alpha \tau +\tilde{\Sigma}_{\kb-\qb}^{(2,2)}(-\zetat+\alpha \tau)] [\zetat-\alpha' \tau' + \tilde{\Sigma}_{\kb'}^{(2,2)}(-\zetat+\alpha' \tau')] (\zetat-\beta \tau -\beta' \tau')}
\end{equation} where the coefficients are those of $\tau$ and $\tau'$ in the {\yct integral in} (\ref{eq282}): $\alpha=(3\gamma/4)^{1/2}\qb\kb$, $\alpha'=-(3\gamma/4)^{1/2}\qb\kb'$, $\beta=-(3\gamma/4)^{1/2}\kb\kb'$, $\beta'=(3\gamma/4)^{1/2}(\kb-\qb)\kb'$. As functions $\tilde{\Sigma}^{(2,2)}$ in the denominator of (\ref{eq308}) have branch cuts in the complex plane, {\yc see section \ref{sec2.2},} Cauchy's theorem no longer allows us to calculate the double integral and we can only integrate once, for example over $\tau'$, restricting ourselves as usual to $\im\zetat>0$.\footnote{The denominator of the integrand in (\ref{eq308}) has a branch cut with respect to variable $\tau$ only in the strict upper half-plane because $\im(\zetat/\alpha)>0$, and with respect to variable $\tau'$ only in the strict lower half-plane because $\im(\zetat/\alpha')<0$. We can therefore integrate on $\tau'$ by the residue theorem, closing the integration path with a large semicircle at infinity in the upper half-plane; this path necessarily surrounds the pole $\tau_0'=(\zetat-\beta \tau)/\beta'$ arising from the third factor in the denominator of (\ref{eq308}) because $\im \tau_0'>0$. The integrand of the result in (\ref{eq310}) this time has a branch cut with respect to variable $\tau$ both in the upper half-plane ($\im(\zetat/\alpha)>0$) and in the lower half-plane ($\im(\zetat/\alpha')<0$), which prevents the integration path from being closed only at infinity.} By inserting the corresponding contribution to $K(\kb,\kb')$ in (\ref{eq242}), we end up with the corrected self-energy of the same form as (\ref{eq303}), now with 
\begin{multline}
\label{eq310}
\tilde{\Sigma}^{(4,2)}_{\mbox{\scriptsize div. {\yct reg.}}}(\qb,\zetat)= \qb\left(\frac{9(1+\Lambda)^2}{8\rho \xi^2}\right)^2 \\
\times \int_{\qb}^\eta \frac{\bar{k}\dd\kb}{(2\pi)^2}
\int_0^{\eta-\kb} \frac{\kb' \dd\kb'}{(2\pi)^2}
\int_{-\infty}^{+\infty} \mathrm{d}\tau 
\frac{(-4\ii\pi/\beta')\bar{k}\kb' (\kb-\qb)(\kb'+\qb)(\kb+\kb')(\bar{n}^{\rm lin}_{k'+q}-\bar{n}^{\rm lin}_{k'+k})(\bar{n}^{\rm lin}_{k-q}-\bar{n}^{\rm lin}_{k})}
{\left[\tilde{\zeta}-\alpha \tau +\tilde{\Sigma}^{(2,2)}_{\kb-\qb}(-\tilde{\zeta}+\alpha \tau)\right]
\left[-\frac{\beta}{\beta'}(\tilde{\zeta}-\alpha'\tau)+\tilde{\Sigma}^{(2,2)}_{\kb'}
\left(\frac{\beta}{\beta'}(\tilde{\zeta}-\alpha'\tau)\right)\right]}
\end{multline}
 knowing that $\alpha'-\beta'=\beta$. The results of this approach are shown as orange stars in Fig.~\ref{fig3}a and as red and black dotted lines in Fig.~\ref{fig3}b. \footnote{Getting them is rather cumbersome, since each evaluation of functions $\tilde{\Sigma}_{\bar{K}}^{(2,2)}$ requires numerical integration over a wavenumber, as in (\ref{eq125}). We have discretized the reduced wavenumbers with a step of $1/64$ (results {\yct in the} figure), but we have verified that a step of $1/96$ leads to very similar results.} The two regularization procedures, atomic physics or condensed matter, ultimately give results close to each other and close to the simulations, which reassures us as to their validity. 

\subsection{Regularized theory in the weakly interacting limit}
\label{sec4.3}

\begin{SCfigure}
\includegraphics[width=8cm,clip=]{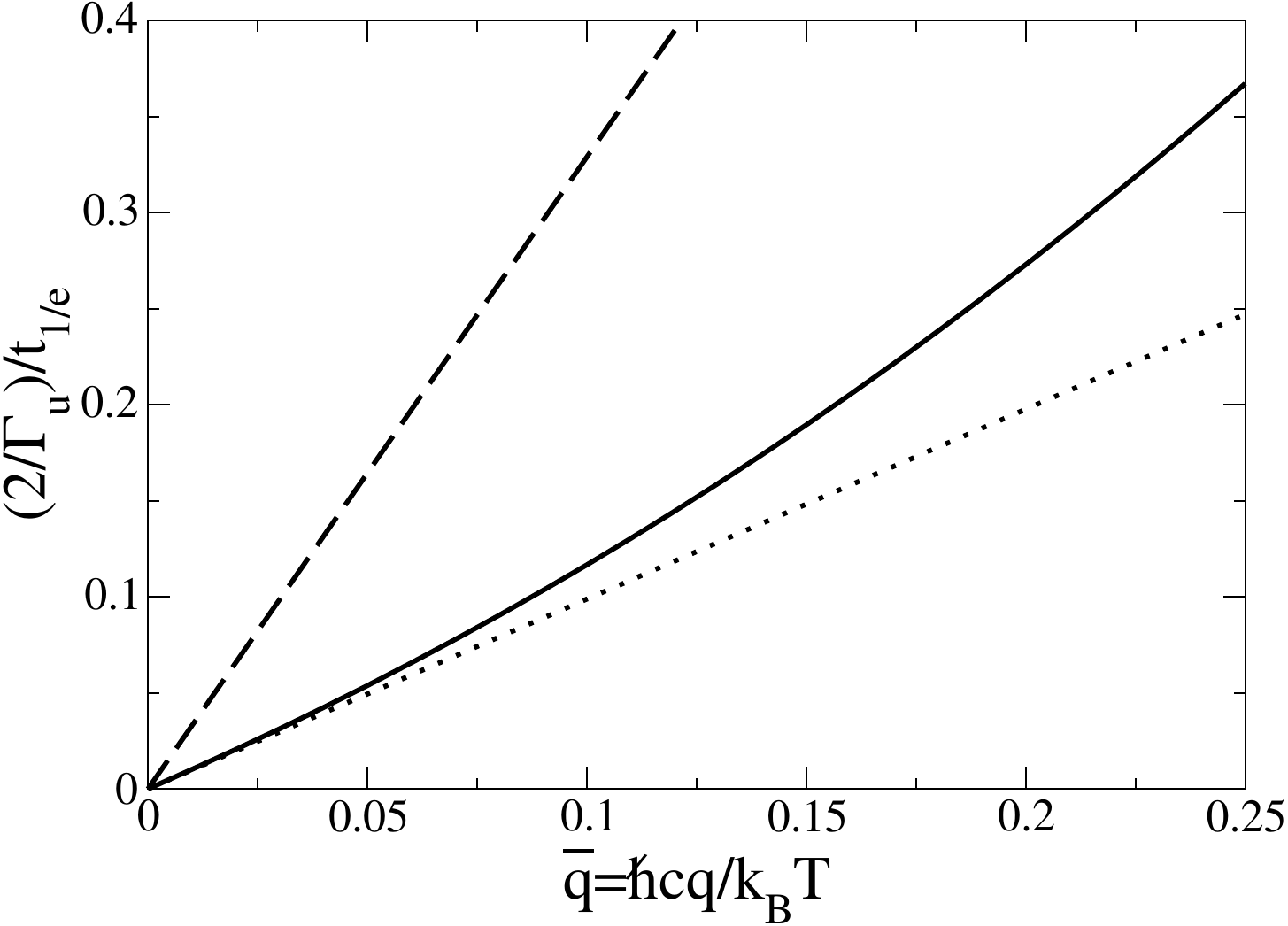}
\caption{In the low-temperature limit $\epsilon=k_BT/mc^2\to 0$ then weakly interacting limit $(\rho\xi^2)^{-1}\to 0$ for the quantum field {\yc of infinite cut-off $\eta$}, inverse of the decay time at $1/\eee$ of signal (\ref{eq012}) for the phonon mode $\qq$ in a two-dimensional superfluid, as a function of the reduced wavenumber $\qb$. Solid line: prediction from non-perturbative theories in section \ref{sec4}, corresponding to the corrected self-energy (\ref{eq311}) in section \ref{sec4.3}. Dotted line: slope at origin (\ref{eq313}). Dashed line: golden rule (\ref{eq033}). The rate $\Gamma_u$ serving as unit is given by (\ref{eq113}).}
\label{fig9}
\end{SCfigure}

Here we calculate the limit of regularized theories when $(\rho\xi^2)^{-1}\to 0$. To do this, we scale variable $\zetat$ and functions $\tilde{\Sigma}$ to their characteristic values $\propto (\rho\xi^2)^{-1}$ by means of {\yct changes} of variables (\ref{eq130}) and (\ref{eq131}), marked with a Czech accent (a factor $\qb$ has been taken out in view of taking the $\qb\to 0$ limit, and all parameters are collected together in the quantity $u\propto\rho\xi^2$ of equation (\ref{eq112})), then {\yct we take} the $u\to+\infty$ limit at fixed $\check{\zeta}$ in equations (\ref{eq303}) and (\ref{eq304}). Contrary to what happened at uncorrected second order in $H_3$ (\ref{eq131}), there remains a dependency in $\check{\zeta}$: \footnote{In two steps, we show that the Green's function associated with self-energy (\ref{eq311}) has, as it should, no pole in the upper complex half-plane. First, we establish the {\yct upper bound} $\im\check{\Sigma}_{{\rm corr}\,\infty}^{(2,2)}(\qb,\check{\zeta})\leq\im\check{\Sigma}_{{\rm corr}\,\infty}^{(2,2)}(\qb,0)\ \forall \check{\zeta}\in\mathbb{C}$, $\im\zetac\geq 0$; to this end, we write the denominator of the integrand in (\ref{eq311}) in the form $x+\textrm{i}y$, with $x\in\mathbb{R}$ and $y\in\mathbb{R}^+$, then use the chain $\im[-1/(x+\textrm{i}y)]=y/(x^2+y^2)\leq 1/y\leq 2/[\kb'(\kb-\qb)(\check{\Gamma}_{\kb-\qb}+\check{\Gamma}_{\kb'})]$. In a second step, we check that $\im\check{\Sigma}_{{\rm corr}\,\infty}^{(2,2)}(\qb,0)<0\ \forall\qb>0$. It follows that equation $\check{\zeta}=\check{\Sigma}_{{\rm corr}\,\infty}^{(2,2)}(\qb,\check{\zeta})$ cannot have a solution $\check{\zeta}$ with a positive imaginary part without leading to the contradiction $\im\check{\zeta}<0$.} 
\begin{equation}\label{eq311}
\boxed{\check{\Sigma}_{\rm corr}^{(2,2)}(\qb,\check{\zeta}) \underset{u \to +\infty}{\stackrel{\im\check{\zeta}\,>\,0}{\longrightarrow}} \check{\Sigma}_{{\rm corr}\,\infty}^{(2,2)}(\qb,\check{\zeta})\equiv-\frac{\ii}{2}\check{\Gamma}_{\qb}-\frac{2}{\qb}\int_{\qb}^\eta \dd\bar{k}\int_0^{\eta-\kb} \dd\kb' \frac{\kb\kb'(\kb-\qb)(\kb'+\qb)(\kb+\kb')(\bar{n}^{\rm lin}_{k'+q}-\bar{n}^{\rm lin}_{k'+k}) (\bar{n}^{\rm lin}_{k-q}-\bar{n}^{\rm lin}_{k})} {\check{\zeta}\qb(\kb+\kb')+\frac{\ii}{2}\kb'(\kb-\qb)\left(\check{\Gamma}_{\kb-\qb}+\check{\Gamma}_{\kb'}\right)}}
\end{equation} Here, $\check{\Gamma}_{\bar{K}}=2\ii\check{\Sigma}_{\bar{K}}^{(2,2)}(\textrm{i}0^+)$ is the golden rule damping rate at leading order in $\epsilon$ and scaled as in (\ref{eq130}), $\check{\Gamma}_{\bar{K}}=8u\Gamma_{\KK}/(3\gamma\epsilon^2\hbar cK)$; we have $\check{\Gamma}_{\bar{K}}=\check{\Gamma}_{\bar{K}}^{{\yct\rm clas}}=2\eta-\bar{K}$ given (\ref{eq113}) in the case of the classical field and $\check{\Gamma}_{\bar{K}}=4\zeta(2)+\bar{K}^2/6$ given (\ref{eq033}) or (\ref{eq141}) in the case of the infinite cut-off quantum field. Starting from the regularized form (\ref{eq310}), we obtain exactly the same expression as (\ref{eq311}).\footnote{Variable changes $\zetat=(3\gamma\qb/8u)\check{\zeta}$ and $\tau=(3\gamma\qb/8u)\check{\tau}$ are carried out simultaneously, {\yct and} the limit $u\to +\infty$ is taken under the integral sign at fixed $\check{\tau}$. Then the branch cuts of functions $\tilde{\Sigma}^{(2,2)}_{\bar{K}}$ are rejected at infinity, these functions are replaced by constants, e.g. $\lim_{u\to+\infty} (8u/3\gamma\qb) \tilde{\Sigma}^{(2,2)}_{\kb-\qb}((3\gamma\qb/8u)(\alpha\check{\tau}-\check{\zeta}))=[(\kb-\qb)/\qb] \check{\Sigma}^{(2,2)}_{\kb-\qb}(\textrm{i}0^-)=[(\kb-\qb)/\qb](\ii/2)\check{\Gamma}_{\kb-\qb}$, and the integral over $\check{\tau}$ can be calculated by Cauchy's formula. We find (\ref{eq311}).} Remarkably, signal (\ref{eq012}) associated with the self-energy (\ref{eq311}) takes real values at all times.\footnote{This results from the property ${\yc\check{\Sigma}(\check{\zeta})^*}=-\check{\Sigma}(-\check{\zeta}^*)$, $\forall\check{\zeta}\in\mathbb{C}\setminus\ii\mathbb{R}^-$, satisfied by function $\check{\Sigma}_{{\rm corr}\,\infty}^{(2,2)}(\qb,\check{\zeta})$ (if we extend the expression in the right-hand side as it is to the case $\im\check{\zeta}<0$).} {\yc When} $(\rho\xi^2)^{-1}\to 0$, therefore, there is persistent disagreement with Fermi's golden rule on the real part of the signal, as predicted in Figure \ref{fig3}b, {\ycd but not on the imaginary part, of zero limit in both theories}. Figure \ref{fig9} shows, as a function of the reduced wavenumber $\qb$, the decay time at $1/\eee$ of the phonon mode $\qq$ deduced from (\ref{eq311}) for quantum field theory (cutoff $\eta\to +\infty$). This is our prediction to leading order in temperature and interaction strength for a two-dimensional superfluid, which replaces and corrects that of Fermi's golden rule shown as a dashed line in Fig.~\ref{fig9}.
 Three simple analytical predictions can be drawn from equation (\ref{eq311}). Firstly, in the $\qb\to 0$ limit, the $\check{\zeta}$ dependence disappears in the right-hand side of (\ref{eq311}) and signal (\ref{eq012}) becomes purely exponential, $s(t)=\exp(-A\check{t})$ with 
\begin{equation}\label{eq312}
A=\frac{1}{2}\check{\Gamma}_0+4\int_0^\eta \mathrm{d}\bar{k}\int_0^{\eta-\kb} \mathrm{d}\kb'
\frac{\kb\kb'(\kb+\kb')(\bar{n}_{k'}^{\rm lin}-\bar{n}_{k'+k}^{\rm lin})\frac{\dd}{\mathrm{d}\kb}\bar{n}_{k}^{\rm lin}}
{\check{\Gamma}_{\kb}+\check{\Gamma}_{\kb'}} =
\left\{ \begin{array}{ll}
[1-4(4 \ln \frac{4}{3} -1)]\eta \simeq 0.4\,\eta & \mbox{(classical field)} \\
& \\
0.988\,618 \ldots & \mbox{(quantum field)}
\end{array}
\right.
\end{equation} If we leave the rescaled time $\check{t}$ (\ref{eq130}), which brings out the quantity $\Gamma_u$ already introduced in (\ref{eq113}) {\yc since $\check{t}=\qb\Gamma_u t/2$}, we find that this corresponds to a corrected damping rate tending towards zero linearly in $\qb$ : 
\begin{equation}\label{eq313}
\boxed{
\Gamma_q^{{\rm corr}\,\infty} \underset{\qb\to 0}{\sim} A \Gamma_u \bar{q}= A \frac{9k_B T\epsilon^2(1+\Lambda)^2}{4\pi\hbar(3\gamma)^{1/2}\rho\xi^2} \bar{q}\sim \left\{ \begin{array}{ll} \displaystyle\frac{A}{\eta} \Gamma_q^{{\yct\rm clas}} \simeq 0.4\,\Gamma_q^{{\yct\rm clas}} & \mbox{(classical field)} \\ \displaystyle\frac{A}{2\zeta(2)} \Gamma_q \simeq 0.3\,\Gamma_q & \mbox{(quantum field)}\end{array} \right.}
\end{equation} shown as a dotted line in Figure \ref{fig9} for the quantum field and in Figure \ref{fig3}a for the classical field. Here, $\Gamma_q^{{\yct\rm clas}}$ and $\Gamma_q$ are the damping rates of the golden rule for the classical field (\ref{eq113}) and for the quantum field (\ref{eq033}). This shows that the slope at the origin of the corrected rate is considerably smaller than that of the golden rule. Then, in the $\qb\to +\infty$ limit, of course for the infinite cut-off quantum field, the integral in the right-hand side of (\ref{eq311}) tends uniformly towards zero in the upper half-plane and the signal decay becomes purely exponential, with a rate equivalent to that of the golden rule: \footnote{This result is also that of zero temperature, which explains why the right-hand side of (\ref{eq315}) is in fact independent of temperature when we return to the non-{\yct scaled} wave number $q$. As it results from the effective Hamiltonian of quantum hydrodynamics, it is then only valid at the leading order in $q\xi$, i.e. in the linear part of the {\yc acoustic branch.}}
\begin{equation}\label{eq315}
\Gamma_q^{{\rm corr}\,\infty} \underset{\qb\to+\infty}{\sim} \Gamma_q \underset{\qb\to+\infty}{\sim} \frac{3 k_B T\epsilon^2(1+\Lambda)^2}{16\pi(3\gamma)^{1/2}\rho\xi^2} \qb^3
\end{equation} Finally, in the limit of long times $\check{t}\to +\infty$ at fixed $\qb$, the signal decay is not exponential but power-law up to a logarithm, taking the same form for the classical and quantum fields, as shown in \ref{app5}: 
\begin{equation}\label{eq314}
s^{(2,2)}_{{\rm corr}\,\infty}(t)\underset{\check{t}\to+\infty}{\sim} \frac{\qb^4(\dd\bar{n}^{\rm lin}_q/\dd\qb)\ln(\check{t}^2)}{\left[\check{\Gamma}_0\check{\Sigma}_{{\rm corr}\,\infty}^{(2,2)}(\qb,0)\check{t}\right]^2}
\end{equation} For $\qb=1/4$, we find numerically from the self-energy (\ref{eq311}) that the asymptotic behavior (\ref{eq314}) is not reached before a {\yc reduced} time $\Gamma_q t/2=100$, for both the quantum and classical fields, which goes far beyond our simulations in Fig.~\ref{fig3}b and prevents us from testing it. 

\section{Conclusion}
\label{sec5}

In this work, we have considered the phonon gas of a two-dimensional superfluid described by quantum hydrodynamics, an effective low-energy theory, with cubic and quartic coupling between phonons for an acoustic branch $\veps_{\qq}$ which is a convex function of wavenumber $q$. Two important parameters characterize the system: the reduced temperature $\epsilon=k_BT/mc^2$ and the {\yct interaction} strength $(\rho \xi^2)^{-1}$ between superfluid particles, where $c$ is the speed of sound and $\rho$ the density of the superfluid in the ground state, $m$ the mass of a particle and $\xi=\hbar/mc$ the relaxation length.  In this system, we have studied the damping of the phonon mode $\qq$ excited by a Bragg pulse in the linear response and collisionless regime, limiting ourselves to the leading order in $\epsilon$ as going beyond this would make no sense with our model Hamiltonian. 

A first unexpected result is that in dimension two, taking the low-temperature limit, $\epsilon\to 0$ at fixed $(\rho \xi^2)^{-1}$, is not enough to bring the phonon gas into the weak-coupling regime. This defeats the pole approximation, Fermi's golden rule and the assumption of exponential decay.  Indeed, as we show by a scaling-law reasoning, the self-energy $\Sigma(\qq,z)$ has a typical value and a scale of variation in energy in the vicinity of $z=\varepsilon_\qq$ both of order $k_BT \epsilon^2$, whereas it would be necessary to have a typical value much lower than the scale of variation {\yc to be able to apply the golden rule}  (as is the case in dimension three {\ycd where the scale of variation remains $k_B T\epsilon^2$ but the typical value is lowered to $k_B T\epsilon^4$ due to a reduction of the density of final states at low energy)}. This first theoretical result is confirmed by classical field simulations. 

A second unexpected result of the two-dimensional case is that, even in the weakly interacting limit $(\rho \xi^2)^{-1}\to 0$ {\yc in the superfluid}, a perturbative calculation does not give access to the self-energy to the leading order in $\epsilon$.  Thus, the calculation of $\Sigma(\qq,z)$ at second order in the cubic phonon coupling does not reproduce the damping of mode $\qq$ observed numerically, and the fourth-order calculation {\yct exhibits} (after taking the $\epsilon\to 0$ limit at fixed $\qb=\hbar c q/k_B T$ and $(\rho \xi^2)^{-1}$) an unphysical divergence in $z=\varepsilon_\qq$ that renders it useless. 

To break this deadlock, guided by physical intuition, we give a finite lifetime to the virtual phonons ({\yc those} present only in the intermediate state of four-phonon collisional processes).  This removes the divergence of the fourth-order contribution by transforming it into a formally second-order contribution, which corrects the second-order calculation. The regularized theory is found to be in good agreement with simulations in the weakly interacting regime.  It makes it possible to take the $(\rho \xi^2)^{-1}\to 0$ limit in the expression of $\Sigma(\qq,z)$ to the leading order in $\epsilon$ and to propose a phonon damping law replacing that of the golden rule. The new law is non-exponential, except in the $\qb\to 0$ limit where it is characterized by the rate (\ref{eq313}), linear in wavenumber like the golden rule but with a coefficient about three times smaller, and in the $\qb\to+\infty$ limit where its rate (\ref{eq315}) reproduces that of the golden rule.

\medskip
\noindent {\bf Acknowledgements~:} Alan Serafin received funding from the European Horizon 2020 research and innovation project macQsimal number 820393. {\ycd This article is part of a special issue dedicated to the CNRS Gold Medal awarded to Jean Dalibard in December 2021. The authors congratulate Jean on his award and thank him for his brilliant lectures at the Collège de France and for all the enriching discussions they had with him on cold atomic gases (Y.C. in particular recalls with emotion his teaching on Fermi's golden rule at the ENS \g{DEA de physique quantique} in 1987-1988).}

\appendix 
\section{Analytic continuation of $\check{\Sigma}^{(2,2)}_{\qb}(\check{\zeta})$ to the lower half-plane and validity conditions of the golden rule}
\label{app1}

To analytically continuate from the upper complex half-plane to the lower half-plane expression (\ref{eq131}) of the self-energy $\check{\Sigma}^{(2,2)}_{\qb}(\check{\zeta})$, we proceed in two steps: (i) we reduce (\ref{eq131}) to a single integral, that of an analytic function of a well-chosen variable $x$ over an interval of $\mathbb{R}$, and (ii) to extend this integral to $\im\zetac<0$, we deform the integration path on $x$ in the complex plane and rotate the branch cuts of the corresponding analytic function. In application, we give conditions for the validity of Fermi's golden rule, which are necessary but not sufficient.

\paragraph{Reduce to a single integral}
 In Landau's contribution to (\ref{eq131}), i.e. {\yc in} the second integral, we perform the change of variable $x=\kb(\kb+\qb)$ or $\kb=(x+\qb^2/4)^{1/2}-\qb/2$, where $x\in\mathbb{R}^+$. In Beliaev's contribution, the integrand is invariant by changing $\kb$ to $\qb-\kb$, which allows us to restrict the integration on $\kb$ to the interval $[0,\qb/2]$ at the cost of multiplying by a factor $2$; we then perform the now injective change of variable $x=\kb(\kb-\qb)$ to $\kb=\qb/2-(\qb^2/4+x)^{1/2}$, where $x\in[-\qb^2/4,0]$; we use the identity $\bar{n}_k^{\rm lin}=-(1+\bar{n}_{-k}^{\rm lin})$ to give the bosonic amplification factor the same structure of difference of occupation numbers as in Landau. In both cases, we arrive at the same integrand, except that Beliaev shows a global factor $(-x)^{1/2}/(\zetac/u-x)^{1/2}$ where Landau contains a factor $x^{1/2}/(x-\zetac/u)^{1/2}$; however, we can give them the common expression $(x-\textrm{i}0^+)^{1/2}/(x-\zetac/u)^{1/2}$ by means of the relations $(-x)^{1/2}=\ii(x-\textrm{i}0^+)^{1/2} \forall x<0$, $x^{1/2}=(x-\textrm{i}0^+)^{1/2} \forall x>0$, $(\zetac/u-x)^{1/2}=\ii(x-\zetac/u)^{1/2}\forall x\in\mathbb{R}$ knowing that $\im(\zetac/u)>0$. We arrive at the single integral we're looking for: 
\begin{equation}\label{eq1001}
\check{\Sigma}^{(2,2)}_{\qb}(\check{\zeta})= \frac{1}{2\ii\qb}\int_{-\qb^2/4}^{+\infty}\mathrm{d}x \frac{x\sqrt{x-\textrm{i}0^+}}{\sqrt{x-\zetac/u}\sqrt{x+\qb^2/4}} \left[\frac{1}{\eee^{\sqrt{x+\qb^2/4}-\qb/2}-1}-\frac{1}{\eee^{\sqrt{x+\qb^2/4}+\qb/2}-1}\right]
\end{equation}
 \begin{SCfigure}
\begin{tikzpicture}[x=6mm,y=6mm]
\draw[thick,->,>=stealth](-7,0)--(7,0);
\node(a)at(7,0)[below]{${\rm Re\,}x$};
\draw[thick,->,>=stealth](0,-5)--(0,6);
\node(a)at(0,6)[right]{${\rm Im\,}x$};
\node(a)at(-3,0)[]{$\bullet$};
\node(a)at(-3,-0.15)[above]{$-\bar{q}^2/4$};
\draw[ultra thick](-7,0)--(-3,0);
\draw[ultra thick](-7,0.7)--(0,0.7);
\node(a)at(0,0.7)[]{$\bullet$};
\node(a)at(0,0.7)[right]{${\rm i} 0^+$};
\draw[dashed,ultra thick](-3,0)--(-7,-4);
\draw[dashed,ultra thick](0,0.7)--(4,4.7);
\draw[ultra thick](-7,2)--(3,2);
\node(a)at(3,2)[]{$\bullet$};
\node(a)at(3,2)[below]{$\check{\zeta}/u$};
\draw[dashed,ultra thick](3,2)--(6,5);
\draw[thick,color=black,->,>=stealth](-3,0)+(-180:1)arc(-180:-135:1);
\node(a)at(-4.5,-0.5)[]{$\alpha$};
\draw[thick,color=black,->,>=stealth](0,0.7)+(180:0.8)arc(180:45:0.8);
\node(a)at(-1.5,1.3)[]{$\alpha-\pi$};
\draw[thick,color=black,->,>=stealth](3,2)+(180:0.8)arc(180:45:0.8);
\node(a)at(3.2,3)[]{$\alpha-\pi$};
\draw[thick,color=blue,->,>=stealth](-3,0-0.1)--(-5,-2-0.1);
\draw[thick,color=blue](-3,0-0.1)--(-7,-4-0.1);
\draw[thick,color=red,->,>=stealth](1,-0.1)+(-180:4)arc(-180:-135:4);
\draw[thick,color=red,->,>=stealth](1,-0.1)+(-135:4)arc(-135:0:4);
\draw[->,>=stealth,color=red,thick](5,-0.1)--(6,-0.1);
\node(a)at(-0.25,0)[below]{$0$};
\end{tikzpicture}
\caption{Graphical illustration of the method for analytically continuating $\check{\Sigma}_{\qb}^{(2,2)}(\zetac)$ from the upper half-plane $\im\zetac>0$ to the lower half-plane $\im\zetac<0$ using integral form (\ref{eq1001}). Red curve: first deformation of the integration path on variable $x$ in the complex plane. {\yct Black h}alf-lines: singularity lines of the integrand of (\ref{eq1001}) considered as a function of $x$, in thick solid line for their original position, in thick dashed line after rotation by an angle $\alpha$ or $\alpha-\pi$ as shown ($\alpha=\pi/4$ on the figure). Black discs: branch points $x_0$ of the corresponding $(x-x_0)^{1/2}$ square root functions. Blue half-line: final integration path on $x$, leading to the result sought (\ref{eq140}).}
\label{fig11}
\end{SCfigure}
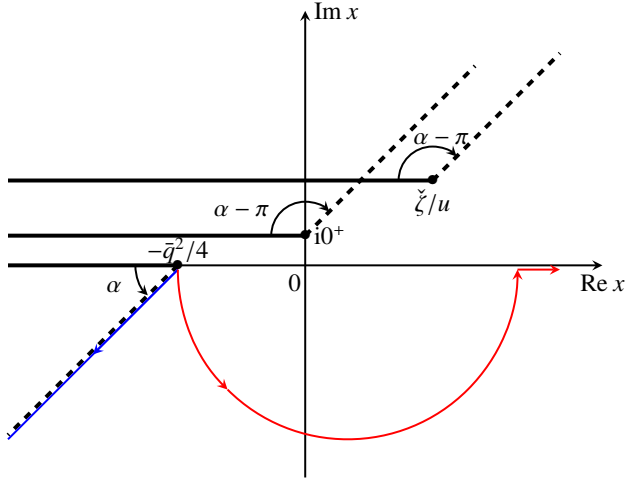
 
\paragraph{Extending it analytically}
 Integrand of (\ref{eq1001}) is an analytic function of $x$ in the complex plane, except for the three singularity lines resulting from the branch cuts of the three square roots involved: $\textrm{i}0^++\mathbb{R}^-$ in the numerator, $\zetac/u+\mathbb{R}^-$ and $-\qb^2/4+\mathbb{R}^-$ in the denominator, shown in black in Figure \ref{fig11}. \footnote{The integrand also has poles $x_n=2\textrm{i}n\pi(\qb+2\textrm{i}n\pi)$, $n\in\mathbb{Z}^*$, coming from the Bose occupation numbers ($x=0$ is not a pole because of the factor $x$ in the numerator). Future continuous deformations of the integration path in (\ref{eq1001}), shown in red and then blue in Fig.~\ref{fig11}, do not pass over any of these poles under the following conditions, which are assumed to be verified in the following (they ensure that the most dangerous pole $x_{n=-1}$ remains to the left of the blue half-line {\yc at an angle $\alpha$}): (i) $-\qb^2/4>\re x_{-1}=-(2\pi)^2$ i.e. $\qb<4\pi$, (ii) $\tan\alpha>2\pi\qb/[(2\pi)^2-\qb^2/4]$.} This allows the integration path to be deformed at will, provided it doesn't cross any singularity of the integrand, see the red line in Figure \ref{fig11}. You'd think this would be enough to make the integrand analytic in $\zetac/u$ in the region between the red line and the real axis; but it's not, because when $\zetac/u$ enters this region, the (horizontal) branch cut attached to it necessarily crosses the red line. So we have to do better, by also deforming the branch cuts of the integrand; the simplest way to do this is to introduce the new $z_\theta^{1/2}\equiv\exp(\ii\theta/2)[\exp(-\ii\theta)z]^{1/2}$ square-root determinations, parameterized by the angle $\theta\in]-\pi,\pi[$ by which the original branch cut was rotated (the $\mathbb{R}^-$ half-line being sent to $\exp(\ii\theta)\mathbb{R}^-$). As long as the branch cuts, in their rotating motion, do not sweep the integration path in red, the integrand does not change on this path and the value of the integrand is preserved. We decide to rotate the $(x+\qb^2/4)^{1/2}$ branch cut by an arbitrary angle $\alpha\in ]0,\pi/2[$, and the other two by an angle $\alpha-\pi\in]-\pi,-\pi/2[$ so that they are parallel but in the opposite direction to the first one, see the dashed lines in Figure \ref{fig11}. Finally, we take as our final integration path the blue half-line in Figure \ref{fig11}, which follows the new branch cut of $(x+\qb^2/4)^{1/2}$\footnote{We could close the integration path with a circular arc of radius $R\to +\infty$, of contribution $\to 0$ since the integrand there is exponentially small in $\sqrt{R}$.} from below and which we parameterize as follows, $x=\exp(\ii\alpha)(-v-\textrm{i}0^+)-\qb^2/4$, $v\in\mathbb{R}^+$. Then 
\begin{equation}\label{eq1002}
(x+\qb^2/4)^{1/2}_\alpha=-\ii\eee^{\ii\alpha/2}\sqrt{v} \ \  ; \ \  (x-\zetac/u)^{1/2}_{\alpha-\pi}=-\ii\eee^{\ii\alpha/2}[v+\eee^{-\ii\alpha}(\qb^2/4+\zetac/u)]^{1/2} \ \  ;\ \  (x-\textrm{i}0^+)^{1/2}_{\alpha-\pi} = -\ii\eee^{\ii\alpha/2} {\yc(}v+\eee^{-\ii\alpha}\qb^2/4{\yc)}^{1/2}
\end{equation} where all $\textrm{i}0^+$ could be omitted without difficulty in the right-hand sides. The complex number $\zetac/u$ can now be made to descend continuously into the lower half-plane, remaining constantly to the right or left of the blue half-line without the latter being crossed by the branch cut attached to $\zetac/u$: we have indeed continuated $\check{\Sigma}_{\qb}^{(2,2)}(\zetac)$ analytically to $\im\zetac<0$. It remains to set $v=\kb^2$, $\kb\in\mathbb{R}^+$, to arrive at the announced result (\ref{eq140}) with $\phi=\alpha-\pi\in]-\pi,-\pi/2[$.

\paragraph{Application}
 Now that we know from equation (\ref{eq1001}) that $\check{\Sigma}^{(2,2)}_{\qb}(\zetac)$ has no branch point at $\zetac=0$, we can calculate its value and derivative at $\zetac=\textrm{i}0^+$. For its value, the simplest way is to make $\zetac$ tend towards zero by positive imaginary parts under the integral sign in form (\ref{eq131}); we integrate over $\kb$ after expanding the occupation numbers in integer series of variable $\exp(-\kb)$, which brings up Bose functions $g_\alpha(z)$ evaluated at $z=1$ or at $z=\exp(-\qb)$ (in the latter case, they cancel out in the final result). For the derivative, the simplest way is to use the analytic continuation (\ref{eq140}), whose integrand shows no singularity on the integration path for $\zetac=0$; we can then differentiate with respect to $\zetac$ and take the limit $\zetac\to 0$ under the integral sign.\footnote{If we stick to form (\ref{eq131}), we can't take the derivative under the integral sign without making a non-absolutely convergent integral appear. We can proceed in a hybrid way after choosing an infinitesimal infrared cut $\kappa$ over $\kb$: in form (\ref{eq131}), limited to $\kb>\kappa$ in each contribution, we differentiate under the integral sign and then let $\zetac$ tend towards $0$ ; in form (\ref{eq1001}), limited to $-\kappa\qb<x<\kappa\qb$, we keep the exact expression of the factor $(x-\textrm{i}0^+)^{1/2}/(x-\zetac/u)^{1/2}$ but expand the rest of the integrand in powers of $x$ (zero order is in fact sufficient), and we analytically calculate the resulting integral for $\zetac=s+\textrm{i}0^+$, where $s$ is a real number (if necessary in terms of hypergeometric functions) then we calculate its derivative with respect to $s$ at $s=0$. The final result is the same.} We finally obtain 
\begin{equation}\label{eq141}
\check{\Sigma}^{(2,2)}_{\qb}(\check{\zeta}=\textrm{i}0^+) = \frac{1}{\ii}\left(\frac{\qb^2}{12}+2\zeta(2)\right) \quad; \quad
\frac{\dd}{\dd\check{\zeta}} \check{\Sigma}^{(2,2)}_{\qb}(\check{\zeta}=\textrm{i}0^+) = \frac{1}{u\qb} \left(\frac{\pi}{2}+\frac{\ii\qb}{4}\right)
\end{equation} The value of the function is simply a rewriting of result (\ref{eq033}) under the effect of the change of variables in (\ref{eq130},\ref{eq131}). Its derivative is involved in a well-known validity condition of the golden rule \cite{CCTbordeaux}, $|\im(\dd/\dd\zeta)\Sigma_{\qq}(\zeta=\textrm{i}0^+)|\ll 1$, which is written here as $1/(4u)\ll 1$ and imposes, unsurprisingly, $u\gg 1$. A more precise condition requires that the residue $Z$ of {\yct the} Green's function, which appears as a factor of the exponential $\exp(-\zetac_{\qb}\check{t})$ in the exact expression (\ref{eq2001}) of the signal $s^{(2,2)}(t)$, be very close to the value one taken in the golden rule, which adds the constraint $|\re(\dd/\dd\zeta)\Sigma_{\qq}(\zeta=\textrm{i}0^+)|\ll 1$, here $\qb\gg \pi/(2 u)$. As shown in Figure \ref{fig1}d, in which $u\simeq 18.14$, these conditions are not sufficient. {\ycd Sections \ref{sec3.3} and \ref{sec4} help to understand why (in essence, a singularity appears at $\zeta=\ii 0^+$ in $\Sigma_\qq^{(4)}(\zeta)$, the fourth-order contribution in $H_3$, making the present second-order calculation irrelevant in the limit $\epsilon\to 0$).}

\section{First deviation of the signal from the golden-rule exponential for the theory at second order in $H_3$}
\label{app2}

Here we establish result (\ref{eq151}) of section \ref{sec2.2}, valid in the weakly interacting limit $u\to+\infty$ {\yc at fixed $\check{t}$ and $Q=u^{1/2}\qb$ as in (\ref{eq150})}, by starting from the writing (\ref{eq132}) of the signal. To this end, guided by Fig.~\ref{fig5}b, we first apply the residue theorem to the contour formed by the union of the path $C_+$, the portions of the circle at infinity and the contour around the branch cut $-Q^2/4-\eee^{\ii(\phi+\pi)}\mathbb{R}^+$: the integral of $\exp(-\ii\check{\zeta}\check{t})\check{\mathcal{G}}_{\qb}^{(2,2)}(\zetac)$ on this contour is $2\ii\pi$ times the residue in $\zetac=\zetac_{\qb}$, the pole of {\yct the} Green's function {\yct analytically continuated} to the lower half-plane by means of (\ref{eq140}). The signal (\ref{eq132}) can therefore be written exactly as the sum of the contributions of the pole $\zetac_{\qb}$ and the contour: 
\begin{equation}\label{eq2001}
s^{(2,2)}(t) = \frac{\exp(-{\yct\ii}\zetac_{\qb}\check{t})}{1-\frac{\dd}{\dd\zetac}\check{\Sigma}^{(2,2)}_{\qb\downarrow}(\zetac_{\qb})}
+\eee^{\textrm{i}Q^2\check{t}/4} \int_0^{\eee^{\ii(\phi+\pi)}(+\infty)} \frac{\mathrm{d}z}{2\ii\pi} \frac{(\check{\Sigma}_- - \check{\Sigma}_+)\exp(\textrm{i}z \check{t})}{(-Q^2/4-z-\check{\Sigma}_-)(-Q^2/4-z-\check{\Sigma}_+)}
\end{equation} {\yc where the} complex variable $z$ goes in a straight line from $0$ to $\exp[\ii(\phi+\pi)](+\infty)$ and we note $\check{\Sigma}_{\pm}\equiv\check{\Sigma}^{(2,2)}_{\qb\downarrow}(-Q^2/4-z+\eee^{\ii(\phi+\pi)}\textrm{i}0^\pm)$ the values of the analytic continuation (\ref{eq140}) of the self-energy immediately to the left (sign $+$) or right (sign $-$) of the branch cut. We then make use of the fundamental property\footnote{The leading term is obtained by writing the difference in the left-hand side of (\ref{eq2003}) as a single integral over $\kb$ by means of (\ref{eq140}), multiplying by $u^{1/2}$, then making the change of variable $\kb=K/u^{1/2}$ before taking the limit $u\to +\infty$ under the integral sign.} 
\begin{equation}\label{eq2003}
\check{\Sigma}^{(2,2)}_{\qb\downarrow}(\zetac)-\check{\Sigma}^{(2,2)}_{\qb\downarrow}(0)\stackrel{Q\,\mbox{\scriptsize fixed}}{\underset{u\to+\infty}{=}} \frac{\eee^{\ii\phi/2}}{\textrm{i}u^{1/2}} \int_0^{+\infty} \mathrm{d}K \left\{\frac{(K^2-\eee^{-\ii\phi}Q^2/4)^{1/2}}{[K^2-\eee^{-\ii\phi}(\zetac+Q^2/4)]^{1/2}}-1\right\}+O(1/u)
\end{equation} In the contribution of the contour to (\ref{eq2001}), at the leading order in $u^{-1/2}$, we can thus replace in the denominator $\check{\Sigma}_\pm$ by $\check{\Sigma}^{(2,2)}_{\qb\downarrow}(0)\to -2\ii\zeta(2)=-\ii\pi^2/3$ according to (\ref{eq141}), and in the numerator the difference $\check{\Sigma}_--\check{\Sigma}_+$ by its asymptotic equivalent, which is written 
\begin{equation}\label{eq2004}
\check{\Sigma}_--\check{\Sigma}_+ \stackrel{Q\,\mbox{\scriptsize fixed}}{\underset{u\to+\infty}{\sim}} \frac{\eee^{\ii\phi/2}}{\textrm{i}u^{1/2}} \int_0^{+\infty} \mathrm{d}K
\left(K^2-\eee^{-\ii\phi}Q^2/4\right)^{1/2}\left[\frac{1}{(K^2-x+\textrm{i}0^-)^{1/2}}-\frac{1}{(K^2-x+\textrm{i}0^+)^{1/2}}\right]
\end{equation} knowing that $z=\eee^{\ii(\phi+\pi)}x$, $x>0$. The integrand in (\ref{eq2004}) is non-zero only for $K<x^{1/2}$, which prompts us to put $K=x^{1/2}\sin\alpha$, $\alpha\in [0,\pi/2]$, especially as this makes the square roots in the denominator disappear: we find $S_{\rm contour}(\check{t})$ in (\ref{eq152}). In the implicit equation defining the pole, rewritten by setting aside its zero-order approximation, $\zetac_{\qb}-\check{\Sigma}^{(2,2)}_{\qb\downarrow}(0)=\check{\Sigma}^{(2,2)}_{\qb\downarrow}(\zetac_{\qb})-\check{\Sigma}^{(2,2)}_{\qb\downarrow}(0)$, it suffices for the right-hand side to use (\ref{eq2003}) with $\zetac=-\ii\pi^2/3$; the convenient choice $\phi={\ycd(-\pi)}^{+}$ (on which the result does not depend) yields $\delta\zetac_{\qb}$ in (\ref{eq154}). Finally, to obtain the first deviation from one of the residue, simply differentiate (\ref{eq2003}) with respect to $\zetac$ (under the integral sign in the right-hand side), replace $\zetac$ by $-\ii\pi^2/3$ and then take $\phi={\ycd(-\pi)}^+$: {\yct one finds} $\delta\check{Z}$ in (\ref{eq154}).

\section{Double Matsubara sums in the theory at fourth order in $H_3$}
\label{app3}

Here we explicitly calculate the Matsubara sums contributing to (\ref{eq212},\ref{eq232}) in section \ref{sec3.1} and show how to obtain their simplified forms (\ref{eq213},\ref{eq233}) in the small-angle limit. The calculation relies on the generalization of relation (\ref{eq122}) to any number $J$ of factors in the summand, which is demonstrated as in footnote \ref{notedemo}: 
\begin{equation}\label{eq3000}
\sum_{n\in\mathbb{Z}} \frac{1}{\prod_{j=1}^{J} (2\ii\pi n-\vepsb_j)} = \sum_{j=1}^{J} \frac{-\bar{n}_{\veps_j}}{\prod_{\stackrel{\mbox{\scriptsize$k\!=\!1$}}{k\neq j}}^{J} (\vepsb_j-\vepsb_k)} \quad\forall J\geq 2
\end{equation}
 
\paragraph{Internal-loop diagrams}
 Let's write the sum appearing in (\ref{eq212}) after outputting a factor $(k_B T)^5$ in the denominator (we have set $\veps_{\kk_j}=k_B T\vepsb_j$) and {\yct expressing} the Matsubara energies in {\yct terms} of the independent variables $\omega=2\ii\pi k_B T n$ and $\omega'=2\ii\pi k_B T n'$ as in equations (\ref{eq210},\ref{eq211}), for a complex energy $z$ temporarily equal to $\Omega=k_B T\bar{\Omega}\in 2\ii\pi k_B T\mathbb{Z}$: 
\begin{equation}\label{eq3001}
S_{B_i}(z)=(k_BT)^{-5}\sum_{n,n'\in\mathbb{Z}} \frac{1}{s_1(\bar{\Omega}-2\ii\pi s_2 n)-\vepsb_1}\frac{1}{2\ii\pi n-\vepsb_2}\frac{1}{2\ii\pi n'-\vepsb_3}\frac{1}{2\ii\pi s_4(s_2 n - s_3 n')-\vepsb_4} \frac{1}{2\ii\pi s_2 s_5 n-\vepsb_5}
\end{equation} Using (\ref{eq122}), let's calculate the sum over $n'$ of the product of the two factors that depend on $n'$, then let's make variable $\bar{\Omega}$ disappear from the occupation numbers as in the first equality of (\ref{eq123}). After the clever change of variable $n\to s_2 n$, which doesn't affect the summation domain since $s_2=\pm 1$, we're left with 
\begin{equation}\label{eq3002}
S_{B_i}(z)= (k_BT)^{-5}\sum_{n\in\mathbb{Z}} \frac{s_1s_2s_4s_5(\bar{n}_{\veps_3}-\bar{n}_{-s_3s_4\veps_4})}{[2\ii\pi n-(\bar{\Omega}-s_1\vepsb_1)](2\ii\pi n-s_2\vepsb_2)(2\ii\pi n -s_5 \vepsb_5)[2\ii\pi n-(s_3\vepsb_3+s_4\vepsb_4)]}
\end{equation} We've taken out all the necessary factors to put the result into a simple form, as in the left-hand side of relation (\ref{eq3000}), which makes this relation easy to use: 
\begin{multline}
\label{eq3003}
S_{B_i}(z)=-(k_B T)^{-2}s_1s_2s_4s_5(\bar{n}_{\veps_3}-\bar{n}_{-s_3s_4\veps_4})\Bigg\{\frac{1}{z-z_2}\left[\frac{\bar{n}_{-s_1\veps_1}}{(z-z_1)(z-z_3)}-\frac{\bar{n}_{s_3\veps_3+s_4\veps_4}}{(z_2-z_1)(z_2-z_3)}\right] \\
+\frac{1}{z_1-z_3}\left[\frac{\bar{n}_{s_2\veps_2}}{(z_1-z)(z_1-z_2)}-\frac{\bar{n}_{s_5\veps_5}}{(z_3-z)(z_3-z_2)}\right]\Bigg\}
\end{multline}
 At this point, we have {\yct analytically continuated the result} for variable $z$ over $\mathbb{C}\setminus\mathbb{R}$ and introduced the energies of the three intermediate states (marked by the red vertical lines on the left-hand side of Figure \ref{fig7}) counted algebraically, $z_1=s_1\veps_1+s_2\veps_2$, $z_2=s_1\veps_1+s_3\veps_3+s_4\veps_4$ and $z_3=s_1\veps_1+s_5\veps_5$. In the case of $s_2=s_5$, we also have $\veps_2=\veps_5$ and $z_1=z_3$, since $\kk_2$ is then equal to $\kk_5$; this reveals a division by zero in the second contribution of (\ref{eq3003}) between curly braces, which we therefore extend by continuity with L'Hospital's rule. After some skilful rearrangement, we arrive at:\footnote{In particular, we use the identity $(z_1-z_2)^{-2}(z-z_2)^{-1}+(z_1-z_2)^{-2}(z_1-z)^{-1}+(z_1-z)^{-2}(z_1-z_2)^{-1}=(z-z_1)^{-2}(z-z_2)^{-1}$.} 
\begin{equation}\label{eq3004}
S_{B_i}^{s_2=s_5}(z)=-(k_B T)^{-2}s_1s_4(\bar{n}_{\veps_3}-\bar{n}_{-s_3s_4\veps_4})\left[\frac{\bar{n}_{-s_1\veps_1}-\bar{n}_{s_2\veps_2}}{(z-z_1)^2(z-z_2)}+\frac{\bar{n}_{s_2\veps_2}-\bar{n}_{s_3\veps_3+s_4\veps_4}}{(z_2-z_1)^2(z-z_2)}+\frac{\bar{n}_{s_2\veps_2}(1+\bar{n}_{s_2\veps_2})}{k_B T(z_2-z_1)(z_1-z)}\right]
\end{equation} Note that the global factor $(k_B T)^{-2}$ in (\ref{eq3003},\ref{eq3004}) simplifies with the factor $(-k_B T)^2$ in (\ref{eq212}). Finally, consider the small-angle limit $\epsilon\to 0$ in (\ref{eq3004}), in which $z$, $z_1$ and $z_2$ are written as $\veps_q+O(k_B T\epsilon^2)$. Between the square brackets, the second contribution is subleading because $s_3\veps_3+s_4\veps_4=s_2\veps_2+z_2-z_1=s_2\veps_2+O(k_B T\epsilon{\yc^2)}$, so its numerator tends to zero; the third is also subleading because its denominator is of order $(k_B T)^3\epsilon^4$ instead of $(k_B T)^3\epsilon^6$ like the others. We find (\ref{eq213}).

\paragraph{Bridge diagrams}
 We proceed as before, with the same notations. The double sum in (\ref{eq232}) is written 
\begin{equation}\label{eq3010}
S_{P_i}(z)=(k_B T)^{-5} \sum_{n,n'\in\mathbb{Z}} \frac{1}{2\ii\pi n-\vepsb_1} \frac{1}{s_2(\bar{\Omega}-2\ii\pi s_1 n)-\vepsb_2}\frac{1}{2\ii\pi n'-\vepsb_3}\frac{1}{s_4(\bar{\Omega}-2\ii\pi s_3 n')-\vepsb_4} \frac{1}{2\ii\pi s_5(s_1 n-s_3 n')-\vepsb_5}
\end{equation} Using relation (\ref{eq3000}) with $J=3$, we calculate the sum over $n'$, perform a partial fraction decomposition of the result for variable $2\ii\pi n$ and then make the change of variable $n\to s_1 n$, so that 
\begin{equation}\label{eq3011}
S_{P_i}(z)=\frac{(k_B T)^{-5}s_1s_2s_4s_5}{\bar{\Omega}-s_3\vepsb_3-s_4\vepsb_4}\sum_{n\in\mathbb{Z}} \frac{1}{(2\ii\pi n-s_1\vepsb_1)[2\ii\pi n-(\bar{\Omega}-s_2\vepsb_2)]} \left[\frac{\bar{n}_{\veps_3}-\bar{n}_{-s_3s_5\veps_5}}{2\ii\pi n-(s_3\veps_3+s_5\veps_5)}+\frac{\bar{n}_{-s_3s_5\veps_5}-\bar{n}_{-s_3s_4\veps_4}}{2\ii\pi n-(\bar{\Omega}+s_5\veps_5-s_4\veps_4)}\right]
\end{equation} We use (\ref{eq3000}) again and then analytically continuate over $z$ to $\mathbb{C}\setminus\mathbb{R}$; there are two terms between the square brackets in (\ref{eq3011}), so two contributions, type I and type II in that order: 
\begin{eqnarray}\label{eq3012}
S_{P_i^{\rm I}}(z)&=&\frac{(k_B T)^{-2}s_1s_2s_4s_5(\bar{n}_{\veps_3}-\bar{n}_{-s_3s_5\veps_5})}{z_2-z}\left[
\frac{\bar{n}_{s_1\veps_1}}{(z_1-z)(z_1-z_4)}+\frac{\bar{n}_{-s_2\veps_2}}{(z-z_1)(z-z_4)}+\frac{\bar{n}_{s_3\veps_3+s_5\veps_5}}{(z_4-z_1)(z_4-z)}\right] \\
\label{eq3013}
S_{P_i^{\rm II}}(z)&=&\frac{(k_B T)^{-2}s_1s_2s_4s_5(\bar{n}_{-s_3s_5\veps_5}-\bar{n}_{-s_3s_4\veps_4})}{z_2-z}\left[
\frac{\bar{n}_{s_1\veps_1}}{(z_1-z)(z_3-z)}+\frac{\bar{n}_{-s_2\veps_2}}{(z-z_1)(z_3-z_1)}+\frac{\bar{n}_{s_5\veps_5-s_4\veps_4}}{(z-z_3)(z_1-z_3)}\right]
\end{eqnarray} with the algebraic energies of the intermediate states marked by the red lines on the right-hand side of Figure \ref{fig7}, $z_1=s_1\veps_1+s_2\veps_2$ and $z_2=s_3\veps_3+s_4\veps_4$ common to both types, $z_4=s_2\veps_2+s_3\veps_3+s_5\veps_5$ for type I and finally $z_3=s_1\veps_1+s_4\veps_4-s_5\veps_5$ for type II. These variables are not independent since $z_1+z_2=z_3+z_4$. It remains to take the limit of small angles $\epsilon\to 0$, in which {\yc $z=\veps_q+O(k_B T \epsilon^2)$ and} $z_j=\veps_q+O(k_B T \epsilon^2)$, $1\leq j\leq 4$. Then $s_3\veps_3+s_5\veps_5=s_1\veps_1+O(k_BT\epsilon^2)$ and we can, at leading order, replace $\bar{n}_{s_3\veps_3+s_5\veps_5}$ by $\bar{n}_{s_1\veps_1}$ in (\ref{eq3012}); grouping by occupation number and simplifying, we find the first contribution to the right-hand side of (\ref{eq233}). Similarly, we replace $\bar{n}_{s_5\veps_5-s_4\veps_4}$ by $\bar{n}_{-s_2\veps_2}$ at leading order in (\ref{eq3013}), which reproduces the second contribution to (\ref{eq233}).

\section{A linear divergence of $\tilde{\Sigma}_{\qb}^{(4,2)}(\zetat)$ away from the origin}
\label{app4}

As we show in this appendix, the $\tilde{\Sigma}_{\qb}^{(4,2)}(\zetat)$ component of the self-energy at fourth-order in $H_3$ and second-order in $\epsilon$, obtained by summing all the contributions (\ref{eq220}), (\ref{eq240}), (\ref{eq242}) of the fourth-order diagrams in section \ref{sec3.1}, diverges linearly when $\zetat\to\zetat_0\equiv -3\gamma\qb^3/32$ (here by positive imaginary parts): 
\begin{equation}\label{eq4000}
\tilde{\Sigma}_{\qb}^{(4,2)}(\zetat) \stackrel{\im\zetat>0}{\underset{\zetat\to \zetat_0}{\sim}} \frac{\qb^3(\bar{n}^{\rm lin}_{q/2}+{1}/{2})}{24\ii\pi^2\gamma(\zetat-\zetat_0)} \left(\frac{9(1+\Lambda)^2}{8\rho\xi^2}\right)^2\left[\int_0^{\qb/2} \dd\kb' \kb'(\kb'-\qb/2)(\bar{n}^{\rm lin}_{k'}-\bar{n}^{\rm lin}_{k'-{q}/{2}})+2\int_{\qb/2}^{\eta} \dd\kb'\kb'(\kb'-\qb/2)(\bar{n}^{\rm lin}_{k'}-\bar{n}^{\rm lin}_{k'-{q}/{2}})\right]
\end{equation} This results from two phenomena. (i) At a fixed reduced wavenumber $\kb$, the angular integral (\ref{eq221}) in inner-loop diagrams diverges at $\zetat=A_1^B$ with exponent $3/2$ because, for this value of $\zetat$, the denominator of the integrand has $\thetat=0$ as quadruple root. To see this quantitatively, we can use the identity $\int_{\mathbb{R}}\dd\thetat/(z-\thetat^2)^2=-\ii\pi\,{\ycd\sig(\im z)}/(2 z^{3/2}), \forall z\in\mathbb{C}\setminus\mathbb{R}$, with $z=(\zetat-A_1^B)/B_1^B$. (ii) This $(\zetat-A_1^B)^{-3/2}$ {\yc divergence} occurs at a position $A_1^B$ depending on $\kb$, as shown in equation (\ref{eq222}), so it is generally washed out by integration over $\kb$ at fixed $\zetat$ in (\ref{eq220}), unless $\zetat$ coincides with a parabolic extremum of $A_1^B(\kb)$ on the integration domain $\mathcal{D}_1^B$. To see this, let's choose an arbitrary reduced wavenumber $\kb_0$ in $\mathcal{D}_1^B$, let's set $\zetat=w+A_1^B(\kb_0)$ and take the limit $w\to 0$ by positive imaginary parts in (\ref{eq220}): the result can only diverge if the value of the integral 
\begin{equation}\label{eq4005}
\sigma_{\kb_0}(w)=\int_{\kb_0-\veps}^{\kb_0+\veps} \frac{\dd\kb}{[w+A_1^B(\kb_0)-A_1^B(\kb)]^{3/2}}
\end{equation} diverges in this limit, where $\veps>0$ is fixed. Since we can take $\veps$ to be as small as we like without changing the conclusion, we can replace $A_1^B(\kb)-A_1^B(\kb_0)$ in (\ref{eq4005}) by its Taylor expansion in $\kb_0$ to the non-zero leading order. If $(\mathrm{d}A_1^B/\mathrm{d}\kb)(\kb=\kb_0)\neq 0$, an expansion to order one is sufficient, and there is no divergence because $\int_{\mathbb{R}} \mathrm{d}\kappa/(w-\kappa)^{3/2}=0$. \footnote{If $\kb_0$ lies on the edge of $\mathcal{D}_1^B$, the neighborhood of $\kb_0$ reduces to $[\kb_0,\kb_0+\veps]$ or $[\kb_0-\veps,\kb_0]$ and there is square-root divergence because $\int_0^{+\infty} \mathrm{d}\kappa/(w-\kappa)^{3/2}=-2/w^{1/2}$. This divergence is subleading and ignored here.} Otherwise, expansion to second order is required, and there is linear divergence $[\zetat-A_1^B(\kb_0)]^{-1}$ as predicted by (\ref{eq4000}) because 
\begin{equation}\label{eq4001}
\int_{\mathbb{R}} \frac{\mathrm{d}\kappa}{(w-\kappa^2)^{3/2}} = \frac{2}{\textrm{i}w} \quad\mbox{and}\quad \int_{\mathbb{R}} \frac{\mathrm{d}\kappa}{(w+\kappa^2)^{3/2}} = \frac{2}{w}\quad\forall w\in \mathbb{C}, \im w>0
\end{equation}
 Let's go into more detail on the example of diagram $B_1$ in Figure \ref{fig8}, in which the signs $s_i$ of the internal line orientations are all $1$, the integration domain on $\kb$ is $\mathcal{D}_1^B=[0,\qb]$ and that on $\kb'$ at fixed $\kb$ is $\mathcal{D}_2^B(\kb)=[0,\kb]$. Note that $A_1^B(\kb)$ reaches its minimum at $\kb_0=\qb/2$, {\yct in the interior of} $\mathcal{D}_1^B$ and therefore with zero derivative, and that this minimum is precisely $\zetat_0=-3\gamma\qb^3/32$, as in (\ref{eq4000}). To obtain an equivalent of $\tilde{\Sigma}_{B_1}^{(4,2)}(\qb,\zetat)$ when $\zetat\to\zetat_0$, we can treat $\thetat^2$, $(\kb-\qb/2)^2$ and $w=\zetat-\zetat_0$ as infinitesimals of the same order. So, in the angular integral (\ref{eq221}), we can replace $\zetat$ by $\zetat_0+\textrm{i}0^+$ and $\thetat^2$ by $0$ in the second factor in the denominator, and everywhere replace $\kb$ by $\kb_0=\qb/2$ except in coefficient $A_1^B(\kb)$, which is written as $A_1^B(\kb)=\tilde{\zeta}_0+(3\gamma\qb/8)(\kb-\qb/2)^2$. After calculating the integral over $\phi$ and then over $\thetat$ using the residue theorem, it remains
\begin{equation}\label{eq4002}
I(\kb,\kb') \sim \frac{-8\pi^2}{(6\gamma\qb^3)^{1/2} \kb'}\frac{1}{[w-(3\gamma/8)\qb(\kb-\qb/2)^2]^{3/2}}
\end{equation} which can be inserted in the integrand of (\ref{eq220}), replacing $\kb$ by $\kb_0=\qb/2$ in all other factors and in the upper bound of integration over $\kb'$. {\yc We then change the order of integration.} In the {\yc now internal} integral over $\kb$ at fixed $\kb'$, it only remains
\begin{equation}\label{eq4003}
\int_0^{\qb} \frac{\mathrm{d}\kb}{[w-(3\gamma/8)\qb(\kb-\qb/2)^2]^{3/2}}\sim \left(\frac{8}{3\gamma\qb}\right)^{1/2}\int_{-\infty}^{+\infty} \frac{\dd\kappa}{(w-\kappa^2)^{3/2}}=\left(\frac{8}{3\gamma\qb}\right)^{1/2} \frac{2}{\textrm{i}w}
\end{equation} where we have made the change of variable $\kb=\qb/2+[8/(3\gamma\qb)]^{1/2}\kappa$, extended the integration bounds over $\kappa$ to $\pm\infty$ without changing the leading behavior and used equation (\ref{eq4001}). We finally obtain the divergent part of diagram $B_1$: 
\begin{equation}\label{eq4004}
\tilde{\Sigma}_{B_1}^{(4,2)}(\qb,\zetat) \stackrel{\im\zetat>0}{\underset{\zetat\to \zetat_0}{\sim}} \frac{\qb^3(\bar{n}^{\rm lin}_{q/2}+{1}/{2})}{24\ii\pi^2\gamma(\zetat-\zetat_0)} \left(\frac{9(1+\Lambda)^2}{8\rho\xi^2}\right)^2\int_0^{\qb/2} \dd\kb' \kb'(\kb'-\qb/2)(\bar{n}^{\rm lin}_{k'}-\bar{n}^{\rm lin}_{k'-{q}/{2}})
\end{equation} Diagram $B_2$ is treated in the same way and gives a similar divergent contribution (with a two times larger prefactor and the integral over $\kb'$ from $\qb/2$ to $\eta$). Diagrams $B_3$ and $B_4$ are subleading in $\epsilon$ because $s_5\neq s_2$ and must be omitted here. Diagrams $B_5$ and $B_6$ lead to a strictly increasing coefficient $A_1^B(\kb)=(3\gamma/8)\qb\kb(\kb+\qb)$ on the integration interval $\kb\in[0,\eta-\qb]$ and therefore cannot contribute to the linear divergence (\ref{eq4000}). Diagrams $B_7$ and $B_8$ have the same problem, with $A_1^B(\kb)=(3\gamma/8)\qb\kb(\kb-\qb)$ and $\mathcal{D}_1^B=[\qb,\eta]>\qb/2$. It remains to combine the contributions of $B_1$ and $B_2$ to obtain (\ref{eq4000}), whose expression between square brackets in the quantum field case tends towards $-\qb(16\pi^2+\qb^2)/48$ when $\eta\to+\infty$. 

Finally, let's explain why bridge diagrams do not contribute to the linear divergence (\ref{eq4000}) on the example of diagram $P_1$ (all $s_i=1$, $\mathcal{D}_1^P=[0,\qb]$ and $\mathcal{D}_2^P(\kb)=[0,\kb]$, type II). As noted in section \ref{sec3.3}, the angular integral (\ref{eq243}) diverges linearly at {\yc some} $\zetat=X_0(\kb,\kb')$ if the cancellation loci of the three factors in the denominator of (\ref{eq243}) intersect at the same point in the $(\tilde{\theta},\tilde{\theta}')$ plane. The triple intersection condition {\ycd in the general case $X_0\neq 0$} is obtained by replacing all coefficients $A_i^P$ by $A_i^P-X_0$ in (\ref{eq281}); the resulting second-degree equation on $X_0$ has the obvious zero solution, to which section \ref{sec3.3} was implicitly restricted, and a second real root, $X_0(\kb,\kb')=-3\gamma \kb(\kb-\qb)\kb'(\kb'-\qb)/(2\qb)$ ; the latter leads to a triple intersection in the plane $(\tilde{\theta},\tilde{\theta}')$, at points with coordinates $\pm[\sqrt{3\gamma}/(2\qb)]((\qb-\kb)(\qb-2\kb'),-(\qb-2\kb)(\qb-\kb'))$, if we add constraints $\kb>\qb/2$ and $\kb'<\qb/2$ to have the right signs in the coefficients of the small deviations $(\tau,\tau')$ at these points (otherwise, as in (\ref{eq282}), there is no divergence). Then, proceeding as in (\ref{eq282}), we obtain the divergent part of the angular integral (for $\zetat\to X_0$ by positive imaginary parts), 
\begin{equation}\label{eq4006}
K_{\rm div}(\kb,\kb') = \frac{-32\pi^2}{3\gamma\qb\kb\kb'(\kb-\kb')[\zetat-X_0(\kb,\kb')]}
\end{equation} to be integrated on $(\kb,\kb')\in[\qb/2,\qb]\times[0,\qb/2]$ after multiplication by the same function of $(\kb,\kb')$ as in (\ref{eq242}). Now $X_0(\kb,\kb')$ has a parabolic minimum, located at $(\kb,\kb')=(\qb/2,\qb/2)$ and precisely equal to $\zetat_0$! The singularity (\ref{eq4006}) will therefore survive the wave number integration if $\zetat\to \zetat_0$ as in (\ref{eq4000}). Let's then perform the change of variables $(\kb=\kappa+\qb/2,\kb'=-\kappa'+\qb/2)$ and treat $w=\zetat-\zetat_0$, $\kappa^2$ and $\kappa'^2$ as infinitesimals of the same order. By writing the function multiplying $[\zetat-X_0(\kb,\kb')]^{-1}$ to the leading order in $\kappa$ and $\kappa'$ and replacing the pole $X_0(\kb,\kb')$ by its quadratic approximation in $\kappa$ and $\kappa'$ near the minimum, we obtain the equivalent 
\begin{equation}\label{eq4008}
\tilde{\Sigma}^{(4,2)}_{P_1^{\rm II}}(\qb,\zetat)\stackrel{\im w>0}{\underset{w\to 0}{\sim}} (\qb/2)^{\ycd 4} \left(\frac{9(1+\Lambda)^2}{8\rho\xi^2}\right)^2 (2\bar{n}^{\rm lin}_{q/2}+1)\frac{(-32\pi^2)}{3\gamma(2\pi)^4} \int_0^{\qb/2}\dd\kappa \int_0^{\qb/2} \dd\kappa' \frac{1}{\kappa+\kappa'} \frac{1}{w-(3\gamma\qb/8)(\kappa^2+\kappa'^2)}
\end{equation} We can extend the integration bounds $\qb/2$ to $+\infty$ without losing equivalence. The double integral is then easily calculated in polar coordinates $(K,\alpha)$, where $K=(\kappa^2+\kappa'^2)^{1/2}$ spans $\mathbb{R}^+$ and $\alpha$ spans $[0,\pi/2]$. We finally find a square-root divergence, $\propto (\zetat-\zetat_0)^{-1/2}$, subleading in (\ref{eq4000}).

\section{The long-time behavior of the signal in the $(\rho\xi^2)^{-1}\to 0$ limit of non-perturbative theories}
\label{app5}

Here we obtain the asymptotic behavior (\ref{eq314}) of the signal associated with the self-energy $\check{\Sigma}_{{\rm corr}\,\infty}^{(2,2)}(\qb,\check{\zeta})$ of equation (\ref{eq311}). To this end, we apply the technique of \ref{app2} by folding back the integration path $C_+$ in (\ref{eq012}) onto the branch cut of $\check{\Sigma}_{{\rm corr}\,\infty}^{(2,2)}(\qb,\check{\zeta})$ located on $\ii\mathbb{R}^{-}$ and described parametrically by $\zeta=-\ii\check{y}$, $\check{y}\geq 0$ (this branch cut is the set of values of $\zetac$ cancelling the denominator of the integrand in (\ref{eq311})). Transposing equation (\ref{eq2001}) to the present case, we obtain 
\begin{equation}\label{eq5001}
s^{(2,2)}_{{\rm corr}\,\infty}(t)\underset{\check{t}\to+\infty}{\sim}\int_0^{+\infty}\frac{\dd\check{y}}{2\pi}\frac{\delta\sigma_{\qb}(\check{y})\exp(-\check{y}\check{t})}{\left[\check{\Sigma}_{{\rm corr}\,\infty}^{(2,2)}(\qb,0)\right]^2}\quad\mbox{where }\quad
\delta\sigma_{\qb}(\check{y})=\check{\Sigma}_{{\rm corr}\,\infty}^{(2,2)}(\qb,0^+-\ii\check{y})-\check{\Sigma}_{{\rm corr}\,\infty}^{(2,2)}(\qb,0^--\ii\check{y})
\end{equation} is the discontinuity of the self-energy across its branch cut at distance $\check{y}$ from the origin. The identity in the sense of distributions $1/(\check{y}+\textrm{i}0^+)=\mathrm{v.p.}(1/\check{y})-\ii\pi\delta(\check{y})$ leads to 
\begin{equation}\label{eq5002}
\delta\sigma_{\qb}(\check{y})=-\frac{4\pi}{\qb^2}\int_{\qb}^{\eta}\mathrm{d}\bar{k}\int_0^{\eta-\kb}\dd\kb' \bar{k}\kb' (\kb-\qb)(\kb'+\qb) (\bar{n}_{k'+q}^{\rm lin}-\bar{n}_{k'+k}^{\rm lin})(\bar{n}_{k-q}^{\rm lin}-\bar{n}_k^{\rm lin})
\delta\left(\check{y}-\kb'(\kb-\qb)\frac{\check{\Gamma}_{\kb-\qb}+\check{\Gamma}_{\kb'}}{2\qb(\kb+\kb')}\right)
\end{equation} In the $\check{t}\to+\infty$ limit, the exponential function in (\ref{eq5001}) restricts the integral to arbitrarily small values of $\check{y}$ and it suffices to know the leading behavior of $\delta\sigma_{\qb}(\check{y})$ when $\check{y}\to 0^+$. Since the reduced rates $\check{\Gamma}_{\bar{K}}$ cannot tend to zero, the argument of the Dirac distribution in (\ref{eq5002}) can be zero in the limit $\check{y}\to 0$ only if $\kb'$ or $\kb-\qb$ is infinitesimal. For an arbitrary choice of a bound $\veps$ that is not infinitesimal but $\ll 1$, we can then restrict the integration domain on $\kb$ and $\kb'$ to the two parts, (i) $\kb\in[\qb,\qb+\veps]$, $\kb'\in[0,\eta-\kb]$ and (ii) $\kb\in [\qb+\veps,\eta-\veps]$, $\kb'\in[0,\veps]$. \footnote{We note in part (ii) that, if $\kb\leq\eta-\veps$, we have $\veps\leq\eta-\kb$ so $\kb'$, which must remain $\leq \eta-\kb$, can indeed go as far as $\veps$. We also check that \g{missing bit} $\kb\in[\eta-\veps,\eta]$, $\kb'\in[0,\veps]$ gives a negligible contribution $O(\check{y}\veps)$ to $\delta\sigma_{\qb}(\check{y})$.} In part (i) of the integration domain, we have $\kb-\qb\leq\veps\ll 1$ so we can replace the various quantities depending on $\kb$ in (\ref{eq5002}) by their Taylor expansion at $\kb=\qb$ to leading order, e.g. $(\kb-\qb)(\bar{n}^{\rm lin}_{k-q}-\bar{n}^{\rm lin}_{k})\simeq 1$, $\bar{n}_{k'+q}^{\rm lin}-\bar{n}_{k'+k}^{\rm lin}\simeq(\kb-\qb)(-\dd/\dd\kb')\bar{n}_{k'+q}^{\rm lin}$ {\yct or} $\check{\Gamma}_{\kb-\qb}\simeq \check{\Gamma}_0$ and the {\yc upper} integration bound on $\kb'$ can be replaced by $\eta-\qb$. The argument of the Dirac distribution in (\ref{eq5002}) becomes an affine function of $\kb-\qb$, making it easy to integrate over $\kb$: 
\begin{equation}\label{eq5003}
\delta\sigma_{\qb}^{\rm (i)}(\check{y}) \simeq -16\pi\qb\check{y}\int_0^{\eta-\bar{q}} \dd\kb'\frac{(\kb'+\qb)^3}{(\check{\Gamma}_0+\check{\Gamma}_{\kb'})^2}\left(-\frac{\dd}{\dd\kb'}\bar{n}_{k'+q}^{\rm lin}\right) Y\left(\veps-\frac{2(\kb'+\qb)\qb\check{y}}{\kb'(\check{\Gamma}_0+\check{\Gamma}_{\kb'})}\right) \simeq -16\pi\qb\check{y}\int_{\qb^2\check{y}/\veps\check{\Gamma}_0}^{\eta-\qb}\frac{\dd\kb'}{\kb'} \frac{(\kb'+\qb)^3}{(\check{\Gamma}_0+\check{\Gamma}_{\kb'})^2} \left(-\frac{\dd}{\dd\kb'}\bar{n}_{k'+q}^{\rm lin}\right)
\end{equation} The form of the {\yct second integral in} (\ref{eq5003}) is based on the following reasoning: if $\check{y}\to 0$ at fixed $\veps$, the argument of the Heaviside function $Y$ in the {\yct first integral} is positive, unless $\kb'\to 0$; at the edge of the support of this function $Y$, we therefore have $\kb'\ll 1$, which allows us to replace in its argument the second term by its leading-order approximation in $\kb'$; this brings up $Y(\veps-\qb^2\check{y}/\kb'\check{\Gamma}_0)$, hence the lower bound announced in the integration over $\kb'$. In part (ii) of the integration domain, we proceed in the same way: since $\kb'\leq\veps\ll 1$, we replace the various quantities in (\ref{eq5002}) by their leading order in $\kb'$. As the argument of the Dirac distribution becomes an affine function of $\kb'$, we integrate over this variable; the resulting function $Y(\veps-\frac{2\qb\kb\check{y}}{(\kb-\qb)(\check{\Gamma}_{\kb-\qb}+\check{\Gamma}_0)})$ is inoperative when $\check{y}\to 0$ because $\kb-\qb$ is here $\geq\veps$, hence 
\begin{equation}\label{eq5004}
\delta\sigma_{\qb}^{\rm (ii)}(\check{y}) \simeq -16\pi\qb\check{y} \int_{\qb+\veps}^{\eta-\veps} \frac{\dd\kb}{\kb-\qb}\frac{\kb^3}{(\check{\Gamma}_{\kb-\qb}+\check{\Gamma}_0)^2} (\bar{n}_q^{\rm lin}-\bar{n}_k^{\rm lin})(\bar{n}_{k-q}^{\rm lin}-\bar{n}_k^{\rm lin})
\end{equation} It remains to take the limit $\veps\to 0$ in results (\ref{eq5003},\ref{eq5004}). To do this, we separate each integrand into its infrared divergent part $\propto 1/\kb'$ or $\propto 1/(\kb-\qb)$ and its regular part.\footnote{To calculate the coefficient of the divergent part $1/(\kb-\qb)$, we use the fact that $\lim_{\kb\to\qb}(\bar{n}_q^{\rm lin}-\bar{n}_k^{\rm lin})(\bar{n}_{k-q}^{\rm lin}-\bar{n}_k^{\rm lin})=(-\dd/\dd\qb)\bar{n}_q^{\rm lin}.$
} In the integral of the regular parts, we can make $\veps$ tend to zero within the integration bounds without damage. In the integral of divergent parts, doing the same gives rise to dangerous contributions $\pm\ln\veps$, but which exactly cancel out in the sum of parts (i) and (ii). Finally, 
\begin{equation}\label{eq5005}
\delta\sigma_{\qb}(\check{y})\underset{\check{y}\to 0^+}{=}\frac{4\pi\qb^4}{\check{\Gamma}_0^2} \left(-\frac{\dd}{\dd\qb}\bar{n}^{\rm lin}_{{\yct q}}\right) \check{y}\ln\check{y} + C_{\qb} \check{y} + O(\check{y}^2\ln\check{y})
\end{equation} For the sake of brevity, we do not give the expression of the coefficient $C_{\qb}$ of the linear part in (\ref{eq5005}), to which our method nevertheless gives access. The form of $O(\ldots)$ in (\ref{eq5005}), on the other hand, has been obtained numerically. Insertion of the leading term $\propto\check{y}\ln\check{y}$ in (\ref{eq5001}) leads to the announced result (\ref{eq314}).



\begin{thebibliography}{99}
\bibitem{LK} L. Landau, I. Khalatnikov, \g{Teoriya vyazkosti Geliya-II}, Zh. Eksp. Teor. Fiz. {\bf 19}, 637 (1949).
\bibitem{SonWingate} D.T. Son, M. Wingate, \href{https://doi.org/10.1016/j.aop.2005.11.001}{\g{General coordinate invariance and conformal invariance in nonrelativistic physics: Unitary Fermi gas}}, Ann. Physics {\bf 321}, 197 (2006).
\bibitem{SDM} S. Van Loon, C.A.R. Sá de Melo, \href{https://doi.org/10.1103/PhysRevLett.131.113001}{\g{Effects of quantum fluctuations on the low-energy collective modes of two-dimensional superfluid Fermi gases from the BCS to the Bose Limit}}, {\ycd Phys. Rev. Lett. {\bf 131}, 113001 (2023)}.
\bibitem{livreK} I. Khalatnikov, \textit{An Introduction to the Theory of Superfluidity} (CRC Press-Taylor \& Francis, Boca Raton, 2018).
\bibitem{Annalen} H. Kurkjian, Y. Castin, A. Sinatra, \href{https://hal.archives-ouvertes.fr/hal-01392846}{\g{Three-phonon and four-phonon interaction processes in a pair-condensed Fermi gas}}, Annalen der Physik {\bf 529}, 1600352 (2017).
\bibitem{Maris} H.J. Maris, \href{https://doi.org/10.1103/RevModPhys.49.341}{Phonon-phonon interactions in liquid helium}, Rev. Mod. Phys. {\bf 49}, 341 (1977).
\bibitem{FW} A.L. Fetter, J.D. Walecka, \textit{Quantum Theory of Many-Particle Systems} (Dover, Mineola, 2003).
\bibitem{Grynberg} D.R. Meacher, D. Boiron, H. Metcalf, C. Salomon, G. Grynberg, \href{https://doi.org/10.1103/PhysRevA.50.R1992}{\g{Method for velocimetry of cold atoms}}, Phys. Rev. A {\bf 50}, R1992(R) (1994).
\bibitem{Ketterle} J. Stenger, S. Inouye, A.P. Chikkatur, D.M. Stamper-Kurn, D.E. Pritchard, W. Ketterle, \href{https://link.aps.org/doi/10.1103/PhysRevLett.82.4569}{\g{Bragg spectroscopy of a Bose-Einstein condensate}}, Phys. Rev. Lett. {\bf 82}, 4569 (1999).
\bibitem{Davidson} J. Steinhauer, R. Ozeri, N. Katz, N. Davidson, \href{https://link.aps.org/doi/10.1103/PhysRevLett.88.120407}{\g{Excitation spectrum of a Bose-Einstein condensate}}, Phys. Rev. Lett. {\bf 88}, 120407 (2002).
\bibitem{Vale} G. Veeravalli, E. Kuhnle, P. Dyke, C.J. Vale, \href{https://link.aps.org/doi/10.1103/PhysRevLett.101.250403}{\g{Bragg spectroscopy of a strongly interacting Fermi gas}}, Phys. Rev. Lett. {\bf 101}, 250403 (2008).
\bibitem{Cartago} A. Sinatra, C. Lobo, Y. Castin, \href{https://doi.org/10.1088/0953-4075/35/17/301}{\g{The truncated Wigner method for Bose condensed gases: limits of validity and applications}}, J. Phys. B {\bf 35}, 3599 (2002).
\bibitem{CCTbordeaux} C. Cohen-Tannoudji, J. Dupont-Roc, G. Grynberg, \textit{Processus d'interaction entre photons et atomes} (EDP Sciences/CNRS Éditions, Paris, 1988).
\bibitem{Barc} {\yct M.A. Escobedo, C. Manuel, \href{https://doi.org/10.1103/PhysRevA.82.023614}{\g{Effective field theory and dispersion law of the phonons of a nonrelativistic superfluid}}, Phys. Rev. A {\bf 82}, 023614 (2010).}
\bibitem{AK} A. Andreev, I.M. Khalatnikov, \href{http://jetp.ras.ru/cgi-bin/dn/e_017_06_1384.pdf}{\g{Sound in Liquid Helium II Near Absolute Zero}}, Zh. Eksp. Teor. Fiz. {\bf 44}, 2058 (1963) [Sov. Phys. JETP {\bf 17}, 1384 (1963)].
\bibitem{CB} Ming-Chiang Chung, A.B. Bhattacherjee, \href{https://doi.org/10.1088/1367-2630/11/12/123012}{\g{Damping in 2D and 3D dilute Bose gases}},  New J. Phys. {\bf 11}, 123012 (2009).
\bibitem{HM} M. Bohlen, L. Sobirey, N. Luick, H. Biss, T. Enss, Th. Lompe, H. Moritz,\href{https://journals.aps.org/prl/abstract/10.1103/PhysRevLett.124.240403}{\g{Sound Propagation and Quantum-Limited Damping in a Two-Dimensional Fermi Gas}}, Phys. Rev. Lett. {\bf 124}, 240403 (2020).
\bibitem{PCZH} P. Christodoulou, M. Gałka, N. Dogra, R. Lopes, J. Schmitt, Z. Hadzibabic, \href{https://doi.org/10.1038/s41586-021-03537-9}{\g{Observation of first and second sound in a BKT superfluid}}, Nature {\bf 594}, 191 (2021).
\bibitem{JDJB} J.L. Ville, R. Saint-Jalm, \'E. Le Cerf, M. Aidelsburger, S. Nascimb\`ene, J. Dalibard, J. Beugnon, \href{https://doi.org/10.1103/PhysRevLett.121.145301}{\g{Sound Propagation in a Uniform Superfluid Two-Dimensional Bose Gas}}, Phys. Rev. Lett. {\bf 121}, 145301 (2018).
\bibitem{Salasnich} A. Cappellaro, F. Toigo, L. Salasnich, \href{https://doi.org/10.1103/PhysRevA.98.043605}{\g{Collisionless dynamics in two-dimensional bosonic gases}}, Phys. Rev. A {\bf 98}, 043605 (2018).
\bibitem{Stringari} Miki Ota, F. Larcher, F. Dalfovo, L. Pitaevskii, N.P. Proukakis, S. Stringari, \href{https://doi.org/10.1103/PhysRevLett.121.145302}{\g{Collisionless Sound in a Uniform Two-Dimensional Bose Gas}}, Phys. Rev. Lett. {\bf 121}, 145302 (2018).
\bibitem{Bere} V.L. Berezinskii, \href{http://jetp.ras.ru/cgi-bin/dn/e_034_03_0610.pdf}{\g{Destruction of long-range order in one-dimensional and two-dimensional systems having a continuous symmetry group II. Quantum systems}}, Zh. Eksp. Teor. Fiz. {\bf 61}, 1144 (1971) [Sov. Phys. JETP {\bf 34}, 610 (1972)].
\bibitem{Nel} D.R. Nelson, J.M. Kosterlitz, \href{https://doi.org/10.1103/PhysRevLett.39.1201}{\g{Universal Jump in the Superfluid Density of Two-Dimensional Superfluids}}, Phys. Rev. Lett. {\bf 39}, 1201 (1977).
\bibitem{SJ} R. Saint-Jalm, \href{https://theses.hal.science/tel-03116129}{\it Exploration de la physique à deux dimensions avec des gaz de Bose dans des potentiels à fond plat: ordre en phase et symétrie dynamique}, PhD thesis of École normale supérieure and PSL University (Paris, October 2019).
\bibitem{broui} H. Kurkjian, Y. Castin, A. Sinatra, \href{https://doi.org/10.1016/j.crhy.2016.02.005}{Brouillage thermique d'un gaz cohérent de fermions}, Comptes Rendus Physique {\bf 17}, 789 (2016).
\bibitem{CD} Y. Castin, R. Dum, \href{https://doi.org/10.1007/s100530050584}{\g{Bose-Einstein condensates with vortices in rotating traps}}, Eur. Phys. J. D {\bf 7}, 399 (1999).
\bibitem{HD} Z. Hadzibabic, J. Dalibard, \href{https://doi.org/10.1393/ncr/i2011-10066-3}{\g{Two-dimensional Bose fluids: An atomic physics perspective}}, Rivista del Nuovo Cimento {\bf 34}, 389 (2011).
\bibitem{Popov} V.N. Popov, {\it Functional Integrals in Quantum Field Theory and Statistical Physics} (Reidel, Dordrecht, 1983).
\bibitem{Morathese} C. Mora, \href{https://theses.hal.science/tel-00005472v1}{\it Gaz de bosons et de fermions condens\'es~: phases de Fulde-Ferrell-Larkin-Ovchinnikov et quasicondensats}, PhD thesis of Universit\'e Paris VI (Paris, March 2004).
\bibitem{MC2} C. Mora, Y. Castin, \href{https://doi.org/10.1103/PhysRevLett.102.180404}{\g{Ground state energy of the two-dimensional weakly interacting Bose gas: First correction beyond Bogoliubov theory}}, Phys. Rev. Lett. {\bf 102}, 180404 (2009). 
\bibitem{Petrov} {\yc D.S. Petrov, G.V. Shlyapnikov, \href{https://doi.org/10.1103/PhysRevA.64.012706}{\g{Interatomic collisions in a tightly confined Bose gas}}, Phys. Rev. A {\bf 64}, 012706 (2001).}
\bibitem{Olshanii} {\yc L. Pricoupenko, M. Olshanii, \href{https://doi.org/10.1088/0953-4075/40/11/009}{\g{Stability of two-dimensional Bose gases in the resonant regime}}, J. Phys. B {\bf 40}, 2065 (2007).}
\bibitem{Ludovic} {\yc L. Pricoupenko, \href{https://doi.org/10.1103/PhysRevA.83.062711}{\g{Isotropic contact forces in arbitrary representation: Heterogeneous few-body problems and low dimensions}}, Phys. Rev. A {\bf 83}, 062711 (2011).}
\bibitem{TheseAlan} A. Serafin, \href{https://theses.hal.science/tel-04002097}{\textit{Deux études de propriétés collectives de systèmes quantiques}}, PhD thesis of Sorbonne University (Paris, December 2022).
\bibitem{NumRec} W.H. Press, S.A. Teukolsky, W.T. Vetterling, B.P.  Flannery, {\it Numerical Recipes} (Cambridge University Press, Cambridge, 1988).
\bibitem{SCE} A. Sinatra, Y. Castin, E. Witkowska, \href{https://doi.org/10.1103/PhysRevA.75.033616}{\g{Nondiffusive phase spreading of a Bose-Einstein condensate at finite temperature}}, Phys. Rev. A {\bf 75}, 033616 (2007).
\end{thebibliography}
\end{document}


\begin{frontmatter}


\selectlanguage{french}
\title{Amortissement des phonons dans un superfluide 2D : insuffisance de la règle d'or de Fermi à basse température / {\it Phonon damping in a 2D superfluid: insufficiency of Fermi's golden rule at low temperature}}


\author{Yvan Castin}
\ead{yvan.castin@lkb.ens.fr}
\author{Alan Serafin}
\ead{a.serafin@protonmail.com}
\author{Alice Sinatra}
\ead{alice.sinatra@lkb.ens.fr}

\address{Laboratoire Kastler Brossel, ENS-Universit\'e PSL, CNRS, Universit\'e Sorbonne et Coll\`ege de France, 24 rue Lhomond, 75231 Paris, France}



\begin{abstract}
Il est en général admis que le gaz de phonons d'un superfluide entre toujours dans un régime de couplage faible à suffisamment basse température, quelle que soit la force des interactions entre les particules sous-jacentes (constitutives du superfluide). Ainsi, dans cette limite, on devrait pouvoir toujours calculer le taux d'amortissement des phonons thermiques en appliquant la règle d'or de Fermi à l'hamiltonien $H_3$ de couplage cubique phonon-phonon tiré de l'hydrodynamique quantique, du moins dans le cas d'une branche acoustique convexe et dans le régime faiblement collisionnel (où la pulsation propre des phonons considérés reste très supérieure au taux de thermalisation du gaz). À l'aide de la méthode des fonctions de Green à $N$ corps, nous prédisons que, contrairement au cas tridimensionnel et de manière inattendue, ceci n'est pas vrai en dimension deux. Nous confirmons cette prédiction par des simulations de champ phononique classique et par une théorie non perturbative en $H_3$, qui régularise l'ordre quatre en donnant à la main une énergie complexe aux phonons virtuels des processus collisionnels à quatre phonons. Pour un fluide en interaction faible et un mode de phonon dans la limite des grandes longueurs d'onde, nous prédisons un taux d'amortissement environ trois fois plus faible que celui de la règle d'or. {\yc Une version multilingue est disponible en fichiers s\'epar\'es sur l'archive ouverte HAL \`a l'adresse \url{https://hal.science/hal-04168815}.}
\\
\noindent{\small{\it Mots-cl\'es~:} gaz de bosons~; superfluide bidimensionnel~; gaz d'atomes froids~; amortissement Belyaev-Landau des phonons~; fonctions de Green \`a $N$ corps ; hydrodynamique quantique}
\\
\vskip 0.25\baselineskip
\noindent{\bf Abstract}
\vskip 0.5\baselineskip
\noindent
\selectlanguage{english}
It is generally accepted that the phonon gas of a superfluid always enters a weak coupling regime at sufficiently low temperatures, whatever the strength of the interactions between the underlying particles (constitutive of the superfluid). Thus, in this limit, we should always be able to calculate the damping rate of thermal phonons by applying Fermi's golden rule to the Hamiltonian {\yc $H_3$} of cubic phonon-phonon coupling taken from quantum hydrodynamics, at least in the case of a convex acoustic branch and in the collisionless regime (where the eigenfrequency of the considered phonons remains much greater than the gas thermalization rate). Using the many-body Green's function method, we predict that, unexpectedly, this is not true in two dimensions, contrary to the three-dimensional case. We confirm this prediction with classical phonon-field simulations and a non-perturbative theory in $H_3$, where the fourth order is regularized by hand, giving a complex energy to the virtual phonons of the four-phonon collisional processes. For a weakly interacting fluid and a phonon mode in the long-wavelength limit, we predict a damping rate about three times lower than that of the golden rule. {\yc A multilingual version is available in separate files on the open archive HAL at \url{https://hal.science/hal-04168815}.}
\\
\noindent{\small {\it Keywords:} Bose gas ; two-dimensional superfluid~; cold atom gas~; Beliaev-Landau phonon damping; many-body Green's functions; quantum hydrodynamics}
\selectlanguage{french}

\noindent 
\vskip 0.5\baselineskip
\end{abstract} 
\end{frontmatter}

\selectlanguage{french}

\tableofcontents


\section{Présentation du problème et raisonnements généraux}
\label{sec0}

Nous considérons en dimension $d\geq 2$ un système spatialement homogène de particules avec des interactions à courte portée, entièrement superfluide dans son état fondamental et préparé à température non nulle mais arbitrairement basse, ce qui permet de limiter son étude à l'ordre dominant en température au moyen d'une théorie effective de basse énergie. Indépendamment de la statistique quantique (bosonique ou fermionique) des particules et de la force de leurs interactions (en phase gazeuse ou liquide), ce système admet une branche d'excitation acoustique, c'est-à-dire de départ linéaire en nombre d'onde (sans bande interdite), dont les quanta sont les phonons. 

Le couplage entre phonons de faible nombre d'onde (d'énergie $<\eta k_B T$, où la température $T\to 0$ puis la coupure $\eta\to+\infty$ à la fin des calculs) se déduit de l'hydrodynamique quantique de Landau et Khalatnikov \cite{LK}, l'hamiltonien effectif prenant la forme\footnote{\label{note1} L'inclusion de termes d'ordre supérieur $H_5, H_6$, etc., n'aurait pas de sens sans des corrections de gradient à la fonctionnelle énergie de l'hydrodynamique quantique \cite{SonWingate}, corrections qui font aussi apparaître le terme de courbure (ici mis à la main) dans l'équation (\ref{eq003}).}
\be
\label{eq001}
H=H_2+H_3+H_4
\ee
où le terme $H_n$ est de degré $n$ en les $\hat{b}_\kk^\dagger$ et $\hat{b}_\kk$, opérateurs bosoniques de création et d'annihilation d'un phonon de vecteur d'onde $\kk$ dans la boîte de quantification $[0,L]^d$ avec des conditions aux limites périodiques. Le terme $H_2$ représente un gaz parfait de phonons de relation de dispersion $\veps_\kk$,
\be
\label{eq002}
H_2=\sum_{\kk\neq\zero} \veps_\kk \hat{b}_\kk^\dagger\hat{b}_\kk
\ee
La forme précise des termes de couplage $H_3$ et $H_4$ dans l'équation (\ref{eq001}) importe peu à ce stade et sera donnée dans la section \ref{sec1}. Le couplage entre phonons rend le système acoustiquement non linéaire et provoque l'amortissement des ondes sonores, objet du présent travail. 

Nous limitons ici notre étude de l'amortissement (i) aux modes de vecteurs d'onde $\qq$ dans le régime faiblement collisionnel $\veps_\qq\gg\hbar\Gamma_{\rm th}$ où $\Gamma_{\rm th}$ est le taux de thermalisation du gaz de phonons ou, ce qui revient au même, le taux d'amortissement des phonons thermiques d'énergie $\veps_\kk\simeq k_B T$, et (ii) au cas d'une branche acoustique de départ convexe en nombre d'onde, que l'on peut limiter à son approximation cubique à basse température:
\be
\label{eq003}
\veps_\kk=\hbar c k \left[1+\frac{\gamma}{8} \left(\frac{\hbar k}{m c}\right)^2\right]\quad\mbox{avec}\quad \gamma>0
\ee
où $m$ est la masse d'une particule, $c$ est la vitesse du son à température nulle et $\gamma$ un paramètre de courbure sans dimension.\footnote{\label{note0} Comme en électrodynamique quantique, les paramètres apparaissant dans l'hamiltonien sont en fait les valeurs nues, différentes des valeurs effectives mesurées dans une expérience, le passage des unes aux autres s'effectuant par une procédure dite de \g{renormalisation}. Il faudrait donc parler dans l'équation (\ref{eq002}) de relation de dispersion nue $\veps_\kk^{(0)}$ et dans l'équation (\ref{eq003}) de vitesse du son nue $c^{(0)}$ et de paramètre de courbure nu $\gamma^{(0)}$. Nous ne le faisons pas ici pour alléger les notations, et parce que cette renormalisation n'intervient pas en dimension $d=2$ à l'ordre dominant en température.}\footnote{Le choix $\gamma>0$ exclut les superfluides de fermions de spin $1/2$ appariés dans la limite, dite BCS, d'une interaction faiblement attractive entre particules \cite{SDM}.} Ces hypothèses autorisent à décrire l'amortissement en termes de processus élémentaires de couplage entre phonons\footnote{Dans le régime opposé $\veps_\qq\ll\hbar\Gamma_{\rm th}$, le système a le temps de relaxer vers une juxtaposition d'états d'équilibre thermodynamique locaux pendant une période d'oscillation de l'onde sonore de vecteur d'onde $\qq$ et l'amortissement à température non nulle se décrit au travers de coefficients de transport ou de viscosité, obtenus par résolution d'équations cinétiques et qui entrent dans les équations aux dérivées partielles sur les densités, les vitesses, l'entropie volumique, etc., d'un modèle à deux fluides \cite{livreK}.} et, comme il est généralement admis, à déduire le taux d'amortissement de la règle d'or de Fermi appliquée au terme $H_3$ de l'hamiltonien.\footnote{Dans le cas $\gamma<0$ d'une branche acoustique de départ concave, les processus à trois phonons induits par $H_3$ à l'ordre un ne conservent pas l'énergie-impulsion, et il faut utiliser une formulation étendue de la règle d'or (tenant compte des processus à quatre phonons induits par $H_3$ à l'ordre deux et par $H_4$ à l'ordre un) pour obtenir un taux d'amortissement non nul, voir les références \cite{LK,Annalen} et notre section \ref{sec3.2}.} En effet, on s'attend à ce qu'à suffisamment basse température, le gaz de phonons entre toujours dans le régime de couplage faible, même si les particules constitutives du système sont en interaction forte (comme dans l'hélium 4 liquide par exemple \cite{Maris}). Nous allons montrer dans cette section, par un raisonnement simple, que ceci est vrai en dimension $d=3$ mais pas en dimension $d=2$. L'amortissement des phonons dans le cas bidimensionnel fera donc l'objet d'une étude plus poussée (au-delà de la règle d'or) dans les sections suivantes. 

Notre raisonnement simple s'appuie sur la méthode des fonctions de Green à $N$ corps \cite{FW}. Pour fixer les idées et la fonction de Green à considérer, nous supposons que le gaz, initialement à l'équilibre thermique pour l'hamiltonien $H$ (opérateur densité $\hat{\sigma}(0)=\hat{\sigma}_{\rm th}\propto \exp(-H/k_B T)$), est excité par la création soudaine d'un état cohérent de phonons dans le mode $\qq$:
\be
\label{eq010}
\hat{\sigma}(0^+) = \hat{U}_{\rm exc}\hat{\sigma}_{\rm th} \hat{U}^\dagger_{\rm exc} \quad\mbox{avec}\quad \hat{U}_{\rm exc}=\eee^{\alpha\hat{b}_\qq^\dagger-\alpha^*\hat{b}_\qq}
\ee
Expérimentalement, cette procédure d'excitation peut être réalisée par une impulsion laser de Bragg de courte durée \cite{Grynberg,Ketterle,Davidson,Vale,Cartago}. On mesure ensuite l'amplitude $\langle\hat{b}_\qq(t)\rangle$ de l'état cohérent au cours du temps, au travers par exemple de la densité du gaz (plus précisément de ses composantes de Fourier de vecteurs d'onde $\pm\qq$). Dans la limite d'une très faible excitation initiale, on entre dans le régime de réponse linéaire et l'on peut formellement développer l'opérateur unitaire d'excitation $\hat{U}_{\rm exc}$ à l'ordre un en $\alpha$ dans l'équation (\ref{eq010}) pour obtenir
\be
\label{eq011}
\langle\hat{b}_\qq(t)\rangle\stackrel{t>0}{\underset{\alpha\to 0}{=}} \alpha \langle[\hat{b}_\qq(t),\hat{b}_\qq^\dagger(0)]\rangle_{\rm th}+O(\alpha^2)
\ee
où $\hat{b}_\qq(t)$ est écrit en représentation de Heisenberg pour l'évolution non perturbée par $\hat{U}_{\rm exc}$, $\hat{b}_{\qq}(t)=\eee^{\ii H t/\hbar} \hat{b}_{\qq}(0)\eee^{-\ii H t/\hbar}$, la moyenne au premier membre est prise dans l'état perturbé $\hat{\sigma}(0^+)$ et celle au second membre dans l'état thermique initial non perturbé $\hat{\sigma}_{\rm th}$, et on a utilisé le fait que $\langle\hat{b}_{\qq}(0)\hat{b}_{\qq}(t)\rangle_{\rm th}=\langle\hat{b}_{\qq}(t)\hat{b}_{\qq}(0)\rangle_{\rm th}=0$ par non-conservation de l'impulsion totale. À l'aide de la section 31 de la référence \cite{FW}, nous relions notre signal à la fonction de Green à deux points $\mG$ {\yc du champ de phonons} et à la fonction énergie propre associée $\Sigma$ (notée $\Sigma^{\star}$ dans \cite{FW}):\footnote{Rappelons qu'à l'origine, $\mG$ et $\Sigma$ sont des fonctions de $(\rr,t)$ et $(\rr',t')$ où $\rr,\rr'$ sont deux vecteurs positions et $t,t'$ deux temps. On les réduit à des fonctions de $(\rr-\rr',t-t')$ par homogénéité spatiale et stationnarité de l'état thermique, dont les transformées de Fourier spatio-temporelles sont des fonctions de $(\kk,\omega)$; le prolongement analytique de ces dernières aux pulsations complexes $z/\hbar$ conduit aux fonctions $\mG(\kk,z)$ et $\Sigma(\kk,z)$ de l'équation (\ref{eq012}).}
\be
\label{eq012}
\boxed{s(t)\equiv \eee^{\ii\veps_\qq t/\hbar}\langle[\hat{b}_\qq(t),\hat{b}_\qq^\dagger(0)]\rangle_{\rm th}\stackrel{t>0}{=} \int_{C_+} \frac{\dd z}{2\ii\pi} \eee^{-\ii (z-\veps_\qq)t/\hbar} \mathcal{G}(\qq,z)\quad\mbox{avec}\quad \mathcal{G}(\qq,z)=\frac{1}{z-\veps_\qq-\Sigma(\qq,z)}}
\ee
où le chemin d'intégration $C_+$ court dans le demi-plan {\yc complexe} supérieur parallèlement à l'axe réel de $+\infty$ à $-\infty$ et où nous avons, par commodité, retiré l'évolution libre due à l'hamiltonien $H_2$. Dans la suite, il sera avantageux de prendre comme nouvelle variable d'énergie complexe l'écart à l'énergie de l'évolution libre, toute confusion étant évitée par mise en indice de la dépendance en vecteur d'onde:
\be
\label{eq013}
\Sigma(\qq,z)=\Sigma_\qq(\zeta)\quad\mbox{avec}\quad\zeta\equiv z-\veps_\qq\quad{\ycd\mbox{d'où}\quad \mathcal{G}(\qq,z)=\frac{1}{\zeta-\Sigma_\qq(\zeta)}}
\ee
À partir de ce formalisme exact, on peut retrouver l'amortissement exponentiel de la règle d'or de Fermi $|s(t)|^2\simeq \exp(-\Gamma_q t)$ en deux étapes. (i) On effectue d'abord l'approximation du pôle \cite{CCTbordeaux} négligeant la dépendance en énergie de la fonction $\Sigma$:
\be
\label{eq014}
\Sigma_\qq(\zeta) \approx \Sigma_\qq(\ii 0^+)\quad \mbox{si bien que}\quad s(t)\simeq s_{\mbox{\scriptsize pôle}}(t)=\exp[-\ii \Sigma_\qq(\ii 0^+)t/\hbar]
\ee
(ii) Comme l'approximation du pôle n'est en général légitime que dans une limite de couplage faible entre les phonons, il suffit de calculer $\Sigma_\qq$ à l'ordre dominant en le couplage, c'est-à-dire à l'ordre deux en $H_3$, mais aussi à l'ordre un en $H_4$ si l'on a besoin de connaître $\re\Sigma_\qq$. La question centrale de cette section est de savoir si le simple fait de prendre une limite de basse température avec mise à l'échelle du nombre d'onde du mode considéré,\footnote{Faut-il le préciser, la force des interactions entre les particules constitutives du fluide ne varie pas dans la prise de limite (\ref{eq020}), si bien que la vitesse du son $c$ reste constante. Par ailleurs, la limite (\ref{eq020}) fait automatiquement entrer le mode $\qq$ dans le régime faiblement collisionnel $\veps_{\qq}\gg \hbar\Gamma_{\rm th}$ car le taux $\Gamma_{\rm th}$ de thermalisation du gaz, que l'on peut estimer par le taux d'amortissement des phonons de nombre d'onde $\approx k_B T/\hbar c$, tend vers zéro plus vite que linéairement en température, comme nous le verrons.}
\be
\label{eq020}
\epsilon \equiv k_B T/mc^2 \to 0 \quad \mbox{à}\quad \qb\equiv \hbar c q/k_B T \quad \mbox{fixé}
\ee
suffit à faire entrer le gaz de phonons dans le régime de couplage faible et à justifier les deux étapes. Nous verrons que la réponse dépend de la dimensionalité du système.

\paragraph{Étape (i)} Pour justifier l'approximation du pôle, nous nous appuyons sur les lois d'échelle décrivant le comportement de la fonction énergie propre à basse température:
\be
\label{eq021}
\Sigma_\qq(\zeta) = k_B T \epsilon^\nu\, \tilde{\Sigma}_{\qb}(\zeta/(k_BT\epsilon^\sigma))
\ee
Après sortie de l'échelle d'énergie $k_B T$, il reste deux exposants $\nu$ et $\sigma$, le premier caractérisant l'ordre de grandeur de la fonction $\Sigma_\qq(\zeta)$, le second sa largeur en énergie autour de $\zeta=0$. Il est entendu que la nouvelle fonction $\tilde{\Sigma}_{\qb}$ admet une limite finie et non nulle lorsque $\epsilon\to 0$. Après report de la forme (\ref{eq021}) dans l'expression (\ref{eq012}) du signal, nous effectuons les changements d'échelle suivants sur les variables conjuguées $\zeta$ et $t$,
\be
\label{eq022}
\zeta\equiv k_B T \epsilon^\nu \zetat \quad ; \quad t\equiv \tilde{t}\,\hbar/(k_B T \epsilon^\nu)
\ee
pour obtenir l'expression réduite
\be
\label{eq023}
\boxed{s(t)=\int_{C_+} \frac{\dd \zetat}{2\ii\pi} \frac{\eee^{-\ii \zetat \tilde{t}}}{\zetat-\tilde{\Sigma}_{\qb}(\epsilon^{\nu-\sigma}\zetat)}}
\ee
dans laquelle il reste à faire tendre $\epsilon$ vers zéro à $\tilde{t}$ fixé. On constate alors que, si $\nu>\sigma$, l'argument de la fonction $\tilde{\Sigma}_{\qb}$ tend vers zéro par parties imaginaires strictement positives lorsque $\epsilon\to 0$ ce qui autorise à le remplacer par $\ii 0^+$: on retrouve l'approximation du pôle (\ref{eq014}), exacte dans cette limite. 

\paragraph{Étape (ii)} Pour savoir s'il suffit de calculer la fonction énergie propre à l'ordre dominant en le couplage entre phonons, gardons pour simplifier seulement le couplage cubique $H_3$ (on peut vérifier que le couplage quartique $H_4$ ne change rien qualitativement), omettons les processus non résonnants qu'elle contient (termes en $\hat{b}^\dagger\hat{b}^\dagger\hat{b}^\dagger$ et $\hat{b}\hat{b}\hat{b}$) de contribution sous-dominante dans la limite $\epsilon\to 0$, et estimons l'ordre de grandeur de sa contribution d'ordre $2n$ à $\Sigma_\qq{\ycd(\ii 0^+)}$ comme suit:\footnote{Comme $H_3$ change la parité du nombre de phonons, les contributions d'ordre impair $2n+1$ sont nulles.}
\be
\label{eq024}
\Sigma^{(2n)}_\qq(\ii 0^+)\approx \int \left(\prod_{i=1}^{n} \dd^d k_i\right) \langle\ |\mathcal{H}_3|\ \rangle \frac{1}{\Delta E_1} \langle\ |\mathcal{H}_3|\ \rangle \ldots \langle\ |\mathcal{H}_3|\ \rangle \frac{1}{\Delta E_{2n-1}} \langle\ |\mathcal{H}_3|\ \rangle
\ee
Justifions brièvement la forme (\ref{eq024}) (le lecteur est renvoyé aux sections \ref{sec2} et \ref{sec3} pour le traitement détaillé des cas $n=1$ et $n=2$). L'ordre $2n$ fait intervenir $2n$ actions de $H_3$ donc $2n$ éléments de matrice $\langle\,|H_3|\,\rangle$, chacune d'elles introduisant un nouveau vecteur d'onde de phonon indépendant $\kk_i$ (termes de Belyaev en $\hat{b}^\dagger\hat{b}^\dagger\hat{b}$ dans $H_3$) ou en faisant disparaître un (termes de Landau en $\hat{b}^\dagger\hat{b}\hat{b}$); les autres vecteurs d'onde de phonons qui apparaissent se déduisent des $\kk_i$ et de $\qq$ par conservation de l'impulsion. Comme les $2n$ actions de $H_3$ doivent conserver le nombre de phonons dans leur ensemble, il doit y avoir autant de processus de Belyaev que de Landau ce qui conduit à exactement $n$ vecteurs d'onde indépendants sur lesquels sommer. À la limite thermodynamique, on exprime $H_3$ en termes de l'hamiltonien volumique tel que $H_3=\mH_3/L^{d/2}$ et on remplace la somme sur les $\kk_i$ par une intégrale. Enfin, entre chaque élément de matrice de $H_3$ apparaît un dénominateur {\yc d'énergie} traduisant la propagation libre d'hamiltonien $H_2$ et donnant la différence d'énergie entre {\yct l'état initial à un phonon $\qq$ et} un état intermédiaire. Chaque $\Delta E_j$ ($1\leq j\leq 2n-1$) peut être exprimé comme la somme algébrique (c'est-à-dire avec un signe plus ou moins) de $j$ défauts d'énergie élémentaires $\veps_\kk+\veps_{\kk'}-\veps_{\kk''}$ correspondant à un processus Belyaev-Landau $\kk+\kk'\leftrightarrow\kk''$. Dans la limite de basse température $\epsilon\to 0$, l'intégrale (\ref{eq024}) est en général dominée par les processus \g{aux petits angles}, où {\it tous} les phonons intermédiaires (ceux de vecteurs d'onde $\kk_i$ et les autres) sont presque colinéaires à et de même sens que $\qq$ \cite{LK,Annalen}; plus précisément, on trouve que les angles $\theta_i$ entre les vecteurs $\qq$ et $\kk_i$ doivent être des $O(\gamma^{1/2}\epsilon)$ pour que les $\Delta E_j$ atteignent leur échelle d'énergie minimale {\yc $k_B T\epsilon^2$} et que l'intégrande soit amplifié par l'effet \g{petit dénominateur}, ce qui conduit au développement du défaut d'énergie élémentaire
\be
\label{eq025}
\veps_\kk+\veps_{\kk'}-\veps_{\kk''=\kk+\kk'}\underset{\epsilon\to 0}{\sim} k_B T\gamma\epsilon^2\left[\frac{\kb\kpb}{\kb+\kpb}\frac{(\theta-\theta')^2}{2\gamma\epsilon^2}-\frac{3}{8}\kb\kpb(\kb+\kpb)\right]
\ee
où $\gamma$ est le paramètre de courbure dans la relation de dispersion (\ref{eq003}), {\yc on a introduit les angles $\theta=\widehat{(\qq,\kk)}$ et $\theta'=\widehat{(\qq,\kk')}$,} et l'adimensionnement de $k$ en $\kb$ par sa valeur typique $k_B T/\hbar c$ est calqué sur l'équation (\ref{eq020}). Les $\Delta E_j$ sont donc {\yc bien} d'ordre $k_B T\epsilon^2$ et chaque élément d'intégration $\dd^dk_i$ dans (\ref{eq024}) écrit en coordonnées polaires ou sphériques fait sortir un facteur $(k_B T/\hbar c)^d (\gamma^{1/2}\epsilon)^{d-1}$. Il reste à donner l'estimation des éléments de matrice de $\mH_3$, que l'on tire aisément de leur expression (\ref{eq102}) à venir (la constante de couplage $\Lambda$ est donnée par l'équation (\ref{eq032})):
\be
\label{eq027}
\langle\ |\mathcal{H}_3|\ \rangle \approx (k\xi k'\xi k''\xi)^{1/2}(1+\Lambda)\frac{mc^2}{\rho^{1/2}} \approx \epsilon^{3/2}(1+\Lambda)\frac{mc^2 \xi^{d/2}}{(\rho\xi^d)^{1/2}}
\ee
où l'on a introduit la densité $\rho$ du fluide et la longueur dite de relaxation ou de corrélation du superfluide, 
\be
\label{eq026}
\xi=\frac{\hbar}{mc}
\ee
Nous aboutissons ainsi à l'estimation
\be
\label{eq028}
\boxed{\Sigma_\qq^{(2n)}(\ii0^+)\underset{\epsilon\to 0}{\approx} k_BT \gamma \epsilon^2 \left[\frac{\epsilon^{2d-4}(1+\Lambda)^2}{\gamma^{(5-d)/2}\rho\xi^d}\right]^n \quad\quad (n>0,d\geq 2)}
\ee
qui désigne la quantité entre crochets comme petit paramètre du développement perturbatif de $\Sigma_\qq$ en le couplage. Le degré en $\epsilon$ de l'ordre dominant non nul, correspondant au choix $n=1$ dans l'équation (\ref{eq028}), donne la valeur de l'exposant $\nu$ dans l'équation (\ref{eq021}):
\be
\label{eq004}
\nu=2d-2
\ee
La valeur du second exposant $\sigma$ s'obtient en généralisant l'estimation (\ref{eq024}) au cas $\zeta\neq 0$, ce qui revient à ajouter $\zeta$ à l'énergie de l'état initial donc à chaque dénominateur d'énergie, $\Delta E_j\to \Delta E_j+\zeta$. On l'a vu, $\Delta E_j\approx k_B T\epsilon^2$, et c'est aussi la largeur de la fonction énergie propre autour de $z=\epsilon_\qq$ (autour de $\zeta=0$), si bien que
\be
\label{eq005}
\sigma=2
\ee
indépendamment de la dimensionalité.

\paragraph{Cas tridimensionnel} En dimension $d=3$, où $\nu=4>\sigma=2$ et où $2d-4=2>0$ entre les crochets de l'équation (\ref{eq028}), les étapes (i) et (ii) sont justifiées et l'amortissement du mode $\qq$ est décrit de manière exacte par la règle d'or de Fermi dans la limite (\ref{eq020}). On trouve alors le comportement dominant à basse température de la partie imaginaire de $\Sigma_\qq(\ii 0^+)$,  
\begin{multline}
\label{eq030}
\im\Sigma_\qq^{d=3}(\ii 0^+) \stackrel{\qb\,\mbox{\scriptsize fix\'e}}{\underset{\epsilon\to 0}{\sim}} -\frac{9 k_B T \epsilon^4(1+\Lambda)^2}{32\pi\rho\xi^3}\Bigg[\int_0^{\qb}\dd\kb\, \kb^2(\qb-\kb)^2(\bar{n}^{\rm lin}_k+1/2)+\int_0^{+\infty}\dd\kb\, \kb^2(\kb+\qb)^2(\bar{n}^{\rm lin}_k-\bar{n}^{\rm lin}_{k+q})\Bigg] \\
=-\frac{9 k_B T \epsilon^4(1+\Lambda)^2}{32\pi\rho\xi^3}\Bigg\{\frac{\qb^5}{60}+48[g_5(1)-g_5(\eee^{-\qb})]-24\qb\, g_4(\eee^{-\qb})+4\qb^2[g_3(1)-g_3(\eee^{-\qb})]\Bigg\}
\end{multline}
correspondant exactement au taux d'amortissement $\Gamma_q$ donné dans la référence \cite{Annalen} (multiplié par l'habituel facteur $-\hbar/2$). Dans l'équation (\ref{eq030}), on a introduit les fonctions de Bose $g_\alpha(z)=\sum_{n=1}^{+\infty} z^n/n^\alpha$ et on a pu remplacer la relation de dispersion $\veps_\kk$ par son approximation linéaire $\hbar c k$ dans les nombres d'occupation, ce qu'indique l'exposant \g{lin}:
\be
\label{eq029}
\bar{n}_k^{\rm lin}=\frac{1}{\exp\kb-1}
\ee
On constate avec satisfaction que le préfacteur dans (\ref{eq030}) est bien celui prédit par (\ref{eq028}) pour $n=1$ (et $d=3$) ; il est en particulier indépendant du paramètre de courbure $\gamma$. Le formalisme des fonctions de Green donne en plus accès à la partie réelle, qui échappait à la référence \cite{Annalen} et qui représente le déplacement d'énergie thermique du phonon $\qq$,\footnote{On omet dans (\ref{eq031}) la composante dite quantique, indépendante des nombres d'occupation, qui est à l'ordre dominant en $\epsilon$ la somme de deux contributions, (i) une contribution mal définie, combinaison linéaire des monômes $q$, $q^3$ et $q^5$ avec des coefficients dépendant de façon divergente de la coupure ultraviolette $\eta$ de l'équation (\ref{eq103}), qui doit donc être réabsorbée dans $\veps_\qq$, c'est-à-dire dans une redéfinition de la relation de dispersion (voir aussi la note \ref{note0}), et (ii) une contribution universelle prédite quantitativement par l'hydrodynamique quantique, $\re\Sigma_{\qq,{\rm quant\ univ}}^{d=3}(\ii 0^+)=[3(1+\Lambda)^2/(160\pi^2)](\hbar^2 q^5/m\rho)\ln(\hbar q/mc)$. Une telle contribution en $q^5\ln q$ était attendue qualitativement dans la référence \cite{SonWingate}. Elle apparaît dans l'équation (35) de la référence \cite{Barc}, avec le bon coefficient numérique mais avec un facteur $c$ en trop.}
\begin{multline}
\label{eq031}
\re\Sigma_{\qq,{\rm th}}^{d=3}(\ii 0^+) \stackrel{\qb\,\mbox{\scriptsize fix\'e}}{\underset{\epsilon\to 0}{\sim}} \frac{k_B T \epsilon^4(1+\Lambda)^2}{2\pi^2\rho\xi^3}
\Bigg\{\qb\zeta(4)\left[-27\ln\frac{\eee(3\gamma)^{1/2}\epsilon}{4}+\frac{2+18\Lambda+3\Lambda_*/2}{(1+\Lambda)^2}\right] \\
+\frac{9}{8} \int_0^{+\infty} \dd\kb\, \kb^2\, \bar{n}^{\rm lin}_k \left[(\kb-\qb)^2\ln|\kb-\qb|-(\kb+\qb)^2\ln(\kb+\qb)\right] \Bigg\}
\end{multline}
où $\zeta$ est la fonction zêta de Riemann, telle que $\zeta(s)=g_s(1)$, et $\zeta(4)=\pi^4/90$. On obtient pour la partie réelle une forme plus compliquée que celle de (\ref{eq028}) car (a) les angles entre $\qq$ et $\kk$ d'ordre $\theta\approx\epsilon^0$ contribuent à la même hauteur que les petits angles $\theta\approx\epsilon$,\footnote{On remarque en effet dans l'intégration angulaire, en $d=3$, que $\sin\theta\,\dd\theta/(\Delta E/k_B T)$ est $\approx \epsilon^0$ aux grands angles et $\approx \epsilon^2/\epsilon^2=\epsilon^0$ aux petits angles, où $\Delta E$ est un dénominateur d'énergie. Ce phénomène ne se produit pas pour la partie imaginaire, compte tenu du Dirac de conservation de l'énergie $\delta(\Delta E)$ et du fait que $\Delta E$ ne s'annule qu'aux petits angles pour des phonons thermiques à basse température. En pratique, pour obtenir (\ref{eq031}), on sépare l'intégrale angulaire en deux intervalles, $\theta\in[0,\alpha]$ et $\theta\in [\alpha,\pi]$ où $\alpha\ll 1$ est fixé. Dans le premier intervalle, on garde le terme de courbure dans la relation de dispersion au dénominateur de l'intégrande mais on remplace le numérateur par son ordre dominant en $\theta$ (ainsi, $\sin\theta\simeq\theta$). Dans le second intervalle, on utilise la relation de dispersion linéaire mais on garde toute la dépendance en $\theta$ des fonctions trigonométriques. Tous calculs faits, la partie basse et la partie haute de l'intégrale comportent une divergence en $\ln\alpha$ qui disparaît exactement dans leur somme.} et (b) l'hamiltonien quartique $H_4$ contribue également, ce qui nous oblige à introduire la seconde constante de couplage $\Lambda_*$ de l'équation (\ref{eq032}). En particulier, on remarque une correction logarithmique à la loi en $k_B T \epsilon^4$. Dans la limite $\qb\to 0$, l'expression (\ref{eq031}) devient $\sim \hbar q \delta c_{\rm th}$, où le coefficient $\delta c_{\rm th}$ est la correction thermique à la vitesse du son {\ycd faiblement collisionnel} dans le fluide. La valeur de $\delta c_{\rm th}$ tirée de la théorie cinétique d'Andreev et Khalatnikov \cite{AK}, voir l'équation (45) de la référence \cite{Barc}, est en bon accord avec la nôtre,\footnote{Il y a accord parfait si, dans l'équation (45) de \cite{Barc}, on remplace $\ln 27$ par $25/6-2C+\ln(3/2)+2\zeta'(4)/\zeta(4)=\ln(26,\!852\ldots)$ où $C=0,\!577\ldots$ est la constante d'Euler. Ce petit écart résulte d'une approximation faite par Andreev et Khalatnikov dans le calcul d'une intégrale, celle de l'équation (13) dans \cite{AK}.} mais celle déduite de l'ordre dominant de la théorie effective des champs, voir l'équation (33) de la référence \cite{Barc}, est en désaccord.

\paragraph{Cas bidimensionnel} En dimension $d=2$, où $\nu=\sigma=2$ et où $2d-4=0$, nous échouons à justifier l'étape (i) et l'étape (ii) à basse température! En particulier, l'expression entre crochets dans l'équation (\ref{eq028}) ne dépend pas de $\epsilon$ et ne devient un petit paramètre que dans la limite $\rho\xi^2\gg 1$ où le superfluide sous-jacent lui-même (et pas simplement le gaz de phonons) entre dans le régime d'interaction faible. Ainsi, en $d=2$, la prise de la limite $\epsilon\to 0$ ne rend exactes ni l'approximation du pôle ni le calcul de $\Sigma_\qq(\zeta)$ à l'ordre deux en le couplage entre phonons. Pour obtenir la bonne expression du signal dans cette limite, il faut calculer $\tilde{\Sigma}_{\qb}(\zetat)$ pour toutes les valeurs de $\zetat$ et à tous les ordres en $H_3$. Donnons quand même comme valeur de référence le comportement à basse température de $\Sigma_\qq(\ii 0^+)$ à l'ordre dominant en le couplage ($n=1$):
\begin{multline} 
\label{eq033}
\Sigma_\qq^{(2)}(\ii 0^+)\stackrel{\qb\,\mbox{\scriptsize fix\'e}}{\underset{\epsilon\to 0}{\sim}} -\ii\frac{9k_B T\epsilon^2(1+\Lambda)^2}{8\pi(3\gamma)^{1/2}\rho\xi^2} \Bigg[\int_0^{\qb}\dd\kb\, \kb(\qb-\kb)(\bar{n}^{\rm lin}_k+1/2)+\int_0^{+\infty}\dd\kb\, \kb (\kb+\qb) (\bar{n}^{\rm lin}_k-\bar{n}^{\rm lin}_{k+q})\Bigg] \\
=\boxed{-\ii\frac{9k_B T\epsilon^2(1+\Lambda)^2}{8\pi(3\gamma)^{1/2}\rho\xi^2} \Bigg[\frac{\qb^3}{12}+2\zeta(2)\qb\Bigg]\equiv -\ii \frac{\hbar\Gamma_q}{2}}
\end{multline}
avec $\zeta(2)=\pi^2/6$. À suffisamment basse température, $\Sigma^{(2)}_\qq(\ii 0^+)$ devient donc imaginaire pur et se réduit à $-\ii\hbar\Gamma_q/2$ où $\Gamma_q$ est le taux d'amortissement de la règle d'or à l'ordre dominant en $\epsilon$ (l'hamiltonien quartique $H_4$ ne fait pas apparaître ici de partie réelle car il est sous-dominant en $\epsilon$, voir la section \ref{sec2}). L'équation (\ref{eq033}) est à notre connaissance originale. Dans les limites $\qb\to 0$ et $\qb\to +\infty$, elle redonne les équations (30) et (22) de la référence \cite{CB} (fois $-\ii\hbar$ compte tenu des conventions de \cite{CB}) qui traite du gaz bidimensionnel de bosons en interaction faible pour lequel $\gamma=1$ (comme dans le spectre de Bogolioubov) et $\Lambda=0$ (comme pour l'équation d'état de champ moyen à température nulle $\mu=\rho g$ utilisée dans \cite{CB}). Dans la suite, nous irons en dimension $d=2$ au-delà de la règle d'or de Fermi avec des simulations numériques de champ classique dans la section \ref{sec1} et avec la méthode des fonctions de Green à des ordres en $H_3$ de plus en plus élevés dans les sections ultérieures: à l'ordre deux dans la section \ref{sec2}, à l'ordre quatre dans la section \ref{sec3} et de manière non perturbative dans la section \ref{sec4} ; à notre surprise, nous y sommes contraints par l'échec de l'estimation (\ref{eq028}) pour $n\geq 2$. {\ycd Nous établissons en effet, par un calcul explicite de l'ordre quatre en $H_3$ dans la section \ref{sec3}, un résultat troublant qui dément (\ref{eq028}) en dimension $d=2$: $\Sigma_{\qq}^{(4)}(\ii 0^+)$ ne tend pas vers zéro comme $k_B T \epsilon^2$ mais comme $k_B T\epsilon^0$ à basse température $\epsilon\to 0$, voir l'équation (\ref{eq296}). Ceci rend en fait inutilisable le développement perturbatif en puissances de $H_3$ dans cette limite, puisque l'ordre quatre devient beaucoup plus grand que l'ordre deux à force des interactions $(\rho\xi^2)^{-1}$ fixée dans le superfluide. Pour obtenir une description opérationnelle de l'amortissement, la section \ref{sec4}, au contraire de l'équation (\ref{eq028}), prend la limite $\epsilon\to 0$ à $\zetat\neq 0$ fixé dans $\tilde{\Sigma}_{\qb}^{(4)}(\zetat)$, régularise le résultat (qui diverge en $\zetat=0$) de manière non perturbative en donnant à la main une durée de vie finie aux phonons intermédiaires des processus collisionnels à quatre phonons et calcule l'intégrale sur $\zetat$ dans l'équation (\ref{eq023}), pour aboutir à un taux d'amortissement variant en $k_B T\epsilon^2$ comme attendu.} Nous concluons dans la section \ref{sec5}.

\section{Expérience numérique dans le modèle de champ classique}
\label{sec1}

Les études expérimentales du son dans des gaz d'atomes froids superfluides bidimensionnels se sont concentrées pour l'instant sur le régime hydrodynamique \cite{HM,PCZH} ou, dans le régime faiblement collisionnel, sur l'intervalle de température $\epsilon=k_B T/mc^2\geq 1$ \cite{JDJB,Salasnich,Stringari}. Pour vérifier qu'en dimension $d=2$, la limite de basse température (\ref{eq020}) ne suffit pas à rendre exacte la décroissance exponentielle prédite par la règle d'or de Fermi, nous effectuons donc dans cette section une étude numérique de l'amortissement des phonons à tous les ordres en le couplage $H_3$ dans le modèle de champ classique. Nous rappelons d'abord l'hamiltonien de l'hydrodynamique quantique de Landau et Khalatnikov. Nous expliquons ensuite comment intégrer numériquement sa version de champ classique et introduisons une fonction de corrélation plus commode à calculer que le signal (\ref{eq012}) mais presque équivalente. Nous analysons ensuite les résultats obtenus pour différentes forces des interactions. 

\subsection{Hamiltonien de l'hydrodynamique quantique}
\label{sec1.1}

L'hydrodynamique quantique est une théorie effective de basse énergie qui permet d'obtenir des prédictions exactes à l'ordre dominant en température, même pour un fluide en interaction forte \cite{LK}. Le fluide est décomposé en petits éléments de taille $b$ tout de même mésoscopique ($b\gg\xi=\hbar/mc$). Chaque élément, de centre $\rr$, est (i) suffisamment grand pour qu'il comporte un grand nombre de particules ($\rho b^d\gg 1$) ce qui permet de lui attribuer un opérateur phase $\hat{\phi}(\rr)$ canoniquement conjugué à son opérateur nombre de particules $\hat{\rho}(\rr)b^d$, $[\hat{\rho}(\rr)b^d,\hat{\phi}({\yct\rr'})]=\ii\delta_{\rr,\rr'},$ et (ii) suffisamment petit pour être considéré aussi bien comme homogène ($b\ll\lambda_{\rm th}$ où $\lambda_{\rm th}=2\pi\hbar c/k_B T$ est la longueur d'onde de de Broglie d'un phonon thermique du fluide) que dans son état fondamental ($k_B T\ll\veps_{\rm exc}^{\rm min}$ où $\veps_{\rm exc}^{\rm min}=\hbar c (2\pi/b)$ est l'énergie d'excitation phononique minimale dans le petit élément).\footnote{La coupure ultraviolette à venir correspond peu ou prou au choix $\pi/b\approx \eta k_B T/\hbar c$. Même dans ce cas, $b$ reste très inférieure à la longueur de cohérence du fluide à suffisamment basse température. C'est évident en $d=3$ à cause de la présence d'un condensat. C'est vrai aussi en $d=2$ où la fonction de cohérence du premier ordre $g_1(\rr)$ décroît à grande distance comme $\exp[mk_B T\ln(\xi/r)/(2\pi\hbar^2\rho_s)]$ où $\rho_s\simeq \rho$ est la densité superfluide \cite{Bere,Nel,SJ}. Ceci rend plausible l'hypothèse $\hat{\vv}=(\hbar/m)\mathbf{grad}\,\hat{\phi}$ utilisée dans la référence \cite{LK} et dans l'équation (\ref{eq050}).} L'hamiltonien est alors somme des énergies cinétiques du centre de masse et de l'énergie interne de chaque petit élément:
\be
\label{eq050}
H_{\rm hydro}=\sum_\rr \frac{1}{2} m \hat{\vv}(\rr)\cdot\hat{\rho}(\rr) b^d\hat{\vv}(\rr)+b^d e_0(\hat{\rho}(\rr))
\ee
où $\hat{\vv}=(\hbar/m)\mathbf{grad}\,\hat{\phi}$ est le champ de vitesse du fluide et $e_0(\rho)$ est la densité volumique d'énergie dans l'état fondamental à la densité $\rho$. À basse température, le champ de densité $\hat{\rho}(\rr)$ fluctue peu et le champ de phase $\hat{\phi}(\rr)$ varie lentement; on les sépare en leurs composantes de Fourier de vecteur d'onde nul $\hat{\rho}_0=\hat{N}/L^d\simeq \rho$ et $\hat{\phi}_0$, et en des fluctuations spatiales $\delta\hat{\rho}(\rr)$ et $\delta\hat{\phi}(\rr)$ admettant les développements modaux {\yc\cite{broui}}
\be
\label{eq052}
\delta\hat{\rho}(\rr)=\frac{1}{L^{d/2}} \sum_{\kk\neq\zero} \left(\frac{\hbar\rho k}{2mc}\right)^{1/2}(\hat{b}_\kk+\hat{b}_{-\kk}^\dagger)\,\eee^{\ii\kk\cdot\rr} \quad ; \quad \delta\hat{\phi}(\rr)=\frac{-\ii}{L^{d/2}}  \sum_{\kk\neq\zero} \left(\frac{mc}{2\hbar\rho k}\right)^{1/2} (\hat{b}_\kk-\hat{b}_{-\kk}^\dagger)\,\eee^{\ii\kk\cdot\rr}
\ee
où les $\hat{b}_\kk^\dagger$ et les $\hat{b}_\kk$ sont les opérateurs de création et d'annihilation d'un phonon de vecteur d'onde $\kk$ dans la boîte de quantification $[0,L]^d$, $c$ est la vitesse du son à température nulle, donnée par $mc^2=\rho(\dd\mu/\dd\rho)$ avec $\mu=\dd e_0/\dd\rho$ le potentiel chimique du fluide dans l'état fondamental. Comme dans toute théorie effective, il faut mettre une coupure ultraviolette dans l'espace des vecteurs d'onde, que nous prenons ici isotrope
\be
\label{eq103}
\hbar c k < \eta k_B T
\ee
et à laquelle les sommes (\ref{eq052}) sont soumises; on fera tendre la température vers zéro comme dans (\ref{eq020}) puis $\eta\to+\infty$ (dans le cas quantique) sans déclencher de divergence si tout va bien. Il reste à développer $H_{\rm hydro}$ en puissances de $\delta\hat{\rho}$ et $\mathbf{grad}\,\delta\hat{\phi}$. L'ordre zéro fournit une constante sans intérêt. L'ordre un est nul par construction.  L'ordre deux, diagonalisé par la transformation (\ref{eq052}), redonne l'hamiltonien $H_2$ de l'équation (\ref{eq002}) pour une relation de dispersion linéaire, la courbure dans l'équation (\ref{eq003}) devant être mise à la main, voir les références \cite{LK,Annalen} et notre note \ref{note1}. Les ordres trois et quatre, donnant le couplage entre phonons dans l'équation (\ref{eq001}), s'écrivent en point de vue $\rr$:
\be
\label{eq104}
H_3=b^d\sum_\rr \frac{1}{2} m\hat{\vv}(\rr)\cdot\delta\hat{\rho}(\rr)\hat{\vv}(\rr)+\frac{1}{6}\frac{\dd^2\mu}{\dd\rho^2}\delta\hat{\rho}(\rr)^3 \quad ; \quad H_4=b^d\sum_\rr\frac{1}{24} \frac{\dd^3\mu}{\dd\rho^{3}}\delta\hat{\rho}(\rr)^4
\ee
sachant que $\mathbf{grad}\,\hat{\phi}=\mathbf{grad}\,\delta\hat{\phi}$. Ils font apparaître les constantes de couplage adimensionnées
\be
\label{eq032}
\Lambda=\rho\frac{\dd^2\mu}{\dd\rho^2}/(3\frac{\dd\mu}{\dd\rho})=\rho^2\frac{\dd^2\mu}{\dd\rho^2}/(3 mc^2)\quad\mbox{et}\quad \Lambda_*=\rho^3\frac{\dd^3\mu}{\dd\rho^3}/(mc^2)
\ee
Pour révéler les processus physiques contenus dans $H_3$ et déjà mentionnés dans la section \ref{sec0}, on injecte dans (\ref{eq104}) les développements modaux (\ref{eq052}) et on regroupe les termes de même type, respectivement Belyaev, Landau et non résonnants:
\be
\label{eq100}
H_3 = {\ycd H_3^{(+)} + H_3^{(-)}} + H_3^{\mbox{\scriptsize (non rés.)}}
\ee
avec
\be
\label{eq101}
H_3^{(-)}= (H_3^{(+)})^\dagger = \frac{1}{2 L^{d/2}}\sum_{\kk,\kk'} \mA(\kk,\kk';\kk+\kk') \hat{b}_{\kk+\kk'}^\dagger \hat{b}_\kk \hat{b}_{\kk'} \quad ; \quad H_3^{\mbox{\scriptsize (non rés.)}} = \frac{1}{6 L^{d/2}} \sum_{\kk,\kk'} \mA(\kk,\kk';-(\kk+\kk')) (\hat{b}_{\kk}\hat{b}_{\kk'}\hat{b}_{-(\kk+\kk')}+\mbox{h.c.})
\ee
et l'amplitude de couplage\footnote{Par choix de notation, notre amplitude $\mA$  diffère d'un facteur deux de celle de la référence \cite{Annalen} {\yct(plus précisément d'un facteur $2mc^2/\rho^{1/2}$)}.}
\be
\label{eq102}
\mA(\kk,\kk';\kk'')=mc^2\left(\frac{\hbar^3kk'k''}{8\rho m^3c^3}\right)^{1/2}\left(3\Lambda+\frac{\kk\cdot\kk'}{k k'}+\frac{\kk\cdot\kk''}{k k''}+\frac{\kk'\cdot\kk''}{k' k''}\right)
\ee
Bien entendu, tous les vecteurs d'onde $\kk$, $\kk'$, $\pm(\kk+\kk')$ apparaissant dans l'hamiltonien sont non nuls et soumis à la condition (\ref{eq103}). L'écriture de $H_4$ en point de vue $\hat{b}$ et $\hat{b}^\dagger$ ne présente guère d'intérêt ici et nous en faisons l'économie.

\subsection{Modèle de champ classique}
\label{sec1.2}

Le modèle est construit en remplaçant les opérateurs $\hat{b}_\kk$ et $\hat{b}_\kk^\dagger$ dans la section \ref{sec1.1} par des amplitudes complexes $b_\kk$ et $b_\kk^*$. Ceci n'a de sens que pour les modes de phonons fortement peuplés, ce qui impose un choix de coupure $\eta\lesssim 1$ dans l'équation (\ref{eq103}); dans les simulations, nous prendrons $\eta=1$. À l'équilibre thermique, les nombres d'occupation des modes ne sont plus donnés par la loi de Bose mais par l'équipartition $\bar{n}_\kk^{\rm clas}=k_B T/\veps_\kk$. 

Limitons notre étude numérique au cas expérimentalement pertinent d'un gaz de bosons sans spin dans le régime quasi 2D où le potentiel chimique $\mu$ dans le plan $xOy$ est inférieur à mais pas très petit devant le quantum d'excitation $\hbar\omega_\perp$ du mouvement selon $Oz$ ; dans ce cas, l'équation d'état prend la forme simple $\mu=\rho g$ avec $g=(8\pi)^{1/2}\hbar^2a_{\rm 3D}/(ma_\perp)$, où $a_{\rm 3D}$ est la longueur de diffusion à 3D et $a_\perp=(\hbar/m\omega_\perp)^{1/2}$ la taille de l'oscillateur harmonique selon $Oz$ \cite{CD,HD}, si bien que $\Lambda=\Lambda_*=0$ dans l'équation (\ref{eq032}) et $H_4\equiv 0$ dans l'équation {\yc(\ref{eq104})}.\footnote{L'équation d'état de Popov d'un gaz de bosons strictement bidimensionnel dans le régime d'interaction faible correspond à une pseudo-constante de couplage dépendant du potentiel chimique, $g(\mu)=4\pi\hbar^2/[m\ln(\veps_0/\mu)]$ où $\veps_0=4\hbar^2/[ma_{\rm 2D}^2\exp(2C+1)]$, $a_{\rm 2D}$ est la longueur de diffusion à 2D et $C=0,\!577\ldots$ est la constante d'Euler \cite{Popov,Morathese}. Dans son domaine de validité \cite{MC2}, elle conduit à des valeurs non nulles mais faibles de $\Lambda$ et $\Lambda_*$, par exemple {\ycd $\Lambda=mg(\mu)/(12\pi\hbar^2)/[1-mg(\mu)/(4\pi\hbar^2)]^2\simeq mg(\mu)/(12\pi\hbar^2) \ll 1$. On raccorde les régimes quasi et strictement 2D en remarquant que $g=g(\mu=0,\!105\,965\ldots\hbar\omega_\perp)$ au vu de l'expression de $a_{\rm 2D}$ en termes de $a_{\rm 3D}$ et $a_\perp$ donnée dans les références \cite{Petrov,Olshanii,Ludovic}.}} 

La méthode de simulation numérique est décrite en détail dans la référence \cite{TheseAlan}. Pour échantillonner l'état initial, on tire les $b_\kk$ selon la distribution thermique (gaussienne) pour l'hamiltonien $H_2$, puis on fait évoluer le champ résultant avec l'hamiltonien complet $H$ pour atteindre le vrai équilibre thermique {\yc par ergodicité}. L'évolution temporelle sur chaque pas $dt$ est séparée en l'évolution due à $H_2$ pendant $dt/2$, l'évolution due à $H_3$ seul pendant $dt$ puis à nouveau l'évolution due à $H_2$ pendant $dt/2$. L'évolution due à $H_2$ se calcule exactement dans l'espace des $\kk$. L'évolution due à $H_3$ est faite dans l'espace des $\rr$\footnote{Comme les équations du mouvement dérivant de $H_3$ sont locales dans l'espace des $\rr$ (ou quasi locales puisqu'elles font intervenir des gradients spatiaux), la complexité sur un pas $dt$ est de degré un en le nombre de points $\mathcal{N}$ sur la grille numérique ; dans l'espace des $\kk$, elles sont fortement non locales (chaque amplitude $b_\kk$ est couplée en gros à toutes les autres) d'où une complexité catastrophique $\propto \mathcal{N}^2$.} par la méthode de Runge-Kutta d'ordre quatre \cite{NumRec}, en pensant à filtrer le champ dans l'espace des $\kk$ comme dans l'équation (\ref{eq103}) après chaque action de $H_3$ dans l'espace des $\rr$ pour s'affranchir des processus violant la coupure qu'elle a induits.\footnote{Le passage de $\rr$ à $\kk$ est effectué par une transformation de Fourier rapide, dans laquelle les vecteurs d'onde décrivent l'ensemble $\mD\cap(2\pi/L)\mathbb{Z}^d$ avec $\mD=[-k_{\rm max},k_{\rm max}[^d$ (chaque composante cartésienne $k_\alpha$ de $\kk$ a un sens modulo $2 k_{\rm max}$, si bien que $\mD$ peut être considéré comme la première zone de Brillouin du réseau spatial $(\pi/k_{\rm max})\mathbb{Z}^d$, qui est précisément celui décrit par $\rr$ {\yc dans les simulations avec} restriction à $[0,L[^d$). Le choix naïf $k_{\rm max}=\eta k_B T/\hbar c$ associé à la coupure (\ref{eq103}) ne convient pas: les processus de $H_3$ faisant sortir les phonons de la première zone de Brillouin correspondante sont {\yc alors} repliés dans cette zone (processus Umklapp) et ne peuvent plus être filtrés. Il faut utiliser au minimum $k_{\rm max}=2\eta k_B T/\hbar c$: le filtrage nous oblige à doubler le nombre de points par dimension.}  Le pas $dt$ doit être assez petit pour que l'énergie totale sur une seule réalisation soit conservée à mieux qu'un pour cent de l'énergie d'interaction typique $\langle |H_3|\rangle$, où $\langle\ldots\rangle$ est la moyenne sur un grand nombre de réalisations, elle-même une très faible fraction de l'énergie totale.

Il serait inefficace numériquement de déterminer l'amortissement des phonons par simulation de l'excitation de Bragg (\ref{eq010}) car ceci obligerait à refaire un grand nombre de simulations indépendantes pour chaque nouvelle valeur du vecteur d'onde $\qq$, tout en s'assurant que l'amplitude $\alpha$ est assez faible pour qu'on soit en régime de réponse linéaire.\footnote{La version de champ classique $\langle\partial b_\qq(t)/\partial b_\qq(0)\rangle_{\rm th}$ de la fonction de corrélation $\langle[\hat{b}_\qq(t),\hat{b}_\qq^\dagger(0)]\rangle_{\rm th}$, qui fait intervenir une dérivée numérique dans le mode particulier $\qq$, est confrontée aux mêmes problèmes.} Nous préférons utiliser la fonction de corrélation suivante, permettant de traiter en parallèle toutes les valeurs possibles de $\qq$ par simple moyenne sur un même jeu de réalisations indépendantes:\footnote{Le passage au troisième membre de l'équation (\ref{eq110}) {\yc se fait au moyen de la relation exacte (31.24) de la référence \cite{FW} mais est ici} approché car nous avons remplacé au dénominateur $\langle\hat{b}_\qq(0)\hat{b}_\qq^\dagger(0)\rangle_{\rm th}$ par $1+\bar{n}_{\qq}$ où $\bar{n}_{\qq}$ est la loi de Bose à l'énergie $\veps_{\qq}$; {\yc cependant, c'est} une excellente approximation pour nos simulations.}
\be
\label{eq110}
s_*(t)\equiv \frac{\eee^{\ii\veps_\qq t/\hbar}\langle\hat{b}_\qq(t)\hat{b}_\qq^\dagger(0)\rangle_{\rm th}}{\langle\hat{b}_\qq(0)\hat{b}_\qq^\dagger(0)\rangle_{\rm th}}\stackrel{t>0}{\simeq} \lim_{\delta\to 0^+} \int_{-\infty}^{+\infty} 
\frac{\dd\veps}{-2\ii\pi} \eee^{-\ii(\veps-\veps_\qq)t/\hbar} \frac{\mathcal{G}(\qq,\veps+\ii\delta)+\left[\mathcal{G}(\qq,\veps+\ii\delta)-\mathcal{G}(\qq,\veps-\ii\delta)\right]/[\exp(\veps/k_B T)-1]}{1+1/[\exp(\veps_\qq/k_B T)-1]}
\ee
Il se trouve que ce signal est en pratique très proche de celui de l'équation (\ref{eq012}), {\yc ce qui autorise dans les simulations à l'identifier à $s(t)$, comme nous le ferons dans la légende des figures:}
\be
\label{eq111}
s_*(t) \simeq s(t)
\ee
En effet, même pour l'interaction la plus forte considérée dans nos simulations, on dispose, pour toute valeur de l'énergie $\veps$, de la propriété $|\im\Sigma(\qq,\veps+\ii 0^+)|\ll \veps_\qq$, si bien que la différence entre les fonctions de Green $\mG(\qq,\veps\pm\ii\delta)$ entre crochets dans l'équation (\ref{eq110}), qui ne peut résulter que d'une partie imaginaire non nulle de la fonction énergie propre dans la limite $\delta\to 0^+$, est très petite sauf dans la largeur de la \g{lorentzienne} $\mathcal{G}(\qq,\veps+\ii 0^+)$ où $|\veps-\veps_\qq|\lesssim |\im\Sigma(\qq,\veps+\ii 0^+)|\ll \veps_\qq, k_B T$. C'est donc une excellente approximation que de remplacer $\veps$ par $\veps_\qq$ dans la loi de Bose au numérateur de l'intégrande dans (\ref{eq110}). Par analyticité dans le demi-plan complexe inférieur, le terme proportionnel à $\mathcal{G}(\qq,z-\ii\delta){\ycd\,\exp(-\ii zt/\hbar)}$ donne alors une contribution nulle à l'intégrale {\ycd pour $t>0$} (on le voit en refermant le chemin d'intégration par un demi-cercle à l'infini et en utilisant le théorème de Cauchy) et on aboutit à (\ref{eq111}). La même conclusion vaut dans le modèle de champ classique, où les deux fonctions exponentielles dans (\ref{eq110}) sont linéarisées.

\subsection{Les résultats numériques}
\label{sec1.3}

\begin{figure}[t]
\begin{center}
\includegraphics[width=6cm,clip=]{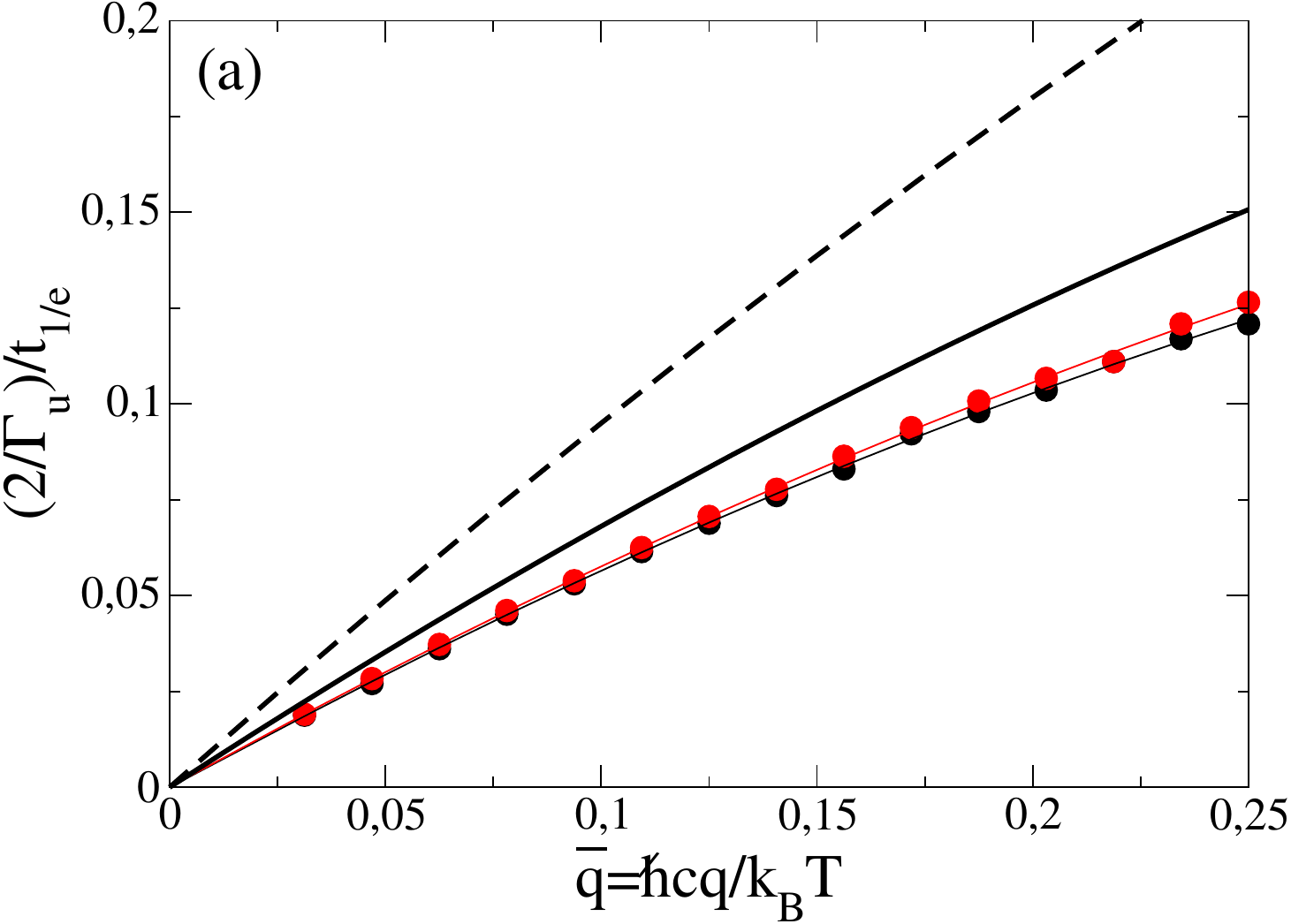} \hspace{1cm} \includegraphics[width=6cm,clip=]{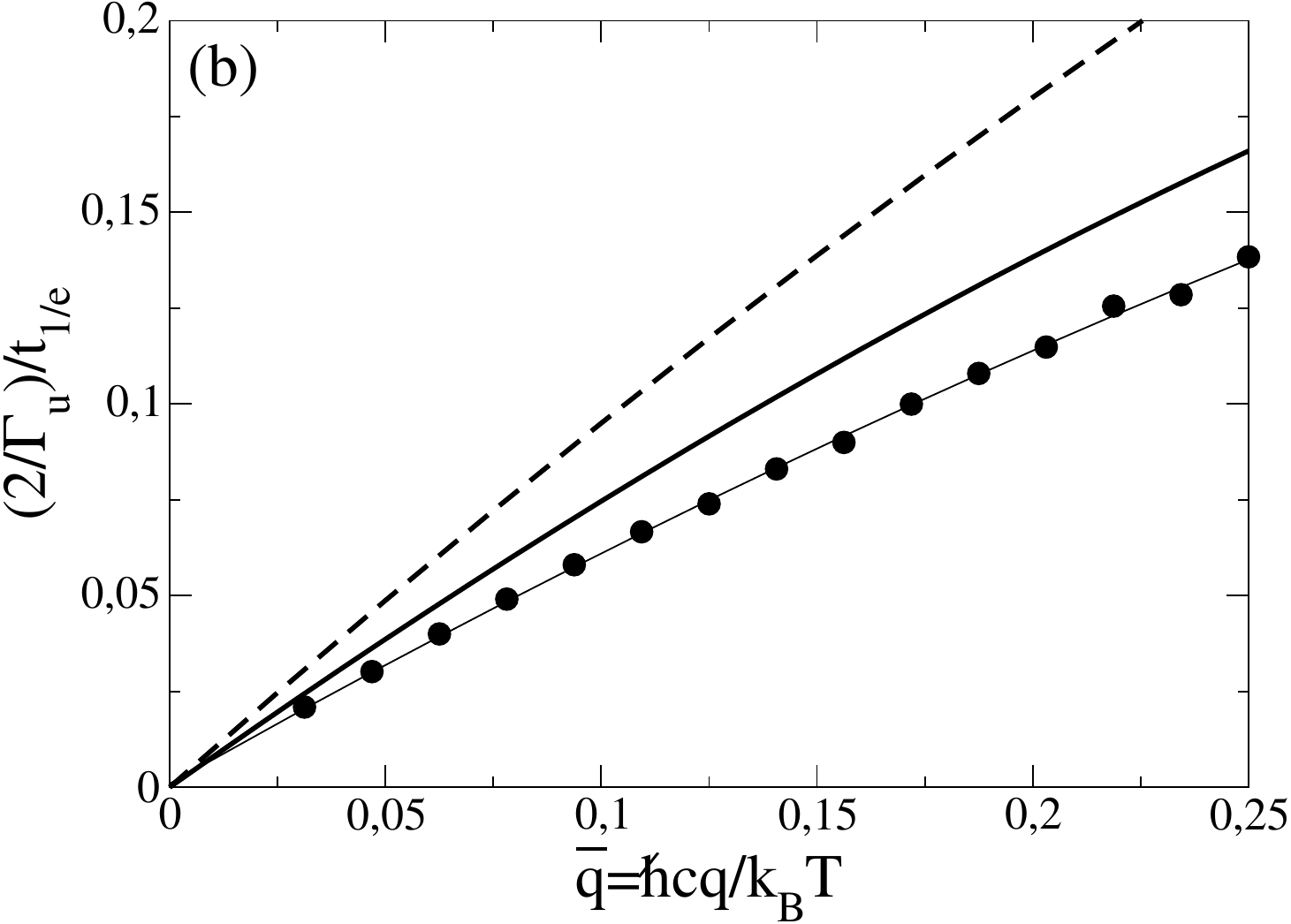} \\
\includegraphics[width=6cm,clip=]{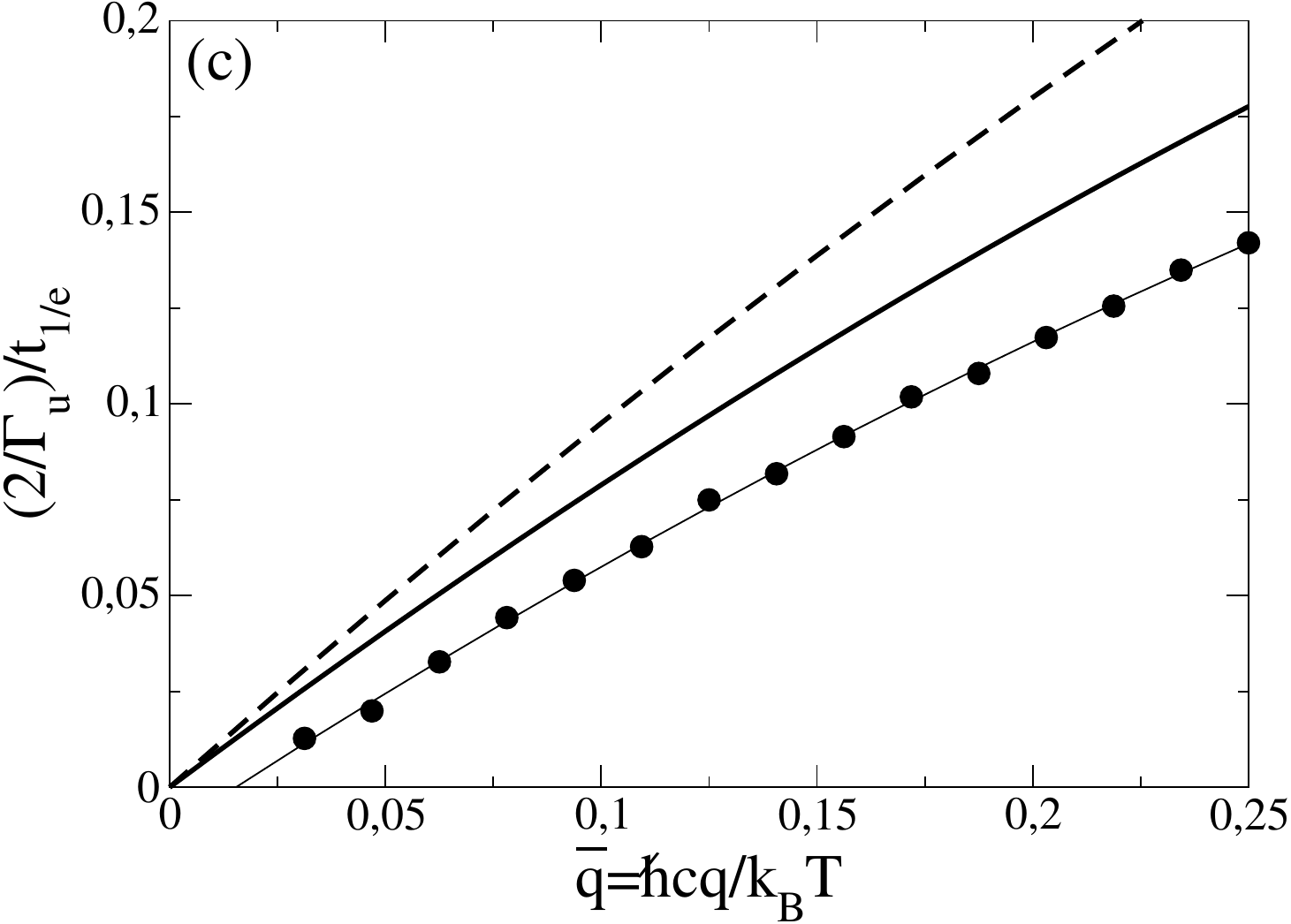} \hspace{1cm} \includegraphics[width=6cm,clip=]{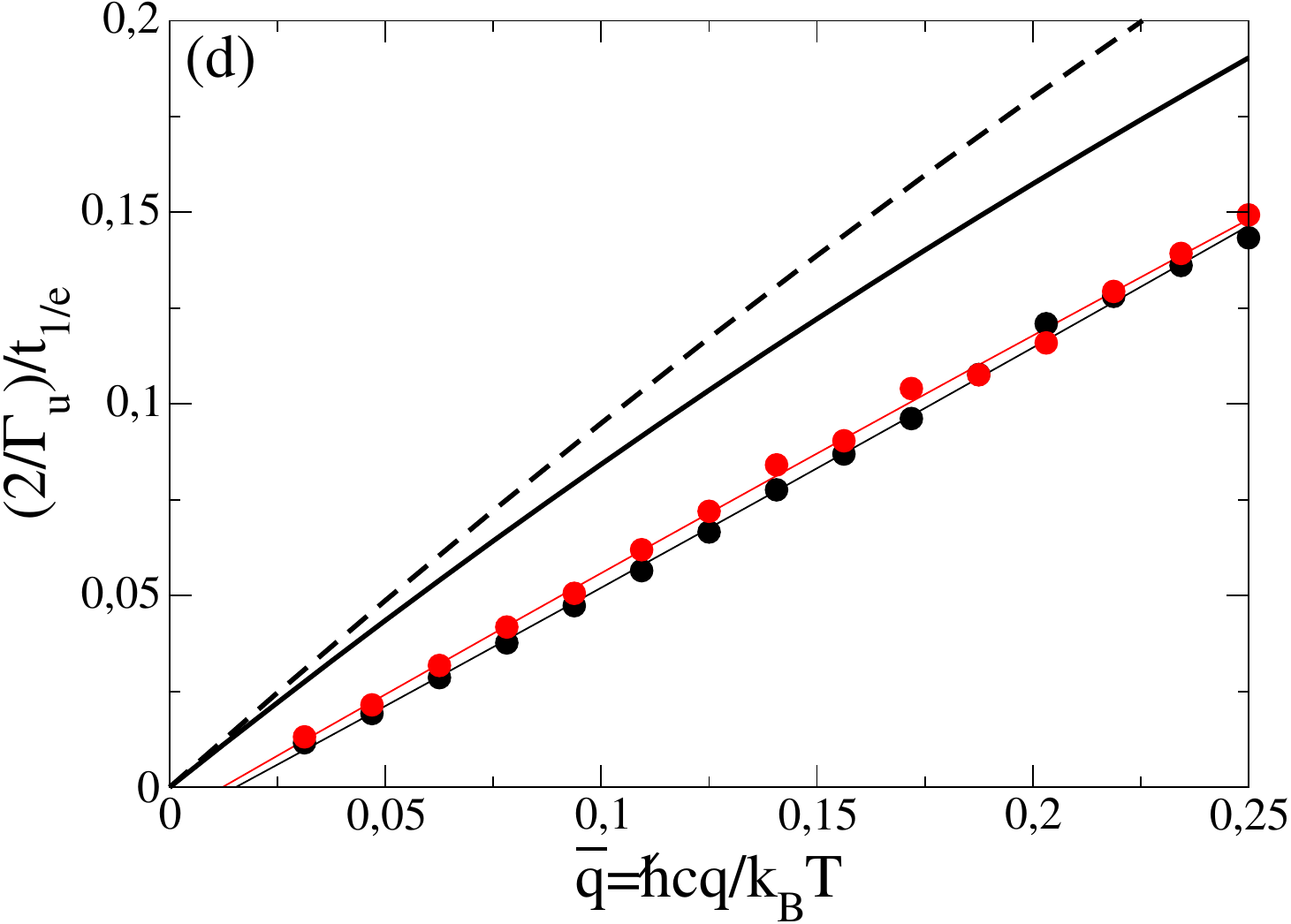}
\end{center}
\caption{Inverse du temps de décroissance à $1/\eee$ du module du signal $s(t)$ (\ref{eq012}) en fonction du nombre d'onde $q$ des modes selon $Ox$, $\qq=\pm q \mathbf{e}_x$, pour différentes forces des interactions (de plus en plus faibles) dans le superfluide {\yct bidimensionnel} de bosons: (a) $(\rho\xi^2)^{-1}=0,\!64$, (b) $(\rho\xi^2)^{-1}=0,\!46$, (c) $(\rho\xi^2)^{-1}=0,\!28$, (d) $(\rho\xi^2)^{-1}=0,\!1$. Disques (superposés à un ajustement quadratique en trait fin): expérience numérique avec $\epsilon=1/2$ (en noir) ou $\epsilon=1/3$ (en rouge). Tireté noir: règle d'or de Fermi (\ref{eq113}) pour le champ phononique classique {\yc dans} la limite $\epsilon\to 0$. Trait plein noir: méthode des fonctions de Green à $N$ corps avec la fonction énergie propre {\yc $\tilde{\Sigma}^{(2,2)}_{\qb}(\zetat)$} (\ref{eq125}) (ordre dominant en $H_3$ et $\epsilon$). Les autres paramètres ont les valeurs fixées suivantes: $\gamma=1$, $\Lambda=\Lambda_*=0$, $\eta=1$. Dans les simulations, la taille du système est assez grande pour que la limite thermodynamique soit atteinte (une largeur de boîte de quantification $L=128\pi\hbar c/k_B T$ suffit). Dans la théorie, les fonctions énergies propres sont remplacées par leurs versions de champ classique pour permettre la comparaison au numérique. Le taux $\Gamma_u$ servant d'unité est celui de l'équation (\ref{eq113}).}
\label{fig1}
\end{figure}

\begin{figure}[t]
\begin{center}
\includegraphics[width=6cm,clip=]{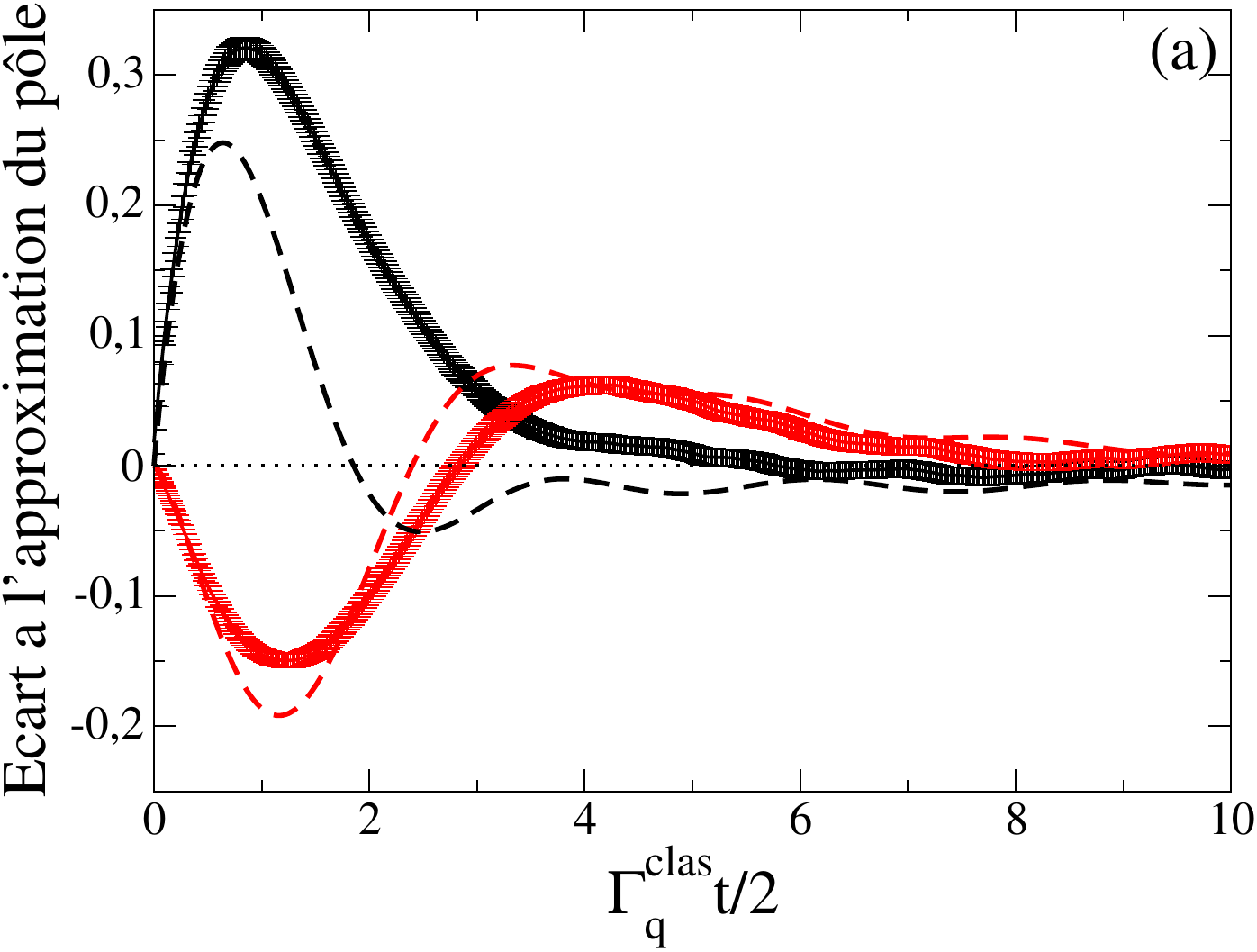}\hspace{1cm} \includegraphics[width=6cm,clip=]{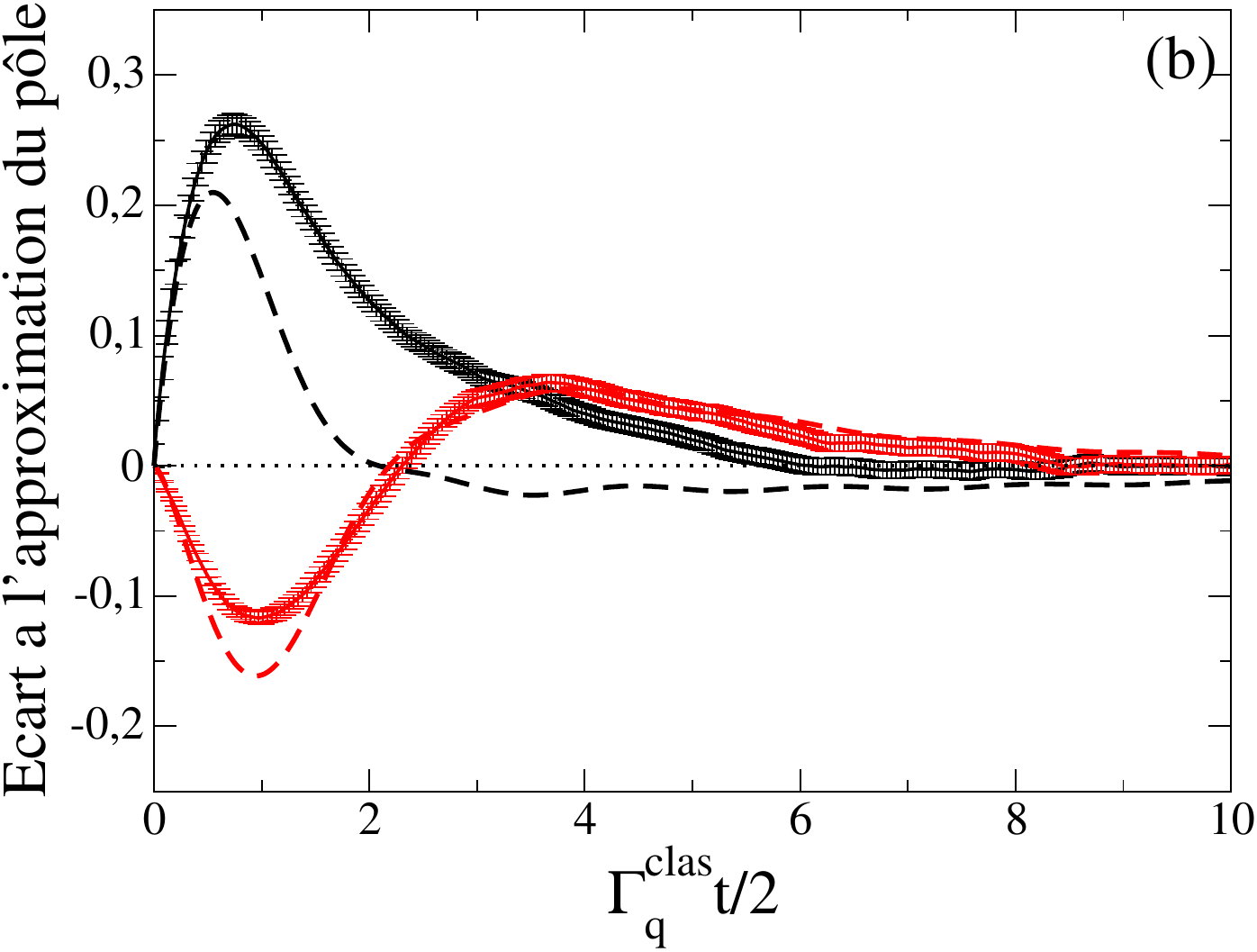} \\
\includegraphics[width=6cm,clip=]{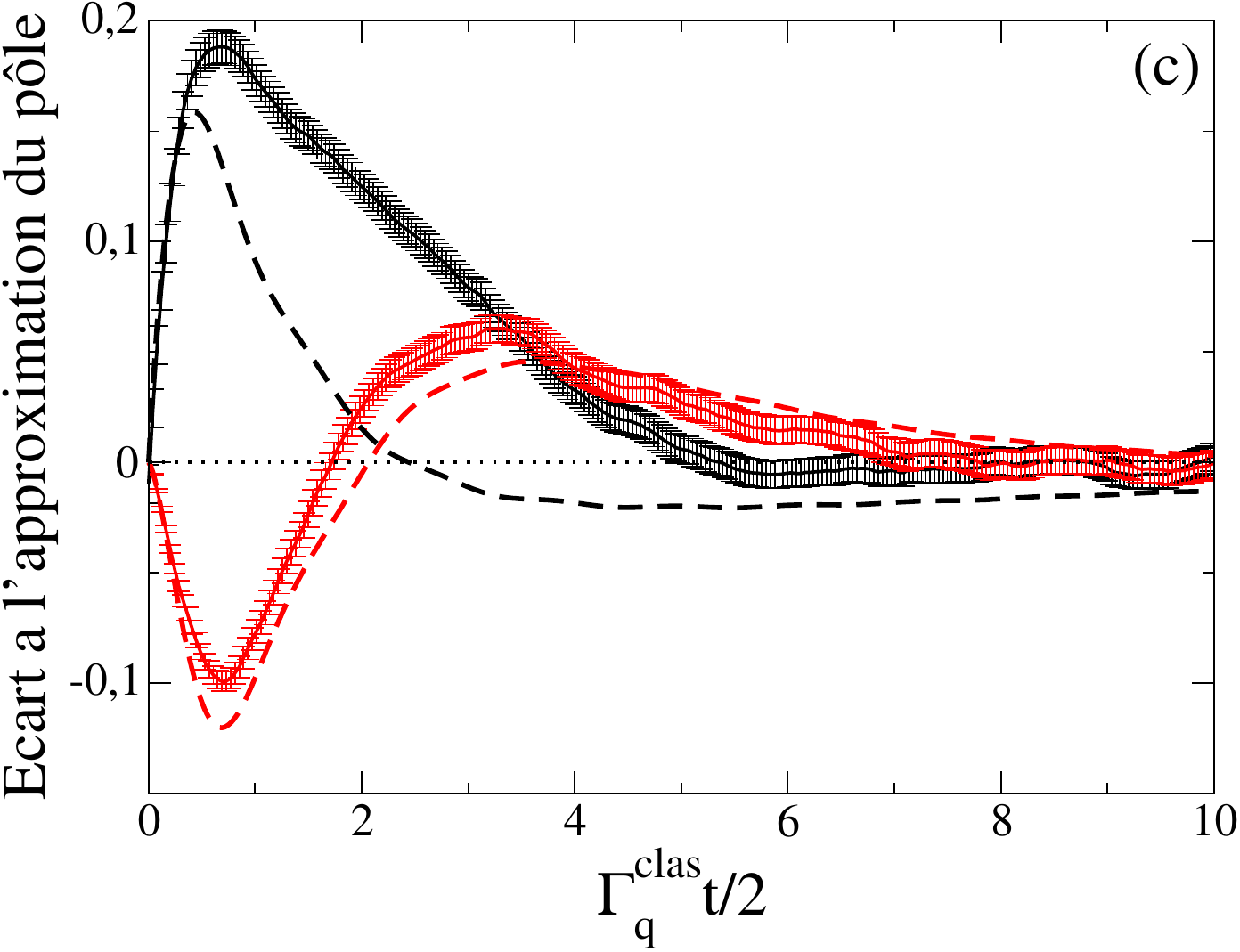}\hspace{1cm} \includegraphics[width=6cm,clip=]{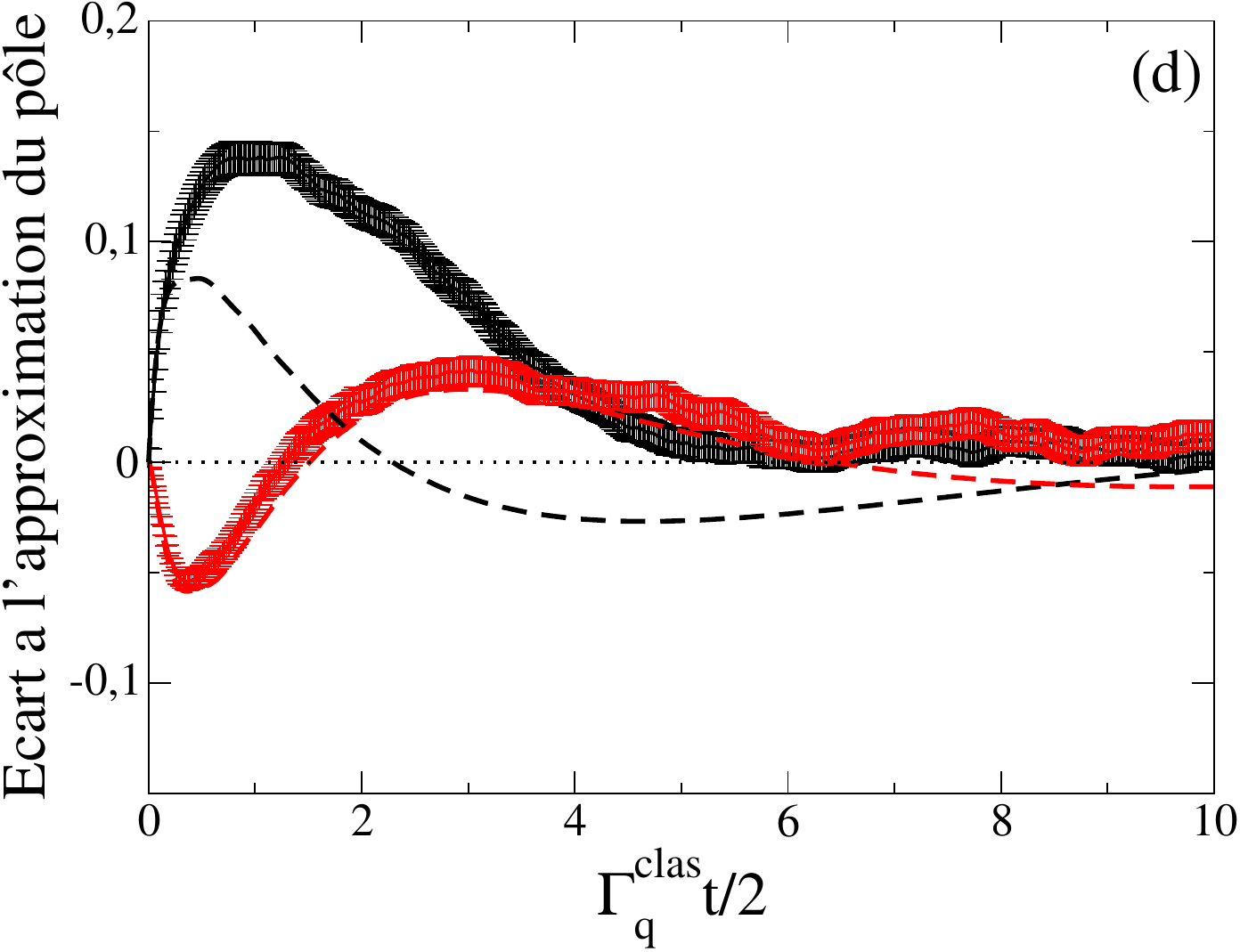}
\end{center}
\caption{Écart {\yc $s(t)-s_{\mbox{\scriptsize pôle}}(t)$} du signal (\ref{eq012}) à l'approximation du pôle (\ref{eq014}) en fonction du temps, pour le mode de nombre d'onde réduit $\qb=1/4$ selon $Ox$ et pour les mêmes forces des interactions que {\yc sur} la figure \ref{fig1}: (a) $(\rho\xi^2)^{-1}=0,\!64$, (b) $(\rho\xi^2)^{-1}=0,\!46$, (c) $(\rho\xi^2)^{-1}=0,\!28$, (d) $(\rho\xi^2)^{-1}=0,\!1$. Courbes avec barres d'erreur: résultats de l'expérience numérique ; tireté : théorie des fonctions de Green à $N$ corps avec la fonction énergie propre $\tilde{\Sigma}^{(2,2)}_{\yc\qb}(\zetat)$ (\ref{eq125}) (ordre dominant en $H_3$ et $\epsilon$) dans sa version de champ classique. La partie réelle de l'écart est représentée en noir, la partie imaginaire en rouge. On a pris $\epsilon=1/2$ dans l'expérience numérique et pour le calcul de $s_{\mbox{\scriptsize pôle}}(t)$ (effectué à l'ordre deux en $H_3$ et à tous les ordres en $\epsilon$) mais on a vérifié qu'on obtient les mêmes résultats (dans les barres d'erreur) pour $\epsilon=1/3$. Les autres paramètres sont fixés à $\gamma=1$, $\Lambda=\Lambda_*=0$, $\eta=1$. Dans les simulations, les tailles choisies assurent l'atteinte de la limite thermodynamique (largeur $L=128\pi\hbar c/k_B T$ pour (a), (b) et (c) ; largeur double pour (d)). Le temps est adimensionné par le demi-taux de la règle d'or en champ classique (\ref{eq113}).}
\label{fig2}
\end{figure}

Sur la figure \ref{fig1}, nous portons l'inverse du temps de décroissance à $1/\eee$ du signal en fonction du nombre d'onde réduit $\qb=\hbar cq/k_B T$ du mode considéré, pour quatre valeurs de la force des interactions repérée par $1/(\rho\xi^2)$ et pour une ou deux valeurs de $\epsilon=k_B T/mc^2$, voir les disques. L'unité de temps choisie est $2\Gamma_u^{-1}$, où la constante $\Gamma_u$ apparaît en préfacteur dans le taux d'amortissement de la règle d'or de Fermi à l'ordre dominant en $\epsilon$, ici pour le champ classique:\footnote{Dans l'expression (\ref{eq033}) pour le champ quantique, on remplace $\bar{n}_k^{\rm lin}$ par $1/\kb$, on néglige le terme $1/2$ devant $\bar{n}_k^{\rm lin}$ dans la première intégrale et on met la coupure $\kb<\eta-\qb$ dans la seconde.}
\be
\label{eq113}
\boxed{\Gamma_q^{\mbox{\scriptsize clas}}=\Gamma_u \qb(\eta-\qb/2) \quad \mbox{avec}\quad \Gamma_u = \frac{9k_B T\epsilon^2(1+\Lambda)^2}{4\pi\hbar(3\gamma)^{1/2}\rho\xi^2}}
\ee
Comme on le voit sur les panneaux (a) et (d) de la figure \ref{fig1}, les résultats ainsi normalisés ne dépendent pas significativement de $\epsilon$ à $\qb$ fixé et la limite $\epsilon\to 0$ semble atteinte ; dans les simulations, l'inverse du temps de décroissance à $1/\eee$ varie donc bien comme $k_B T\epsilon^2$ à basse température, comme le prédisaient la règle d'or et plus généralement l'estimation (\ref{eq028}) à tous les ordres en $H_3$. En revanche, la dépendance en $\qb$ du temps d'amortissement est fort mal décrite par la règle d'or, en tireté sur la figure \ref{fig1}: comme le prévoyait le raisonnement général de la section \ref{sec0}, la limite $\epsilon\to 0$ ne suffit donc pas à 2D à faire entrer le gaz de phonons dans un régime de couplage faible. Ce que la section \ref{sec0} ne disait pas, cependant, c'est que la règle d'or reste insuffisante même si le superfluide sous-jacent entre dans le régime d'interaction faible $(\rho\xi^2)^{-1}\to 0$, voir le panneau (d) où un calcul perturbatif à l'ordre dominant en $H_3$ devrait a priori être légitime; le départ à faible $\qb$ prédit par (\ref{eq113}) y est d'ailleurs particulièrement mauvais. 

Pour rendre l'insuffisance de la règle d'or encore plus criante, nous représentons sur la figure \ref{fig2}, en fonction du temps, l'écart du signal numérique à l'approximation du pôle (\ref{eq014}) limitée à l'ordre deux en $H_3$ {\yc mais prise à tous les ordres en $\epsilon$} [voir l'équation (\ref{eq124})] et traduite en champ classique (si l'on était à l'ordre dominant en $\epsilon$, {\yc ce serait simplement} l'écart à l'exponentielle $\exp(-\Gamma_q^{\rm clas}t/2)$ de la règle d'or), pour les mêmes forces d'interaction que précédemment mais pour une valeur fixée du nombre d'onde réduit, $\qb=1/4$, suffisamment grande pour rendre aisée l'atteinte de la limite thermodynamique dans les simulations.\footnote{Plus $\qb$ est petit, plus le taux d'amortissement du mode est faible et plus la granularité des niveaux d'énergie du gaz de phonons dans la boîte de quantification se fait sentir, et plus il faut augmenter la taille du système pour atteindre la limite thermodynamique.} Là aussi, la limite $\epsilon\to 0$ est atteinte dans les simulations (lorsque le temps $t$ est mis à l'échelle comme dans (\ref{eq022}) avec $\nu=2d-2=2$, ici en l'exprimant en unités de $2/\Gamma_q^{\rm clas}$), l'écart {\yc attendu} à la règle d'or est bien là et persiste même dans la limite d'interaction faible $(\rho\xi^2)^{-1}\to 0$, du moins sur la partie réelle.\footnote{Si l'on passe du panneau (c) au panneau (d) sur la figure \ref{fig2}, la force des interactions est divisée par $\simeq 3$, la partie imaginaire est divisée par $\simeq 2$ en valeur absolue maximale donc pourrait bien tendre vers zéro, alors que la partie réelle ne décroît que d'un facteur $\simeq 1,\!3$.}

\section{Étude à l'ordre dominant en le couplage phonon-phonon}
\label{sec2}

Les simulations de champ classique de la section \ref{sec1} ont établi l'insuffisance de la règle d'or de Fermi dans la description de l'amortissement des phonons en dimension $d=2$; nous ne l'avons pas montré ici, mais outre le désaccord sur le temps de décroissance à $1/\eee$, elles révèlent, en échelle logarithmique, une décroissance non exponentielle du signal $s(t)$ \cite{TheseAlan}. Dans cette section, nous incriminons l'approximation faite dans l'étape (i) de la section \ref{sec0}, et y renonçons en gardant toute la dépendance en énergie de la fonction énergie propre dans l'expression intégrale (\ref{eq012}) de $s(t)$, en continuant cependant à calculer $\Sigma_\qq(\zeta)$ à l'ordre dominant en le couplage phonon-phonon et en $\epsilon$ (comme dans l'étape (ii) de la section \ref{sec0}). Après un calcul diagrammatique de $\Sigma^{(2)}$, nous obtenons quelques prédictions analytiques mises sous une forme universelle puis nous comparons aux simulations.

\subsection{Calcul de la fonction énergie propre}
\label{sec2.1}

Dans l'hamiltonien $H$ de l'équation (\ref{eq001}), le couplage cubique $H_3$ semble prépondérant par rapport au couplage quartique $H_4$, puisqu'il apparaît à l'ordre $\delta\hat{\rho}^3$ plutôt que $\delta\hat{\rho}^4$ dans le développement de l'hamiltonien hydrodynamique, voir la section \ref{sec1.1}. Cependant, comme il ne conserve pas le nombre de phonons, il doit être traité dans le calcul de $\Sigma_\qq(\zeta)$ au minimum au second ordre de la théorie des perturbations, alors que $H_4$ donne une contribution non nulle dès le premier ordre, à première vue aussi importante que celle de $H_3$.\footnote{Là où apparaît dans la contribution de $H_3$ {\yc à $\Sigma_\qq(\ii 0^+)$} le rapport $\langle\ |\mH_3|\ \rangle^2/\Delta E\approx \delta\hat{\rho}^{2\times 3-2}$, intervient dans celle de $H_4$ l'élément de matrice $\langle\ |\mH_4|\ \rangle\approx \delta\hat{\rho}^4$, si l'on utilise les notations de l'équation (\ref{eq024}) et le fait que le dénominateur $\Delta E$, constitué de différences d'énergie de l'hamiltonien quadratique $H_2$, est $\approx\delta\hat{\rho}^2$.} Le calcul complet s'impose.

Au premier ordre de la théorie des perturbations, il n'est point besoin d'utiliser de diagramme de Feynman, la contribution à $\Sigma_\qq(\zeta)$ étant indépendante de la variable $\zeta$ et s'obtenant par la procédure simple suivante, justifiable par la méthode de l'équation pilote de la référence \cite{SCE}. On écrit l'opérateur de la perturbation, ici $H_4$, en point de vue $\hat{b},\hat{b}^\dagger$ et dans l'ordre normal en s'aidant des relations de commutation bosoniques (tous les $\hat{b}^\dagger$ mis à gauche, tous les $\hat{b}$ mis à droite). Les termes ne conservant pas le nombre de phonons ne contribuent pas. Les termes en $\hat{b}^\dagger\hat{b}$, {\yc quadratiques}, sont absorbés dans une redéfinition des énergies propres nues $\veps_\kk$ des modes (comme nous y préparait la note \ref{note0}). Enfin, on isole dans les \g{vrais} termes de couplage quartique {\yc restants} les contributions en $\hat{b}^\dagger_\qq\hat{b}^\dagger_\kk\hat{b}_\kk\hat{b}_\qq$ dont la moyenne thermique sur les modes $\kk\neq\qq$ donne le déplacement d'énergie du mode $\qq$. Tout ceci se résume par l'expression compacte
\be
\label{eq120}
\Sigma_\qq^{(H_4)}(\zeta)=\int\frac{\dd^dk}{(2\pi)^d} \langle\qq,\kk|\mathcal{H}_4^{\rm vrai}
|\qq,\kk\rangle \bar{n}_k = \frac{\hbar^2q\Lambda_*}{4m\rho}\int\frac{\dd^dk}{(2\pi)^d} k\bar{n}_k \underset{\epsilon\to 0}{\approx} \frac{\bar{q}k_B T\epsilon^{d+1}}{\rho\xi^d}
\ee
où $\mH_4$ est l'hamiltonien volumique tel que $H_4=\mH_4/L^{d}$, la constante de couplage $\Lambda_*$ est celle de l'équation (\ref{eq032}), nous avons déduit l'élément de matrice dans l'état à deux phonons de l'expression (\ref{eq104}) de $H_4$ {\yc et} (\ref{eq052}) de $\delta\hat{\rho}(\rr)$, et nous avons obtenu le comportement d'échelle dans la limite (\ref{eq020}) au quatrième membre de l'équation (\ref{eq120}) comme dans la section \ref{sec0}. Comme nous le voyons sur la contribution (\ref{eq028}) de $H_3$ écrite pour $n=1$, la contribution de $H_4$ en dimension $d=2$ (mais pas en dimension trois) est sous-dominante et peut donc être négligée lorsque $\epsilon\to 0$. C'est ce que nous ferons désormais.

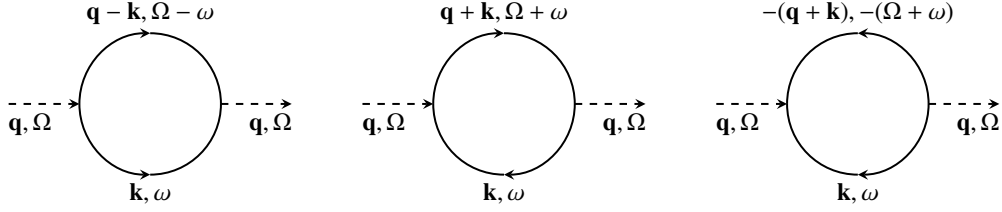
\begin{figure}[t]
\begin{center}
\begin{tikzpicture}
\draw[thick,dashed,->,>=stealth](0,0)--(1,0);
\draw[thick,->,>=stealth](1,0)arc(-180:-90:1);
\draw[thick,->,>=stealth](1,0)arc(180:90:1);
\draw[thick](2,-1)arc(-90:0:1);
\draw[thick](2,1)arc(90:0:1);
\draw[thick,dashed,->,>=stealth](3,0)--(4,0);
\node(a)at(0.3,0)[below]{$\vec{q},\Omega$};
\node(a)at(3.7,0)[below]{$\vec{q},\Omega$};
\node(a)at(2,1)[above]{$\vec{q}-\vec{k},\Omega-\omega$};
\node(a)at(2,-1)[below]{$\vec{k},\omega$};
\draw[thick,dashed,->,>=stealth](0+5,0)--(1+5,0);
\draw[thick](1+5,0)arc(-180:-90:1);
\draw[thick,->,>=stealth](1+5,0)arc(180:90:1);
\draw[thick,->,>=stealth](3+5,0)arc(0:-90:1);
\draw[thick](2+5,1)arc(90:0:1);
\draw[thick,dashed,->,>=stealth](3+5,0)--(4+5,0);
\node(a)at(0.3+5,0)[below]{$\vec{q},\Omega$};
\node(a)at(3.7+5,0)[below]{$\vec{q},\Omega$};
\node(a)at(2+5,1)[above]{$\vec{q}+\vec{k},\Omega+\omega$};
\node(a)at(2+5,-1)[below]{$\vec{k},\omega$};
\draw[thick,dashed,->,>=stealth](0+10,0)--(1+10,0);
\draw[thick](1+10,0)arc(-180:-90:1);
\draw[thick](1+10,0)arc(180:90:1);
\draw[thick,->,>=stealth](3+10,0)arc(0:-90:1);
\draw[thick,->,>=stealth](3+10,0)arc(0:90:1);
\draw[thick,dashed,->,>=stealth](3+10,0)--(4+10,0);
\node(a)at(0.3+10,0)[below]{$\vec{q},\Omega$};
\node(a)at(3.7+10,0)[below]{$\vec{q},\Omega$};
\node(a)at(2+10,1)[above]{$-(\vec{q}+\vec{k}),-(\Omega+\omega)$};
\node(a)at(2+10,-1)[below]{$\vec{k},\omega$};
\end{tikzpicture}
\end{center}
\caption{La fonction énergie propre $\Sigma(\qq,z)$ est donnée à l'ordre deux en $H_3$ par des diagrammes à une boucle, correspondant aux processus de Belyaev, de Landau et non résonnants (de gauche à droite sur la figure). Les variables $\Omega$ et $\omega$ sont les énergies de Matsubara associées aux vecteurs d'onde entrant $\qq$ et interne $\kk$. Les lignes internes (d'orientations quelconques et sur lesquelles il faut sommer) sont en trait plein, les lignes externes (dont l'orientation est fixée par le choix de la fonction de Green) sont en tireté.}
\label{fig6}
\end{figure}

Au second ordre en $H_3$, il vaut mieux avoir recours à la méthode de la référence \cite{FW}.\footnote{Nos états de phonons ayant directement la symétrie bosonique, nous pouvons faire l'économie des lignes d'interaction directe et d'échange introduites dans \cite{FW}.} Cette méthode {\yc fournit} un développement diagrammatique de la fonction énergie propre $\Sigma(\qq,z)$ pour des valeurs particulières de la variable énergie complexe, $z=\Omega\in 2\ii\pi k_B T \mathbb{Z}$, où $\Omega$ est appelée énergie de Matsubara,\footnote{Petite entorse à la notation habituelle \cite{FW}, nos variables de Matsubara $\Omega$, $\omega$, etc., ne sont pas des fréquences réelles mais des énergies imaginaires pures, c'est-à-dire que nous y incorporons un facteur $\ii\hbar$ supplémentaire.} que l'on complète dans un deuxième temps par un prolongement analytique à $\mathbb{C}\setminus\mathbb{R}$ \cite{FW}. Ici, comme sur la figure \ref{fig6}, chaque diagramme doit comporter deux sommets (nous sommes au second ordre), chaque sommet voyant se rejoindre trois lignes de phonons ($H_3$ est cubique en les $\hat{b},\hat{b}^\dagger$). La seule topologie possible est à une boucle et, puisqu'il faut considérer comme identiques deux boucles se déduisant l'une de l'autre par rotation d'angle $\pi$ autour leur axe, les quatre orientations possibles des deux lignes internes ne conduisent qu'à trois diagrammes distincts, représentés dans l'ordre Belyaev, Landau et non résonnant sur la figure \ref{fig6} (pour reprendre la typologie de l'équation (\ref{eq100})). Compte tenu de la conservation de l'énergie de Matsubara et de l'impulsion à chaque sommet (la somme des quantités entrantes doit égaler celle des quantités sortantes), il reste une seule énergie de Matsubara $\omega$ et un seul vecteur d'onde $\kk$ indépendants, sur lesquels il faut sommer. Les règles de Feynman {\yc de} la section 25 de la référence \cite{FW} conduisent alors pour notre gaz de phonons de potentiel chimique nul à
\be
\label{eq121}
\Sigma^{(2)}(\qq,\Omega) = -k_B T \sum_{\omega\in 2\ii\pi k_B T\mathbb{Z}}\int\frac{\dd^dk}{(2\pi)^d} \Bigg\{\frac{(1/2)|\langle\kk,\qq-\kk|\mathcal{H}_3|\qq\rangle|^2}{(\omega-\veps_\kk)(\Omega-\omega-\veps_{\qq-\kk})}+\frac{|\langle\kk+\qq|\mathcal{H}_3|\kk,\qq\rangle|^2}{(\omega-\veps_\kk)(\Omega+\omega-\veps_{\kk+\qq})}+\frac{(1/2)|\langle 0|\mathcal{H}_3|\qq,\kk,-(\kk+\qq)\rangle|^2}{(\omega-\veps_\kk)(-\Omega-\omega-\veps_{-(\kk+\qq)})}
\Bigg\}
\ee
{\yc dans la limite thermodynamique avec $\mathcal{H}_3=H_3 L^{d/2}$, $|0\rangle$ le vide de phonons et,} pour chaque contribution, un domaine d'intégration en $\kk$ assurant que les vecteurs d'onde mis en jeu $\kk$, $\qq-\kk$, etc., satisfont à la coupure (\ref{eq103}). Sont inclus aussi les facteurs de symétrie $1/2$ lorsque les deux branches de la boucle ont la même orientation et sont donc physiquement identiques. Les sommes de Matsubara se calculent au moyen de l'expression générale \footnote{\label{notedemo} La démonstration se fait très bien par décomposition du sommande en éléments simples vis-à-vis de la variable $n$ et utilisation de l'identité $\lim_{N\to +\infty} \sum_{n=-N}^{N} 1/(2\ii\pi n-\vepsb)=-(\bar{n}_\veps+1/2)\ \forall \veps\in{\yc\mathbb{C}\setminus 2\ii\pi\mathbb{Z}}$ résultant des propriétés de la fonction digamma $\psi(z)$ {\yc comme la formule de réflexion}.}
\be
\label{eq122}
\sum_{n\in\mathbb{Z}} \frac{1}{(2\ii\pi n-\vepsb_1)(2\ii\pi n-\vepsb_2)} = -\frac{\bar{n}_{\veps_1}-\bar{n}_{\veps_2}}{\vepsb_1-\vepsb_2}
\ee
où les énergies réelles {\ycd ou même complexes} $\veps_j$ sont deux à deux distinctes, on a posé $\vepsb=\veps/k_B T$ et $\bar{n}_\veps=1/(\exp\vepsb-1)$. On utilise ensuite le fait que $\Omega/k_B T\in 2\ii\pi\mathbb{Z}$ pour faire disparaître cette variable de la loi de Bose, et la relation $\bar{n}_{-\veps}=-(1+\bar{n}_\veps)$ pour se ramener à de vrais nombres d'occupation des modes (à une énergie positive). Ainsi, par exemple:
\be
\label{eq123}
\bar{n}_{\Omega-\veps_{\qq-\kk}}=\bar{n}_{-\veps_{\qq-\kk}}=-(1+\bar{n}_{\qq-\kk})
\ee
L'intégrande devient une fonction rationnelle de la variable discrète $\Omega$, dont le prolongement analytique à $\mathbb{C}\setminus\mathbb{R}$ s'effectue par la simple substitution $\Omega\to z$. Il reste
\begin{multline}
\label{eq124}
\Sigma_\qq^{(2)}(\zeta)=\int\frac{\dd^dk}{(2\pi)^d}\Bigg[\frac{(1/2)|\langle\kk,\qq-\kk|\mathcal{H}_3|\qq\rangle|^2}{\zeta+\veps_\qq-(\veps_\kk+\veps_{\qq-\kk})}(\bar{n}_\kk+\bar{n}_{\qq-\kk}+1)+\frac{|\langle\kk+\qq|\mathcal{H}_3|\kk,\qq\rangle|^2}{\zeta+\veps_\qq+\veps_\kk-\veps_{\qq+\kk}}(\bar{n}_\kk-\bar{n}_{\qq+\kk})\\
-\frac{(1/2)|\langle 0|\mathcal{H}_3|\qq,\kk,-(\kk+\qq)\rangle|^2}{\zeta+\veps_\qq+\veps_{-(\kk+\qq)}+\veps_\kk}(\bar{n}_\kk+\bar{n}_{-(\qq+\kk)}+1)\Bigg\}
\end{multline}
Il reste à passer à la limite $\epsilon\to 0$ à $\qb$ et $\zetat$ fixés comme dans les équations (\ref{eq020},\ref{eq021},\ref{eq022}) avec, rappelons-le, des exposants $\nu=\sigma=2$ en dimension $d=2$. Au travers de l'effet \g{petit dénominateur} décrit dans la section \ref{sec0}, chaque contribution à l'intégrale (\ref{eq124}) est dominée par les configurations où les trois vecteurs d'onde au dénominateur d'énergie sont quasi colinéaires et de même sens (à un angle $O(\epsilon)$ près). Dans la première contribution, il suffit d'utiliser le développement (\ref{eq025}) avec $\kk'=\qq-\kk$ (avoir $\kk$ et $\kk'$ quasi colinéaires et de même sens implique alors $k<q$); dans la seconde, on utilise (\ref{eq025}) avec $\kk'=\qq$ ce qui n'impose rien sur $k$; dans la troisième, on a affaire à une somme à coefficients positifs de trois énergies de phonons, qui ne peut donc jamais être un $O(k_B T\epsilon^2$) (elle est toujours $\approx k_B T\epsilon^0$) si bien que la contribution correspondante est sous-dominante et négligeable à basse température.\footnote{Après ces développements, l'intégrale angulaire est lorentzienne donc s'effectue très bien avec la formule de Cauchy en refermant le chemin d'intégration par un demi-cercle infini dans le plan complexe, après avoir effectué le changement de variable $\theta=\epsilon\tilde{\theta}$ {\yc sur l'angle $\theta$ entre $\qq$ et $\kk$} et avoir envoyé les bornes d'intégration $\pm\pi/\epsilon$ sur $\tilde{\theta}$ à l'infini. {\ycd Dans la première contribution, on a au préalable utilisé le comportement limite $\theta'\sim-k\theta/(q-k)$ de l'angle $\theta'$ entre $\qq$ et $\kk'$.}} Il reste, en respectant l'ordre des contributions dans (\ref{eq124}), et en se limitant au demi-plan {\yc complexe} supérieur strict comme dans l'expression (\ref{eq012}) du signal:\footnote{\label{notegen} La valeur de $\tilde{\Sigma}_{\qb}^{(2,2)}(\zetat)$ dans le demi-plan {\yc complexe} inférieur strict s'en déduit par la relation générale valable à tous les ordres en le couplage entre phonons, ${\yc\Sigma_\qq(\zeta)^*}=\Sigma_\qq(\zeta^*)$.}
\be
\label{eq125}
\boxed{\tilde{\Sigma}_{\qb}^{(2,2)}(\zetat)\stackrel{\im\zetat>0}{=}\frac{9(1+\Lambda)^2(2\qb)^{1/2}}{64\ii\pi\rho\xi^2} \Bigg\{\int_0^{\qb} \dd\kb\, \frac{[\kb(\qb-\kb)]^{3/2}(1+\bar{n}_k^{\rm lin}+\bar{n}_{q-k}^{\rm lin})}{\left[(3\gamma/8)\qb\kb(\qb-\kb)+\tilde{\zeta}\right]^{1/2}} +2 \int_0^{\eta-\qb} \dd\kb \frac{[\kb(\qb+\kb)]^{3/2}(\bar{n}^{\rm lin}_k-\bar{n}^{\rm lin}_{k+q})}{\left[(3\gamma/8)\qb\kb(\kb+\qb)-\tilde{\zeta}\right]^{1/2}}\Bigg\}}
\ee
où les nombres d'occupation repérés par l'exposant \g{lin} sont ceux (\ref{eq029}) des modes de phonons pour la relation de dispersion linéarisée. Au premier membre de (\ref{eq125}), le second entier $2$ dans l'exposant rappelle que $\Sigma^{(2)}$ est calculé à l'ordre deux en $\epsilon$ ; les tildes sur $\Sigma$ et $\zeta$ rappellent qu'un facteur $k_BT\epsilon^2$ a été sorti de ces quantités comme dans (\ref{eq021},\ref{eq022}). Enfin, malgré la menace toujours présente des fluctuations quantiques du champ, on peut prendre la limite d'une coupure $\eta$ infinie sans déclencher de divergence ultraviolette, donc sans avoir à renormaliser quelque quantité que ce soit. Alors, si l'on fait tendre $\zetat$ vers zéro par parties imaginaires positives dans (\ref{eq125}), on retrouve l'équation (\ref{eq033}).

\subsection{Études analytiques: universalité à un paramètre}
\label{sec2.2}

De façon remarquable, il est possible, par un simple changement de variable, de faire en sorte qu'il ne reste plus, dans la théorie à l'ordre dominant {\yc en $\epsilon$ et en le couplage} traitée dans cette section \ref{sec2}, qu'un seul paramètre $u$ contenant tous les paramètres physiques comme la force des interactions $1/\rho\xi^2$, le paramètre de courbure $\gamma$ et la constante de couplage $\Lambda$, à savoir\footnote{Pour qu'il en aille de même dans le modèle de champ classique, il faudrait de plus incorporer le paramètre de coupure $\eta$ dans $u$, c'est-à-dire introduire la quantité $u_{\rm clas}=\eta u$. C'est inutile ici car nous avons pris $\eta=1$ dans les simulations. À titre de curiosité, signalons pour le champ classique et dans la limite $\qb\ll 1$ que le signal $s^{(2,2)}(t)$ défini plus bas cesse de commencer à la valeur $1$ en $t=0$ (comme il se doit physiquement) si $u_{\rm clas}<1$, ce que nous relions mathématiquement à l'apparition d'un pôle de la fonction de Green sur l'axe réel. Par conséquent, il serait intéressant de savoir si tous les systèmes physiques bidimensionnels (même en interaction très forte, pourvu qu'elle soit à courte portée) conduisent à $u>1$.}
\be
\label{eq112}
u=\frac{\pi\rho\xi^2\gamma^{3/2}}{\sqrt{3}(1+\Lambda)^2}
\ee
En effet, les nouvelles définitions repérées par un accent tchèque,
\be
\label{eq130}
\check{t}=\frac{3\gamma\qb}{8u}\tilde{t}=\frac{3\gamma\epsilon^2 c q t}{8u}\quad ; \quad \zetac=\frac{8u}{3\gamma\qb}\zetat=\frac{8u\zeta}{3\gamma\epsilon^2\hbar c q} 
\ee
conduisent à une nouvelle fonction énergie propre et à une nouvelle expression du signal ne mettant visiblement en jeu que $u$ et les variables réduites $\qb$ et $\check{t}$:
\be
\label{eq131}
\check{\Sigma}^{(2,2)}_{\qb}(\check{\zeta}) = \frac{8u}{3\gamma\qb} \tilde{\Sigma}^{(2,2)}_{\qb}(\zetat) \stackrel{\im\zetac>0}{=} \frac{1}{\ii} \int_0^{\qb} \frac{\dd\kb}{2\qb} \frac{[\kb(\qb-\kb)]^{3/2} (1+\bar{n}^{\rm lin}_k+\bar{n}^{\rm lin}_{q-k})}{[\kb(\qb-\kb)+\check{\zeta}/u]^{1/2}}+\frac{1}{\ii} \int_0^{+\infty} \frac{\dd\kb}{\qb}\frac{[\kb(\qb+\kb)]^{3/2} (\bar{n}^{\rm lin}_k-\bar{n}^{\rm lin}_{q+k})}{[\kb(\qb+\kb)-\check{\zeta}/u]^{1/2}}
\ee
et
\be
\label{eq132}
s^{(2,2)}(t) = \int_{C_+} \frac{\dd\check{\zeta}}{2\ii\pi} \eee^{-\ii\check{\zeta}\check{t}} \check{\mathcal{G}}_{\qb}^{(2,2)}(\zetac)  \quad\mbox{avec}\quad \check{\mathcal{G}}_{\qb}^{(2,2)}(\zetac)=\frac{1}{\check{\zeta}-\check{\Sigma}^{(2,2)}_{\qb}(\check{\zeta})}
\ee
Nous parlons dans ce cas d'universalité à un paramètre.

\begin{figure}[t]
\begin{center}
\begin{tikzpicture}[x=8mm,y=8mm]
\draw[thick,->,>=stealth](0,0)--(7,0);
\draw[thick,->,>=stealth,color=black,->,>=stealth](7,0.3)--(4,0.3);
\draw[thick,color=black](4,0.3)--(0,0.3);
\node(a)at(5,0.3)[above,color=black]{$C_+$};
\node(a)at(7,0)[below]{${\rm Re\,}\check{\zeta}$};
\draw[thick,->,>=stealth](3,-1)--(3,3);
\node(a)at(3,3)[above]{${\rm Im\,}\check{\zeta}$};
\node(a)at(0.2,3)[above]{$(a)$};
\node(a)at(0.8,0.3)[above]{$-u\bar{q}^2/4$};
\draw[dashed,ultra thick](1,0)--(7,0);
\node(a)at(1,0)[]{$\bullet$};
\draw[thick,color=white](3,-4.25)--(3,-1);
\end{tikzpicture}
\quad
\begin{tikzpicture}[x=8mm,y=8mm]
\draw[thick,->,>=stealth](-1,0)--(7,0);
\draw[thick,->,>=stealth,color=black,->,>=stealth](7,0.3)--(4,0.3);
\draw[thick,color=black](4,0.3)--(-1,0.3);
\node(a)at(5,0.3)[above,color=black]{$C_+$};
\node(a)at(7,0)[below]{$\re\check{\zeta}$};
\draw[thick,->,>=stealth](3,-1)--(3,3);
\node(a)at(3,3)[above]{$\im\check{\zeta}$};
\node(a)at(-0.8,3)[above]{$(b)$};
\node(a)at(0.8,0.3)[above]{$-u\bar{q}^2/4$};
\draw[dashed,ultra thick](1,0)--(-1,-2);
\node(a)at(1,0)[]{$\bullet$};
\draw[thick,color=blue,->,>=stealth](3,0.3)+(180:4)arc(-180:-155:4);
\draw[thick,color=blue,->,>=stealth](3,0.3)+(-150:4)arc(-150:-60:4);
\draw[thick,color=blue](3,0.3)+(-60:4)arc(-60:0:4);
\draw[thick,color=red,->,>=stealth](-0.5-0.08,-1.5+0.16)--(1-0.08,0+0.16);
\draw[thick,color=red,->,>=stealth](1+0.05,0-0.2)--(-0.5+0.05,-1.5-0.2);
\draw[thick,color=red](1,0)+(120:0.18)arc(120:-48:0.18);
\draw[thick,->,>=stealth](1,0)+(-0:0.5)arc(0:-135:0.5);
\node(a)at(1.3,-0.7)[]{$\phi$};
\node(a)at(4,-1)[ultra thick,color=red]{$\times$};
\node(a)at(4,-1.2)[ultra thick,color=black,below]{$\check{\zeta}_{\bar{q}}$};
\end{tikzpicture}
\end{center}
\caption{(a) Singularités de la fonction de Green $\check{\mG}^{(2,2)}_{\qb}(\zetac)$ dans le plan complexe et (b) chemin d'intégration suivi dans le calcul du signal (\ref{eq132}) pour obtenir le premier écart (\ref{eq151}) à l'approximation du pôle. On a effectué le changement de variable (\ref{eq130}) $\zetac=8(z-\veps_\qq)u/(3\gamma\hbar c q \epsilon^2)$ pour révéler une universalité à un paramètre $u$ (\ref{eq112}). Ligne horizontale orientée noire: chemin d'intégration originel dans (\ref{eq132}). Disque noir: point de branchement de $\check{\zeta}\mapsto\check{\mG}^{(2,2)}_{\qb}(\zetac)$. Demi-droite en tireté: ligne de coupure originelle de $\check{\zeta}\mapsto\check{\mG}^{(2,2)}_{\qb}(\zetac)$ dans (a), et tournée d'un angle $\phi\in]-\pi,-\pi/2[$ dans (b) par prolongement analytique de $\check{\zeta}\mapsto\check{\mG}^{(2,2)}_{\qb}(\zetac)$ du demi-plan supérieur au demi-plan inférieur, ce qui a fait apparaître le pôle $\check{\zeta}_{\qb}$ (croix rouge). Dans (b), pour appliquer le théorème de Cauchy, on referme $C_+$ par deux portions de cercle à l'infini (en bleu) et un lacet (en rouge) qui contourne la ligne de coupure déplacée.}
\label{fig5}
\end{figure}

Effectuons le prolongement analytique de la fonction (\ref{eq131}) du demi-plan supérieur au demi-plan inférieur à travers sa ligne de coupure sur l'axe réel.\footnote{Il y a ici une subtilité mathématique que le lecteur voudra bien nous pardonner. Implicitement, à partir de l'expression (\ref{eq131}) de $\check{\Sigma}_{\qb}^{(2,2)}(\zetac)$, valable pourtant seulement pour $\im\zetac>0$, nous définissons une fonction dans tout $\mathbb{C}\setminus\mathbb{R}$ en appliquant (\ref{eq131}) telle quelle pour $\im\zetac<0$; la ligne de coupure $[-u\qb^2/4,+\infty[$ mentionnée dans le texte et représentée sur la figure \ref{fig5}a correspond à cette fonction plutôt qu'à la vraie fonction énergie propre $\check{\Sigma}_{\qb}^{(2,2)}(\check{\zeta})$ dont la ligne de coupure est $\mathbb{R}$ tout entier et qui serait dès lors moins commode à prolonger analytiquement du demi-plan supérieur au demi-plan inférieur pour le même résultat final (on voit que c'est $\mathbb{R}$ tout entier (i) en considérant le dénominateur d'énergie $\zeta+\Delta E$ dans la contribution de Landau de l'équation (\ref{eq124}) et en vérifiant que $\Delta E$ décrit $\mathbb{R}$ lorsque $\theta/\epsilon$ et $\kb$ décrivent $\mathbb{R}^+$ dans (\ref{eq025}) avec $\kk'=\qq$, ou (ii) en utilisant la relation $\ii z^{1/2}=-\sig(\im z)(-z)^{1/2}$, $\forall z\in\mathbb{C}\setminus\mathbb{R}$, pour faire disparaître le facteur $\ii$ dans (\ref{eq131}) et obtenir une expression physiquement correcte sur tout $\mathbb{C}\setminus\mathbb{R}$ car satisfaisant à $\Sigma_{\qq}(\zeta)^*=\Sigma_{\qq}(\zeta^*)$, dont on montre ensuite que la ligne de coupure est $\mathbb{R}$).} Ceci va permettre, comme dans la référence \cite{CCTbordeaux}, de séparer le signal en la contribution d'un lacet autour d'une ligne de coupure déplacée (contribution décroissant en loi de puissance aux temps longs) et la contribution d'un pôle dans la fonction de Green ainsi prolongée (contribution décroissant exponentiellement en temps), voir notre figure \ref{fig5}b, ce qui améliore la compréhension physique. Il faut d'abord identifier les points de branchement (singularités immuables) de $\check{\Sigma}_{\qb}^{(2,2)}(\check{\zeta})$ sur l'axe réel. Comme les valeurs de $\check{\zeta}$ annulant l'argument de la racine carrée au dénominateur décrivent l'intervalle $[-u\qb^2/4,0]$ lorsque $\kb$ décrit $[0,\qb]$ dans la première contribution à (\ref{eq131}), et décrivent $[0,+\infty[$ lorsque $\kb$ décrit $\mathbb{R}^+$ dans la seconde, on pourrait s'attendre à avoir deux points de branchement, $\check{\zeta}=-u\qb^2/4$ et $\check{\zeta}=0$.\footnote{Par exemple, $\zetac\mapsto\int_a^b\dd x/(\zetac-x)^{1/2}=2(\zetac-b)^{1/2}-2(\zetac-a)^{1/2}$ admet les réels $a$ et $b$ comme points de branchement.} En réalité, par compensation subtile entre les deux contributions, seul $\zetac=-u\qb^2/4$ subsiste ; de ce point part la ligne de coupure $[-u\qb^2/4,+\infty[$ comme sur la figure \ref{fig5}a.\footnote{\label{note30} Si l'on revient à la variable $\zetat$, le point de branchement se trouve en $-3\gamma\qb^3/32$.} On peut alors faire tourner cette dernière d'un angle $\phi \in ]-\pi,-\pi/2[$ pour l'amener dans le troisième quadrant, par prolongement analytique de $\check{\Sigma}^{(2,2)}_{\qb}(\check{\zeta})$ du demi-plan supérieur au demi-plan inférieur, la nouvelle ligne de coupure étant la demi-droite $-u\qb^2/4+\exp(\ii\phi)\mathbb{R}^+$, voir la figure \ref{fig5}b. La fonction ainsi prolongée s'écrit, comme le montre l'\ref{app1} et en convenant que $z^{3/2}=z\sqrt{z}$:
\be
\label{eq140}
\check{\Sigma}^{(2,2)}_{\qb\downarrow}(\check{\zeta})= \frac{\eee^{3\ii\phi/2}}{\ii} \int_0^{+\infty} \frac{\dd\kb}{\qb} 
\frac{(\kb^2-\eee^{-\ii\phi}\qb^2/4)^{3/2}}{[\kb^2-\eee^{-\ii\phi}(\qb^2/4+\zetac/u)]^{1/2}} \Bigg[
\frac{1}{\exp(\eee^{\ii\phi/2}\kb-\qb/2)-1}-\frac{1}{\exp(\eee^{\ii\phi/2}\kb+\qb/2)-1} \Bigg]
\ee
En particulier, elle est analytique sur un voisinage de $\zetac=0$ (voir l'\ref{app1}, qui donne aussi sa valeur et celle de sa dérivée en ce point). En revanche, la fonction de Green $\check{\mG}_{\qb\downarrow}^{(2,2)}(\zetac)$ correspondante admet un pôle $\check{\zeta}_{\qb}$ dans le quatrième quadrant comme nous l'avons vérifié numériquement.\footnote{Si l'angle $\phi$ est trop proche de $-\pi/2$ (ce que nous éviterons dans la suite) et si $\qb$ est assez faible, un second pôle de partie imaginaire $<0$, d'interprétation physique incertaine, peut apparaître à gauche de la ligne de coupure déplacée.} 

\begin{SCfigure}
\includegraphics[width=10cm,clip=]{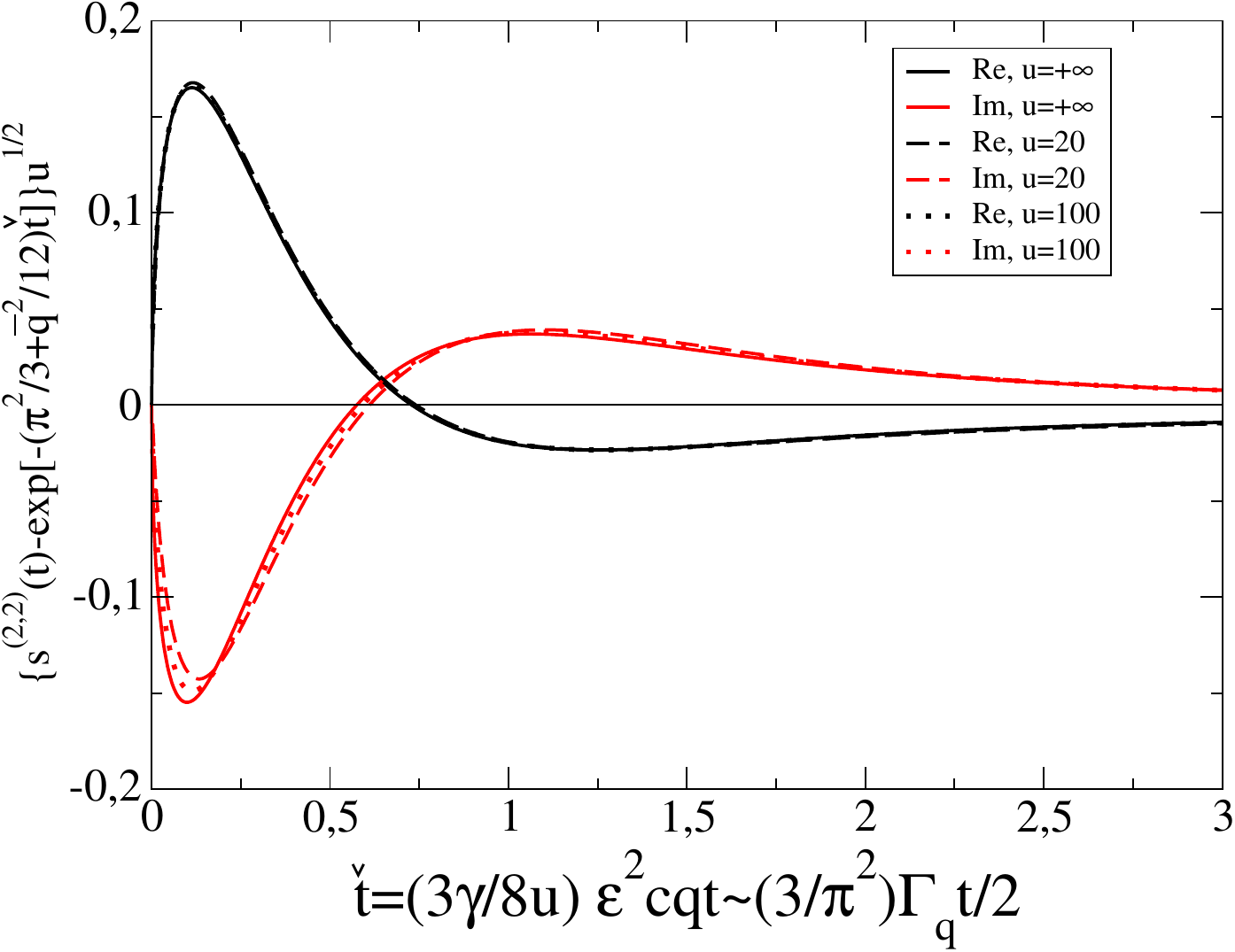}
\caption{Dans la théorie quantique et pour une fonction énergie propre calculée à l'ordre deux en $H_3$ et à l'ordre deux en $\epsilon=k_B T/mc^2$, écart du signal (\ref{eq012}) à {\yc l'exponentielle $\exp(-\Gamma_qt/2)$ de} la règle d'or de Fermi multiplié par $u^{1/2}$ (le paramètre universel $u$ est donné par (\ref{eq112}), {\yc le taux $\Gamma_q$ par (\ref{eq033})}) et représenté en fonction du temps réduit (\ref{eq130}) en partie réelle et partie imaginaire, pour des valeurs croissantes de $u$ (le superfluide bidimensionnel sous-jacent est en interaction de plus en plus faible) mais pour un nombre d'onde mis à l'échelle comme dans (\ref{eq150}) {\yc constant}, $Q=1$. Courbes pour des valeurs finies de $u$: obtenues par intégration numérique de l'expression (\ref{eq132}). Courbes pour $u=+\infty$: tirées de la prédiction analytique (\ref{eq151}) (elles correspondent simplement à $\re S(\check{t})$ et $\im S(\check{t})$).}
\label{fig4}
\end{SCfigure}

Plutôt que d'évaluer numériquement les contributions du lacet et du pôle au signal, cherchons maintenant à extraire analytiquement de la forme (\ref{eq140}) la \g{substantifique moelle} de l'ordre deux en $H_3$, en remarquant par exemple qu'il est illusoire de calculer la position du pôle $\zetac_{\qb}$ à tous les ordres en $1/u$ puisque la fonction énergie propre est calculée ici seulement à l'ordre un en $1/u$. En revanche, se limiter à l'ordre zéro en $1/u$, ce qui reviendrait à prendre la limite $u\to+\infty$ à $\qb$ fixé, serait trop pauvre, car le signal se réduirait alors à l'exponentielle de la règle d'or de Fermi. On le voit bien sur l'écriture (\ref{eq131}), la nouvelle fonction énergie propre dépend de $u$ seulement au travers du rapport $\zetac/u$ ; aussi tend-elle vers sa valeur en $\ii 0^+$ lorsqu'on fait tendre $u$ vers $+\infty$ sous le signe intégral ($\zetac$ et $\check{t}$ étant fixés dans (\ref{eq132})). Un résultat plus intéressant s'obtient en prenant la limite $u\to+\infty$ à
\be
\label{eq150}
Q=u^{1/2}\qb \quad\mbox{et}\quad \check{t}\quad\mbox{fixés}
\ee
de façon que le point de branchement sur la figure \ref{fig5} reste fixe au lieu d'être rejeté à l'infini (ce qui ferait effectivement disparaître la contribution du lacet). On trouve alors une correction d'ordre $u^{-1/2}$ {\yc au signal de} la règle d'or de Fermi ; comme $u^{-1}\ll u^{-1/2}$ il semble entendu que l'on puisse garder cette correction sans avoir à calculer la fonction énergie propre à l'ordre suivant (l'ordre quatre) en $H_3$, {\yc même si nous verrons} dans la section \ref{sec4} qu'il en va autrement. Reportons les détails techniques dans l'\ref{app2} et donnons directement le résultat {\yc dans la théorie quantique}:
\be
\label{eq151}
s^{(2,2)}(t) \stackrel{Q\,\mbox{\scriptsize et}\,\check{t}\,\mbox{\scriptsize fixés}}{\underset{u\to+\infty}{=}}\eee^{-\pi^2\check{t}/3} + \frac{S(\check{t})}{u^{1/2}}+O(1/u) \quad\mbox{avec}\quad S(\check{t})=S_{\rm lacet}(\check{t})+S_{\mbox{\scriptsize pôle}}(\check{t})
\ee
La première contribution est l'exponentielle réelle de la règle d'or dans la limite $\qb\to 0$. La seconde est partagée en les contributions du lacet et du pôle établies dans l'\ref{app2}:
\be
\label{eq152}
S_{\rm lacet}(\check{t})= -\frac{\eee^{\ii Q^2\check{t}/4}}{\pi}\int_0^{\eee^{\ii(\phi+\pi)}(+\infty)} \dd z \frac{\eee^{\ii z \check{t}}\int_0^{\pi/2} \dd\alpha \left(z\sin^2\alpha+\frac{Q^2}{4}\right)^{1/2}}{(\frac{Q^2}{4}+z-\ii\frac{\pi^2}{3})^2} \quad ; \quad S_{\mbox{\scriptsize pôle}}(\check{t})=\eee^{-\pi^2\check{t}/3} (\delta \check{Z}-\ii \delta \check{\zeta}_{\qb}\check{t})
\ee
où l'angle $\phi$ est choisi librement dans $]-\pi,-\pi/2[$ (le résultat (\ref{eq152}) n'en dépend pas), $\delta\check{\zeta}_{\qb}$ est le premier écart du pôle à l'approximation du pôle $-\ii\pi^2/3$ (multiplié par $u^{1/2}$) et $\delta\check{Z}$ est le premier écart à un du résidu de $\check{\mG}_{\qb\downarrow}^{(2,2)}(\zetac)$ (multiplié lui aussi par $u^{1/2}$) dans un développement en puissances de $u^{-1/2}$:\footnote{Comme le signal (\ref{eq012}) commence exactement à un, il faut que $S(0)=0$ donc que la contribution du lacet pour $\check{t}=0$ compense exactement celle de l'écart à un du résidu du pôle. La vérification est non triviale. L'idée directrice est de ramener $S_{\rm lacet}(0)$ à une intégrale simple en intervertissant l'intégration sur $z$ et sur $\alpha$ dans (\ref{eq152}) (on connaît en effet une primitive de $(z+a)^{1/2}/(z+b)^2$ où $a$ et $b$ sont deux constantes). {\yc Par ailleurs, on remarque que $\re\delta\check{Z}>0$ pour tout $Q$: la première correction en $u^{-1/2}$ rend le résidu $\check{Z}$ de module $>1$ donc le pôle $\check{\zeta}_{\qb}$ de poids spectral $>1$.}}
\be
\label{eq154}
\delta\check{\zeta}_{\qb} = \int_0^{+\infty} \dd K \left[1-\frac{(K^2+Q^2/4)^{1/2}}{(K^2-\ii\frac{\pi^2}{3}+\frac{Q^2}{4})^{1/2}}\right]
\quad;\quad 
\delta \check{Z}=\frac{1}{2} \int_0^{+\infty} \dd K \frac{(K^2+Q^2/4)^{1/2}}{(K^2-\ii\frac{\pi^2}{3}+\frac{Q^2}{4})^{3/2}} 
\ee
Sur la figure \ref{fig4}, nous représentons la fonction $S(\check{t})$ de l'équation (\ref{eq151}) pour $Q=1$, ainsi que les prédictions tirées numériquement de $\check{\Sigma}_{\qb}^{(2,2)}{\yc(\check{\zeta})}$ pour des valeurs finies de $u$. On constate que $u=20$ a déjà atteint le régime asymptotique; on s'attend donc à ce qu'il en aille de même pour les paramètres des simulations de champ classique de la figure \ref{fig2}d, qui correspondent à $u\simeq 18,\!14$ et $Q\simeq 1,\!06$. Ainsi, la fonction $S(\check{t})$ n'est pas qu'une curiosité mathématique et ses prédictions ne sont pas dénuées d'intérêt pratique. Gardons-en deux, dues au lacet: le comportement aux temps courts est en racine carrée, ce qui rend compte du départ rapide du signal sur la figure \ref{fig4}, et le comportement aux temps longs est sinusoïdal amorti comme l'inverse du temps, son raccordement forcément continu à celui aux temps courts expliquant les changements de signe sur cette même figure,\footnote{Si $\check{t}\to +\infty$, l'intégrale donnant $S_{\rm lacet}(\check{t})$ est dominée par les valeurs de $z$ proches de zéro et on peut remplacer $z$ par $0$ dans l'intégrande, partout sauf dans $\exp(\ii z\check{t})$. Si $\check{t}\to 0^+$, l'intégrale donnant $\exp(\ii Q^2\check{t}/4)S_{\rm lacet}(\check{t})-S_{\rm lacet}(0)$ est dominée par les valeurs de $z$ proches de l'infini, donc se réduit à l'intégrale de Fresnel $(-1/\pi)\int_0^{+\infty} \dd z [\exp(\ii z\check{t})-1]/z^{3/2}$ dont la valeur est connue (on a pris $\phi+\pi=0^+$ pour simplifier).}\footnote{\label{note2} On peut obtenir une expression plus explicite du signal dans la limite $Q\to 0$: $S(\check{t})=\left(\frac{\check{t}}{\pi}\right)^{1/2}\eee^{-\ii\pi/4}-\frac{\sqrt{3}}{2\pi} \eee^{-\ii\pi/4}\left(\frac{2\pi^2\check{t}}{3}-1\right)\erfi(\sqrt{\pi^2\check{t}/3})\eee^{-\pi^2\check{t}/3}$, où $\erfi$ est la fonction d'erreur imaginaire, ce qui admet le même comportement aux temps courts que dans (\ref{eq155}) mais le nouveau comportement aux temps longs $S(\check{t}) \underset{\check{t}\to +\infty}{\sim} - \frac{\eee^{-\ii\pi/4}}{2\pi^{1/2}} \left(\frac{3}{\pi^2}\right)^2 \frac{1}{\check{t}^{3/2}}$. Comme la fonction $\erfi$ est, malgré son nom, à valeurs réelles sur l'axe réel, on trouve que $\re S(\check{t})=-\im S(\check{t})$ dans cette limite, une propriété dont il reste une trace sur la figure \ref{fig4}.}
\be
\label{eq155}
S(\check{t}) \underset{\check{t}\to 0^+}{\sim} 2\eee^{-\ii\pi/4} (\check{t}/\pi)^{1/2} \quad\quad ; \quad\quad
S(\check{t})\underset{\check{t}\to +\infty}{\sim} \frac{-\ii Q/4}{(\frac{Q^2}{4}-\ii\frac{\pi^2}{3})^2} \frac{\eee^{\ii Q^2\check{t}/4}}{\check{t}}
\ee

\subsection{Comparaison aux simulations de champ classique}
\label{sec2.3}

Afin de comparer aux simulations de la section \ref{sec1}, nous transposons les traitements quantiques des sections \ref{sec2.1} et \ref{sec2.2} au champ classique: dans l'équation (\ref{eq125}), il faut garder le paramètre de coupure $\eta$ fini (il vaut un dans les simulations), négliger le terme $1$ devant les nombres d'occupation $\bar{n}^{\rm lin}_k$ et remplacer ceux-ci par la loi d'équipartition classique $1/\kb$. À titre de test, notons que la valeur correspondante de $\Sigma_\qq^{(2,2)}(\ii 0^+)$ s'écrit alors $-\ii\hbar\Gamma_q^{\mbox{\scriptsize clas}}/2$ où $\Gamma_q^{\mbox{\scriptsize clas}}$ est le taux d'amortissement (\ref{eq113}) comme il se doit. Dans la suite, nous gardons bien sûr toute la dépendance de $\tilde{\Sigma}_\qq^{(2,2)}(\zetat)$ en $\zetat$ dans le calcul du signal $s^{(2,2)}(t)$ par fonction de Green.

Les prédictions correspondantes pour le temps de décroissance à $1/\eee$ sont représentées en trait plein noir épais sur chaque panneau de la figure \ref{fig1}. Elles sont dans tous les cas plus proches des simulations que la règle d'or mais, paradoxalement, l'amélioration est la plus nette dans le régime d'interaction forte (a) et devient très modeste dans le régime d'interaction faible (d). Cette dégradation était cependant prévisible par le raisonnement précédant l'équation (\ref{eq150}): si $\rho\xi^2$ augmente à $\qb$ fixé, le calcul par fonction de Green à l'ordre deux en $H_3$ doit se réduire à la règle d'or donc s'écarter des simulations.

L'analyse de la figure \ref{fig2} montrant la dépendance temporelle du signal nous donne un indice supplémentaire sur ce qui se passe lorsqu'on réduit la force des interactions, $(\rho \xi^2)^{-1}\to 0$. Les prédictions tirées de $\tilde{\Sigma}_\qq^{(2,2)}(\zetat)$ sont cette fois-ci représentées en tireté (celles de la règle d'or seraient identiquement nulles). Elles se rapprochent de manière remarquable des simulations pour la partie imaginaire de l'écart $s(t)-s_{\mbox{\scriptsize pôle}}(t)$ (le tireté rouge se confond presque avec les résultats numériques sur le panneau (d)) mais elles présentent un désaccord persistant, et même croissant en valeurs relatives, sur la partie réelle, qui explique l'échec précédent sur le temps de décroissance à $1/\eee$.

L'anomalie que représente la partie réelle sur la figure \ref{fig2}d est criante si l'on se souvient des prédictions analytiques de la figure \ref{fig4} (même si elles sont pour le champ quantique): la partie réelle et la partie imaginaire, pour les valeurs de $u$ et de $Q$ correspondantes ($u\simeq 18,\!14$ et $Q\simeq 1,\!06$, rappelons-le), devraient être plus ou moins symétriques l'une de l'autre par rapport à l'axe des abscisses (voir aussi la note \ref{note2}), ce qui est loin d'être le cas dans les simulations.

En conclusion, le calcul du signal avec la fonction énergie propre $\tilde{\Sigma}^{(2,2)}_{\yc\qb}(\zetat)$, par la seule faute de la partie réelle, échoue à rendre compte du premier écart à la règle d'or dans la limite $u\to +\infty$ où il avait toutes les chances d'exceller.

\section{Étude à l'ordre quatre en $H_3$}
\label{sec3}

Pour comprendre l'écart persistant sur la partie réelle du signal entre les simulations numériques de la section \ref{sec1} et l'étude théorique à l'ordre deux en $H_3$ de la section \ref{sec2}, même quand le superfluide bidimensionnel entre dans la limite d'interaction faible $(\rho\xi^2)^{-1}\to 0$, nous calculons ici la fonction énergie propre à l'ordre quatre en $H_3$, soit au moyen de l'habituel formalisme diagrammatique dans la section \ref{sec3.1}, soit par un raisonnement à la Landau dans la section \ref{sec3.2}, qui apporte un éclairage physique intéressant et aboutit à une expression plus compacte. Nous montrons cependant, dans la section \ref{sec3.3}, que le résultat à l'ordre quatre est inutilisable dans la limite de basse température $\epsilon=k_B T/mc^2\to 0$ qui nous intéresse ici.

\subsection{Calcul avec le formalisme diagrammatique de la référence \cite{FW}}
\label{sec3.1}

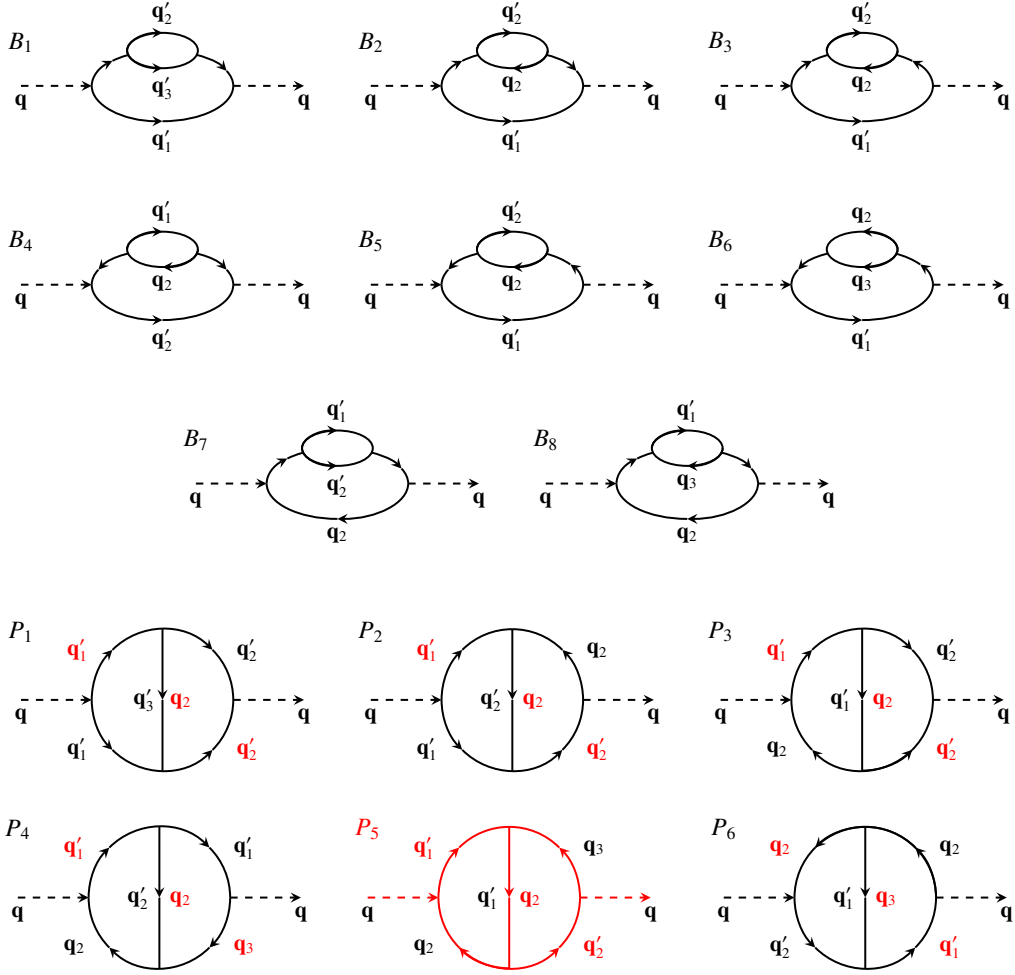
\begin{figure}[p]
\begin{center}
\begin{tikzpicture}[x=10mm,y=5mm]
\draw[thick,->,>=stealth](1,0)arc(180:135:1);
\draw[thick,->,>=stealth](2,0)+(135:1)arc(135:25:1);
\draw[thick](2,0)+(25:1)arc(25:0:1);
\draw[thick,dashed,->,>=stealth](3,0)--(4,0);
\node(a)at(0,0.7)[above]{$B_1$};
\draw[thick,dashed,->,>=stealth](0,0)--(1,0);
\draw[thick,->,>=stealth](1,0)arc(-180:-90:1);
\draw[thick](2,0)+(-90:1)arc(-90:0:1);
\draw[thick,fill=white] (2,1) circle (0.5);
\draw[thick,->,>=stealth](2,1)+(180:0.5)arc(180:90:0.5);
\draw[thick,->,>=stealth](2,1)+(-180:0.5)arc(-180:-90:0.5);
\node(a)at(0,0)[below]{\small $\vec{q}$};
\node[below](b)at(2,-1) {\small $\vec{q}_1^\prime$};
\node[](b)at(0.8,0.7) {};
\node[above](b)at(2,1.5) {\small $\vec{q}_2^\prime$};
\node[below](b)at(2,0.5) {\small $\vec{q}_3^\prime$};
\node[](b)at(3.2,0.7) {};
\node(a)at(4,0)[below]{\color{black} {\small $\vec{q}$}};
\end{tikzpicture}
\quad
\begin{tikzpicture}[x=10mm,y=5mm]
\draw[thick,->,>=stealth](1,0)arc(180:135:1);
\draw[thick,->,>=stealth](2,0)+(135:1)arc(135:25:1);
\draw[thick](2,0)+(25:1)arc(25:0:1);
\draw[thick,dashed,->,>=stealth](3,0)--(4,0);
\node(a)at(0,0.7)[above]{$B_2$};
\draw[thick,dashed,->,>=stealth](0,0)--(1,0);
\draw[thick,->,>=stealth](1,0)arc(-180:-90:1);
\draw[thick](2,0)+(-90:1)arc(-90:0:1);
\draw[thick,fill=white] (2,1) circle (0.5);
\draw[thick,->,>=stealth](2,1)+(180:0.5)arc(180:90:0.5);
\draw[thick,->,>=stealth](2,1)+(0:0.5)arc(0:-90:0.5);
\node(a)at(0,0)[below]{\small $\vec{q}$};
\node[below](b)at(2,-1) {\small $\vec{q}_1^\prime$};
\node[](b)at(0.8,0.7) {};
\node[below](b)at(2,0.5) {\small $\vec{q}_2$};
\node[above](b)at(2,1.5) {\small $\vec{q}_2^\prime$};
\node[](b)at(3.2,0.7) {};
\node(a)at(4,0)[below]{\color{black} {\small $\vec{q}$}};
\end{tikzpicture}
\quad
\begin{tikzpicture}[x=10mm,y=5mm]
\draw[thick,->,>=stealth](1,0)arc(180:135:1);
\draw[thick](2,0)+(135:1)arc(135:45:1);
\draw[thick,->,>=stealth](2,0)+(0:1)arc(0:45:1);
\draw[thick,dashed,->,>=stealth](3,0)--(4,0);
\node(a)at(0,0.7)[above]{$B_3$};
\draw[thick,dashed,->,>=stealth](0,0)--(1,0);
\draw[thick,->,>=stealth](1,0)arc(-180:-90:1);
\draw[thick](2,0)+(-90:1)arc(-90:0:1);
\draw[thick,fill=white] (2,1) circle (0.5);
\draw[thick,->,>=stealth](2,1)+(180:0.5)arc(180:90:0.5);
\draw[thick,->,>=stealth](2,1)+(0:0.5)arc(0:-90:0.5);
\node(a)at(0,0)[below]{\small $\vec{q}$};
\node[below](b)at(2,-1) {\small $\vec{q}_1^\prime$};
\node[](b)at(0.8,0.7) {};
\node[below](b)at(2,0.5) {\small $\vec{q}_2$};
\node[above](b)at(2,1.5) {\small $\vec{q}_2^\prime$};
\node[](b)at(3.2,0.7) {};
\node(a)at(4,0)[below]{\color{black} {\small $\vec{q}$}};
\end{tikzpicture}

\bigskip

\begin{tikzpicture}[x=10mm,y=5mm]
\draw[thick](1,0)arc(180:155:1);
\draw[thick,->,>=stealth](2,0)+(90:1)arc(90:155:1);
\draw[thick,->,>=stealth](2,0)+(90:1)arc(90:25:1);
\draw[thick](2,0)+(25:1)arc(25:0:1);
\draw[thick,dashed,->,>=stealth](3,0)--(4,0);
\node(a)at(0,0.7)[above]{$B_4$};
\draw[thick,dashed,->,>=stealth](0,0)--(1,0);
\draw[thick,->,>=stealth](1,0)arc(-180:-90:1);
\draw[thick](2,0)+(-90:1)arc(-90:0:1);
\draw[thick,fill=white] (2,1) circle (0.5);
\draw[thick,->,>=stealth](2,1)+(180:0.5)arc(180:90:0.5);
\draw[thick,->,>=stealth](2,1)+(0:0.5)arc(0:-90:0.5);
\node(a)at(0,0)[below]{\small $\vec{q}$};
\node[below](b)at(2,-1) {\small $\vec{q}_2^\prime$};
\node[](b)at(0.8,0.7) {};
\node[below](b)at(2,0.5) {\small $\vec{q}_2$};
\node[above](b)at(2,1.5) {\small $\vec{q}_1^\prime$};
\node[](b)at(3.2,0.7) {};
\node(a)at(4,0)[below]{\color{black} {\small $\vec{q}$}};
\end{tikzpicture}
\quad
\begin{tikzpicture}[x=10mm,y=5mm]
\draw[thick](1,0)arc(180:155:1);
\draw[thick,->,>=stealth](2,0)+(90:1)arc(90:155:1);
\draw[thick](2,0)+(35:1)arc(35:90:1);
\draw[thick,->,>=stealth](2,0)+(0:1)arc(0:35:1);
\draw[thick,dashed,->,>=stealth](3,0)--(4,0);
\node(a)at(0,0.7)[above]{$B_5$};
\draw[thick,dashed,->,>=stealth](0,0)--(1,0);
\draw[thick,->,>=stealth](1,0)arc(-180:-90:1);
\draw[thick](2,0)+(-90:1)arc(-90:0:1);
\draw[thick,fill=white] (2,1) circle (0.5);
\draw[thick,->,>=stealth](2,1)+(180:0.5)arc(180:90:0.5);
\draw[thick,->,>=stealth](2,1)+(0:0.5)arc(0:-90:0.5);
\node(a)at(0,0)[below]{\small $\vec{q}$};
\node[below](b)at(2,-1) {\small $\vec{q}_1^\prime$};
\node[](b)at(0.8,0.7) {};
\node[below](b)at(2,0.5) {\small $\vec{q}_2$};
\node[above](b)at(2,1.5) {\small $\vec{q}_2^\prime$};
\node[](b)at(3.2,0.7) {};
\node(a)at(4,0)[below]{\color{black} {\small $\vec{q}$}};
\end{tikzpicture}
\quad
\begin{tikzpicture}[x=10mm,y=5mm]
\draw[thick](1,0)arc(180:155:1);
\draw[thick,->,>=stealth](2,0)+(90:1)arc(90:155:1);
\draw[thick](2,0)+(35:1)arc(35:90:1);
\draw[thick,->,>=stealth](2,0)+(0:1)arc(0:35:1);
\draw[thick,dashed,->,>=stealth](3,0)--(4,0);
\node(a)at(0,0.7)[above]{$B_6$};
\node[below](b)at(2,-1) {\small $\vec{q}_1^\prime$};
\draw[thick,dashed,->,>=stealth](0,0)--(1,0);
\draw[thick,->,>=stealth](1,0)arc(-180:-90:1);
\draw[thick](2,0)+(-90:1)arc(-90:0:1);
\draw[thick,fill=white] (2,1) circle (0.5);
\draw[thick,->,>=stealth](2,1)+(0:0.5)arc(0:90:0.5);
\draw[thick,->,>=stealth](2,1)+(0:0.5)arc(0:-90:0.5);
\node(a)at(0,0)[below]{\small $\vec{q}$};
\node[below](b)at(2,-1) {};
\node[](b)at(0.8,0.7) {};
\node[below](b)at(2,0.5) {\small $\vec{q}_3$};
\node[above](b)at(2,1.5) {\small $\vec{q}_2$};
\node[](b)at(3.2,0.7) {};
\node(a)at(4,0)[below]{\color{black} {\small $\vec{q}$}};
\end{tikzpicture}

\bigskip
\begin{tikzpicture}[x=10mm,y=5mm]
\draw[thick,->,>=stealth](1,0)arc(180:135:1);
\draw[thick,->,>=stealth](2,0)+(135:1)arc(135:25:1);
\draw[thick](2,0)+(25:1)arc(25:0:1);
\draw[thick,dashed,->,>=stealth](3,0)--(4,0);
\node(a)at(0,0.7)[above]{$B_7$};
\draw[thick,dashed,->,>=stealth](0,0)--(1,0);
\draw[thick,->,>=stealth](2,0)+(0:1)arc(0:-90:1);
\draw[thick](2,0)+(-180:1)arc(-180:-90:1);
\draw[thick,fill=white] (2,1) circle (0.5);
\draw[thick,->,>=stealth](2,1)+(180:0.5)arc(180:90:0.5);
\draw[thick,->,>=stealth](2,1)+(-180:0.5)arc(-180:-90:0.5);
\node(a)at(0,0)[below]{\small $\vec{q}$};
\node[below](b)at(2,-1) {\small $\vec{q}_2$};
\node[](b)at(0.8,0.7) {};
\node[below](b)at(2,0.5) {\small $\vec{q}_2^\prime$};
\node[above](b)at(2,1.5) {\small $\vec{q}_1^\prime$};
\node[](b)at(3.2,0.7) {};
\node(a)at(4,0)[below]{\color{black} {\small $\vec{q}$}};
\end{tikzpicture}
\quad
\begin{tikzpicture}[x=10mm,y=5mm]
\draw[thick,->,>=stealth](1,0)arc(180:135:1);
\draw[thick,->,>=stealth](2,0)+(135:1)arc(135:25:1);
\draw[thick](2,0)+(25:1)arc(25:0:1);
\draw[thick,dashed,->,>=stealth](3,0)--(4,0);
\node(a)at(0,0.7)[above]{$B_8$};
\node[above](b)at(2,1.5) {\small $\vec{q}_1^\prime$};
\draw[thick,dashed,->,>=stealth](0,0)--(1,0);
\draw[thick,->,>=stealth](2,0)+(0:1)arc(0:-90:1);
\draw[thick](2,0)+(-180:1)arc(-180:-90:1);
\draw[thick,fill=white] (2,1) circle (0.5);
\draw[thick,->,>=stealth](2,1)+(180:0.5)arc(180:90:0.5);
\draw[thick,->,>=stealth](2,1)+(0:0.5)arc(0:-90:0.5);
\node(a)at(0,0)[below]{\small $\vec{q}$};
\node[below](b)at(2,-1) {\small $\vec{q}_2$};
\node[](b)at(0.8,0.7) {};
\node[below](b)at(2,0.5) {\small $\vec{q}_3$};
\node[above](b)at(2,1.5) {};
\node[](b)at(3.2,0.7) {};
\node(a)at(4,0)[below]{\color{black} {\small $\vec{q}$}};
\end{tikzpicture}

\bigskip
\bigskip
\begin{tikzpicture}
\draw[thick,->,>=stealth](1,0)arc(180:135:1);
\draw[thick,->,>=stealth](2,0)+(135:1)arc(135:45:1);
\draw[thick](2,0)+(45:1)arc(45:0:1);
\draw[thick,dashed,->,>=stealth](3,0)--(4,0);
\node(a)at(0,0.7)[above]{$P_1$};
\draw[thick,dashed,->,>=stealth](0,0)--(1,0);
\node(a)at(0,0)[below]{\color{black} {\small $\vec{q}$}};
\node(a)at(0.8,0.7)[]{\small  {\color{red} $\vec{q}_1^\prime$}};
\node(a)at(3.2,0.7)[]{\small {\color{black} $\vec{q}_2^\prime$} };
\node(a)at(0.8,-0.7)[]{\small {\color{black} $\vec{q}_1^\prime$} };
\node(a)at(3.2,-0.7)[]{\small  {\color{red} $\vec{q}_2^\prime$}};
\node(a)at(2,0)[]{\small {\color{black} $\vec{q}_3^\prime$}\:\: {\color{red} $\vec{q}_2$}};
\node(a)at(4,0)[below]{\color{black} {\small $\vec{q}$}};
\draw[thick,->,>=stealth](1,0)arc(-180:-135:1);
\draw[thick,->,>=stealth](2,0)+(-135:1)arc(-135:-45:1);
\draw[thick](2,0)+(-45:1)arc(-45:0:1);
\draw[thick,->,>=stealth](2,1)--(2,0);
\draw[thick](2,0)--(2,-1);
\end{tikzpicture}
\quad
\begin{tikzpicture}
\draw[thick,->,>=stealth](1,0)arc(180:135:1);
\draw[thick](2,0)+(135:1)arc(135:45:1);
\draw[thick,->,>=stealth](2,0)+(0:1)arc(0:45:1);
\draw[thick,dashed,->,>=stealth](3,0)--(4,0);
\node(a)at(0,0.7)[above]{$P_2$};
\draw[thick,dashed,->,>=stealth](0,0)--(1,0);
\node(a)at(0,0)[below]{\color{black} {\small $\vec{q}$}};
\node(a)at(0.8,0.7)[]{\small  {\color{red} $\vec{q}_1^\prime$}};
\node(a)at(3.2,0.7)[]{\small {\color{black} $\vec{q}_2$}};
\node(a)at(0.8,-0.7)[]{\small {\color{black} $\vec{q}_1^\prime$}};
\node(a)at(3.2,-0.7)[]{\small  {\color{red} $\vec{q}_2^\prime$}};
\node(a)at(2,0)[]{\small {\color{black} $\vec{q}_2^\prime$} \:\: {\color{red} $\vec{q}_2$}};
\node(a)at(4,0)[below]{\color{black} {\small $\vec{q}$}};
\draw[thick,->,>=stealth](1,0)arc(-180:-135:1);
\draw[thick,->,>=stealth](2,0)+(-135:1)arc(-135:-45:1);
\draw[thick](2,0)+(-45:1)arc(-45:0:1);
\draw[thick,->,>=stealth](2,1)--(2,0);
\draw[thick](2,0)--(2,-1);
\end{tikzpicture}
\quad
\begin{tikzpicture}
\draw[thick,->,>=stealth](1,0)arc(180:135:1);
\draw[thick,->,>=stealth](2,0)+(135:1)arc(135:45:1);
\draw[thick](2,0)+(45:1)arc(45:0:1);
\draw[thick,dashed,->,>=stealth](3,0)--(4,0);
\node(a)at(0,0.7)[above]{$P_3$};
\draw[thick,dashed,->,>=stealth](0,0)--(1,0);
\node(a)at(0,0)[below]{\color{black} {\small $\vec{q}$}};
\node(a)at(0.8,0.7)[]{\small {\color{red} $\vec{q}_1^\prime$}};
\node(a)at(3.2,0.7)[]{\small {\color{black} $\vec{q}_2^\prime$}};
\node(a)at(0.8,-0.7)[]{\small {\color{black} $\vec{q}_2$}};
\node(a)at(3.2,-0.7)[]{\small  {\color{red} $\vec{q}_2^\prime$}};
\node(a)at(2,0)[]{\small {\color{black} $\vec{q}_1^\prime$} \:\: {\color{red} $\vec{q}_2$}};
\node(a)at(4,0)[below]{\color{black} {\small $\vec{q}$}};
\draw[thick](1,0)arc(-180:-135:1);
\draw[thick,->,>=stealth](2,0)+(-45:1)arc(-45:-135:1);
\draw[thick](2,0)+(0:1)arc(0:-45:1);
\draw[thick,->,>=stealth](2,0)+(-90:1)arc(-90:-45:1);
\draw[thick,->,>=stealth](2,1)--(2,0);
\draw[thick](2,0)--(2,-1);
\end{tikzpicture}

\bigskip
\begin{tikzpicture}
\draw[thick,->,>=stealth](1,0)arc(180:135:1);
\draw[thick,->,>=stealth](2,0)+(135:1)arc(135:45:1);
\draw[thick](2,0)+(45:1)arc(45:0:1);
\draw[thick,dashed,->,>=stealth](3,0)--(4,0);
\node(a)at(0,0.7)[above]{$P_4$};
\node(a)at(0.8,0.7)[]{\small  {\color{red} $\vec{q}_1^\prime$}};
\draw[thick,dashed,->,>=stealth](0,0)--(1,0);
\node(a)at(0,0)[below]{\color{black} {\small $\vec{q}$}};
\node(a)at(0.8,0.7)[]{\small };
\node(a)at(3.2,0.7)[]{\small {\color{black} $\vec{q}_1^\prime$}};
\node(a)at(0.8,-0.7)[]{\small {\color{black} $\vec{q}_2$}};
\node(a)at(3.2,-0.7)[]{\small {\color{red} $\vec{q}_3$}};
\node(a)at(2,0)[]{\small {\color{black} $\vec{q}_2^\prime$} \:\: {\color{red} $\vec{q}_2$}};
\node(a)at(4,0)[below]{\color{black} {\small $\vec{q}$}};
\draw[thick](1,0)arc(-180:-135:1);
\draw[thick,->,>=stealth](2,0)+(-45:1)arc(-45:-135:1);
\draw[thick,->,>=stealth](2,0)+(0:1)arc(0:-45:1);
\draw[thick,->,>=stealth](2,1)--(2,0);
\draw[thick](2,0)--(2,-1);
\end{tikzpicture}
\quad
{\color{red}
\begin{tikzpicture}
\draw[thick,->,>=stealth](1,0)arc(180:135:1);
\draw[thick](2,0)+(135:1)arc(135:45:1);
\draw[thick,->,>=stealth](2,0)+(0:1)arc(0:45:1);
\draw[thick,dashed,->,>=stealth](3,0)--(4,0);
\node(a)at(0,0.7)[above]{$P_5$};
\draw[thick,dashed,->,>=stealth](0,0)--(1,0);
\node(a)at(0,0)[below]{\color{black} {\small $\vec{q}$}};
\node(a)at(0.8,0.7)[]{\small  {\color{red} $\vec{q}_1^\prime$}};
\node(a)at(3.2,0.7)[]{\small {\color{black} $\vec{q}_3$}};
\node(a)at(0.8,-0.7)[]{\small {\color{black} $\vec{q}_2$}};
\node(a)at(3.2,-0.7)[]{\small  {\color{red} $\vec{q}_2^\prime$}};
\node(a)at(2,0)[]{\small {\color{black} $\vec{q}_1^\prime$} \:\: {\color{red} $\vec{q}_2$}};
\node(a)at(4,0)[below]{\color{black} {\small $\vec{q}$}};
\draw[thick](1,0)arc(-180:-135:1);
\draw[thick,->,>=stealth](2,0)+(-90:1)arc(-90:-135:1);
\draw[thick,->,>=stealth](2,0)+(-135:1)arc(-135:-45:1);
\draw[thick](2,0)+(-45:1)arc(-45:0:1);
\draw[thick,->,>=stealth](2,1)--(2,0);
\draw[thick](2,0)--(2,-1);
\end{tikzpicture}
}
\quad
\begin{tikzpicture}
\draw[thick](1,0)arc(180:135:1);
\draw[thick,->,>=stealth](2,0)+(90:1)arc(90:135:1);
\draw[thick](2,0)+(135:1)arc(135:45:1);
\draw[thick,->,>=stealth](2,0)+(0:1)arc(0:45:1);
\draw[thick](2,1)arc(90:0:1);
\draw[thick,dashed,->,>=stealth](3,0)--(4,0);
\node(a)at(0,0.7)[above]{$P_6$};
\node(a)at(3.2,-0.7)[]{\small  {\color{red} $\vec{q}_1^\prime$}};
\draw[thick,dashed,->,>=stealth](0,0)--(1,0);
\node(a)at(0,0)[below]{\color{black} {\small $\vec{q}$}};
\node(a)at(0.8,0.7)[]{\small {\color{red} $\vec{q}_2$}};
\node(a)at(3.2,0.7)[]{\small {\color{black} $\vec{q}_2$}};
\node(a)at(0.8,-0.7)[]{\small {\color{black} $\vec{q}_2^\prime$}};
\node(a)at(3.2,-0.7)[]{\small };
\node(a)at(2,0)[]{\small {\color{black} $\vec{q}_1^\prime$} \:\: {\color{red} $\vec{q}_3$}};
\node(a)at(4,0)[below]{\color{black} {\small $\vec{q}$} };
\draw[thick,->,>=stealth](1,0)arc(-180:-135:1);
\draw[thick,->,>=stealth](2,0)+(-135:1)arc(-135:-45:1);
\draw[thick](2,0)+(-45:1)arc(-45:0:1);
\draw[thick,->,>=stealth](2,1)--(2,0);
\draw[thick](2,0)--(2,-1);
\end{tikzpicture}
\end{center}
\caption{Les contributions à la fonction énergie propre $\Sigma(\qq,z)$ d'ordre quatre en $H_3^{(+)}+H_3^{(-)}$ sont données par huit diagrammes à boucles internes (numérotés de $B_1$ à $B_8$) et par six diagrammes à pont (numérotés de $P_1$ à $P_6$).  Les vecteurs d'onde $\qq_i$ $(2\leq i\leq n$) et $\qq'_j$ ($1\leq j\leq n'$) près des lignes internes sont ceux des phonons entrants et sortants des processus collisionnels $n\to n'$ à quatre phonons dans l'interprétation physique à la Landau de la section \ref{sec3.2} (pour les diagrammes à pont: vecteurs d'onde des contributions de type I en noir, de type II en rouge, au sens de l'équation {\yct(\ref{eq233})}) ; pour alléger, les vecteurs d'onde des phonons virtuels ne sont pas indiqués (ils s'en déduisent par conservation de l'impulsion à chaque sommet). {\yc Les contributions des processus non résonnants de $H_3$, sous-dominantes dans la limite $\epsilon\to 0$, ne sont pas représentées. En revanche, une fois cette limite prise, le diagramme $P_5$ (plus précisément sa contribution de type II, d'où la couleur rouge) diverge linéairement en $\zetat\equiv (z-\veps_\qq)/k_B T\epsilon^2=0$ ce qui fait apparaître un pôle non physique dans la fonction de Green, voir la section \ref{sec3.3}.}
}
\label{fig8}
\end{figure}

\paragraph{Les deux topologies} Nous omettons directement l'hamiltonien quartique $H_4$ dans l'équation (\ref{eq001}) et les termes non résonnants $H_3^{\mbox{\yct\scriptsize (non\, r\'es.)}}$ dans l'hamiltonien cubique (\ref{eq100}), car leurs contributions à $\Sigma^{(4)}$ sont sous-dominantes dans la limite $\epsilon\to 0$ (c'est évident pour $H_3^{\mbox{\yct\scriptsize (non\, r\'es.)}}$, qui ne bénéficie pas de l'effet \g{petit dénominateur}, et ceci sera justifié pour $H_4$ dans la section \ref{sec3.2}). L'hamiltonien de couplage se réduit donc aux termes $H_3^{(+/-)}$ donnés dans l'équation (\ref{eq101}). Dans les diagrammes de Feynman d'ordre quatre correspondants, chacun des quatre sommets est le point de rencontre de trois lignes de phonons, ce qui conduit à deux topologies possibles, (i) à boucle interne et (ii) à pont, comme le montre la figure \ref{fig8}. Dans la première topologie, on lève la dégénérescence en mettant arbitrairement la boucle interne dans la branche supérieure, et on se souvient (comme dans la section \ref{sec2}) que l'échange des deux lignes de la boucle interne conduit au même diagramme ; {\yc après orientation des lignes internes,} il reste huit diagrammes distincts. Dans la seconde topologie, on lève la dégénérescence en orientant la ligne du pont vers le bas ; il reste six diagrammes distincts. En réalité, dans la terminologie de la figure \ref{fig8}, les contributions des diagrammes $B_3$ et $B_4$ à $\Sigma^{(4)}$ se doublent, ainsi que celles de $P_2$ et $P_3$ et que celles de $P_4$ et $P_6$.\footnote{Il existe en effet une symétrie non triviale reliant certains diagrammes entre eux. Pour le voir, appliquons à chaque diagramme une rotation d'axe vertical d'angle $\pi$ puis changeons l'orientation de toutes les lignes (internes et externes); ceci ne change pas la valeur de la contribution du diagramme à $\Sigma^{(4)}$. La plupart des diagrammes sont ainsi envoyés sur eux-mêmes, à l'exception des paires $\{B_3,B_4\}$, $\{P_2,P_3\}$ et $\{P_4,P_6\}$ dont les deux partenaires sont simplement échangés (on se souviendra, dans le cas des diagrammes à pont, qu'une rotation d'axe horizontal d'angle $\pi$, qui permet de rétablir l'orientation originelle de la ligne verticale, ne change pas la valeur du diagramme). On en déduit que $\Sigma_{B_3}^{{\yc(4)}}(\qq,z)=\Sigma_{B_4}^{{\yc(4)}}(\qq,z)$, $\Sigma_{P_2}^{{\yc(4)}}(\qq,z)=\Sigma_{P_3}^{{\yc(4)}}(\qq,z)$ et $\Sigma_{P_4}^{{\yc(4)}}(\qq,z)=\Sigma_{P_6}^{{\yc(4)}}(\qq,z)$ comme annoncé.} Il ne reste que onze diagrammes véritablement indépendants.


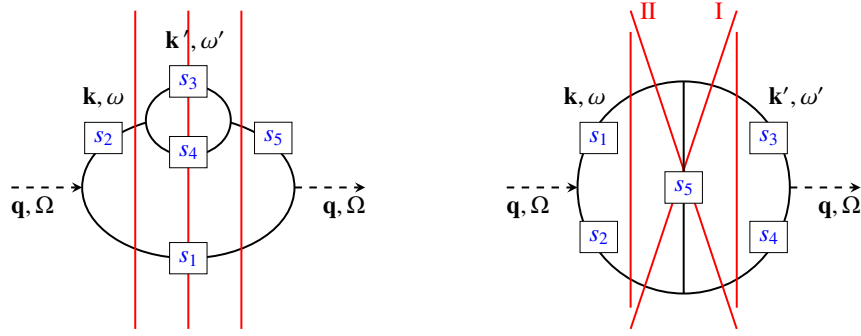
\begin{figure}[t]
\begin{center}
\begin{tikzpicture} 
\draw[thick,dashed,->,>=stealth](-0.5,0)--(0.5,0);
\draw[thick] (2,0) ellipse (1.5 and 1);
\draw[thick,fill=white] (2,0.95) ellipse (0.6 and 0.5);
\draw[thick,red](2,-2)--(2,2.5);
\draw[thick,dashed,->,>=stealth](3.5,0)--(4.5,0);
\node(a)at(-0.2,0)[below]{$\vec{q},\Omega$};
\node(a)at(4.2,0)[below]{$\vec{q},\Omega$};
\node(a)at(2.1,1.8)[above]{$\vec{k\,}^\prime,\omega^\prime$};
\node[draw,rectangle,fill=white,text=blue](b)at(2,-1) {$s_1$};
\node[draw,rectangle,fill=white,text=blue](b)at(2,0.5) {$s_4$};
\node[draw,rectangle,fill=white,text=blue](b)at(2,1.5) {$s_3$};
\node[draw,rectangle,fill=white,text=blue](b)at(0.8,0.7) {$s_2$};
\node(a)at(0.8,1.0)[above]{$\vec{k},\omega$};
\node[draw,rectangle,fill=white,text=blue](b)at(3.2,0.7) {$s_5$};
\draw[thick,red](1.25,-2)--(1.25,2.5); 
\draw[thick,red](2.75,-2)--(2.75,2.5);
\draw[thick,dashed,->,>=stealth](-0.5+7,0)--(0.5+7,0);
\draw[thick] (2+7,0) circle (1.5 );
\draw[thick,dashed,->,>=stealth](3.5+7,0)--(4.5+7,0);
\node(a)at(-0.2+7,0)[below]{$\vec{q},\Omega$};
\node(a)at(4.2+7,0)[below]{$\vec{q},\Omega$};
\node[draw,rectangle,fill=white,text=blue](b)at(0.8+7,0.7) {$s_1$};
\node[draw,rectangle,fill=white,text=blue](b)at(3.2+7,0.7) {$s_3$};
\node[draw,rectangle,fill=white,text=blue](b)at(0.8+7,-0.7) {$s_2$};
\node[draw,rectangle,fill=white,text=blue](b)at(3.2+7,-0.7) {$s_4$};
\draw[thick,red](1.25+7,-1.7)--(1.25+7,2.2);
\draw[thick,red](2.75+7,-1.7)--(2.75+7,2.2);
\draw[thick,red](1.25+7,-2)--(2.75+7,2.5);
\draw[thick,red](2.75+7,-2)--(1.25+7,2.5);
\node[text=red](b)at(2.5+7,2.5) {${\rm I}$};
\node[text=red](b)at(1.5+7,2.5) {${\rm II}$};
\node(a)at(0.6+7,1)[above]{$\vec{k},\omega$};
\node(a)at(3.6+7,1)[above]{$\vec{k}^\prime,\omega'$};
\draw[thick](2+7,-1.5)--(2+7,1.5); 
\node[draw,rectangle,fill=white,text=blue](b)at(2+7,0) {$s_5$};
\end{tikzpicture}
\end{center}
\caption{Dans le développement de la fonction énergie propre $\Sigma(\qq,z)$, formes génériques des diagrammes d'ordre quatre en $H_3$ à boucle interne (partie gauche) ou à pont (partie droite). Les boîtes contiennent les orientations possibles $s_j=\pm 1$ des lignes internes et indiquent leur numérotation $j$. Compte tenu de la conservation de l'énergie-impulsion de Matsubara en chaque sommet, il n'y a que deux vecteurs d'onde $\kk$, $\kk'$ et deux énergies de Matsubara $\omega$, $\omega'$ internes indépendants ; les autres sont donnés par les équations (\ref{eq211}) (cas à boucle interne) et (\ref{eq231}) (cas à pont). Les droites rouges sont une construction graphique permettant de retrouver les dénominateurs d'énergie dans les formes simplifiées (\ref{eq213}) et (\ref{eq233}) des sommes doubles de Matsubara, qui valent à l'ordre dominant en $\epsilon$ lorsque tous les vecteurs d'onde internes font un petit angle $O(\epsilon)$ avec $\qq$. Dans la forme simplifiée (\ref{eq213}), la quantité $\zeta_1$ correspond aux deux droites rouges externes du diagramme à boucle interne (en effet, $s_2=s_5$ à l'ordre dominant en $\epsilon$) et $\zeta_2$ correspond à la droite rouge centrale. Dans la forme simplifiée (\ref{eq233}), la quantité $\zeta_1$ correspond à la droite rouge verticale gauche du diagramme à pont, $\zeta_2$ à l'autre droite rouge verticale, $\zeta_3$ à la droite {\yc oblique} numérotée II et $\zeta_4$ à la droite {\yc oblique} numérotée I.}
\label{fig7}
\end{figure}

\paragraph{Calcul d'un diagramme générique à boucle interne}
Il est commode de numéroter les lignes internes de $j=1$ à $5$ comme sur la partie gauche de la figure \ref{fig7}. Chaque ligne a une orientation $s_j\in\{-1;1\}$ (de gauche à droite si $s_j=1$, de droite à gauche si $s_j=-1$), un vecteur d'onde $\kk_j$ et une énergie de Matsubara $\omega_j\in 2\ii\pi k_B T \mathbb{Z}$. Sur la figure \ref{fig8}, on a fait le choix d'avoir toujours $s_3\geq s_4$ afin d'éviter le double comptage de diagramme évoqué plus haut. On choisit comme quadrivecteurs indépendants ceux des lignes 2 et 3, que l'on renomme comme suit: 
\be
\label{eq210}
(\kk_2,\omega_2)=(\kk,\omega) \quad ; \quad (\kk_3,\omega_3)=(\kk',\omega')
\ee
Les autres s'en déduisent par conservation de l'énergie-impulsion en chaque sommet. Par exemple, la somme des vecteurs d'onde qui arrivent à un sommet donné est égale à la somme des vecteurs d'onde qui en partent, ce qui donne:
\be
\label{eq211}
\kk_1=s_1(\qq-s_2\kk) \quad ; \quad \kk_4=s_4(s_2\kk-s_3\kk') \quad ; \quad \kk_5=s_2 s_5 \kk
\ee
On a les mêmes relations sur les énergies de Matsubara. Compte tenu des règles énoncées dans la section 25 de la référence \cite{FW}, la contribution à la fonction énergie propre du diagramme générique à boucle interne $B_i$ vaut:
\be
\label{eq212}
\Sigma_{B_i}^{(4)}(\qq,z) = \frac{(-k_B T)^2}{1+\delta_{s_3,s_4}} \int\frac{\dd^dk}{(2\pi)^d}\int\frac{\dd^dk'}{(2\pi)^d}
\mA(\kk_1,\kk_2;\qq)\mA(\kk_3,\kk_4;\kk_2)\mA(\kk_3,\kk_4;\kk_5)\mA(\kk_1,\kk_5;\qq)
\sum_{\omega,\omega'\in 2\ii\pi k_B T\mathbb{Z}}\frac{1}{\prod_{j=1}^{5} (\omega_j-\veps_{\kk_j})}
\ee
où l'on a temporairement $z=\Omega\in 2\ii\pi k_B T\mathbb{Z}$ (voir la section \ref{sec2.1}). Pour écrire les amplitudes de transition en chaque sommet, on a utilisé l'invariance de la fonction $\mA$ de l'équation (\ref{eq102}) par permutation de ses trois arguments et donc le fait que les éléments de matrice $\langle \kk_2,\kk_3|\mathcal{H}_3|\kk_1\rangle$, $\langle \kk_1,\kk_3|\mathcal{H}_3|\kk_2\rangle$ et $\langle \kk_1,\kk_2|\mathcal{H}_3|\kk_3\rangle$ sont tous égaux à $\mA(\kk_2,\kk_3;\kk_1)$. Le dénominateur du facteur de symétrie {\yc devant les intégrales} vaut $2$ si les deux lignes de la boucle interne ont la même orientation (pour compenser le double comptage des configurations ne différant que par l'échange des quadrivecteurs $(\kk_3,\omega_3)$ et $(\kk_4,\omega_4)$ et dans ce cas physiquement identiques); sinon, ces deux lignes sont discernables et le facteur de symétrie vaut $1$. La somme double de Matsubara est calculée explicitement dans l'\ref{app3}, qui montre aussi que son expression se simplifie à la limite des petits angles $O(\epsilon)\to 0$ entre les vecteurs d'onde, où l'on se restreint à $s_2=s_5$ (sinon, la contribution du diagramme est sous-dominante en $\epsilon$, voir l'équation (\ref{eq215}) plus bas):
\be
\label{eq213}
\sum_{\omega,\omega'\in 2\ii\pi k_B T\mathbb{Z}}\frac{(k_B T)^2}{\prod_{j=1}^{5}(\omega_j-\veps_{\kk_j})} \stackrel{\mbox{\scriptsize diag.}\, B}{\underset{\epsilon\to 0}{\sim}} -s_1 s_4 \frac{(\bar{n}_{\veps_3}-\bar{n}_{-s_3s_4\veps_4})(\bar{n}_{-s_1\veps_1}-\bar{n}_{s_2\veps_2})}{(\zeta-\zeta_1)^2(\zeta-\zeta_2)}
\ee
avec la variable $\zeta=z-\veps_\qq$ et des $\zeta_j$ tous d'ordre $k_B T\epsilon^2$,
\be
\label{eq214}
\zeta_1\equiv s_1\veps_{\kk_1}+s_2\veps_{\kk_2}-\veps_\qq \quad ; \quad \zeta_2\equiv s_1\veps_{\kk_1}+s_3\veps_{\kk_3}+s_4\veps_{\kk_4}-\veps_{\qq}
\ee
La dépendance en $\zeta$ de la forme (\ref{eq213}) au second membre est facile à interpréter physiquement puisque $\zeta_1+\veps_\qq$ et $\zeta_2+\veps_\qq$ sont les énergies des états intermédiaires repérés par les droites verticales sur la partie gauche de la figure \ref{fig7}: en effet, ce sont les sommes des énergies des lignes internes $j$ coupées par chaque droite verticale comptées algébriquement, c'est-à-dire pondérées par les orientations $s_j$ (si $s_j=-1$, le phonon sur la ligne $j$ remonte le temps et son énergie doit être comptée négativement). Ceci valide l'estimation (\ref{eq024}) en en reproduisant les dénominateurs d'énergie. 

Pour passer à la limite $\epsilon\to 0$ dans (\ref{eq212}), on met directement les angles à zéro dans chaque amplitude de couplage $\mA$ (ce qui fait {\yc apparaître} un facteur $3(1+\Lambda)$), puis on développe les dénominateurs d'énergie de (\ref{eq213}) sur le modèle de (\ref{eq025}) ; pour que l'effet \g{petit dénominateur} joue à plein, il faut que les vecteurs internes $\kk_j$ fassent tous un petit angle $O(\epsilon)$ avec le vecteur d'onde $\qq$ entrant, ce qui impose\footnote{Quand $\kk_2=\kk$ et $\kk_3=\kk'$ font un angle nul avec $\qq$, les autres vecteurs d'onde $\kk_j$ donnés par (\ref{eq211}) doivent satisfaire à la condition $\kk_j\cdot\qq=k_j q$.}
\be
\label{eq215}
\sig(q-s_2k)=s_1 \quad ; \quad \sig(s_2k-s_3k')=s_4 \quad ; \quad s_2s_5=1
\ee
Les deux premières conditions dans (\ref{eq215}) restreignent le domaine de variation des nombres d'onde $k$ et $k'$. La dernière, qui en est indépendante, élimine les diagrammes $B_3$ et $B_4$, de contributions sous-dominantes. En se souvenant des définitions (\ref{eq022},\ref{eq004}) de $\zetat$ et (\ref{eq020}) de $\qb$, on arrive en dimension $d=2$ au résultat (pour $\im\zetat\neq 0$ mais de signe quelconque):
\be
\label{eq220}
\tilde\Sigma^{(4,2)}_{B_i}(\qb,\tilde{\zeta})=\frac{\qb}{1+\delta_{s_3,s_4}}\left(\frac{9(1+\Lambda)^2}{8\rho\xi^2}\right)^2\int_{\mathcal{D}_1^B}\frac{\kb\dd \kb}{(2\pi)^2} \int_{\mathcal{D}_2^B(\kb)} \frac{\kb'\dd \kb'}{(2\pi)^2} \kb^2\kb'(\qb-s_2\kb)(s_2\kb-s_3\kb')(\bar{n}^{\rm lin}_{k'}-\bar{n}^{\rm lin}_{k'-s_2 s_3 k})(\bar{n}^{\rm lin}_{s_2k}-\bar{n}^{\rm lin}_{s_2k-q}) I(\kb,\kb')
\ee
avec {\yc les nombres d'occupation (\ref{eq029}) étendus aux nombres d'onde négatifs et} l'intégrale angulaire\footnote{Dans l'intégrale, avant de prendre la limite $\epsilon\to 0$, on a posé $\tilde{\theta}=\theta/\epsilon$ et $\tilde{\phi}=(\theta-\theta')/\epsilon$, où $\theta$ est l'angle entre $\qq$ et $\kk$, $\theta'$ l'angle entre $\qq$ et $\kk'$.}
\be
\label{eq221}
I(\kb,\kb')=\int_{\mathbb{R}} \dd\tilde\theta \int_{\mathbb{R}} \dd\tilde\phi \frac{1}{[\tilde{\zeta}-(A_1^B+B_1^B\tilde\theta^2)]^{2}[\tilde{\zeta}-(A^B_2+B^B_1\tilde\theta^2+B_2^B\tilde\phi^2)]}
\ee
qui fait intervenir les coefficients
\be
\label{eq222}
A_1^B=\frac{3\gamma}{8}(\kb-s_2 \qb)\kb \qb \quad ; \quad A_2^B=\frac{3\gamma}{8}\kb \kb' (s_2\kb'-s_3 \kb) +A_1^{B} \quad ; \quad B_1^B=\frac{-\kb\qb/2}{\kb-s_2 \qb} \quad ; \quad B_2^B=\frac{-\kb \kb'/2}{s_2\kb'-s_3 \kb}
\ee
et les domaines d'intégration {\yc$\mathcal{D}_1^B$} sur $\kb$, puis {\yc$\mathcal{D}_2^{B}(\kb)$} sur $\kb'$ à $\kb$ fixé, satisfaisant aux conditions (\ref{eq215}) et de coupure (\ref{eq103}) $\kb_j<\eta$, $1\leq j\leq 5$. Au premier membre de (\ref{eq220}), le second entier dans l'exposant rappelle {\yc à quel ordre en $\epsilon$ la fonction} $\Sigma^{(4)}$ est calculé {\yc(ici à l'ordre deux)} et le tilde sur $\Sigma$ signifie comme dans (\ref{eq021}) que {\yc le résultat} a été divisé par $k_BT\epsilon^2$. Enfin, on vérifie au cas par cas que l'expression (\ref{eq220}) admet une limite finie lorsque la coupure $\eta\to+\infty$ (pour le champ quantique).\footnote{Par exemple, dans le cas {\yc du diagramme} $B_8$ ($s_1=s_4=-1$, $s_2=s_3=s_5=1$), les domaines {\yc$\mathcal{D}_1^B=[\qb,\eta]$ et $\mathcal{D}_2^B(\kb)=[\kb,\eta]$} sont non bornés lorsque $\eta\to +\infty$ mais les facteurs d'amplification bosoniques $(\bar{n}_{k'}^{\rm lin}-\bar{n}_{k'-k}^{\rm lin})(\bar{n}_k^{\rm lin}-\bar{n}_{k-q}^{\rm lin})=O[\exp(-\kb-\kb'')]$ sont à décroissance rapide en les deux variables d'intégration découplées $(\kb,\kb''=\kb'-\kb)$ et écrasent toute divergence en loi de puissance qui pourrait provenir des autres facteurs ; dans le cas de $B_7$ ($s_1=-1$, $s_2=s_3=s_4=s_5=1$), $\mathcal{D}_1^B=[\qb,\eta]$ et $\mathcal{D}_2^B(\kb)=[0,\kb]$ restent non bornés et les facteurs bosoniques $(1+\bar{n}_{k'}^{\rm lin}+\bar{n}_{k-k'}^{\rm lin})(\bar{n}_k^{\rm lin}-\bar{n}_{k-q}^{\rm lin})=O(\exp(-\kb))$ ne sont plus à décroissance rapide qu'en la variable $\kb$ mais, comme $\kb'<\kb$, l'intégrale sur $\kb'$ à $\kb$ fixé est au plus un $O(\kb^\alpha)$, $\alpha>0$, et l'intégrale sur $\kb$ converge si $\eta\to +\infty$.}

\paragraph{Calcul d'un diagramme générique à pont} Le début ressemble beaucoup au cas précédent. On numérote les lignes internes comme sur la partie droite de la figure \ref{fig7}, avec la même convention de signe, sauf pour la ligne verticale (le \g{pont}), de signe $s_5=1$ lorsqu'elle orientée vers le bas (ce à quoi nous nous sommes limités sur la figure \ref{fig8} pour éviter un double comptage). On garde comme quadrivecteurs indépendants
\be
\label{eq230}
(\kk_1,\omega_1)=(\kk,\omega) \quad ;\quad (\kk_3,\omega_3)=(\kk',\omega')
\ee
les autres s'en déduisant par conservation de l'énergie-impulsion, par exemple pour les vecteurs d'onde:
\be
\label{eq231}
\kk_2=s_2(\qq-s_1\kk) \quad ;\quad \kk_4=s_4(\qq-s_3\kk') \quad ;\quad \kk_5=s_5(s_1\kk-s_3\kk')
\ee
La contribution du diagramme $P_i$ à la fonction énergie propre s'écrit alors \cite{FW}:
\be
\label{eq232}
\Sigma_{P_i}^{(4)}(\qq,z) = (-k_B T)^2\int\frac{\dd^dk}{(2\pi)^d}\int\frac{\dd^dk'}{(2\pi)^d}
\mA(\kk_1,\kk_2;\qq)\mA(\kk_3,\kk_5;\kk_1)\mA(\kk_2,\kk_5;\kk_4)\mA(\kk_3,\kk_4;\qq)
\sum_{\omega,\omega'\in 2\ii\pi k_B T\mathbb{Z}}\frac{1}{\prod_{j=1}^{5} (\omega_j-\veps_{\kk_j})}
\ee
Nul besoin ici de diviser par un facteur de symétrie. Le calcul explicite de la somme double de Matsubara est donné dans l'\ref{app3}, qui établit aussi son expression simplifiée à la limite des petits angles $O(\epsilon)$ entre les vecteurs d'onde: 
\be
\label{eq233}
\sum_{\omega,\omega'\in 2\ii\pi k_B T\mathbb{Z}}\frac{(k_B T)^2}{\prod_{j=1}^{5}(\omega_j-\veps_{\kk_j})} \stackrel{\mbox{\scriptsize diag.}\, P}{\underset{\epsilon\to 0}{\sim}} 
-s_1 s_2 s_4 s_5 \frac{(\bar{n}_{-s_2\veps_2}-\bar{n}_{s_1\veps_1})(\bar{n}_{\veps_3}-\bar{n}_{-s_3s_5\veps_5})}
{(\zeta-\zeta_1)(\zeta-\zeta_2)(\zeta-\zeta_4)}
-s_1 s_2 s_4 s_5 \frac{(\bar{n}_{-s_3s_5\veps_5}-\bar{n}_{-s_3s_4\veps_4})(\bar{n}_{s_1\veps_1}-\bar{n}_{-s_2\veps_2})}
{(\zeta-\zeta_1)(\zeta-\zeta_2)(\zeta-\zeta_3)}
\ee
avec la variable $\zeta$ et des $\zeta_j$ tous d'ordre $k_BT\epsilon^2$,
\be
\label{eq238}
\zeta_1\equiv s_1\veps_{\kk_1}+s_2\veps_{\kk_2}-\veps_\qq\quad ; \quad \zeta_2=s_3\veps_{\kk_3}+s_4\veps_{\kk_4}-\veps_\qq \quad ; \quad \zeta_3\equiv s_1\veps_{\kk_1}+s_4\veps_{\kk_4}-s_5\veps_{\kk_5}-\veps_{\qq} \quad ;\quad \zeta_4\equiv s_2\veps_{\kk_2}+s_3\veps_{\kk_3}+s_5\veps_{\kk_5}-\veps_{\qq}
\ee
Il y a cette fois deux contributions au second membre, dites de type I et de type II dans cet ordre. Les quantités $\zeta_j$ sont des différences d'énergie entre états intermédiaires et état initial, comme dans le cas à boucle interne, sauf qu'il faut introduire quatre droites pour les repérer, voir la partie droite de la figure \ref{fig7}: deux droites verticales pour $\zeta_1$ et $\zeta_2$, deux droites obliques pour $\zeta_3$ et $\zeta_4$. En effet, du point de vue de la droite oblique marquée I sur la figure, la ligne interne du pont est orientée de gauche à droite si $s_5=1$, et $\zeta_4$ doit comporter un signe plus devant $s_5\veps_{\kk_5}$ dans l'équation (\ref{eq238}); pour la droite oblique II, c'est le contraire, d'où le signe moins devant $s_5\veps_{\kk_5}$ dans $\zeta_3$.

Lorsque $\epsilon\to 0$, la limite des petits angles dans les dénominateurs d'énergie et dans les amplitudes de couplage $\mA$ se prend comme précédemment, si ce n'est qu'il faut distinguer les deux types. Pour la contribution de type I, on obtient (pour $\im\zetat\neq 0$ mais de signe quelconque)
\be
\label{eq240}
\tilde\Sigma^{(4,2)}_{P_{i}^{\rm I}}(\qb,\tilde{\zeta})=\qb\left(\frac{9(1+\Lambda)^2}{8\rho\xi^2}\right)^2\int_{\mathcal{D}_1^P} \frac{\kb\dd \kb}{(2\pi)^2} \int_{\mathcal{D}_2^P(\kb)} \frac{\kb'\dd \kb'}{(2\pi)^2} \kb \kb'(\qb-s_1\kb)(\qb-s_3\kb')(\kb-s_1s_3\kb')(\bar{n}^{\rm lin}_{k'}-\bar{n}^{\rm lin}_{k'-s_1 s_3 k})(\bar{n}^{\rm lin}_{s_1k}-\bar{n}^{\rm lin}_{s_1k-q}) J(\kb,\kb')
\ee
avec l'intégrale angulaire
\be
\label{eq241}
J(\kb,\kb')=\int_{\mathbb{R}} \dd\tilde\theta \int_{\mathbb{R}} \dd\tilde\theta' \frac{1}{[\tilde{\zeta}-(A_1^P+B_1^P\thetat^2)][\tilde{\zeta}-(A_2^P+B_2^P\thetat'^2)][\tilde{\zeta}-(A_4^P+B_1^P\thetat^2+B_4^P(\thetat-\thetat')^2)]}
\ee
Pour celle de type II, il vient (toujours pour $\im\zetat\neq 0$ mais de signe quelconque)
\be
\label{eq242}
\tilde\Sigma^{(4,2)}_{P_{i}^{\rm II}}(\qb,\tilde{\zeta})=\qb\left(\frac{9(1+\Lambda)^2}{8\rho\xi^2}\right)^2\int_{\mathcal{D}_1^P} \frac{\kb\dd \kb}{(2\pi)^2} \int_{\mathcal{D}_2^P(\kb)} \frac{\kb'\dd \kb'}{(2\pi)^2} \kb \kb'(\qb-s_1\kb)(\qb-s_3\kb')(\kb-s_1s_3\kb')(\bar{n}^{\rm lin}_{k'-s_3q}-\bar{n}^{\rm lin}_{k'-s_1 s_3 k})(\bar{n}^{\rm lin}_{s_1k}-\bar{n}^{\rm lin}_{s_1k-q}) K(\kb,\kb')
\ee
avec l'intégrale angulaire
\be
\label{eq243}
K(\kb,\kb')=\int_{\mathbb{R}} \dd\tilde\theta \int_{\mathbb{R}} \dd\tilde\theta' \frac{1}{[\tilde{\zeta}-(A_1^P+B_1^P\thetat^2)][\tilde{\zeta}-(A_2^P+B_2^P\thetat'^2)][\tilde{\zeta}-(A_3^P+B_2^P\thetat'^2+B_3^P(\thetat-\thetat')^2)]}
\ee
Indépendamment du type, les coefficients valent
\be
\label{eq244}
A_1^P=\frac{3\gamma}{8}(\kb-s_1\qb)\kb\qb \quad ; \quad A_2^P=\frac{3\gamma}{8}(\kb'-s_3\qb)\kb'\qb \quad ; \quad A_4^P=A_1^P+\frac{3\gamma}{8}\kb \kb' (s_1\kb'-s_3\kb) \quad ; \quad A_3^P=A_1^P+A_2^P-A_4^P
\ee
\be
\label{eq245}
B_1^P=\frac{-\kb\qb/2}{\kb-s_1\qb} \quad ; \quad B_2^P=\frac{-\kb'\qb/2}{\kb'-s_3\qb} \quad ; \quad B_4^P=\frac{-\kb\kb'/2}{s_1\kb'-s_3\kb} \quad ; \quad B_3^P=-B_4^P
\ee
Les domaines d'intégration $\mathcal{D}_1^P$ et $\mathcal{D}_2^P(\kb)$ sont eux aussi indépendants du type; ils tiennent compte de la coupure (\ref{eq103}) sur tous les vecteurs d'onde internes $\kk_j$, $1\leq j\leq 5$, ainsi que des conditions 
\be
\label{eq239}
\sig(q-s_1k)=s_2 \quad ;\quad \sig(q-s_3k')=s_4 \quad;\quad \sig(s_1k-s_3k')=s_5
\ee
assurant que $\kk_2$, $\kk_4$ et $\kk_5$ sont colinéaires à et de même sens que $\qq$ dès que $\kk$ et $\kk'$ le sont. Pour toutes les orientations $s_j$ prises sur la figure \ref{fig8}, il y a compatibilité entre les conditions (\ref{eq239}) donc aucun des diagrammes $P_i$ n'est sous-dominant en $\epsilon$. Enfin, on vérifie l'existence d'une limite finie dans (\ref{eq240},\ref{eq242}) lorsque la coupure $\eta\to +\infty$ (pour le champ quantique).

\subsection{Calcul par un raisonnement à la Landau}
\label{sec3.2}

Le calcul de $\Sigma^{(4)}$ comme dans la référence \cite{FW}, bien que systématique et rigoureux, ne dit pas explicitement quels processus collisionnels entre phonons se cachent derrière chaque diagramme de la figure \ref{fig8}. Nous développons ici une méthode plus heuristique, inspirée de la référence \cite{LK} et valable à l'ordre dominant en $\epsilon$, qui répond à la question, donne le résultat sous une forme compacte (plutôt qu'éclatée en une dizaine de contributions intégrales) et permet de montrer facilement que la contribution de $H_4$ est sous-dominante dans la limite $\epsilon\to 0$.

La référence \cite{LK} explique, pour l'hamiltonien (\ref{eq001}), comment calculer par une règle d'or de Fermi étendue le taux d'amortissement du mode de phonon $\qq$ dû à des collisions $n\to n'$, $n$ étant le nombre de phonons entrants $\qq_i$ (ils comportent un et un seul phonon de vecteur d'onde $\qq$) et $n'$ le nombre de phonons sortants $\qq'_j$ (ils ne comportent aucun phonon de vecteur d'onde $\qq$). Les cas ordinaires des processus de Belyaev et de Landau correspondent à $(n,n')=(1,2)$ et $(n,n')=(2,1)$, ils sont à $n+n'=3$ phonons ; ici, il s'agira de processus à quatre phonons, comme $(n,n')=(2,2)$. On commence par écrire des équations cinétiques sur les nombres d'occupation des phonons, provisoirement supposés hors d'équilibre, en comptant positivement le processus direct $n\to n'$ (qui tend à vider le mode $\qq$) et négativement le processus inverse $n'\to n$ (qui tend à le remplir): 
\begin{multline}
\label{eq250}
\frac{\dd}{\dd t} {n}_\qq = -\frac{1}{(n-1)!}\int\left(\prod_{i=2}^{n}\frac{\dd^dq_i}{(2\pi)^d}\right)\frac{1}{n'!}\int\left(\prod_{j=1}^{n'}\frac{\dd^dq'_j}{(2\pi)^d}\right)\frac{2\pi}{\hbar} \mA^2_{\ii\to \ff} \delta(E_\ii-E_\ff) (2\pi)^d \delta\left(\sum_{i=1}^{n}\qq_i-\sum_{j=1}^{n'}\qq'_j\right)\\
\times\left[\prod_{i=1}^{n}n_{\qq_i}\prod_{j=1}^{n'}(1+n_{\qq'_j})-\prod_{i=1}^{n} (1+n_{\qq_i}) \prod_{j=1}^{n'}n_{\qq'_j}\right]
\end{multline}
On a pris la limite thermodynamique, et on a divisé par des nombres de symétrie tenant compte de l'invariance du processus par permutation des $n-1$ derniers vecteurs d'onde entrants (on convient que $\qq_1=\qq$) et par permutation des $n'$ vecteurs d'onde sortants. On reconnaît dans (\ref{eq250}) le Dirac de conservation de l'énergie, le facteur $2\pi/\hbar$ et le carré d'une amplitude de transition $\mA_{\ii\to\ff}$ entre état initial $|\ii\rangle=|(\qq_i)_{1\leq i\leq n}\rangle$ et état final $|\ff\rangle=|(\qq'_j)_{1\leq j\leq n'}\rangle$ typiques de la règle d'or, ainsi qu'un Dirac de conservation de l'impulsion et les facteurs d'amplification bosoniques. Le principe de microréversibilité a permis de factoriser l'amplitude au carré du processus direct. 

On se place ensuite dans le cadre de la procédure d'excitation (\ref{eq010}), où tous les modes de phonons sont à l'équilibre thermique à l'exception du mode $\qq$, d'où la substitution $n_\kk\to \bar{n}_\kk$, $\kk\neq\qq$, et on linéarise (\ref{eq250}) autour de $n_\qq=\bar{n}_\qq$ pour une  faible excitation, ce qui donne l'équation de retour à l'équilibre $(\dd/\dd t)(n_\qq-\bar{n}_\qq)|_{t=0^+}=-\Gamma_\qq^{n\to n'}(n_\qq-\bar{n}_\qq)$ avec le taux
\begin{multline}
\label{eq251}
\Gamma_\qq^{n\to n'}=\frac{1}{(n-1)!}\int\left(\prod_{i=2}^{n}\frac{\dd^dq_i}{(2\pi)^d}\right)\frac{1}{n'!}\left(\int\prod_{j=1}^{n'}\frac{\dd^dq'_j}{(2\pi)^d}\right)\frac{2\pi}{\hbar} \mA^2_{\ii\to \ff} \delta(E_\ii-E_\ff) (2\pi)^d \delta\left(\sum_{i=1}^{n}\qq_i-\sum_{j=1}^{n'}\qq'_j\right) \\ {\yct\times}\left[\prod_{i=2}^{n}\bar{n}_{\qq_i}\prod_{j=1}^{n'}(1+\bar{n}_{\qq'_j})-\prod_{i=2}^{n} (1+\bar{n}_{\qq_i}) \prod_{j=1}^{n'}\bar{n}_{\qq'_j}\right]
\end{multline}
où, au contraire de l'équation (\ref{eq250}), tous les produits sur les vecteurs d'onde entrants excluent le mode $\qq$ donc commencent à $i=2$. La différence avec la règle d'or réside dans le calcul de l'amplitude $\mA_{\ii\to\ff}$. D'habitude, on a simplement $\mA_{\ii\to\ff}=\langle\ff|\mH_3|\ii\rangle$. La référence \cite{LK} s'occupe justement du cas utile ici de collisions à quatre phonons, où $\mA_{\ii\to\ff}$ s'obtient en traitant {\yc$H_3=\mathcal{H}_3/L^{d/2}$} au second ordre et {\yc$H_4=\mathcal{H}_4/L^d$} au premier ordre de la théorie des perturbations:
\be
\label{eq252}
\mA_{\ii\to\ff}=\langle\ff|\mH_4|\ii\rangle + \sum_{|\lambda \rangle} \frac{\langle\ff|\mH_3|\lambda\rangle\langle\lambda|\mH_3|\ii\rangle}{E_\ii-E_\lambda}
\ee
et la somme porte sur tous les états de Fock de phonons intermédiaires possibles $|\lambda\rangle$. 
Un simple comptage de puissances comme dans la section \ref{sec0} montre que la contribution de $\mH_4$ à (\ref{eq252}) est sous-dominante dans la limite des petits angles $\epsilon\to 0$;\footnote{Au vu des équations (\ref{eq052},\ref{eq104}), la contribution de $\mH_4$ à (\ref{eq252}) est d'ordre $(mc^2/\rho)\epsilon^2$ ; au vu des équations (\ref{eq025},\ref{eq027}), celle de $\mH_3$ est d'ordre $(mc^2/\rho)\epsilon^0$.} nous l'omettons donc dans la suite. Au dénominateur d'énergie, nous avons choisi comme énergie de référence celle $E_\ii$ de l'état initial plutôt que celle $E_\ff$ de l'état final, car nous parlons d'une transition $\ii\to\ff$, comme le rappelle le choix de $\ff$ en cro- et $\ii$ en -chet dans l'élément de matrice au numérateur, ce qui ne constitue pas une raison physique sérieuse. Heureusement, ceci n'a aucune importance puisque l'amplitude est utilisée dans (\ref{eq250}) sur la couche d'énergie ($E_\ii=E_\ff$): le résultat est invariant par échange des rôles de $\ii$ et $\ff$ dans n'importe quel terme de la somme au second membre de (\ref{eq252}):
\be
\label{eq253}
\frac{\langle\ff|\mH_3|\lambda\rangle\langle\lambda|\mH_3|\ii\rangle}{E_\ii-E_\lambda} \mbox{(sens direct)} \longleftrightarrow \frac{\langle\ii|\mH_3|\lambda\rangle\langle\lambda|\mH_3|\ff\rangle}{E_\ff-E_\lambda} \mbox{(sens inverse)}
\ee
Le taux total d'amortissement s'obtient en sommant $\Gamma_\qq^{n\to n'}$ sur les processus collisionnels $n\to n'$ pertinents à l'ordre considéré. À partir des règles évidentes suivantes, (i) le diagramme donnant l'amplitude de transition $\mA_{\ii\to\ff}$ est connexe, c'est-à-dire qu'il n'y a pas de découplage possible du processus $n\to n'$ en deux processus collisionnels indépendants, (ii) tout état intermédiaire (entre actions successives de $\mH_3$) doit différer de l'état initial et de l'état final, on trouve à l'ordre quatre en $H_3^{(+)}+H_3^{(-)}$ comme seules possibilités les processus $1\to 3$, $2\to 2$ et $3\to 1$:\footnote{Dans la référence \cite{LK}, {\yc contrairement à notre cas,} la branche acoustique est de départ concave {\yc($\gamma<0$ dans l'équation (\ref{eq003}))}: seul le processus $2\to 2$ conserve l'énergie-impulsion et donne une contribution non nulle à (\ref{eq254}); cependant, elle a été mal calculée, et l'erreur a été corrigée dans la référence \cite{Annalen}.}
\be
\label{eq254}
\Gamma^{(4)}_\qq=\Gamma_\qq^{1\to 3}+\Gamma_\qq^{2\to 2}+\Gamma_\qq^{3\to 1}
\ee

De manière heuristique, nous généralisons ce formalisme au calcul de la contribution d'ordre quatre en $H_3^{(+)}+H_3^{(-)}$ à la fonction énergie propre du mode $\qq$. Nous admettons, comme pour le taux d'amortissement, qu'elle est la somme des contributions des processus collisionnels $n\to n'$ susmentionnés:
\be
\label{eq255}
\Sigma^{(4)}(\qq,z)=\Sigma_{1\to 3}(\qq,z)+\Sigma_{2\to 2}(\qq,z)+\Sigma_{3\to 1}(\qq,z)
\ee
Sachant que le Dirac de conservation de l'énergie dans (\ref{eq251}) provient en définitive de l'identité 
\be
\label{eq256}
\lim_{\delta\to 0^+} \im \frac{1}{E_\ii+\ii\delta-E_\ff}=-\pi\delta(E_\ii-E_\ff)\quad,
\ee
que la variable déplacée $\zeta=z-\veps_\qq$ joue le rôle de $\ii\delta$ dans l'approximation du pôle (\ref{eq014}) {\yc et qu'une fonction énergie propre diffère d'un taux par un facteur $-\ii\hbar/2$}, nous arrivons sans hésitation à la forme
\begin{multline}
\label{eq257}
\Sigma_{n\to n'}^{(4)}(\qq,z)=\frac{1}{(n-1)!} \int\left(\prod_{i=2}^{n}\frac{\dd^dq_i}{(2\pi)^d}\right)\frac{1}{n'!} \int\left(\prod_{j=1}^{n'} \frac{\dd^dq'_j}{(2\pi)^d}\right) \frac{\mathcal{A}_{\ii\to\ff}^2(\qq,z)}{\zeta+E_\ii-E_\ff}(2\pi)^d \delta\left(\sum_{i=1}^{n}\qq_i-\sum_{j=1}^{n'}\qq'_j\right) \\
\times\left[\prod_{i=2}^{n}\bar{n}_{\qq_i}\prod_{j=1}^{n'}(1+\bar{n}_{\qq'_j})-\prod_{i=2}^{n} (1+\bar{n}_{\qq_i}) \prod_{j=1}^{n'}\bar{n}_{\qq'_j}\right]
\end{multline}
Le seul piège réside dans la construction de la dépendance en $\zeta$ de l'amplitude de transition {\yc$\mA_{\ii\to\ff}$}. Le choix de $E_\ii$ ou $E_\ff$ comme énergie de référence n'est maintenant plus indifférent. Choisir $E_\ii$ comme au premier membre de (\ref{eq253}), c'est prendre le point de vue d'une transition $\ii\to\ff$ donc parcourir le temps dans le sens ordinaire ; le bon chemin d'intégration est précisément $C_+$, celui décrit par la variable $\zeta$ dans (\ref{eq012}) ; il faut donc remplacer $E_\ii$ par $\zeta+E_\ii$ dans le dénominateur d'énergie interne à {\yc$\mA_{\ii\to\ff}$}. En revanche, choisir $E_\ff$ comme au second membre de (\ref{eq253}) revient à prendre le point de vue inverse d'une transition $\ff\to\ii$ donc à remonter le temps; le bon chemin d'intégration de la fonction de Green est alors $C_-$ \cite{CCTbordeaux}, qui court dans le demi-plan inférieur parallèlement à l'axe réel de $-\infty$ à $+\infty$ et qui est décrit par la variable $-\zeta$ dans (\ref{eq012}) ; il faut donc remplacer $E_\ff$ par $-\zeta+E_\ff$ dans le dénominateur d'énergie interne à {\yc$\mA_{\ii\to\ff}$}. 

Pour trancher selon l'état intermédiaire $|\lambda\rangle$, il faut s'appuyer sur la propriété suivante. Par construction, la fonction $\Sigma^{(4)}(\qq,z)$ dépend du mode au travers de son vecteur d'onde $\qq$ mais pas de son énergie $\veps_\qq$ (lorsqu'on utilise comme variable l'énergie complexe originelle $z$), et ne peut donc pas comporter de dénominateur d'énergie contenant $\veps_\qq$. C'est le contraire qui se produit si l'on utilise la variable déplacée $\zeta$. Ceci veut dire que, si dans le sens direct de l'amplitude de transition, l'action sur $|\ii\rangle$ de l'opérateur $\mH_3^{(\pm)}$ ne fait pas disparaître le phonon de vecteur d'onde $\qq$, {\ycd son énergie $\veps_\qq$ apparaît dans $E_\lambda$ donc disparaît dans $\zeta+E_\ii-E_\lambda$, et} il faut utiliser le sens inverse de l'amplitude de transition, cette règle étant à appliquer état intermédiaire par état intermédiaire. D'où notre prescription finale sur l'amplitude de transition à inclure dans (\ref{eq257}):\footnote{{\ycd En réalité, les processus collisionnels $n\to n'$ à $n$ phonons entrants $\qq_i$ et $n'$ phonons sortants $\qq'_j$ se produisent dans un bain thermique, et les modes des phonons virtuels $\kk$ (ni entrants ni sortants) ne sont pas initialement vides. Il faudrait donc tenir compte des facteurs d'amplification bosoniques correspondants $(n_\kk+1)^{1/2}$ et $n_\kk^{1/2}$ dans l'expression (\ref{eq258}) de l'amplitude de transition, et inclure la contribution des nouveaux processus autorisés, dans lesquels un phonon virtuel $\kk$ est annihilé avant d'être recréé (ce qui est interdit dans le vide). Cependant, ces deux modifications se compensent exactement et l'expression (\ref{eq258}) reste inchangée. Vérifions-le sur un terme de la première somme dans (\ref{eq258}). Pour cela, décrivons plus finement la transition correspondante $|\ii\rangle\to|\lambda\rangle\to|\ff\rangle$ en notant $|\ii\rangle=|\ii_1\rangle\otimes|\ii_2\rangle\otimes|n_\kk:\kk\rangle$ et $|\ff\rangle=|\ff_1\rangle\otimes|\ff_2\rangle\otimes|n_\kk:\kk\rangle$, où $|\ii_1\rangle$ ($|\ii_2\rangle$) sont les photons entrants annihilés par la première (la seconde) action de $\mathcal{H}_3$, $|\ff_1\rangle$ ($|\ff_2\rangle$) sont les phonons sortants créés par la première (la seconde) action de $\mathcal{H}_3$ et $|n_\kk:\kk\rangle$ l'état de Fock à $n_\kk$ quanta dans le mode du phonon virtuel, si bien que l'état intermédiaire s'écrit $|\lambda\rangle=|\ff_1\rangle\otimes|\ii_2\rangle\otimes|n_\kk+1:\kk\rangle$
(nous ne tenons pas compte des autres modes, spectateurs, ni de l'occupation thermique des modes entrants et sortants puisque leurs facteurs bosoniques sont inclus dans (\ref{eq250},\ref{eq257})). Dans l'exemple pris, $\qq\notin|\lambda\rangle$ donc $\qq\in|\ii_1\rangle$.
Les composantes de $\mathcal{H}_3$ assurant les transitions $|\ii\rangle\to|\lambda\rangle$ et $|\lambda\rangle\to|\ff\rangle$ s'écrivent respectivement $W_1=\mathcal{A}_1\hat{b}^\dagger_\kk(\prod_{\qq'_j\in|\ff_1\rangle} \hat{b}^\dagger_{\qq'_j})(\prod_{\qq_i\in|\ii_1\rangle} \hat{b}_{\qq_i})$ et $W_2=\mathcal{A}_2(\prod_{\qq'_j\in|\ff_2\rangle} \hat{b}^\dagger_{\qq'_j})(\prod_{\qq_i\in|\ii_2\rangle} \hat{b}_{\qq_i})\hat{b}_\kk$. 
En présence du bain thermique, (i) le numérateur $\langle\ff|\mH_3^{(\pm)}|\lambda\rangle\langle\lambda|\mH_3^{(\pm)}|\ii\rangle=\langle\ff|W_2|\lambda\rangle\langle\lambda|W_1|\ii\rangle$ devient $\mathcal{A}_1\mathcal{A}_2(1+n_\kk)$, (ii) le nouveau processus associé, correspondant à l'ordre d'action inverse de $W_1$ et $W_2$, est de numérateur $\langle\ff|W_1|\lambda'\rangle\langle\lambda'|W_2|\ii\rangle=\mathcal{A}_1\mathcal{A}_2n _\kk$, (iii) le nouvel état intermédiaire $|\lambda'\rangle=|\ii_1\rangle\otimes|\ff_2\rangle\otimes|n_\kk-1:\kk\rangle$ contient le phonon $\qq$ donc son dénominateur d'énergie doit être pris dans la seconde somme de (\ref{eq258}), et se trouve être l'exact opposé du dénominateur de l'ancien processus, le calcul donnant $-\zeta+E_\ff-E_{\lambda'}=-(\zeta+E_\ii-E_\lambda)$. En définitive, les contributions linéaires en $n_\kk$ de l'ancien et du nouveau processus se compensent exactement dans l'amplitude de transition $\mathcal{A}_{\ii\to\ff}(\qq,z)$, ce qui autorise à oublier le fond thermique comme il est fait dans (\ref{eq258}). Un raisonnement similaire, mais limité aux processus $2\to 2$ sur la couche d'énergie, figure déjà dans la référence \cite{Annalen}.}}
\be
\label{eq258}
\boxed{\mathcal{A}_{\ii\to\ff}(\qq,z) = \sum_{|\lambda\rangle\, / \, \qq\notin|\lambda\rangle} \frac{\langle \ff| \mH_3^{(\pm)} |\lambda\rangle \langle\lambda| \mH_3^{(\pm)}|\ii\rangle}{\zeta+E_\ii-E_\lambda} 
+
\sum_{|\lambda\rangle\, / \, \qq\in|\lambda\rangle} \frac{\langle \ii| \mH_3^{(\pm)} |\lambda\rangle \langle\lambda| \mH_3^{(\pm)}|\ff\rangle}{-\zeta+E_\ff-E_\lambda}} 
\ee
De l'expression (\ref{eq258}) nous tirons l'amplitude de transition pour le processus $1\to 3$ (sachant que $\qq'_1+\qq'_2+\qq'_3=\qq$): 
\begin{multline}
\label{eq260}
\mathcal{A}_{\qq\to\qq'_1,\qq'_2,\qq'_3}(\qq,\zeta)= \frac{\mA(\qq'_1,\qq'_2;\qq'_1+\qq'_2)\mA(\qq'_1+\qq'_2,\qq'_3;\qq)}{\zeta+\epsilon_\qq-(\epsilon_{\qq'_1+\qq'_2}+\epsilon_{\qq'_3})} +\frac{\mA(\qq'_2,\qq'_3;\qq'_2+\qq'_3)\mA(\qq'_2+\qq'_3,\qq'_1;\qq)}{\zeta+\epsilon_\qq-(\epsilon_{\qq'_2+\qq'_3}+\epsilon_{\qq'_1})} \\ +\frac{\mA(\qq'_1,\qq'_3;\qq'_1+\qq'_3)\mA(\qq'_1+\qq'_3,\qq'_2;\qq)}{\zeta+\epsilon_\qq-(\epsilon_{\qq'_1+\qq'_3}+\epsilon_{\qq'_2})}
\end{multline}
puis pour le processus $2\to 2$ (sachant que $\qq'_1+\qq'_2=\qq+\qq_2$):
\begin{multline}
\label{eq262}
\mathcal{A}_{\qq,\qq_2\to\qq'_1,\qq'_2}(\qq,\zeta)= \frac{\mA(\qq'_1,\qq'_2;\qq'_1+\qq'_2)\mA(\qq,\qq_2;\qq+\qq_2)}{\zeta+\epsilon_{\qq}+\epsilon_{\qq_2}-\epsilon_{\qq+\qq_2}} -\frac{\mA(\qq'_1-\qq,\qq'_2;\qq_2)\mA(\qq'_1-\qq,\qq;\qq'_1)}{\zeta+\epsilon_{\qq}+\epsilon_{\qq'_1-\qq}-\epsilon_{\qq'_1}} \\ +\frac{\mA(\qq'_2-\qq_2,\qq_2;\qq'_2)\mA(\qq'_2-\qq_2,\qq'_1;\qq)}{\zeta+\epsilon_{\qq}-(\epsilon_{\qq'_1}+\epsilon_{\qq'_2-\qq_2})} +\frac{\mA(\qq-\qq'_2,\qq_2;\qq'_1)\mA(\qq-\qq'_2,\qq'_2;\qq)}{\zeta+\epsilon_{\qq}-(\epsilon_{\qq'_2}+\epsilon_{\qq-\qq'_2})}-\frac{\mA(\qq'_2-\qq,\qq'_1;\qq_2)\mA(\qq'_2-\qq,\qq;\qq'_2)}{\zeta+\epsilon_{\qq}+\epsilon_{\qq'_2-\qq}-\epsilon_{\qq'_2}}
\end{multline}
et enfin pour le processus $3\to 1$ (sachant que $\qq_1'=\qq+\qq_2+\qq_3$):
\begin{multline}
\label{eq264}
\mathcal{A}_{\qq,\qq_2,\qq_3\to \qq_1'}(\qq,\zeta)= -\frac{\mA(\qq_2,\qq_3;\qq_2+\qq_3)\mA(\qq_2+\qq_3,\qq;\qq_1')}{\zeta+\epsilon_{\qq}+\epsilon_{\qq_2+\qq_3}-\epsilon_{\qq_1'}}+\frac{\mA(\qq+\qq_2,\qq_3;\qq_1')\mA(\qq,\qq_2;\qq+\qq_2)}{\zeta+\epsilon_{\qq}+\epsilon_{\qq_2}-\epsilon_{\qq+\qq_2}}\\ +\frac{\mA(\qq+\qq_3,\qq_2;\qq_1')\mA(\qq,\qq_3;\qq+\qq_3)}{\zeta+\epsilon_{\qq}+\epsilon_{\qq_3}-\epsilon_{\qq+\qq_3}}
\end{multline}
Dans ces équations (\ref{eq260},\ref{eq262},\ref{eq264}), les trois contributions où il a fallu choisir le sens inverse dans (\ref{eq253}), c'est-à-dire prendre $E_\ff$ plutôt que $E_\ii$ comme énergie de référence, sont faciles à repérer car elles commencent par un signe moins. 

Notre prescription (\ref{eq257},{\yc\ref{eq258}}) donne des résultats simples mais en général différents de ceux, exacts, de la théorie diagrammatique à l'ordre quatre, compte tenu des expressions des sommes doubles de Matsubara dans l'\ref{app3}. Cependant, de façon remarquable, 
une fois que ces sommes ont été remplacées par leurs expressions simplifiées (\ref{eq213},\ref{eq233}), ce qui est légitime aux petits angles, elle est en accord parfait avec la valeur de $\Sigma^{(4)}$ prédite par les diagrammes de la figure \ref{fig8}; on le vérifie à l'aide de la table de correspondance \ref{table1}, obtenue au vu des amplitudes de couplage dans (\ref{eq212},\ref{eq232}) et des dénominateurs d'énergie dans (\ref{eq213},\ref{eq233}), et qui permet d'associer à chaque ligne interne sur cette figure un phonon entrant $\qq_i$, sortant $\qq'_j$ ou virtuel (combinaison linéaire des précédents) d'un processus $n\to n'$ donc de donner un sens physique clair au diagramme correspondant.\footnote{{\ycd La vérification est facilitée par le résultat suivant: selon notre prescription à la Landau, chaque contribution diagrammatique aux sommes de Matsubara (\ref{eq213},\ref{eq233}) vaut $\sigma/\prod_j (\zeta-\zeta_j)$ avec $\sigma=s(f_1 f_2-f_3f_4)$, où (i) le signe global $s$ est le produit des orientations $s_i$ des lignes de phonons virtuels (en effet, si un phonon virtuel remonte le temps, son énergie apparaît avec un signe plus au dénominateur dans le facteur $(\zeta-\zeta_j)$ correspondant, donc il contribue au travers de la seconde somme de (\ref{eq258}) et il faut inclure un signe moins global), (ii) chaque $f_k$ est le produit des facteurs d'amplification bosoniques des lignes reliées au sommet $k$ (la ligne $i$ contribue à $f_k$ par un facteur $\bar{n}_{\veps_i}$ si elle va vers le sommet, ou un facteur $1+\bar{n}_{\veps_i}$ si elle s'en éloigne, ce qui s'écrit $s_i(1+\bar{n}_{s_i\veps_i})$ si elle est à droite du sommet et $s_i\bar{n}_{s_i\veps_i}$ si elle est à gauche), à l'exclusion cependant des lignes externes et des lignes de phonons virtuels, dont les nombres d'occupation n'apparaissent pas dans la prescription (\ref{eq257},\ref{eq258}). Les sommets des diagrammes sont numérotés de gauche à droite ; dans le cas des diagrammes à pont, il faut lever la dégénérescence de numérotation des sommets internes en faisant tourner la ligne de phonon verticale d'un petit angle vers la gauche (type I, orientation effective $s_5$) ou vers la droite (type II, orientation effective $-s_5$). On trouve ainsi, en sortant tous les signes, $\sigma_B=s_1s_2s_3s_4s_5[(1+\bar{n}_{s_1\veps_1})(1+\bar{n}_{s_3\veps_3})(1+\bar{n}_{s_4\veps_4})-\bar{n}_{s_1\veps_1}\bar{n}_{s_3\veps_3}\bar{n}_{s_4\veps_4}]$ pour les diagrammes à boucle interne ($s=s_2s_5$), $\sigma_{P^{\rm I}}=s_1s_2s_3s_4s_5[(1+\bar{n}_{s_2\veps_2})(1+\bar{n}_{s_3\veps_3})(1+\bar{n}_{s_5\veps_5})-\bar{n}_{s_2\veps_2} \bar{n}_{s_3\veps_3} \bar{n}_{s_5\veps_5}]$ et $\sigma_{P^{\rm II}}=s_1s_2s_3s_4(-s_5)[(1+\bar{n}_{s_1\veps_1})(1+\bar{n}_{s_4\veps_4})(1+\bar{n}_{-s_5\veps_5})-\bar{n}_{s_1\veps_1} \bar{n}_{s_4\veps_4} \bar{n}_{-s_5\veps_5}]$ pour les diagrammes à pont ($s=s_1s_4$ pour le type I, $s=s_2s_3$ pour le type II). Dans la limite $\epsilon\to 0$, on retrouve alors exactement les formes simplifiées (\ref{eq213},\ref{eq233}) si l'on utilise la conservation de l'énergie au deuxième sommet, respectivement $s_2\veps_2=s_3\veps_3+s_4\veps_4$, $s_1\veps_1=s_3\veps_3+s_5\veps_5$ et $s_2\veps_2=s_4\veps_4+(-s_5)\veps_5$, au travers de l'identité $\bar{n}_{a+b}=\bar{n}_a\bar{n}_b/(1+\bar{n}_a+\bar{n}_b)$, ainsi que l'égalité $s_2=s_5$ dans les diagrammes à boucle interne.}} En définitive, la prescription (\ref{eq257},{\yc\ref{eq258}}) conduit en $d=2$ à une expression rigoureusement équivalente mais beaucoup plus compacte de $\Sigma^{(4)}$ à l'ordre dominant en $\epsilon$:
\be
\label{eq269}
\boxed{
\tilde{\Sigma}_{\qb}^{(4,2)}(\tilde{\zeta})=\frac{1}{3}\int\!\!\!\!\int_{\mathbb{T}_{\qb}} \frac{\qb'_1\dd \qb'_1}{(2\pi)^2} \frac{\qb'_2\dd \qb'_2}{(2\pi)^2} \Phi(\qb'_1,\qb'_2) + \int\!\!\!\!\int_{\mathbb{R}^{+2}\setminus\mathbb{T}_{\qb}} \frac{\qb'_1\dd \qb'_1}{(2\pi)^2} \frac{\qb'_2\dd \qb'_2}{(2\pi)^2} \Phi(\qb'_1,\qb'_2) + \int\!\!\!\!\int_{\mathbb{R}^{+2}}\frac{\qb_2\dd \qb_2}{(2\pi)^2}\frac{\qb_3\dd \qb_3}{(2\pi)^2} \Phi(-\qb_2,-\qb_3)}
\ee
où nous avons respecté l'ordre des processus $1\to 3$, $2\to 2$ et $3\to 1$, $\mathbb{T}_{\qb}$ est le triangle $\{(\qb'_1,\qb'_2)\in\mathbb{R}^{+2}\;|\; \qb'_1+\qb'_2\leq \qb\}$  et, comme précédemment, le tilde sur $\Sigma_{\qb}$ indique une division par $k_B T\epsilon^2$. Une seule et même fonction intervient dans l'intégrande des contributions $n\to n'$, seul le domaine d'intégration change avec le processus considéré:\footnote{En ce qui concerne les facteurs bosoniques, on le montre en faisant bon usage de la relation $\bar{n}_{-k}^{\rm lin}=-(1+\bar{n}_k^{\rm lin})$, $\forall k\in\mathbb{R}^*$.}
\begin{multline}
\label{eq270}
\Phi(\qb'_1,\qb'_2)=\frac{1}{2} \qb \left(\frac{9(1+\Lambda)^2}{8\rho\xi^2}\right)^2 \qb'_1 \qb'_2 (\qb'_1+\qb'_2-\qb) [(1+\bar{n}^{\rm lin}_{q'_1}+\bar{n}^{\rm lin}_{q'_2})\bar{n}^{\rm lin}_{q'_1+q'_2-q}-\bar{n}^{\rm lin}_{q'_1}\bar{n}^{\rm lin}_{q'_2}] \\ \times \int_{\mathbb{R}}\dd\thetat_1\int_{\mathbb{R}}\dd\thetat_2\frac{\left\{\frac{\qb'_1+\qb'_2}{\zetat+\qb(\qb'_1+\qb'_2)(\qb'_1+\qb'_2-\qb)\left[\frac{(\qb'_1\thetat_1+\qb'_2\thetat_2)^2}{2(\qb'_1+\qb'_2-\qb)^2(\qb'_1+\qb'_2)^2}-\frac{3\gamma}{8}\right]}+\frac{\qb-\qb'_1}{\zetat+\qb \qb'_1(\qb-\qb'_1)\left[-\frac{\thetat_1^2}{2(\qb'_1-\qb)^2}+\frac{3\gamma}{8}\right]}+\frac{\qb-\qb'_2}{\zetat+\qb \qb'_2(\qb-\qb'_2)\left[-\frac{\thetat_2^2}{2(\qb'_2-\qb)^2}+\frac{3\gamma}{8}\right]}\right\}^2}{\zetat - \frac{\qb'_1(\qb'_2-\qb)\thetat_1^2+\qb'_2(\qb'_1-\qb)\thetat_2^2-2\qb'_1 \qb'_2\thetat_1 \thetat_2}{2(\qb'_1+\qb'_2-\qb)}+\frac{3\gamma}{8} (\qb'_1+\qb'_2)(\qb'_1-\qb)(\qb'_2-\qb)}
\end{multline}

\begin{table}[t]
\centerline{\small
\begin{tabular}{|c|c|c|c|c|c|}
$B_1$ & tous les carrés de $1\to 3$ & $P_{1}^{\rm I}$  &  tous les rectangles de $1\to 3$ & $P_{1}^{\rm II}$ & $3^{\rm e}$ terme $\times$ $4^{\rm e}$ terme de $2\to 2$\\
$B_2$ & carrés des $3^{\rm e}$ et $4^{\rm e}$ termes de $2\to 2$ & $P_{2}^{\rm I}$ & $1^{\rm er}$ terme $\times$ $3^{\rm e}$ terme de $2\to 2$ & $P_{2}^{\rm II}$ & $3^{\rm e}$ terme $\times$ $5^{\rm e}$ terme de $2\to 2$\\
$B_3$ & $2^{\rm e}$ terme $\times$ $3^{\rm e}$ terme de $2\to 2$ & $P_{3}^{\rm I}$ & $1^{\rm er}$ terme $\times$ $4^{\rm e}$ terme de $2\to 2$ & $P_{3}^{\rm II}$ & $2^{\rm e}$ terme $\times$ $4^{\rm e}$ terme de $2\to 2$\\
$B_4$ & $4^{\rm e}$ terme $\times$ $5^{\rm e}$ terme de $2\to 2$ & $P_{4}^{\rm I}$ & $1^{\rm er}$ terme $\times$ $2^{\rm e}$ terme de $2\to 2$ & $P_{4}^{\rm II}$ & $1^{\rm er}$ terme $\times$ $3^{\rm e}$ terme de $3\to 1$\\
$B_5$ & carrés des $2^{\rm e}$ et $5^{\rm e}$ termes de $2\to 2$ & $P_{5}^{\rm I}$ & $2^{\rm e}$ terme $\times$ $3^{\rm e}$ terme de $3\to 1$ & $P_{5}^{\rm II}$ & $2^{\rm e}$ terme $\times$ $5^{\rm e}$ terme de $2\to 2$\\
$B_6$ & carré du $1^{\rm er}$ terme de $3\to 1$ & $P_{6}^{\rm I}$ & $1^{\rm er}$ terme $\times$ $5^{\rm e}$ terme de $2\to 2$ & $P_{6}^{\rm II}$ & $1^{\rm er}$ terme $\times$ $2^{\rm e}$ terme de $3\to 1$\\
$B_7$ & carré du $1^{\rm er}$ terme de $2\to 2$ & & & & \\
$B_8$ & carrés des $2^{\rm e}$ et $3^{\rm e}$ termes de $3\to 1$ & & & & 
\end{tabular}
}
\caption{Dans le calcul de la fonction énergie propre à l'ordre quatre en $H_3$, correspondance entre les diagrammes de Feynman de la figure \ref{fig8} et les termes carrés ou rectangles (doubles produits) des quantités $\mA^2_{\ii\to\ff}(\qq,z)$ intervenant dans la prescription à la Landau (\ref{eq257}); les différents termes des amplitudes de transition $\mA_{\ii\to\ff}(\qq,z)$ auxquels il est fait référence sont ceux des équations (\ref{eq260},\ref{eq262},\ref{eq264}) écrits dans cet ordre. Pour les diagrammes à pont, il a fallu distinguer entre type I et type II comme dans (\ref{eq233}). La correspondance vaut après intégration sur les vecteurs d'onde comme dans (\ref{eq257}), qui fait par exemple que les carrés de chacun des termes de (\ref{eq260}), qui diffèrent par une permutation des $\qq_j'$, ou que les doubles produits des $2^{\rm e}$ et $4^{\rm e}$ termes et des $5^{\rm e}$ et $3^{\rm e}$ termes de (\ref{eq262}), qui sont envoyés l'un sur l'autre par échange de $\qq_1'$ et $\qq_2'$, donnent la même contribution.}
\label{table1}
\end{table}

\subsection{Échec de l'ordre quatre dans la limite $\epsilon=k_B T/mc^2\to 0$}
\label{sec3.3}

\paragraph{Une divergence inattendue} En première application simple du développement de la fonction énergie propre à l'ordre quatre en $H_3$, nous avions décidé de calculer numériquement\footnote{Les intégrales angulaires sont calculées analytiquement, seule l'intégration sur les nombres d'onde est faite par ordinateur.} $\tilde{\Sigma}_{\qb}^{(4,2)}(\ii 0^+)$ pour voir si, par une meilleure approximation de $\Sigma_{\qq}(\ii 0^+)$ dans l'équation (\ref{eq014}), cela rapprochait l'approximation du pôle des résultats numériques en réduisant l'écart sur la figure \ref{fig2}d. À notre grande surprise, nous avons découvert que $\tilde{\Sigma}_{\yc\qb}^{(4,2)}(\ii\delta)$ diverge comme $1/\ii\delta$ lorsque $\delta\to 0^+$! Nous avons ensuite confirmé ce résultat analytiquement, en montrant que\footnote{Pour faciliter le calcul des intégrales angulaires, nous avons établi (\ref{eq280}) d'abord dans le cas où $\zetat\to 0$ par parties imaginaires $>0$, puis nous avons étendu au cas $\im\zetat<0$ comme dans la note \ref{notegen}.}
\be
\label{eq280}
\boxed{\tilde{\Sigma}_{\qb}^{(4,2)}(\zetat)\underset{\zetat\to 0}{\sim} \frac{-2}{3\pi^2\gamma\zetat} \left(\frac{9(1+\Lambda)^2}{8\rho\xi^2}\right)^2 \int_{\qb}^{\eta} \kb\dd\kb \int_0^{\eta-\kb} \kb'\dd\kb' (\kb-\qb)(\qb+\kb')(\bar{n}_{k-q}^{\rm lin}-\bar{n}_{k}^{\rm lin})(\bar{n}_{k'+q}^{\rm lin}-\bar{n}_{k'+k}^{\rm lin})}
\ee
Cette divergence provient de l'intégrale angulaire $K(\kb,\kb')$ (\ref{eq243}) du diagramme à pont $P_5$. Analysons donc le comportement de (\ref{eq243}) dans la limite $\zetat\to 0$ pour des orientations quelconques $s_j$ des lignes internes. S'il est divergent, il est nécessairement déterminé par un voisinage du lieu d'annulation du dénominateur de l'intégrande dans le plan $(\thetat,\thetat')$ pour $\zetat=0$. Comme le paramètre de courbure $\gamma$ est ici $>0$, ce lieu est la réunion des deux droites verticales $\thetat=\veps(-A_1^P/B_1^P)^{1/2}$ ($\veps=\pm 1$) où s'annule le premier facteur dans le dénominateur, des deux droites horizontales $\thetat'=\veps'(-A_2^P/B_2^P)^{1/2}$ ($\veps'=\pm 1$) où s'annule le deuxième facteur, et de la conique annulant le troisième facteur. Le malheur est que la conique passe toujours par les points d'intersection $P_{\veps,\veps'}\equiv(\veps(-A_1^P/B_1^P)^{1/2},\veps'(-A_2^P/B_2^P)^{1/2})$ des droites horizontales et verticales, {\yc quels} que soient les nombres d'onde $\kb$, $\kb'$ et $\qb$, à la seule condition que $\veps\veps'=s_2 s_4$. Ceci résulte de l'identité sur les coefficients (\ref{eq244},\ref{eq245}):
\be
\label{eq281}
4A_1^P A_2^P B_1^P B_2^P (B_3^P)^2 = [A_2^P B_1^P(B_2^P+B_3^P)+A_1^P B_2^P B_3^P-A_3^P B_1^P B_2^P]^2
\ee
Comme l'intégrale angulaire est bidimensionnelle, une éventuelle partie divergente $K_{\rm div}(\kb,\kb')$ de $K(\kb,\kb')$ ne peut provenir que d'un petit voisinage des points de triple intersection $P_{s_2,s_4}$ et $P_{-s_2,-s_4}$, sur lequel nous remplaçons chaque facteur au dénominateur par son approximation linéaire en $\tau=\thetat-\veps(-A_1^P/B_1^P)^{1/2}$ et $\tau'=\thetat'-\veps'(-A_2^P/B_2^P)^{1/2}$. Il reste
\be
\label{eq282}
K_{\rm div}(\kb,\kb')=\int_{\mathbb{R}^2} \frac{2\dd \tau\dd \tau'}{[\zetat-(\frac{3\gamma}{4})^{1/2}s_1\qb\kb \tau][\zetat-(\frac{3\gamma}{4})^{1/2}s_3\qb\kb' \tau']\{\zetat-(\frac{3\gamma}{4})^{1/2}s_3\kb'[s_1\kb \tau +(\qb-s_1\kb) \tau']\}}=- \frac{8\pi^2(1-s_2)(1-s_1s_3)}{3\gamma\zetat \qb\kb\kb' (s_1\kb-s_3\kb')}
\ee
où nous avons pu étendre l'intégrale à $\mathbb{R}^2$ tout entier puisqu'elle converge à l'infini, et avons pu nous limiter au point $P_{s_2,s_4}$ (au prix de l'ajout d'un facteur $2$ global) par invariance par parité $(\thetat,\thetat')\to (-\thetat,-\thetat')$. Au troisième membre de (\ref{eq282}), on a calculé explicitement l'intégrale sur $\tau'$ puis sur $\tau$ par la méthode des résidus (il faut refermer le chemin d'intégration par un demi-cercle à l'infini dans le demi-plan supérieur ou inférieur contenant le nombre minimal de pôles au vu des contraintes (\ref{eq239})). La partie divergente est donc non nulle seulement si $s_2=s_1s_3=-1$, c'est-à-dire si la branche inférieure gauche de la boucle à pont est orientée vers la gauche et si les deux branches supérieures ont des orientations opposées, ce qui sélectionne comme annoncé le seul diagramme $P_5$ sur la figure \ref{fig8}. Le report de la valeur correspondante de $K_{\rm div}(\kb,\kb')$ dans (\ref{eq242}) redonne l'équation (\ref{eq280}). La même technique appliquée à la contribution de type I des diagrammes à pont conduit dans (\ref{eq241}) à {\yc une partie divergente de la forme}\footnote{En effet, les coefficients satisfont aussi à la relation obtenue en échangeant les indices $1$ et $2$ et {\yct en} remplaçant l'indice $3$ par l'indice $4$ dans (\ref{eq281}).}\footnote{Dans le cas des diagrammes à boucle interne, l'existence d'un point d'annulation triple est moins surprenante puisque le dénominateur de (\ref{eq221}) comporte un trinôme au carré. En fait, il y en a quatre, qui jouent des rôles équivalents. Pour les choix de signes $\veps=s_1 s_2$ et $\veps'=s_2 s_3 s_4$, les coefficients de $\tau$ et $\tau'$ dans les formes linéarisées sont tous négatifs; en les mettant à $-1$ et en prenant $\zetat=\ii\delta$ pour simplifier, on tombe sur $\int_{[-1,1]^2} 4\dd \tau\dd \tau'/[(\ii\delta-\tau)^2 (\ii\delta -\tau-\tau')]$, qui est un $O(1)$ lorsque $\delta\to 0^+$.}
\be
\label{eq283}
J_{\rm div}(\kb,\kb')=\int_{\mathbb{R}^2} \frac{2\dd \tau\dd \tau'}{[\zetat-(\frac{3\gamma}{4})^{1/2}s_1\qb\kb \tau][\zetat-(\frac{3\gamma}{4})^{1/2}s_3\qb\kb' \tau']\{\zetat-(\frac{3\gamma}{4})^{1/2}s_1\kb[(\qb-s_3\kb')\tau +s_3\kb'\tau']\}}=\frac{8\pi^2(1-s_1)(1-s_2s_4)}{3\gamma\zetat \qb\kb\kb'(s_1\kb-s_3\kb')}
\ee
et conclut à l'absence de divergence ({\yc pour avoir $J_{\rm div}\neq 0$,} il faudrait que la branche supérieure gauche de la boucle à pont soit orientée vers la gauche, et que les deux branches inférieures soient d'orientations opposées, ce qui ne se produit pas sur la figure \ref{fig8}).

\paragraph{Implications physiques} À partir du développement de la fonction énergie propre à l'ordre quatre en $H_3$, et toujours à l'ordre dominant (deux) en $\epsilon$, nous formons une nouvelle approximation de la fonction de Green, 
\be
\label{eq289}
\tilde{\mathcal{G}}_{\qb}^{(\leq 4,2)}(\zetat)=\frac{1}{\zetat-\tilde{\Sigma}_{\qb}^{(2,2)}(\zetat)-\tilde{\Sigma}_{\qb}^{(4,2)}(\zetat)}
\ee
On pourrait croire qu'elle est plus précise et plus performante que celle de la section \ref{sec2}. Malheureusement, à cause de la divergence (\ref{eq280}), il n'en est rien: dans le régime d'interaction faible, pourtant le plus favorable, elle admet un pôle $\zetat^{n\phi}_{\qb}$ dans le demi-plan supérieur (et son complexe conjugué dans le demi-plan inférieur), en l'absence, faut-il le préciser, de tout prolongement analytique, d'où l'exposant $n\phi$ (\g{non physique}). Ceci viole {\yc la} propriété fondamentale d'analyticité de la fonction de Green sur $\mathbb{C}\setminus\mathbb{R}$ \cite{FW}, et conduit par exemple à un signal divergeant exponentiellement en temps dans (\ref{eq012}) si le chemin $C_+$ passe au-dessus de $\zetat^{n\phi}_{\qb}$, ou ne commençant pas à un s'il passe en dessous. Pour montrer l'existence de ce pôle, réduisons l'équation implicite
\be
\label{eq290}
\zetat_{\qb}^{n\phi} = \tilde{\Sigma}_{\qb}^{(2,2)}(\zetat_{\qb}^{n\phi})+\tilde{\Sigma}_{\qb}^{(4,2)}(\zetat_{\qb}^{n\phi})
\ee
à son ordre dominant en supposant que $\zetat_{\qb}^{n\phi}$ tend vers zéro comme $(\rho\xi^2)^{-1}$ (par parties imaginaires positives). Puisque $\tilde{\Sigma}_{\qb}^{(2,2)}(\zetat)$ est déjà d'ordre un en $(\rho\xi^2)^{-1}$, et que c'est une fonction dérivable de $\zetat$ en $\ii 0^+$ (voir l'\ref{app1}), on peut la remplacer par sa valeur en ce point. Dans le cas de $\tilde{\Sigma}_{\qb}^{(4,2)}(\zetat)$, il faut garder sa partie divergente (\ref{eq280}), le préfacteur d'ordre deux $(\rho\xi^2)^{-2}$ qu'elle contient étant remonté à l'ordre un par division par $\zetat$, mais on peut négliger sa partie régulière, d'ordre deux près de $\ii 0^+$. En effectuant les changements de variable universalisants (\ref{eq112},\ref{eq130},\ref{eq131}) de la section \ref{sec2}, il reste
\be
\label{eq291}
\zetac^{n\phi}_{\qb} = \check{\Sigma}_{\qb}^{(2,2)}(\ii 0^+)-\frac{(2/\qb^{2})\int_{\qb}^{\eta}\kb\,\dd\kb \int_0^{\eta-\kb}\kb'\dd\kb' (\kb-\qb)(\kb'+\qb)({\yc\bar{n}_{k-q}^{\rm lin}-\bar{n}_{k}^{\rm lin})(\bar{n}_{k'+q}^{\rm lin}-\bar{n}_{k'+k}^{\rm lin})}}{\zetac^{n\phi}_{\qb}}
\ee
où le premier terme au second membre figure dans (\ref{eq141}). Comme l'intégrale au numérateur est $>0$, l'équation (\ref{eq291}) sur $\zetac^{n\phi}_{\qb}$ admet une solution imaginaire pure dans le demi-plan supérieur.\footnote{Posons $\zetac^{n\phi}_{\qb}=\ii y$. Alors {\yc(\ref{eq291}) s'écrit sous la forme} $y=-2B+A/y$ où les coefficients $A$ et $B$ sont réels $>0$. L'équation du second degré sur $y$ qui en résulte admet une (et une seule) racine $>0$, $y_0=A/[(A+B^2)^{1/2}+B]$. Lorsque $\qb\to 0$, $B$ a une limite finie et non nulle au vu de (\ref{eq141}), alors que $A$ diverge comme $1/\qb$, si bien que $\zetac^{n\phi}_{\qb}$ diverge comme $\qb^{-1/2}$ et $\zetat_{\qb}^{n\phi}$ tend vers zéro comme $\qb^{1/2}$. Ainsi, la condition $\im\zetat^{n\phi}_{\qb}\lll|\tilde{\Sigma}_{\qb}^{(2,2)}(\ii 0^+)|=O(\qb)$, qui rend le pôle non physique sans effet sur le signal dans sa mi-largeur temporelle, ne peut être satisfaite à nombre d'onde $\qb\lesssim 1$. Elle peut l'être à grand nombre d'onde $\qb\gg 1$, où $B\approx \qb^2$ au vu de (\ref{eq141}) et $A\sim 2\exp(-\qb)$.} Pour être complets, signalons que $\tilde{\Sigma}_{\qb}^{(4,2)}(\zetat)$ admet une autre divergence linéaire sur l'axe réel, en $\zetat_0=-3\gamma\qb^3/32$,\footnote{Remarquons que $\zetat_0$ est aussi le point de branchement de $\tilde{\Sigma}_{\qb}^{(2,2)}(\zetat)$, voir la section \ref{sec2.1} et la note \ref{note30}.} comme le montre l'\ref{app4}, moins dommageable car n'impliquant pas l'existence d'un pôle non physique de la fonction de Green dans la limite d'interaction faible (il suffit que l'on ait $|\tilde{\Sigma}_{\qb}^{(2,2)}(\ii 0^+)|\lll|\re\zetat_0|$ pour que cette divergence ait peu d'effet sur l'amortissement du signal dans sa mi-largeur). 

\paragraph{Applicabilité de l'ordre quatre} L'apparition d'un pôle non physique oblige-t-elle à remettre en question la validité du développement diagrammatique et par là, celle de la méthode des fonctions de Green à $N$ corps ? Non. La théorie était prévue pour être utilisée dans la limite d'interaction faible $(\rho\xi^2)^{-1}\to 0$ à température réduite fixée $\epsilon=k_B T/mc^2$. Nous avons cependant pris la limite $\epsilon\to 0$ à ordre de développement fixé en $(\rho\xi^2)^{-1}$, ce qui échoue en dimension $d=2$. Montrons que tout va bien, c'est-à-dire que le pôle non physique disparaît, lorsque $\rho\xi^2\epsilon^2$ est assez grand (à $\qb$ fixé). 

À cette fin, voyons comment l'analyse locale conduisant à (\ref{eq282}) est modifiée lorsque $\epsilon$ est petit mais pas infinitésimal, en nous limitant directement au diagramme $P_5$, donc en prenant $s_1=s_4=s_5=1$, $s_2=s_3=-1$. Si $\zetat=0$, le dénominateur d'énergie de type II dans (\ref{eq233}), qui est aussi celui de (\ref{eq282}), s'écrit $(-\zeta_1)(-\zeta_2)(-\zeta_3)$, où les défauts d'énergie $\zeta_i$ sont ceux de l'équation (\ref{eq238}). Le premier facteur $-\zeta_1=-(\veps_{\kk}-\veps_{\kk-\qq}-\veps_{\qq})$ est fonction du seul angle $\theta=\epsilon\thetat$ entre $\qq$ et $\kk$, et continue à s'annuler sur deux droites verticales dans le plan $(\thetat,\thetat')$, à des abscisses légèrement modifiées, ce qui change de manière négligeable (par un facteur $1+O(\epsilon)^2$) le coefficient de $\tau$ dans (\ref{eq282}); le deuxième facteur $-\zeta_2=-(-\veps_{\kk'}+\veps_{\qq+\kk'}-\veps_{\qq})$ est fonction du seul angle $\theta'=\epsilon\thetat'$ entre $\qq$ et $\kk'$, et continue à s'annuler sur deux droites horizontales, légèrement déplacées, avec les mêmes conséquences négligeables sur le coefficient de $\tau'$ dans (\ref{eq282}). En revanche, le lieu d'annulation du troisième facteur, à savoir $-\zeta_3=-(\veps_{\kk}+\veps_{\qq+\kk'}-\veps_{\kk+\kk'}-\veps_{\qq})$, n'est plus exactement une conique dans le plan $(\thetat,\thetat')$ et passe maintenant un peu à côté des points d'intersection $P_{-1,1}$ et $P_{1,-1}$ des droites horizontales et verticales: ceci change très légèrement les coefficients de la combinaison linéaire de $\tau$ et $\tau'$ dans (\ref{eq282}), mais surtout ajoute, à côté de $\zetat$, la valeur de $-\zetat_3$ en $P_{-1,1}$ désormais non nulle:
\be
\label{eq292}
(-\zetat_3)_{P_{-1,1}} \underset{\epsilon\to 0}{\sim} -\frac{27}{64}\gamma^2\epsilon^2\kb\kb'(\kb-\qb)(\kb+\kb')(\kb'+\qb)
\ee
Nous recalculons en conséquence le troisième membre de (\ref{eq282}):
\be
\label{eq293}
K_{\rm div}^{P_5}(\kb,\kb')=-\frac{32\pi^2}{3\gamma\qb\kb\kb'(\kb+\kb')} \frac{1}{\zetat-\frac{27}{64}\gamma^2\epsilon^2\qb\kb\kb'(\kb-\qb)(\kb'+\qb)}
\ee
La nouvelle équation sur le pôle non physique
\be
\label{eq294}
\zetac^{n\phi}_{\qb} = \check{\Sigma}_{\qb}^{(2,2)}(\ii 0^+)-\frac{2}{\qb^{2}}\int_{\qb}^{\eta}\kb\,\dd\kb \int_0^{\eta-\kb}\kb'\dd\kb' \frac{(\kb-\qb)(\kb'+\qb) {\yc(\bar{n}_{k-q}^{\rm lin}-\bar{n}_{k}^{\rm lin})(\bar{n}_{k'+q}^{\rm lin}-\bar{n}_{k'+k}^{\rm lin})}}{\zetac^{n\phi}_{\qb}-\frac{9}{8}\gamma u \epsilon^2\kb\kb'(\kb-\qb)(\kb'+\qb)}
\ee
cesse d'admettre une solution dans le demi-plan supérieur lorsque $\gamma u\epsilon^2>h(\qb)$, où $u$ est le paramètre universel (\ref{eq112}) et $h(\qb)$ est fonction exclusivement de $\qb$ (et de la coupure $\eta$ si on ne la fait pas tendre vers l'infini dans la théorie quantique), comme on le voit en prenant formellement la limite $\gamma u \epsilon^2\to +\infty$ dans (\ref{eq294}). Pour le champ classique, nous trouvons numériquement que $h(\qb)$ est une fonction décroissante de $\qb$; le problème d'applicabilité de l'ordre quatre pour de faibles valeurs de $\epsilon$ subsiste donc dans la limite $\qb\to 0$.\footnote{{\ycd Pour le champ classique ($\eta=1$), on a l'équivalent $h(\qb)\sim \frac{8}{9}(\pi^2/2\qb)^{4/3}$ lorsque $\qb\to 0$ (en pratique $\qb^{1/3}\ll 1$). Ceci s'obtient en écrivant le second membre de (\ref{eq294}) à l'ordre dominant en $\qb$ (en particulier, on remplace $\qb$ par zéro dans la première borne d'intégration et dans le dénominateur de l'intégrande, et on remplace le numérateur par son approximation linéaire en $\qb$) puis en montrant que l'intégrale double est dominée par un voisinage de $(\kb,\kb')=(0,0)$ de rayon $h(\qb)^{-1/4}\ll 1$ lorsque $\qb\to 0$, ce qui autorise à remplacer les bornes supérieures d'intégration par $+\infty$. On aboutit à une intégrale double calculable analytiquement et à l'équation $\zetac^{n\phi}_{\qb}=-\ii+(1+\ii)\pi^2/[2\qb(\frac{9}{8}\gamma u\epsilon^2)^{3/4}(\zetac^{n\phi}_{\qb})^{1/4}]$. Il reste à traduire le fait que le pôle non physique est de partie imaginaire nulle au seuil de disparition.}}  {\ycd Pour le voir plus simplement, nous pouvons évaluer  $\check{\Sigma}_{\qb}^{(4)}(\ii 0^+)$ (contribution d'ordre 4 en $H_3$ considérée à tous les ordres en $\epsilon$) dans la limite $\epsilon\to 0$. Il suffit pour cela de remplacer $\check{\zeta}_{\qb}^{n\phi}$ par $\ii 0^+$ dans l'intégrale double au second membre de (\ref{eq294}), que l'on sait alors calculer explicitement pour le champ quantique de coupure $\eta\to+\infty$:
\be
\label{eq296}
\check{\Sigma}_{\qb}^{(4)}(\ii 0^+)\underset{\epsilon\to 0}{\sim} \frac{16 g_2(\eee^{-\qb})}{9\gamma u \epsilon^2 \qb^2} \quad\mbox{d'où}\quad \Sigma_{\qb}^{(4)}(\ii 0^+)\underset{\epsilon\to 0}{\sim} k_B T \left[\frac{(1+\Lambda)^2}{\gamma^{3/2}\rho\xi^2}\right]^2 \frac{2 g_2(\eee^{-\qb})}{\pi^2\qb}
\ee
À basse température $\epsilon\ll 1$, l'ordre quatre en $H_3$ (\ref{eq296}) n'est donc une petite correction à l'ordre deux (\ref{eq141}) que si $\gamma u\epsilon^2$ est beaucoup plus grand qu'une certaine fonction de $\qb$, à savoir $1/\qb^2$ à faible $\qb$.\footnote{{\yc Le résultat (\ref{eq296}) n'est pas en contradiction avec l'équation (\ref{eq280}) car l'ordre des limites $\zetat\to 0$ et $\epsilon\to 0$ n'est pas le même.}}}

\paragraph{Conclusion} À nombre d'onde réduit $\qb$ fixé du mode de phonon {\yc dont on étudie l'amortissement}, la condition d'applicabilité de la théorie diagrammatique à l'ordre quatre en $H_3$ n'est pas seulement que le paramètre universel $u$ soit assez grand (régime d'interaction faible attendu) mais aussi que le produit ${\yc\gamma}u\epsilon^2$ soit assez élevé. Elle présente donc une incompatibilité de fond avec la limite étudiée dans cette article, celle de basse température $\epsilon\to 0$ à force des interactions fixée. Ainsi, pour les simulations numériques de la figure \ref{fig2}, où $\qb=1/4$ et $\gamma=1$, nous trouvons, en suivant par la méthode de Newton jusqu'à disparition la racine de l'équation (\ref{eq294}) écrite en champ classique, qu'il faudrait avoir $u\epsilon^2>10,\!6$ soit $\rho\xi^2>23,\!4$ pour $\epsilon=1/2$ et $\rho\xi^2>52,\!6$ pour $\epsilon=1/3$. Or $\rho\xi^2$ vaut au plus $10$ sur la figure \ref{fig2} ; la théorie à l'ordre quatre développée dans cette section est donc inapplicable.

\section{Théories non perturbatives en $H_3$}
\label{sec4}

Comme nous l'avons vu dans la section \ref{sec3}, la théorie à l'ordre quatre en $H_3$ n'est pas utilisable en dimension $d=2$ dans la limite de basse température $\epsilon=k_B T/mc^2\to 0$. Pour rendre compte de l'écart entre la théorie à l'ordre deux en $H_3$ et les simulations sur le temps de décroissance à $1/\eee$ du mode de phonon $\qq$ et sur la partie réelle du signal associé (\ref{eq012}) dans le régime d'interaction faible $(\rho\xi^2)^{-1}\ll 1$, voir les figures \ref{fig1}d et \ref{fig2}d, nous proposons ici une théorie non perturbative guidée par des considérations physiques. En bref, nous {\yc omettons la partie finie mais} nous régularisons la partie divergente de l'ordre quatre de la fonction énergie propre en donnant une durée de vie finie aux phonons dits virtuels (ou intermédiaires) des processus collisionnels à quatre phonons ; ceci fait apparaître une contribution qui se révèle être d'ordre deux en $H_3$ et qui vient donc corriger l'expression de l'énergie propre à l'ordre dominant en la force des interactions. Nous présentons une régularisation de type physique atomique dans la section \ref{sec4.1} (les phonons virtuels $\KK$ acquièrent une énergie complexe $\veps_\KK\to \veps_\KK +\Sigma_\KK^{(2,2)}(\ii 0^+)=\veps_\KK-(\ii\hbar/2)\Gamma_\KK$) et une régularisation de type matière condensée dans la section \ref{sec4.2} (on garde dans l'énergie complexe des phonons virtuels la dépendance en fréquence de la fonction énergie propre $\Sigma_\KK^{(2,2)}(\zeta)$). Enfin, dans la section \ref{sec4.3}, nous donnons la nouvelle forme de la théorie dans la limite $(\rho\xi^2)^{-1}\to 0$, commune aux deux régularisations et qui remplace avec succès la règle d'or de Fermi.

\subsection{Régularisation de type physique atomique}
\label{sec4.1}

Nous tirons de la section \ref{sec3.3} l'enseignement que l'ajout d'une partie complexe infinitésimale $\ii 0^+$ à $\zeta=z-\veps_\qq$, très habituel dans la méthode des fonctions de Green, ne suffit pas à donner une valeur finie à $\tilde{\Sigma}_\qq^{(4,2)}(\zeta=0)$. Or, les collisions entre les phonons, qui les rendent instables, fournissent naturellement une contribution imaginaire non infinitésimale à côté de $\zeta$ dans les dénominateurs d'énergie. En suivant la prescription de la physique atomique, nous donnons donc aux phonons virtuels une énergie complexe
\be
\label{eq300}
\veps_\KK\to\veps_\KK+\Sigma_\KK^{(2,2)}(\ii 0^{\pm})=\veps_\KK\mp \frac{\ii\hbar}{2}\Gamma_\KK
\ee
où $\Gamma_\KK$ est le taux d'amortissement des phonons $\KK$ de la règle d'or de Fermi à l'ordre deux en $\epsilon$, voir l'équation (\ref{eq033}). Les phonons virtuels, par définition ni entrants ni sortants dans la formulation de Landau de la section \ref{sec3.2}, sont ceux dont les vecteurs d'onde ne sont pas indiqués sur {\yc les diagrammes de} la figure \ref{fig8}.  Le signe à prendre dans (\ref{eq300}) devant $\Gamma_\KK$ ou dans l'argument de $\Sigma_\KK$ est celui du haut si le phonon descend le temps (ligne interne orientée vers la droite) ou celui du bas si le phonon remonte le temps (ligne interne orientée vers la gauche).\footnote{Dans la formulation de Landau, on choisit $-\ii(\hbar/2)\Gamma_\KK$ dans la première somme de l'équation (\ref{eq258}) et son opposé dans la deuxième somme, pour les raisons expliquées après (\ref{eq257}) {\yc (dans la deuxième somme, les amplitudes de transition sont retournées temporellement),} ce qui conduit au même résultat {\yc que dans la formulation diagrammatique}.} En revanche, nous ne donnons pas une énergie complexe aux phonons entrants ou sortants car ceci remettrait en question la conservation de l'énergie du gaz lors des processus collisionnels. Par conséquent, il n'y a pas de modification à apporter à $\Sigma_{\qq}(\zeta)$ à l'ordre deux en $H_3$, qui ne comporte aucun phonon virtuel. 

\begin{figure}[t]
\includegraphics[width=8cm,clip=]{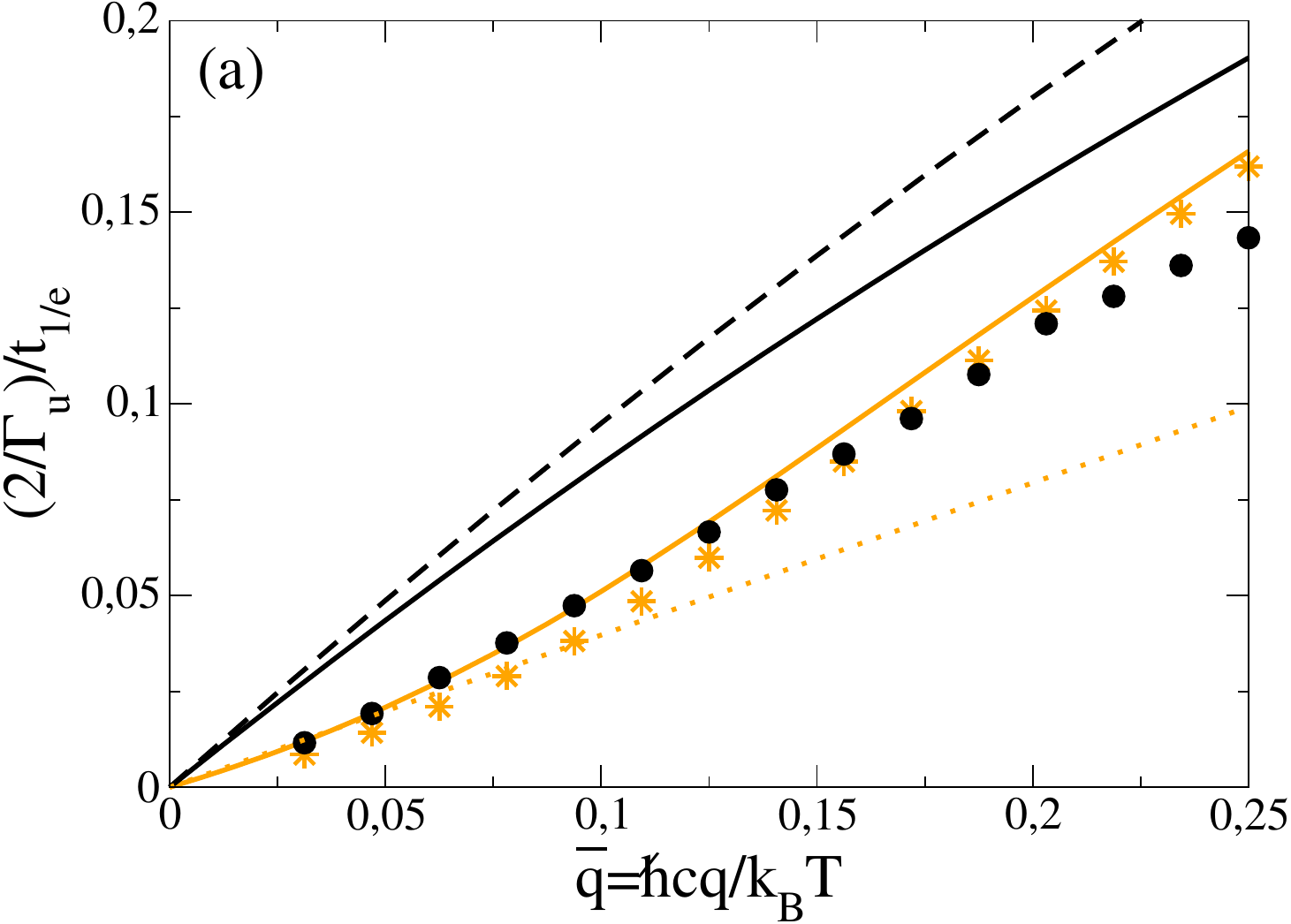} \includegraphics[width=8cm,clip=]{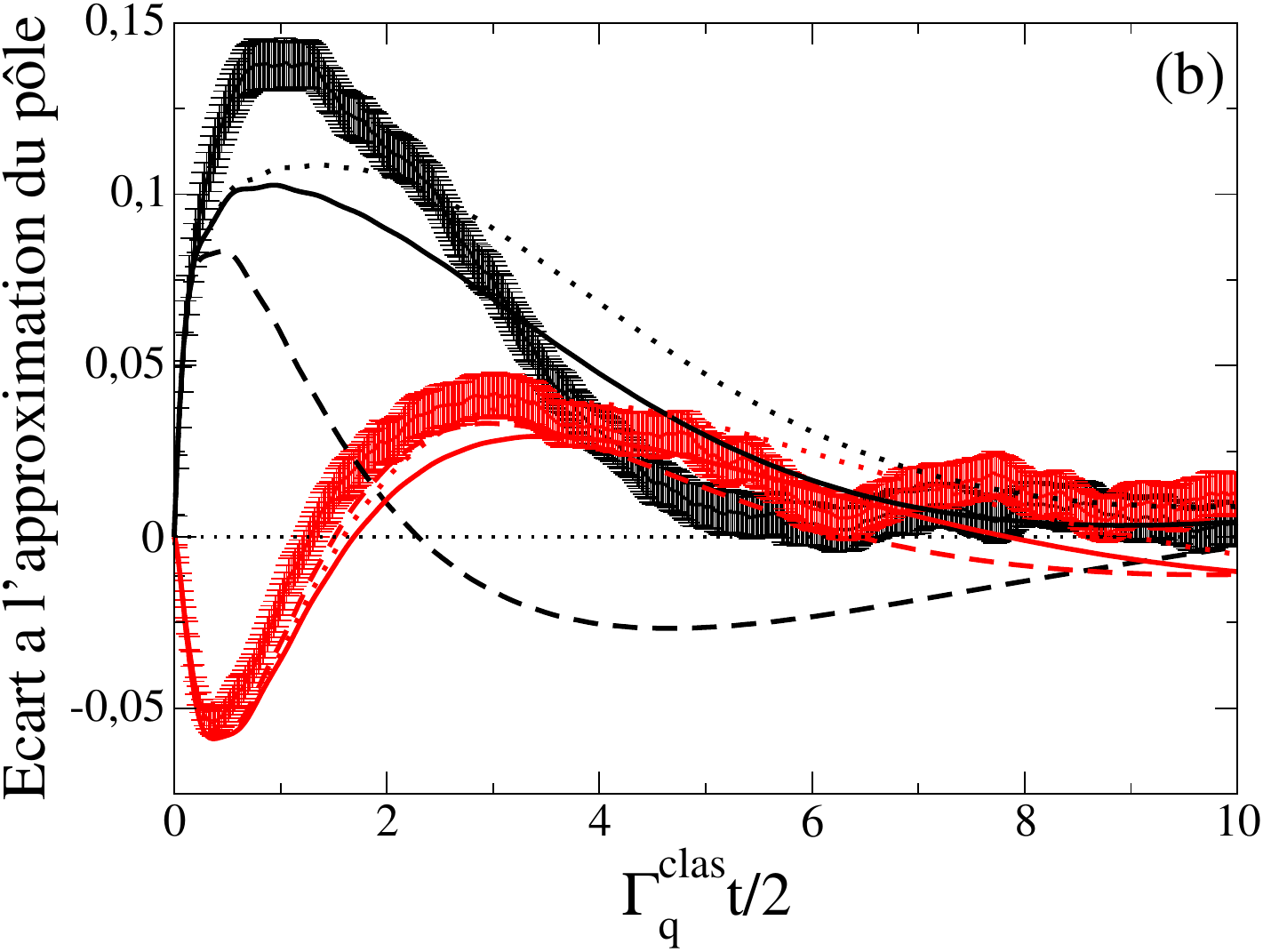}
\caption{(a) Inverse du temps de décroissance à $1/\eee$ du module du signal $s(t)$ (\ref{eq012}) en fonction du nombre d'onde $q$ des modes selon $Ox$, pour les paramètres de la figure \ref{fig1}d, en particulier $(\rho\xi^2)^{-1}=0,\!1$ (c'est la force des interactions la plus faible de la figure \ref{fig1}). Disques noirs: expérience numérique avec $\epsilon=1/2$. Tireté noir: règle d'or de Fermi (\ref{eq113}) pour le champ phononique classique et dans la limite $\epsilon\to 0$. Trait plein noir: méthode des fonctions de Green à $N$ corps avec la fonction énergie propre $\tilde{\Sigma}^{(2,2)}_{\yc\qb}(\zetat)$ (\ref{eq125}) (ordre dominant en $H_3$ et $\epsilon$). Trait plein orange: idem avec la fonction énergie propre $\tilde{\Sigma}^{(2,2)}_{\mbox{\tiny corr}}(\qb,\zetat)=\tilde{\Sigma}^{(2,2)}_{\yc\qb}(\zetat)+\tilde{\Sigma}^{(4,2)}_{\mbox{\tiny div. rég.}}(\qb,\zetat)$ de l'équation (\ref{eq303}) (on ne garde que la partie divergente de l'ordre quatre en $H_3$ et on la régularise selon la prescription de la physique atomique en donnant à la main l'énergie complexe de la règle d'or de Fermi aux phonons virtuels). Étoiles orange: idem avec la composante $\tilde{\Sigma}^{(4,2)}_{\mbox{\tiny div. rég.}}(\qb,\zetat)$ de la fonction énergie propre (\ref{eq310}) obtenue par une régularisation de type physique de la matière condensée (on donne aux phonons virtuels $\KK$ une énergie complexe $\Sigma^{(2,2)}_{\KK}(\zeta_{\KK})$ dépendant de la fréquence c'est-à-dire de la variable $\zeta$ dans (\ref{eq012})). Pointillé orange: tangente à l'origine (\ref{eq313}) pour la limite $(\rho\xi^2)^{-1}\to 0$ des théories régularisées. Toutes les fonctions énergies propres sont écrites dans leurs versions de champ classique pour permettre la comparaison au numérique. Le taux $\Gamma_u$ servant d'unité est celui de l'équation (\ref{eq113}) donnant $\Gamma_q^{{\yct\rm clas}}$. (b) Écart {\yc $s(t)-s_{\mbox{\scriptsize pôle}}(t)$} du signal à l'approximation du pôle pour les paramètres de la figure \ref{fig2}d, en particulier {\yc$\qb=1/4$ et} $(\rho\xi^2)^{-1}=0,\!1$. Courbes avec barres d'erreur: expérience numérique. Tireté: méthode des fonctions de Green avec la fonction énergie propre $\tilde{\Sigma}^{(2,2)}_{\yc\qb}(\zetat)$. Trait plein: idem en prenant la fonction énergie propre $\tilde{\Sigma}^{(2,2)}_{\mbox{\tiny corr}}(\qb,\zetat)$ de l'équation (\ref{eq303}). Pointillé: idem en utilisant l'expression (\ref{eq310}) pour la composante $\tilde{\Sigma}^{(4,2)}_{\mbox{\tiny div. rég.}}(\qb,\zetat)$ de la fonction énergie propre $\tilde{\Sigma}^{(2,2)}_{\yc\rm corr}(\qb,\zetat)$. La partie réelle du signal est en noir, la partie imaginaire en rouge.}
\label{fig3}
\end{figure}

Nous appliquons la procédure (\ref{eq300}) à la partie divergente de $\tilde{\Sigma}^{(4,2)}_{\qb}$, qui provient de la contribution de type II du diagramme $P_5$, et nous omettons toutes les autres contributions à $\tilde{\Sigma}^{(4,2)}_{\qb}$, qui se révèlent être des $O((\rho\xi^2)^{-2})$ et sous-dominantes lorsque $(\rho\xi^2)^{-1}\to 0$.\footnote{En l'absence de divergence en $\zeta=0$, l'ajout d'une partie imaginaire à $\veps_\KK$ ne change pas l'ordre dominant du diagramme, qui reste d'ordre quatre en $H_3$.} Comme le montre la figure \ref{fig8}, les phonons virtuels sont alors {\ycd $\kk_2=\kk-\qq$} et $\kk_3=\kk'$ dans les notations de la figure \ref{fig7}. Tous deux remontent le temps ($s_2=s_3=-1$), si bien que les quantités $\zeta_1$ et $\zeta_2$ dans (\ref{eq233}), {\ycd qui s'écrivent $\zeta_1=\veps_\kk-\veps_{\kk-\qq}-\veps_\qq$ et $\zeta_2=-\veps_{\kk'}+\veps_{\kk'+\qq}-\veps_{\qq}$,} et les coefficients $A_1^P$ et $A_2^P$ dans (\ref{eq243}) sont changés comme suit, 
\be
\label{eq301}
\zeta_1\to\zeta_1-\ii\frac{\hbar}{2}\Gamma_{\yc\kk-\qq} \quad ;\quad \zeta_2\to\zeta_2-\ii\frac{\hbar}{2}\Gamma_{\kk'} \quad ; \quad A_1^P\to A_1^P-\frac{\ii}{2}\tilde{\Gamma}_{\kb-\qb} \quad ; \quad A_2^P\to A_2^P-\frac{\ii}{2}\tilde{\Gamma}_{\kb'}
\ee
l'ordre des facteurs au dénominateur de (\ref{eq243}) respectant celui de (\ref{eq233}) ; ici $\tilde{\Gamma}_{\bar{K}}=\hbar\Gamma_{\KK}/k_BT\epsilon^2$ dans l'esprit de l'équation (\ref{eq022}). {\yc En revanche, la quantité $\zeta_3$ ne met en jeu aucun phonon virtuel et n'est pas modifiée.} L'ancienne partie divergente de l'intégrale angulaire (\ref{eq282}) devient pour $\im\zetat>0$:\footnote{On continue à linéariser autour du même point de triple intersection des lieux d'annulation en regroupant les corrections complexes avec la variable $\zetat$ si bien qu'au second membre de (\ref{eq282}), on remplace $\zetat$ par $\zetat+(\ii/2)\tilde{\Gamma}_{\kb-\qb}$ dans le premier facteur et par $\zetat+(\ii/2)\tilde{\Gamma}_{\kb'}$ dans le deuxième facteur au dénominateur.}
\be
\label{eq302}
K_{\mbox{\scriptsize div. rég.}}^{P_5^{\rm II}}(\kb,\kb')=-\frac{32\pi^2}{3\gamma\qb\kb\kb'} \frac{1}{\zetat(\kb+\kb')+\frac{\ii}{2}\kb'\tilde{\Gamma}_{\kb-\qb}+\frac{\ii}{2}(\kb-\qb)\tilde{\Gamma}_{\kb'}}
\ee
Après remplacement de $K(\kb,\kb')$ par (\ref{eq302}) dans (\ref{eq242}) et ajout de la contribution d'ordre deux en $H_3$, nous aboutissons à la prescription non perturbative pour la fonction énergie propre corrigeant le résultat de la section \ref{sec2}:
\be
\label{eq303}
\boxed{\tilde{\Sigma}_{\rm corr}^{(2,2)}(\qb,\zetat) = \tilde{\Sigma}_{\qb}^{(2,2)}(\zetat) + \tilde{\Sigma}^{(4,2)}_{\mbox{\scriptsize div. rég.}}(\qb,\zetat)}
\ee
où la première contribution est celle de l'équation (\ref{eq125}) et la seconde, régularisation de la partie divergente de $\tilde{\Sigma}_{\qb}^{(4,2)}(\zetat)$, s'écrit, toujours pour $\im\zetac>0$:
\be
\label{eq304}
\boxed{\tilde{\Sigma}^{(4,2)}_{\mbox{\scriptsize div. rég.}}(\qb,\zetat)=-\frac{2}{3\pi^2\gamma} \left(\frac{9(1+\Lambda)^2}{8\rho \xi^2}\right)^2
\int_{\qb}^\eta \dd\kb \int_0^{\eta-\kb} \dd\kb' \frac{\kb \kb' (\kb-\qb)(\kb'+\qb)(\kb+\kb')}
{\tilde{\zeta}(\kb+\kb')+\frac{\ii}{2}\kb'\tilde{\Gamma}_{\kb-\qb}+\frac{\ii}{2}(\kb-\qb)\tilde{\Gamma}_{\kb'}} 
(\bar{n}^{\rm lin}_{k'+q}-\bar{n}^{\rm lin}_{k'+k})(\bar{n}^{\rm lin}_{k-q}-\bar{n}^{\rm lin}_{k})}
\ee
En reportant la fonction (\ref{eq303}) dans l'équation (\ref{eq023}) (avec $\nu=\sigma$ en dimension $d=2$) dans sa version de champ classique, nous obtenons pour la force des interactions la plus faible un bien meilleur accord avec les simulations numériques, voir le trait plein orange sur la figure \ref{fig3}a et le trait plein noir sur la figure \ref{fig3}b, à comparer respectivement au trait plein noir sur la figure \ref{fig3}a et au tireté noir sur la figure \ref{fig3}b correspondant à l'ordre deux en $H_3$ non corrigé. 

\subsection{Régularisation de type matière condensée}
\label{sec4.2}

On sait bien en physique de la matière condensée que l'énergie complexe d'une quasi-particule dans un système en interaction dépend de la fréquence à laquelle on la regarde, c'est-à-dire de la fréquence à laquelle on mesure les fonctions de réponse. C'est d'ailleurs l'essence de ce qu'est la fonction $\Sigma$ dans (\ref{eq012}). Il faut donc remplacer dans (\ref{eq300}) l'énergie complexe fixée $\mp(\ii\hbar/2)\Gamma_\KK$ du phonon virtuel {\yc$\KK$} par une fonction de {\yc la variable d'intégration} $\zeta$:
\be
\label{eq305} 
\veps_\kk\to \veps_\KK+\Sigma_\KK^{(2)}(\pm(\zeta+\Delta E)-\veps_\KK)
\ee
où il faut prendre le signe $+$ si le phonon $\KK$ descend le temps, et le signe $-$ s'il le remonte. La quantité $\Delta E$ s'obtient en réinterprétant le dénominateur d'énergie auquel participe le phonon $\KK$ dans $\Sigma^{(4)}$ comme celui du propagateur libre de ce phonon à l'énergie complexe $z_{\KK}=\zeta+\Delta E$. Ainsi, si le phonon $\KK$ descend le temps (signe $s_\KK=+1$ devant $\veps_\KK$ dans (\ref{eq238})), il contribuera à $\Sigma^{(4)}$ dans les formes simplifiées (\ref{eq213},\ref{eq233}) au travers du facteur $1/(\zeta+\Delta E-\veps_\KK)$, qui n'est autre que son propagateur non perturbé à l'énergie complexe $z_\KK$ donc à l'énergie complexe redéfinie comme dans (\ref{eq013}) $\zeta_\KK=z_\KK-\veps_\KK$. Nous proposons donc de le remplacer par la fonction de Green (\ref{eq012}) du phonon $\KK$ {\yc écrite comme dans (\ref{eq013})}, avec $\Sigma_{\KK}$ calculé à l'ordre deux en $H_3$:
\be
\label{eq306}
\frac{1}{\zeta+\Delta E-\veps_\KK}\to \frac{1}{\zeta+\Delta E-\veps_\KK-\Sigma_\KK^{(2)}(\zeta+\Delta E-\veps_\KK)}
\ee
ce qui explique la substitution (\ref{eq305}). Si le phonon virtuel remonte le temps (signe $s_\KK=-1$), nous nous ramenons au cas précédent par la chaîne de transformations
\be
\label{eq307}
\frac{1}{\zeta+\Delta E+\veps_\KK} = -\frac{1}{-(\zeta+\Delta{\ycd E})-\veps_\KK} \longrightarrow -\frac{1}{-(\zeta+\Delta E)-\veps_\KK-\Sigma_\KK^{(2)}(-(\zeta+\Delta E{\yc)-}\veps_\KK)} = \frac{1}{\zeta+\Delta E+\veps_\KK+\Sigma_\KK^{(2)}(-(\zeta+\Delta E{\yc)-}\veps_\KK)}
\ee
 en accord là aussi avec (\ref{eq305}). En pratique, comme dans la section \ref{sec4.1}, on garde seulement la partie divergente de $\tilde{\Sigma}^{(4,2)}_{\qb}$ (provenant de la contribution de type II du diagramme $P_5$), dans laquelle les phonons virtuels {\yc remontent tous deux le temps et} sont de vecteur d'onde $\KK=\kk-\qq$ (avec $\Delta E=\veps_{\qq}-\veps_\kk$ {\ycd de façon que $\zeta-\zeta_1=\zeta+\Delta E+\veps_{\kk-\qq}$}) ou $\KK=\kk'$ (avec $\Delta E=\veps_{\qq}-\veps_{\kk'+\qq}$ {\ycd de façon que $\zeta-\zeta_2=\zeta+\Delta E+\veps_{\kk'}$}). Après régularisation, la partie divergente (\ref{eq282}) de l'intégrale angulaire devient
\be
\label{eq308}
K_{\mbox{\scriptsize div. rég.}}^{P_5^{\rm II}}(\kb,{\yct\kb'}) = \int_{\mathbb{R}^2} \frac{2\dd \tau \dd \tau'}{[\zetat-\alpha \tau +\tilde{\Sigma}_{\kb-\qb}^{(2,2)}(-\zetat+\alpha \tau)] [\zetat-\alpha' \tau' + \tilde{\Sigma}_{\kb'}^{(2,2)}(-\zetat+\alpha' \tau')] (\zetat-\beta \tau -\beta' \tau')}
\ee
où les coefficients sont ceux de $\tau$ et $\tau'$ au second membre de (\ref{eq282}): $\alpha=(3\gamma/4)^{1/2}\qb\kb$, $\alpha'=-(3\gamma/4)^{1/2}\qb\kb'$, $\beta=-(3\gamma/4)^{1/2}\kb\kb'$, $\beta'=(3\gamma/4)^{1/2}(\kb-\qb)\kb'$. Comme les fonctions $\tilde{\Sigma}^{(2,2)}$ au dénominateur de (\ref{eq308}) admettent des lignes de coupure dans le plan complexe, {\yc voir la section \ref{sec2.2},} le théorème de Cauchy ne permet plus de calculer l'intégrale double et l'on peut seulement intégrer une fois, par exemple sur $\tau'$, en se limitant comme d'habitude à $\im\zetat>0$.\footnote{Le dénominateur de l'intégrande dans (\ref{eq308}) admet une ligne de coupure vis-à-vis de la variable $\tau$ uniquement dans le demi-plan supérieur strict car $\im(\zetat/\alpha)>0$, et vis-à-vis de la variable $\tau'$ uniquement dans le demi-plan inférieur strict car $\im(\zetat/\alpha')<0$. On peut donc intégrer sur $\tau'$ par le théorème des résidus en refermant le chemin d'intégration par un grand demi-cercle à l'infini dans le demi-plan supérieur ; ce chemin entoure nécessairement le pôle $\tau_0'=(\zetat-\beta \tau)/\beta'$ issu du troisième facteur au dénominateur de (\ref{eq308}) car $\im \tau_0'>0$. L'intégrande du résultat (\ref{eq310}) admet cette fois une ligne de coupure vis-à-vis de la variable $\tau$ aussi bien dans le demi-plan supérieur ($\im(\zetat/\alpha)>0$) que dans le demi-plan inférieur ($\im(\zetat/\alpha')<0$) ce qui empêche de refermer le chemin d'intégration seulement à l'infini.}  
En reportant la contribution correspondante à $K(\kb,\kb')$ dans (\ref{eq242}), nous aboutissons à la fonction énergie propre corrigée de même forme que (\ref{eq303}), avec cette fois
\begin{multline}
\label{eq310}
\tilde{\Sigma}^{(4,2)}_{\mbox{\scriptsize div. rég.}}(\qb,\zetat)= \qb\left(\frac{9(1+\Lambda)^2}{8\rho \xi^2}\right)^2 \\
\times \int_{\qb}^\eta \frac{\kb \dd\kb}{(2\pi)^2}
\int_0^{\eta-\kb} \frac{\kb' \dd\kb'}{(2\pi)^2}
\int_{-\infty}^{+\infty} \dd \tau 
\frac{(-4\ii\pi/\beta')\kb \kb' (\kb-\qb)(\kb'+\qb)(\kb+\kb')(\bar{n}^{\rm lin}_{k'+q}-\bar{n}^{\rm lin}_{k'+k})(\bar{n}^{\rm lin}_{k-q}-\bar{n}^{\rm lin}_{k})}
{\left[\tilde{\zeta}-\alpha \tau +\tilde{\Sigma}^{(2,2)}_{\kb-\qb}(-\tilde{\zeta}+\alpha \tau)\right]
\left[-\frac{\beta}{\beta'}(\tilde{\zeta}-\alpha'\tau)+\tilde{\Sigma}^{(2,2)}_{\kb'}
\left(\frac{\beta}{\beta'}(\tilde{\zeta}-\alpha'\tau)\right)\right]}
\end{multline}
sachant que $\alpha'-\beta'=\beta$. Les résultats de cette approche sont représentés par des étoiles orange sur la figure \ref{fig3}a et par des pointillés rouge et noir sur la figure \ref{fig3}b. \footnote{Leur obtention est assez lourde puisque chaque évaluation des fonctions $\tilde{\Sigma}_{\bar{K}}^{(2,2)}$ requiert une intégration numérique sur un nombre d'onde comme dans (\ref{eq125}). Nous avons discrétisé les nombres d'onde réduits avec un pas $1/64$ (résultats sur la figure) mais nous avons vérifié qu'un pas $1/96$ conduit à des résultats très proches.} Les deux procédures de régularisation, de type physique atomique ou matière condensée, donnent en définitive des résultats proches entre eux et proches des simulations, ce qui nous rassure sur leur validité. 

\subsection{Théorie régularisée dans la limite d'interaction faible}
\label{sec4.3}

\begin{SCfigure}
\includegraphics[width=8cm,clip=]{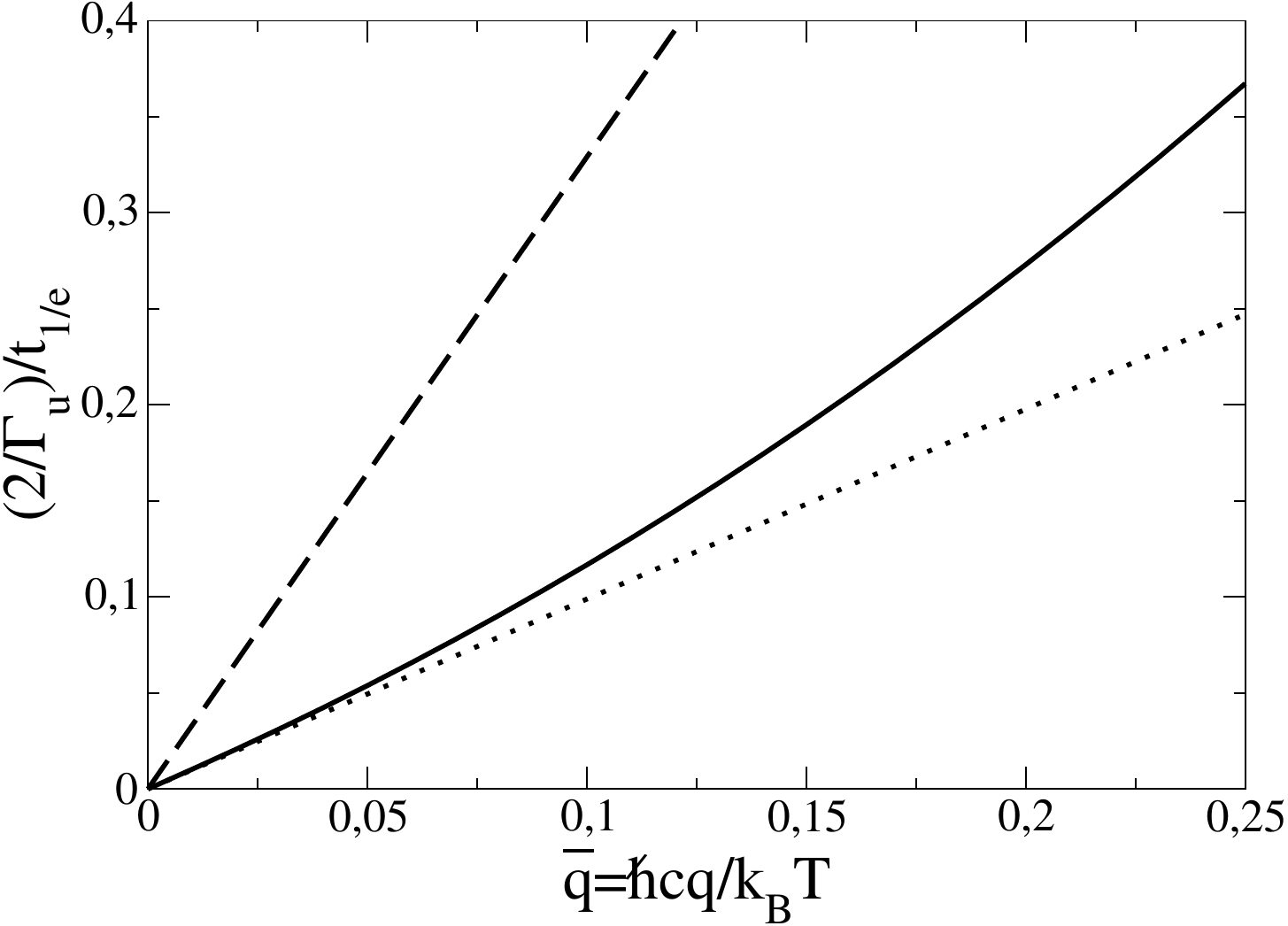}
\caption{Dans la limite de basse température $\epsilon=k_BT/mc^2\to 0$ puis d'interaction faible $(\rho\xi^2)^{-1}\to 0$ pour le champ quantique {\yc de coupure $\eta$ infinie}, inverse du temps de décroissance à $1/\eee$ du signal (\ref{eq012}) pour le mode de phonon $\qq$ dans un superfluide bidimensionnel, en fonction du nombre d'onde réduit $\qb$. Trait plein : prédiction des théories non perturbatives de la section \ref{sec4}, correspondant à la fonction énergie propre corrigée (\ref{eq311}) de la section \ref{sec4.3}. Pointillé: pente à l'origine (\ref{eq313}). Tireté: règle d'or (\ref{eq033}). Le taux $\Gamma_u$ servant d'unité est donné par (\ref{eq113}).}
\label{fig9}
\end{SCfigure}

Nous calculons ici la limite des théories régularisées lorsque $(\rho\xi^2)^{-1}\to 0$. Pour ce faire, nous mettons la variable $\zetat$ et les fonctions $\tilde{\Sigma}$ à l'échelle de leurs valeurs caractéristiques $\propto (\rho\xi^2)^{-1}$ au moyen des {\yct changements de variables} (\ref{eq130}) et (\ref{eq131}), repérés par un accent tchèque (un facteur $\qb$ a été sorti en prévision d'un passage à la limite $\qb\to 0$, et tous les paramètres sont rassemblés dans la quantité $u\propto\rho\xi^2$ de l'équation (\ref{eq112})), puis nous passons à la limite $u\to+\infty$ à $\check{\zeta}$ fixé dans les équations (\ref{eq303}) et (\ref{eq304}). Contrairement à ce qui se passait à l'ordre deux en $H_3$ non corrigé (\ref{eq131}), il reste cette fois une dépendance en $\check{\zeta}$:\footnote{On montre en deux étapes que la fonction de Green associée à la fonction énergie propre (\ref{eq311}) n'admet, comme il se doit, aucun pôle dans le demi-plan complexe supérieur. Dans un premier temps, on établit la majoration $\im\check{\Sigma}_{{\rm corr}\,\infty}^{(2,2)}(\qb,\check{\zeta})\leq\im\check{\Sigma}_{{\rm corr}\,\infty}^{(2,2)}(\qb,0)\ \forall \check{\zeta}\in\mathbb{C}$, $\im\zetac\geq 0$; à cette fin, on écrit le dénominateur de l'intégrande dans (\ref{eq311}) sous la forme $x+\ii y$, avec $x\in\mathbb{R}$ et $y\in\mathbb{R}^+$, puis on utilise la chaîne $\im[-1/(x+\ii y)]=y/(x^2+y^2)\leq 1/y\leq 2/[\kb'(\kb-\qb)(\check{\Gamma}_{\kb-\qb}+\check{\Gamma}_{\kb'})]$. Dans un second temps, on vérifie que $\im\check{\Sigma}_{{\rm corr}\,\infty}^{(2,2)}(\qb,0)<0\ \forall\qb>0$. Il en résulte que l'équation $\check{\zeta}=\check{\Sigma}_{{\rm corr}\,\infty}^{(2,2)}(\qb,\check{\zeta})$ ne peut avoir de solution $\check{\zeta}$ de partie imaginaire positive sans conduire à la contradiction $\im\check{\zeta}<0$.}
\be
\label{eq311}
\boxed{\check{\Sigma}_{\rm corr}^{(2,2)}(\qb,\check{\zeta}) \underset{u \to +\infty}{\stackrel{\im\check{\zeta}\,>\,0}{\longrightarrow}} \check{\Sigma}_{{\rm corr}\,\infty}^{(2,2)}(\qb,\check{\zeta})\equiv-\frac{\ii}{2}\check{\Gamma}_{\qb}-\frac{2}{\qb}\int_{\qb}^\eta \dd\kb \int_0^{\eta-\kb} \dd\kb' \frac{\kb\kb'(\kb-\qb)(\kb'+\qb)(\kb+\kb')(\bar{n}^{\rm lin}_{k'+q}-\bar{n}^{\rm lin}_{k'+k}) (\bar{n}^{\rm lin}_{k-q}-\bar{n}^{\rm lin}_{k})} {\check{\zeta}\qb(\kb+\kb')+\frac{\ii}{2}\kb'(\kb-\qb)\left(\check{\Gamma}_{\kb-\qb}+\check{\Gamma}_{\kb'}\right)}}
\ee
Ici, $\check{\Gamma}_{\bar{K}}=2\ii\check{\Sigma}_{\bar{K}}^{(2,2)}(\ii 0^+)$ est le taux d'amortissement de la règle d'or à l'ordre dominant en $\epsilon$ et adimensionné comme dans (\ref{eq130}), $\check{\Gamma}_{\bar{K}}=8u\Gamma_{\KK}/(3\gamma\epsilon^2\hbar cK)$ ; on a $\check{\Gamma}_{\bar{K}}=\check{\Gamma}_{\bar{K}}^{{\yct\rm clas}}=2\eta-\bar{K}$ au vu de (\ref{eq113}) dans le cas du champ classique et $\check{\Gamma}_{\bar{K}}=4\zeta(2)+\bar{K}^2/6$ au vu de (\ref{eq033}) ou (\ref{eq141}) dans le cas du champ quantique de coupure infinie.
On obtient exactement la même expression que (\ref{eq311}) en partant de la forme régularisée (\ref{eq310}).\footnote{On effectue simultanément les changements de variables $\zetat=(3\gamma\qb/8u)\check{\zeta}$ et $\tau=(3\gamma\qb/8u)\check{\tau}$, puis on passe à la limite $u\to +\infty$ sous le signe intégral à $\check{\tau}$ fixé. Alors les lignes de coupure des fonctions $\tilde{\Sigma}^{(2,2)}_{\bar{K}}$ sont rejetées à l'infini, ces fonctions sont remplacées par des constantes, par exemple $\lim_{u\to+\infty} (8u/3\gamma\qb) \tilde{\Sigma}^{(2,2)}_{\kb-\qb}((3\gamma\qb/8u)(\alpha\check{\tau}-\check{\zeta}))=[(\kb-\qb)/\qb] \check{\Sigma}^{(2,2)}_{\kb-\qb}(\ii 0^-)=[(\kb-\qb)/\qb](\ii/2)\check{\Gamma}_{\kb-\qb}$, et l'intégrale sur $\check{\tau}$ peut être calculée par la formule de Cauchy. On retrouve (\ref{eq311}).} De façon remarquable, le signal (\ref{eq012}) associé à la fonction énergie propre (\ref{eq311}) prend des valeurs réelles à tout temps.\footnote{Ceci résulte de la propriété ${\yc\check{\Sigma}(\check{\zeta})^*}=-\check{\Sigma}(-\check{\zeta}^*)$, $\forall \check{\zeta}\in\mathbb{C}\setminus\ii\mathbb{R}^-$, satisfaite par la fonction $\check{\Sigma}_{{\rm corr}\,\infty}^{(2,2)}(\qb,\check{\zeta})$ (si on étend l'expression au second membre telle quelle au cas $\im\check{\zeta}<0$).} Il y a donc bien, {\yc lorsque} $(\rho\xi^2)^{-1}\to 0$, un désaccord persistant avec la règle d'or de Fermi sur la partie réelle du signal, comme le laissait prévoir la figure \ref{fig3}b, {\ycd mais pas sur la partie imaginaire, de limite nulle dans les deux théories}. La figure \ref{fig9} représente, en fonction du nombre d'onde réduit $\qb$, le temps de décroissance à $1/\eee$ du mode de phonon $\qq$ déduit de (\ref{eq311}) pour la théorie de champ quantique (coupure $\eta\to +\infty$). C'est notre prédiction à l'ordre dominant à basse température et en la force des interactions pour un superfluide bidimensionnel, qui remplace et corrige celle de la règle d'or de Fermi représentée en tireté sur la figure \ref{fig9}.

Trois prédictions analytiques simples peuvent être tirées de l'équation (\ref{eq311}). D'abord, dans la limite $\qb\to 0$, la dépendance en $\check{\zeta}$ disparaît au second membre de (\ref{eq311}) et le signal (\ref{eq012}) devient purement exponentiel, $s(t)=\exp(-A\check{t})$ avec
\be
\label{eq312}
A=\frac{1}{2}\check{\Gamma}_0+4\int_0^\eta \dd \kb \int_0^{\eta-\kb} \dd \kb'
\frac{\kb\kb'(\kb+\kb')(\bar{n}_{k'}^{\rm lin}-\bar{n}_{k'+k}^{\rm lin})\frac{\dd}{\dd \kb}\bar{n}_{k}^{\rm lin}}
{\check{\Gamma}_{\kb}+\check{\Gamma}_{\kb'}} =
\left\{ \begin{array}{ll}
[1-4(4 \ln \frac{4}{3} -1)]\eta \simeq 0,\!4\,\eta & \mbox{(champ classique)} \\
& \\
0,\!988\,618 \ldots & \mbox{(champ quantique)}
\end{array}
\right.
\ee
Si l'on quitte le temps adimensionné $\check{t}$ (\ref{eq130}), ce qui fait apparaître la quantité $\Gamma_u$ déja introduite dans (\ref{eq113}) {\yc puisque $\check{t}=\qb\Gamma_u t/2$}, on trouve que ceci correspond à un taux d'amortissement corrigé tendant vers zéro linéairement en $\qb$ :
\be
\label{eq313}
\boxed{
\Gamma_q^{{\rm corr}\,\infty} \underset{\qb\to 0}{\sim} A \Gamma_u \qb = A \frac{9k_B T\epsilon^2(1+\Lambda)^2}{4\pi\hbar(3\gamma)^{1/2}\rho\xi^2} \qb \sim \left\{ \begin{array}{ll} \displaystyle\frac{A}{\eta} \Gamma_q^{{\yct\rm clas}} \simeq 0,\!4\,\Gamma_q^{{\yct\rm clas}} & \mbox{(champ classique)} \\ \displaystyle\frac{A}{2\zeta(2)} \Gamma_q \simeq 0,\!3\,\Gamma_q & \mbox{(champ quantique)}\end{array} \right. }
\ee
représenté en pointillé sur la figure \ref{fig9} pour le champ quantique et sur la figure \ref{fig3}a pour le champ classique. Ici, $\Gamma_q^{{\yct\rm clas}}$ et $\Gamma_q$ sont les taux d'amortissement de la règle d'or pour le champ classique (\ref{eq113}) et pour le champ quantique (\ref{eq033}). On voit donc que la pente à l'origine du taux corrigé est considérablement plus faible que celle de la règle d'or. Ensuite, dans la limite $\qb\to +\infty$, bien entendu pour le champ quantique de coupure infinie, l'intégrale au second membre de (\ref{eq311}) tend uniformément vers zéro dans le demi-plan supérieur et la décroissance du signal devient purement exponentielle, avec un taux équivalent à celui de la règle d'or: \footnote{Ce résultat est aussi celui de la température nulle, ce qui explique pourquoi le troisième membre de (\ref{eq315}) est en fait indépendant de la température lorsqu'on revient au nombre d'onde $q$ non adimensionné. Comme il résulte de l'hamiltonien effectif de l'hydrodynamique quantique, il ne vaut alors qu'à l'ordre dominant en $q\xi$ c'est-à-dire dans la partie linéaire {\yc de la branche acoustique}.}
\be
\label{eq315}
\Gamma_q^{{\rm corr}\,\infty} \underset{\qb\to+\infty}{\sim} \Gamma_q \underset{\qb\to+\infty}{\sim} \frac{3 k_B T\epsilon^2(1+\Lambda)^2}{16\pi(3\gamma)^{1/2}\rho\xi^2} \qb^3
\ee
Enfin, dans la limite des temps longs $\check{t}\to +\infty$ à $\qb$ fixé, la décroissance du signal n'est pas exponentielle mais en loi de puissance à logarithme près, prenant même forme pour le champ classique et pour le champ quantique, comme le montre l'\ref{app5}:
\be
\label{eq314}
s^{(2,2)}_{{\rm corr}\,\infty}(t)\underset{\check{t}\to+\infty}{\sim} \frac{\qb^4(\dd\bar{n}^{\rm lin}_q/\dd\qb)\ln(\check{t}^2)}{\left[\check{\Gamma}_0\check{\Sigma}_{{\rm corr}\,\infty}^{(2,2)}(\qb,0)\check{t}\right]^2}
\ee
Pour $\qb=1/4$, on trouve numériquement à partir de la fonction énergie propre (\ref{eq311}) que le comportement asymptotique (\ref{eq314}) n'est pas atteint avant un temps {\yc réduit} $\Gamma_q t/2=100$, aussi bien pour le champ quantique que pour le champ classique, ce qui va bien au-delà de nos simulations de la figure \ref{fig3}b et nous empêche de le tester. 

\section{Conclusion}
\label{sec5}

Dans ce travail, nous avons considéré le gaz de phonons d'un superfluide bidimensionnel décrit par l'hydrodynamique quantique, une théorie effective de basse énergie, avec un couplage cubique et quartique entre phonons et une branche acoustique $\veps_{\qq}$ fonction convexe du nombre d'onde $q$. Deux paramètres importants caractérisent le système : la température réduite $\epsilon=k_BT/mc^2$ et la force des interactions $(\rho \xi^2)^{-1}$ entre particules du superfluide, où $c$ est la vitesse du son et $\rho$ la densité du superfluide dans l'état fondamental, $m$ la masse d'une particule et $\xi=\hbar/mc$ la longueur de relaxation. Dans ce système, nous avons étudié en régime faiblement collisionnel l'amortissement du mode de phonon $\qq$ excité par une impulsion de Bragg en réponse linéaire, en nous limitant à l'ordre dominant en $\epsilon$ car aller au-delà n'aurait pas de sens avec notre hamiltonien modèle. 

Un premier résultat inattendu est qu'en dimension deux, la prise de limite de basse température, $\epsilon\to 0$ à $(\rho \xi^2)^{-1}$ fixé, ne suffit pas à faire entrer le gaz de phonons en régime de couplage faible. Ceci met en échec l'approximation du pôle, la règle d'or de Fermi et l'hypothèse d'une décroissance exponentielle. En effet, comme nous le montrons par un raisonnement de loi d'échelle, la fonction énergie propre $\Sigma(\qq,z)$ a une valeur typique et une échelle de variation en énergie au voisinage de $z=\varepsilon_\qq$ toutes les deux d'ordre $k_BT \epsilon^2$, alors qu'il faudrait avoir une valeur typique très inférieure à l'échelle de variation {\yc pour pouvoir appliquer la règle d'or} (comme c'est le cas en dimension trois, {\ycd où l'échelle de variation reste $k_BT\epsilon^2$ mais où la valeur typique est abaissée à $k_B T\epsilon^4$ par réduction de la densité d'états finals à basse énergie)}. Ce premier résultat théorique est confirmé par des simulations de champ classique. 

Un second résultat inattendu du cas bidimensionnel est que, même dans la limite d'interaction faible $(\rho \xi^2)^{-1}\to 0$ {\yc au sein du superfluide}, un calcul perturbatif ne donne pas accès à la fonction énergie propre à l'ordre dominant en $\epsilon$. Ainsi, le calcul de $\Sigma(\qq,z)$ à l'ordre deux en le couplage cubique entre phonons ne reproduit pas l'amortissement du mode $\qq$ observé numériquement et le calcul à l'ordre quatre présente (après prise de limite $\epsilon\to 0$ à $\qb=\hbar c q/k_B T$ et $(\rho \xi^2)^{-1}$ fixés) une divergence non physique en $z=\varepsilon_\qq$ qui le rend inutilisable. 

Pour sortir de cette impasse, guidés par l'intuition physique, nous donnons une durée de vie finie aux phonons virtuels ({\yc ceux} présents  seulement dans l'état intermédiaire des processus collisionnels à quatre phonons).  Ceci supprime la divergence dans l'ordre quatre en la transformant en une contribution formellement du second ordre, qui {\yc vient corriger} le calcul à l'ordre deux. La théorie régularisée se trouve être en bon accord avec les simulations  dans le régime d'interaction faible.  Elle permet de prendre la limite $(\rho \xi^2)^{-1}\to 0$ dans l'expression de $\Sigma(\qq,z)$ à l'ordre dominant en $\epsilon$ et de proposer une loi d'amortissement des phonons remplaçant celle de la règle d'or. La nouvelle loi est non exponentielle, sauf dans la limite $\qb\to 0$ où elle est caractérisé par le taux (\ref{eq313}), linéaire en nombre d'onde comme la règle d'or mais avec un coefficient environ trois fois plus faible, et dans la limite $\qb\to+\infty$ où son taux (\ref{eq315}) reproduit celui de la règle d'or.\\

\noindent {\bf Remerciements~:} Alan Serafin a bénéficié d'un financement par le projet européen de recherche et d'innovation Horizon 2020 macQsimal numéro 820393. {\ycd Cet article fait partie d'un dossier thématique consacré à la médaille d'or du CNRS reçue par Jean Dalibard en décembre 2021. Les auteurs félicitent Jean pour son prix et le remercient pour ses cours lumineux au Collège de France et pour toutes les discussions enrichissantes qu'ils ont pu avoir avec lui sur les gaz d'atomes froids (Y.C. en particulier se souvient avec émotion de son enseignement sur la règle d'or de Fermi au DEA de physique quantique de l'ENS en 1987-1988).}

\appendix
\section{Prolongement analytique de $\check{\Sigma}^{(2,2)}_{\qb}(\check{\zeta})$ au demi-plan inférieur et conditions de validité de la règle d'or}
\label{app1}

Pour prolonger analytiquement du demi-plan complexe supérieur au demi-plan inférieur l'expression (\ref{eq131}) de la fonction énergie propre $\check{\Sigma}^{(2,2)}_{\qb}(\check{\zeta})$, nous procédons en deux étapes: (i) nous réduisons (\ref{eq131}) à une intégrale unique, celle d'une fonction analytique d'une variable $x$ bien choisie sur un intervalle de $\mathbb{R}$, et (ii) pour étendre cette intégrale à $\im\zetac<0$, nous déformons le chemin d'intégration sur $x$ dans le plan complexe et faisons tourner les lignes de coupure de la fonction analytique en question. En application, nous donnons des conditions de validité de la règle d'or de Fermi, nécessaires mais cependant pas suffisantes.

\paragraph{Se ramener à une seule intégrale} Dans la contribution de Landau à (\ref{eq131}), c'est-à-dire {\yc dans} la seconde intégrale, nous effectuons le changement de variable $x=\kb(\kb+\qb)$ soit $\kb=(x+\qb^2/4)^{1/2}-\qb/2$, où $x\in\mathbb{R}^+$. Dans la contribution de {\yc Belyaev}, l'intégrande est invariant par changement de $\kb$ en $\qb-\kb$, ce qui autorise à limiter l'intégration sur $\kb$ à l'intervalle $[0,\qb/2]$ au prix d'une multiplication par un facteur $2$; nous effectuons alors le changement de variable devenu injectif $x=\kb(\kb-\qb)$ soit $\kb=\qb/2-(\qb^2/4+x)^{1/2}$, où $x\in[-\qb^2/4,0]$; nous utilisons au préalable l'identité $\bar{n}_k^{\rm lin}=-(1+\bar{n}_{-k}^{\rm lin})$ pour donner au facteur d'amplification bosonique la même structure de différence de nombres d'occupation que dans Landau. Nous {\yc arrivons} dans les deux cas au même intégrande, si ce n'est que {\yc Belyaev} fait apparaître un facteur global $(-x)^{1/2}/(\zetac/u-x)^{1/2}$ là où Landau contient un facteur $x^{1/2}/(x-\zetac/u)^{1/2}$ ; on peut leur donner cependant l'expression commune $(x-\ii 0^+)^{1/2}/(x-\zetac/u)^{1/2}$ au moyen des relations $(-x)^{1/2}=\ii(x-\ii 0^+)^{1/2} \forall x<0$, $x^{1/2}=(x-\ii 0^+)^{1/2} \forall x>0$, $(\zetac/u-x)^{1/2}=\ii(x-\zetac/u)^{1/2}\forall x\in\mathbb{R}$ sachant que $\im(\zetac/u)>0$. Nous aboutissons à l'intégrale unique cherchée:
\be
\label{eq1001}
\check{\Sigma}^{(2,2)}_{\qb}(\check{\zeta})= \frac{1}{2\ii\qb}\int_{-\qb^2/4}^{+\infty}\dd x \frac{x\sqrt{x-\ii 0^+}}{\sqrt{x-\zetac/u}\sqrt{x+\qb^2/4}} \left[\frac{1}{\eee^{\sqrt{x+\qb^2/4}-\qb/2}-1}-\frac{1}{\eee^{\sqrt{x+\qb^2/4}+\qb/2}-1}\right]
\ee

\begin{SCfigure}
\begin{tikzpicture}[x=6mm,y=6mm]
\draw[thick,->,>=stealth](-7,0)--(7,0);
\node(a)at(7,0)[below]{${\rm Re\,}x$};
\draw[thick,->,>=stealth](0,-5)--(0,6);
\node(a)at(0,6)[right]{${\rm Im\,}x$};
\node(a)at(-3,0)[]{$\bullet$};
\node(a)at(-3,-0.15)[above]{$-\bar{q}^2/4$};
\draw[ultra thick](-7,0)--(-3,0);
\draw[ultra thick](-7,0.7)--(0,0.7);
\node(a)at(0,0.7)[]{$\bullet$};
\node(a)at(0,0.7)[right]{${\rm i} 0^+$};
\draw[dashed,ultra thick](-3,0)--(-7,-4);
\draw[dashed,ultra thick](0,0.7)--(4,4.7);
\draw[ultra thick](-7,2)--(3,2);
\node(a)at(3,2)[]{$\bullet$};
\node(a)at(3,2)[below]{$\check{\zeta}/u$};
\draw[dashed,ultra thick](3,2)--(6,5);
\draw[thick,color=black,->,>=stealth](-3,0)+(-180:1)arc(-180:-135:1);
\node(a)at(-4.5,-0.5)[]{$\alpha$};
\draw[thick,color=black,->,>=stealth](0,0.7)+(180:0.8)arc(180:45:0.8);
\node(a)at(-1.5,1.3)[]{$\alpha-\pi$};
\draw[thick,color=black,->,>=stealth](3,2)+(180:0.8)arc(180:45:0.8);
\node(a)at(3.2,3)[]{$\alpha-\pi$};
\draw[thick,color=blue,->,>=stealth](-3,0-0.1)--(-5,-2-0.1);
\draw[thick,color=blue](-3,0-0.1)--(-7,-4-0.1);
\draw[thick,color=red,->,>=stealth](1,-0.1)+(-180:4)arc(-180:-135:4);
\draw[thick,color=red,->,>=stealth](1,-0.1)+(-135:4)arc(-135:0:4);
\draw[->,>=stealth,color=red,thick](5,-0.1)--(6,-0.1);
\node(a)at(-0.25,0)[below]{$0$};
\end{tikzpicture}
\caption{Illustration graphique de la méthode de prolongement analytique de $\check{\Sigma}_{\qb}^{(2,2)}(\zetac)$ du demi-plan supérieur $\im\zetac>0$ au demi-plan inférieur $\im\zetac<0$ à partir de la forme intégrale (\ref{eq1001}). Courbe rouge: première déformation du chemin d'intégration sur la variable $x$ dans le plan complexe. Demi-droites {\yct noires} : lignes de singularité de l'intégrande de (\ref{eq1001}) considéré comme une fonction de $x$, en trait plein épais pour leur position originelle, en tireté épais après rotation d'un angle $\alpha$ ou $\alpha-\pi$ comme indiqué ($\alpha=\pi/4$ sur la figure). Disques noirs: points de branchement $x_0$ des fonctions racines carrées $(x-x_0)^{1/2}$ correspondantes. Demi-droite bleue: chemin d'intégration définitif sur $x$, conduisant au résultat cherché (\ref{eq140}).}
\label{fig11}
\end{SCfigure}
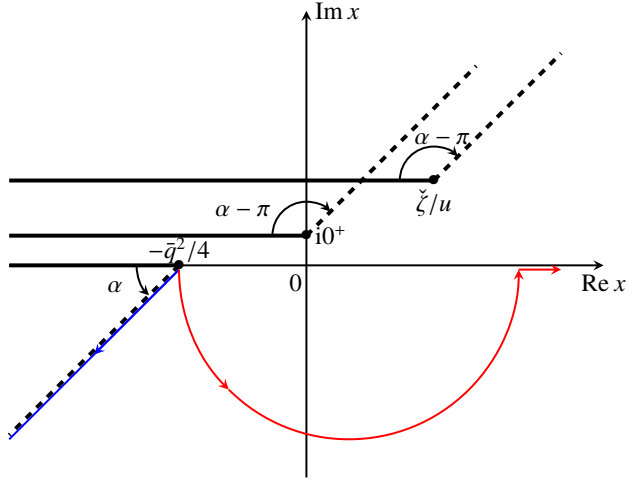

\paragraph{La prolonger analytiquement} L'intégrande de (\ref{eq1001}) est une fonction analytique de $x$ dans le plan complexe, si l'on excepte les trois lignes de singularités issues des lignes de coupure des trois racines carrées mises en jeu: $\ii 0^++\mathbb{R}^-$ au numérateur, $\zetac/u+\mathbb{R}^-$ et $-\qb^2/4+\mathbb{R}^-$ au dénominateur, en noir sur la figure \ref{fig11}.\footnote{L'intégrande admet également des pôles $x_n=2\ii n\pi(\qb+2\ii n\pi)$, $n\in\mathbb{Z}^*$, provenant des nombres d'occupation de Bose ($x=0$ n'est pas un pôle à cause du facteur $x$ au numérateur). Les déformations continues à venir du chemin d'intégration dans (\ref{eq1001}), en rouge puis en bleu sur la figure \ref{fig11}, ne passent sur aucun de ces pôles aux conditions suivantes, supposées vérifiées dans la suite (elles assurent que le pôle le plus dangereux $x_{n=-1}$ reste à gauche de la demi-droite bleue {\yc d'angle $\alpha$}): (i) $-\qb^2/4>\re x_{-1}=-(2\pi)^2$ c'est-à-dire $\qb<4\pi$, (ii) $\tan\alpha>2\pi\qb/[(2\pi)^2-\qb^2/4]$.} Ceci permet de déformer à volonté le chemin d'intégration, pourvu qu'il ne croise aucune singularité de l'intégrande, voir la ligne rouge sur la figure \ref{fig11}. On pourrait croire que ceci suffit à rendre l'intégrale analytique en $\zetac/u$ dans la région comprise entre la ligne rouge et l'axe réel ; il n'en est rien car, lorsque $\zetac/u$ entre dans cette région, la ligne de coupure (horizontale) qui lui est attachée croise nécessairement la ligne rouge. Il faut donc faire mieux, en déformant aussi les lignes de coupure de l'intégrande ; le plus simple est d'introduire les nouvelles déterminations $z_\theta^{1/2}\equiv\exp(\ii\theta/2)[\exp(-\ii\theta)z]^{1/2}$ de la racine carrée, paramétrées par l'angle $\theta\in]-\pi,\pi[$ dont on a fait tourner la ligne de coupure originelle (la demi-droite $\mathbb{R}^-$ étant envoyée sur $\exp(\ii\theta)\mathbb{R}^-$). Tant que les lignes de coupure, dans leur mouvement tournant, ne balaient pas le chemin d'intégration en rouge, l'intégrande ne change pas sur ce chemin et la valeur de l'intégrale est préservée. Nous décidons de faire tourner la ligne de coupure de $(x+\qb^2/4)^{1/2}$ d'un angle arbitraire $\alpha\in ]0,\pi/2[$, et les deux autres d'un angle $\alpha-\pi\in]-\pi,-\pi/2[$ pour qu'elles soient parallèles mais de direction opposée à la première, voir les lignes en tireté sur la figure \ref{fig11}. Enfin, nous prenons comme chemin d'intégration définitif la demi-droite en bleu sur la figure \ref{fig11}, qui suit par en-dessous la nouvelle ligne de coupure de $(x+\qb^2/4)^{1/2}$,\footnote{On a pu refermer par un arc de cercle de rayon $R\to +\infty$, de contribution $\to 0$ puisque l'intégrande y est exponentiellement petit en $\sqrt{R}$.} et que nous paramétrons comme suit, $x=\exp(\ii\alpha)(-v-\ii 0^+)-\qb^2/4$, $v\in\mathbb{R}^+$. Alors
\be
\label{eq1002}
(x+\qb^2/4)^{1/2}_\alpha=-\ii\eee^{\ii\alpha/2}\sqrt{v} \ \  ; \ \  (x-\zetac/u)^{1/2}_{\alpha-\pi}=-\ii\eee^{\ii\alpha/2}[v+\eee^{-\ii\alpha}(\qb^2/4+\zetac/u)]^{1/2} \ \  ;\ \  (x-\ii 0^+)^{1/2}_{\alpha-\pi} = -\ii\eee^{\ii\alpha/2} {\yc(}v+\eee^{-\ii\alpha}\qb^2/4{\yc)}^{1/2}
\ee
où tous les $\ii 0^+$ ont pu être omis sans mal dans les seconds membres. On peut désormais faire descendre le nombre complexe $\zetac/u$ continûment dans le demi-plan inférieur en restant constamment à droite ou à gauche de la demi-droite bleue sans que cette dernière ne soit croisée par la ligne de coupure attachée à $\zetac/u$: on a bien prolongé analytiquement $\check{\Sigma}_{\qb}^{(2,2)}(\zetac)$ à $\im\zetac<0$. Il reste à poser $v=\kb^2$, $\kb\in\mathbb{R}^+$, pour arriver au résultat annoncé (\ref{eq140}) avec $\phi=\alpha-\pi\in]-\pi,-\pi/2[$.

\paragraph{Application} Maintenant que nous savons, au vu de l'équation (\ref{eq1001}), que $\check{\Sigma}^{(2,2)}_{\qb}(\zetac)$ n'a pas de point de branchement en $\zetac=0$, nous pouvons calculer sa valeur et sa dérivée en $\zetac=\ii 0^+$. Pour sa valeur, le plus simple est faire tendre $\zetac$ vers zéro par parties imaginaires $>0$ sous le signe intégral dans la forme (\ref{eq131}); on intègre sur $\kb$ après développement des nombres d'occupation en série entière de la variable $\exp(-\kb)$, ce qui fait apparaître des fonctions de Bose $g_\alpha(z)$ prises en $z=1$ ou $z=\exp(-\qb)$, les secondes ayant le bon goût de se compenser dans le résultat final. Pour la dérivée, le plus simple est d'utiliser le prolongement analytique (\ref{eq140}), dont l'intégrande ne présente aucune singularité sur le chemin d'intégration pour $\zetac=0$; on peut alors dériver par rapport à $\zetac$ puis prendre la limite $\zetac\to 0$ sous le signe intégral.\footnote{Si l'on s'en tient à la forme (\ref{eq131}), on ne peut pas dériver sous le signe intégral sans faire apparaître une intégrale non absolument convergente. On peut procéder de manière hybride après le choix d'une coupure infrarouge infinitésimale $\kappa$ sur $\kb$: dans la forme (\ref{eq131}), limitée à $\kb>\kappa$ dans chaque contribution, on dérive sous le signe intégral puis on fait tendre $\zetac$ vers $0$ ; dans la forme (\ref{eq1001}), limitée à $-\kappa\qb<x<\kappa\qb$, on garde l'expression exacte du facteur $(x-\ii 0^+)^{1/2}/(x-\zetac/u)^{1/2}$ mais on développe le reste de l'intégrande en puissances de $x$ (l'ordre zéro suffit en fait), et on calcule analytiquement l'intégrale résultante pour $\zetac=s+\ii 0^+$, $s$ réel (au besoin en termes de fonctions hypergéométriques) puis sa dérivée par rapport à $s$ en $s=0$. Le résultat final est le même.} Nous obtenons finalement
\be
\label{eq141}
\check{\Sigma}^{(2,2)}_{\qb}(\check{\zeta}=\ii 0^+) = \frac{1}{\ii}\left(\frac{\qb^2}{12}+2\zeta(2)\right) \quad; \quad
\frac{\dd}{\dd\check{\zeta}} \check{\Sigma}^{(2,2)}_{\qb}(\check{\zeta}=\ii 0^+) = \frac{1}{u\qb} \left(\frac{\pi}{2}+\frac{\ii\qb}{4}\right)
\ee
La valeur de la fonction n'est qu'une réécriture du résultat (\ref{eq033}) sous l'effet des changements de variable dans (\ref{eq130},\ref{eq131}). Sa dérivée intervient dans une condition de validité bien connue de la règle d'or \cite{CCTbordeaux}, $|\im(\dd/\dd\zeta)\Sigma_{\qq}(\zeta=\ii 0^+)|\ll 1$, qui s'écrit ici $1/(4u)\ll 1$ et impose, sans surprise, $u\gg 1$. Une condition plus précise exige que le résidu $Z$ de la fonction de Green, qui apparaît en facteur de l'exponentielle $\exp(-\zetac_{\qb}\check{t})$ dans l'expression exacte (\ref{eq2001}) du signal $s^{(2,2)}(t)$, soit très proche de la valeur un prise dans la règle d'or, ce qui ajoute la contrainte $|\re(\dd/\dd\zeta)\Sigma_{\qq}(\zeta=\ii 0^+)|\ll 1$, ici $\qb\gg \pi/(2 u)$. Comme le montre la figure \ref{fig1}d, sur laquelle $u\simeq 18,\!14$, ces conditions ne sont pas suffisantes. {\ycd Les sections \ref{sec3.3} et \ref{sec4} permettent de comprendre pourquoi (en substance, une singularité apparaît en $\zeta=\ii 0^+$ dans la contribution $\Sigma_\qq^{(4)}(\zeta)$ d'ordre 4 en $H_3$, ce qui rend le présent calcul à l'ordre 2 non pertinent dans la limite $\epsilon\to 0$)}.

\section{Premier écart du signal à l'exponentielle de la règle d'or pour la théorie à l'ordre deux en $H_3$}
\label{app2}

Nous établissons ici le résultat (\ref{eq151}) de la section \ref{sec2.2}, valable dans la limite d'interaction faible $u\to+\infty$ {\yc à $\check{t}$ et $Q=u^{1/2}\qb$ fixés comme dans (\ref{eq150})}, en partant de l'écriture (\ref{eq132}) du signal. À cette fin, guidés par la figure \ref{fig5}b, nous appliquons d'abord le théorème des résidus au contour formé de la réunion du chemin $C_+$, des portions de cercle à l'infini et du lacet autour de la ligne de coupure $-Q^2/4-\eee^{\ii(\phi+\pi)}\mathbb{R}^+$: l'intégrale de $\exp(-\ii\check{\zeta}\check{t})\check{\mathcal{G}}_{\qb}^{(2,2)}(\zetac)$ sur ce contour vaut $2\ii\pi$ fois le résidu en $\zetac=\zetac_{\qb}$, pôle de la fonction de Green prolongée analytiquement au demi-plan inférieur au moyen de (\ref{eq140}). Le signal (\ref{eq132}) s'écrit donc de manière exacte comme somme des contributions du pôle $\zetac_{\qb}$ et du lacet:
\be
\label{eq2001}
s^{(2,2)}(t) = \frac{\exp(-{\yct\ii}\zetac_{\qb}\check{t})}{1-\frac{\dd}{\dd\zetac}\check{\Sigma}^{(2,2)}_{\qb\downarrow}(\zetac_{\qb})}
+\eee^{\ii Q^2\check{t}/4} \int_0^{\eee^{\ii(\phi+\pi)}(+\infty)} \frac{\dd z}{2\ii\pi} \frac{(\check{\Sigma}_- - \check{\Sigma}_+)\exp(\ii z \check{t})}{(-Q^2/4-z-\check{\Sigma}_-)(-Q^2/4-z-\check{\Sigma}_+)}
\ee
{\yc où la} variable complexe $z$ va en ligne droite de $0$ à $\exp[\ii(\phi+\pi)](+\infty)$ et l'on note $\check{\Sigma}_{\pm}\equiv\check{\Sigma}^{(2,2)}_{\qb\downarrow}(-Q^2/4-z+\eee^{\ii(\phi+\pi)}\ii 0^\pm)$ les valeurs du prolongement analytique (\ref{eq140}) de la fonction énergie propre immédiatement à gauche (signe $+$) ou à droite (signe $-$) de la ligne de coupure. Nous faisons ensuite bon usage de la propriété fondamentale\footnote{Le terme dominant s'obtient en écrivant la différence au premier membre de (\ref{eq2003}) comme une seule intégrale sur $\kb$ au moyen de (\ref{eq140}), en la multipliant par $u^{1/2}$, puis en faisant le changement de variable $\kb=K/u^{1/2}$ avant de passer à la limite $u\to +\infty$ sous le signe intégral.}
\be
\label{eq2003}
\check{\Sigma}^{(2,2)}_{\qb\downarrow}(\zetac)-\check{\Sigma}^{(2,2)}_{\qb\downarrow}(0)\stackrel{Q\,\mbox{\scriptsize fix\'e}}{\underset{u\to+\infty}{=}} \frac{\eee^{\ii\phi/2}}{\ii u^{1/2}} \int_0^{+\infty} \dd K \left\{\frac{(K^2-\eee^{-\ii\phi}Q^2/4)^{1/2}}{[K^2-\eee^{-\ii\phi}(\zetac+Q^2/4)]^{1/2}}-1\right\}+O(1/u)
\ee
Dans la contribution du lacet à (\ref{eq2001}), à l'ordre dominant en $u^{-1/2}$, nous pouvons ainsi remplacer au dénominateur $\check{\Sigma}_\pm$ par $\check{\Sigma}^{(2,2)}_{\qb\downarrow}(0)\to -2\ii\zeta(2)=-\ii\pi^2/3$ d'après (\ref{eq141}), et au numérateur la différence $\check{\Sigma}_--\check{\Sigma}_+$ par son équivalent, qui s'écrit
\be
\label{eq2004}
\check{\Sigma}_--\check{\Sigma}_+ \stackrel{Q\,\mbox{\scriptsize fix\'e}}{\underset{u\to+\infty}{\sim}} \frac{\eee^{\ii\phi/2}}{\ii u^{1/2}} \int_0^{+\infty} \dd K
\left(K^2-\eee^{-\ii\phi}Q^2/4\right)^{1/2}\left[\frac{1}{(K^2-x+\ii 0^-)^{1/2}}-\frac{1}{(K^2-x+\ii 0^+)^{1/2}}\right]
\ee
sachant que $z=\eee^{\ii(\phi+\pi)}x$, $x>0$. L'intégrande dans (\ref{eq2004}) est non nul seulement pour $K<x^{1/2}$, ce qui incite à poser $K=x^{1/2}\sin\alpha$, $\alpha\in [0,\pi/2]$, d'autant plus que ceci fait disparaître les racines carrées au dénominateur: nous retrouvons $S_{\rm lacet}(\check{t})$ dans (\ref{eq152}). Dans l'équation implicite définissant le pôle, réécrite en mettant à part son approximation d'ordre zéro, $\zetac_{\qb}-\check{\Sigma}^{(2,2)}_{\qb\downarrow}(0)=\check{\Sigma}^{(2,2)}_{\qb\downarrow}(\zetac_{\qb})-\check{\Sigma}^{(2,2)}_{\qb\downarrow}(0)$, il suffit au second membre d'utiliser (\ref{eq2003}) avec $\zetac=-\ii\pi^2/3$; le choix commode $\phi={\ycd(-\pi)}^{+}$ (dont le résultat ne dépend pas) redonne $\delta\zetac_{\qb}$ dans (\ref{eq154}). Enfin, pour obtenir le premier écart à un du résidu, il suffit de dériver (\ref{eq2003}) par rapport à $\zetac$ (sous le signe intégral au second membre), de remplacer $\zetac$ par $-\ii\pi^2/3$ puis de prendre $\phi={\ycd(-\pi)}^+$~: {\yct on retrouve} $\delta\check{Z}$ dans (\ref{eq154}).

\section{Les sommes doubles de Matsubara dans la théorie à l'ordre quatre en $H_3$}
\label{app3}

Nous calculons ici explicitement les sommes de Matsubara contribuant à (\ref{eq212},\ref{eq232}) dans la section \ref{sec3.1} et montrons comment obtenir leurs formes simplifiées (\ref{eq213},\ref{eq233}) dans la limite des petits angles. Le calcul repose sur la généralisation de la relation (\ref{eq122}) à un nombre quelconque $J$ de facteurs dans le sommande, qui se démontre comme dans la note \ref{notedemo}:
\be
\label{eq3000}
\sum_{n\in\mathbb{Z}} \frac{1}{\prod_{j=1}^{J} (2\ii\pi n-\vepsb_j)} = \sum_{j=1}^{J} \frac{-\bar{n}_{\veps_j}}{\prod_{\stackrel{\mbox{\scriptsize$k\!=\!1$}}{k\neq j}}^{J} (\vepsb_j-\vepsb_k)} \quad\forall J\geq 2
\ee

\paragraph{Diagrammes à boucle interne} Écrivons la somme apparaissant dans (\ref{eq212}) après sortie d'un facteur $(k_B T)^5$ au dénominateur (on a posé $\veps_{\kk_j}=k_B T\vepsb_j$) et élimination des énergies de Matsubara au profit des variables indépendantes $\omega=2\ii\pi k_B T n$ et $\omega'=2\ii\pi k_B T n'$ sur le modèle des équations (\ref{eq210},\ref{eq211}), pour une énergie complexe $z$ valant temporairement $\Omega=k_B T\bar{\Omega}\in 2\ii\pi k_B T\mathbb{Z}$:
\be
\label{eq3001}
S_{B_i}(z)=(k_BT)^{-5}\sum_{n,n'\in\mathbb{Z}} \frac{1}{s_1(\bar{\Omega}-2\ii\pi s_2 n)-\vepsb_1}\frac{1}{2\ii\pi n-\vepsb_2}\frac{1}{2\ii\pi n'-\vepsb_3}\frac{1}{2\ii\pi s_4(s_2 n - s_3 n')-\vepsb_4} \frac{1}{2\ii\pi s_2 s_5 n-\vepsb_5}
\ee
Calculons au moyen de (\ref{eq122}) la somme sur $n'$ du produit des deux facteurs qui en dépendent, puis faisons disparaître la variable $\bar{\Omega}$ des nombres d'occupation comme dans la première égalité de (\ref{eq123}). Après le changement de variable astucieux $n\to s_2 n$, qui n'affecte pas le domaine de sommation puisque $s_2=\pm 1$, il reste
\be
\label{eq3002}
S_{B_i}(z)= (k_BT)^{-5}\sum_{n\in\mathbb{Z}} \frac{s_1s_2s_4s_5(\bar{n}_{\veps_3}-\bar{n}_{-s_3s_4\veps_4})}{[2\ii\pi n-(\bar{\Omega}-s_1\vepsb_1)](2\ii\pi n-s_2\vepsb_2)(2\ii\pi n -s_5 \vepsb_5)[2\ii\pi n-(s_3\vepsb_3+s_4\vepsb_4)]}
\ee
Nous avons pris soin de sortir tous les facteurs nécessaires pour mettre le résultat sous une forme simple, comme au premier membre de la relation (\ref{eq3000}), ce qui permet d'utiliser facilement cette dernière:
\begin{multline}
\label{eq3003}
S_{B_i}(z)=-(k_B T)^{-2}s_1s_2s_4s_5(\bar{n}_{\veps_3}-\bar{n}_{-s_3s_4\veps_4})\Bigg\{\frac{1}{z-z_2}\left[\frac{\bar{n}_{-s_1\veps_1}}{(z-z_1)(z-z_3)}-\frac{\bar{n}_{s_3\veps_3+s_4\veps_4}}{(z_2-z_1)(z_2-z_3)}\right] \\
+\frac{1}{z_1-z_3}\left[\frac{\bar{n}_{s_2\veps_2}}{(z_1-z)(z_1-z_2)}-\frac{\bar{n}_{s_5\veps_5}}{(z_3-z)(z_3-z_2)}\right]\Bigg\}
\end{multline}
À ce stade, nous avons prolongé analytiquement en la variable $z$ sur $\mathbb{C}\setminus\mathbb{R}$ et introduit les énergies des trois états intermédiaires (repérés par les droites verticales rouges sur la partie gauche de la figure \ref{fig7}) comptées algébriquement, $z_1=s_1\veps_1+s_2\veps_2$, $z_2=s_1\veps_1+s_3\veps_3+s_4\veps_4$ et $z_3=s_1\veps_1+s_5\veps_5$. Dans le cas $s_2=s_5$, on a aussi $\veps_2=\veps_5$ et $z_1=z_3$ puisque $\kk_2$ est alors égal à $\kk_5$ ; ceci fait apparaître une division par zéro dans la seconde contribution de (\ref{eq3003}) entre accolades, que nous prolongeons donc par continuité avec la règle de l'Hôpital. Après d'habiles réarrangements, il vient:\footnote{Nous utilisons en particulier l'identité $(z_1-z_2)^{-2}(z-z_2)^{-1}+(z_1-z_2)^{-2}(z_1-z)^{-1}+(z_1-z)^{-2}(z_1-z_2)^{-1}=(z-z_1)^{-2}(z-z_2)^{-1}$.}
\be
\label{eq3004}
S_{B_i}^{s_2=s_5}(z)=-(k_B T)^{-2}s_1s_4(\bar{n}_{\veps_3}-\bar{n}_{-s_3s_4\veps_4})\left[\frac{\bar{n}_{-s_1\veps_1}-\bar{n}_{s_2\veps_2}}{(z-z_1)^2(z-z_2)}+\frac{\bar{n}_{s_2\veps_2}-\bar{n}_{s_3\veps_3+s_4\veps_4}}{(z_2-z_1)^2(z-z_2)}+\frac{\bar{n}_{s_2\veps_2}(1+\bar{n}_{s_2\veps_2})}{k_B T(z_2-z_1)(z_1-z)}\right]
\ee
Remarquons que le facteur global $(k_B T)^{-2}$ de (\ref{eq3003},\ref{eq3004}) se simplifie avec le facteur $(-k_B T)^2$ dans (\ref{eq212}). Pour terminer, prenons la limite des petits angles $\epsilon\to 0$ dans (\ref{eq3004}), dans laquelle $z$, $z_1$ et $z_2$ s'écrivent $\veps_q+O(k_B T\epsilon^2)$. Entre les crochets, la deuxième contribution est sous-dominante parce que $s_3\veps_3+s_4\veps_4=s_2\veps_2+z_2-z_1=s_2\veps_2+O(k_B T\epsilon{\yc^2)}$, si bien que son numérateur tend vers zéro; la troisième l'est aussi car son dénominateur est d'ordre $(k_B T)^3\epsilon^4$ au lieu de $(k_B T)^3\epsilon^6$ comme les autres. Nous retrouvons (\ref{eq213}).

\paragraph{Diagrammes à pont} Nous procédons comme précédemment, avec les mêmes notations. La somme double dans (\ref{eq232}) s'écrit
\be
\label{eq3010}
S_{P_i}(z)=(k_B T)^{-5} \sum_{n,n'\in\mathbb{Z}} \frac{1}{2\ii\pi n-\vepsb_1} \frac{1}{s_2(\bar{\Omega}-2\ii\pi s_1 n)-\vepsb_2}\frac{1}{2\ii\pi n'-\vepsb_3}\frac{1}{s_4(\bar{\Omega}-2\ii\pi s_3 n')-\vepsb_4} \frac{1}{2\ii\pi s_5(s_1 n-s_3 n')-\vepsb_5}
\ee
Au moyen de la relation (\ref{eq3000}) avec $J=3$, nous calculons la somme sur $n'$, décomposons le résultat en éléments simples pour la variable $2\ii\pi n$ puis effectuons le changement de variable $n\to s_1 n$, si bien que
\be
\label{eq3011}
S_{P_i}(z)=\frac{(k_B T)^{-5}s_1s_2s_4s_5}{\bar{\Omega}-s_3\vepsb_3-s_4\vepsb_4}\sum_{n\in\mathbb{Z}} \frac{1}{(2\ii\pi n-s_1\vepsb_1)[2\ii\pi n-(\bar{\Omega}-s_2\vepsb_2)]} \left[\frac{\bar{n}_{\veps_3}-\bar{n}_{-s_3s_5\veps_5}}{2\ii\pi n-(s_3\veps_3+s_5\veps_5)}+\frac{\bar{n}_{-s_3s_5\veps_5}-\bar{n}_{-s_3s_4\veps_4}}{2\ii\pi n-(\bar{\Omega}+s_5\veps_5-s_4\veps_4)}\right]
\ee
Nous utilisons à nouveau (\ref{eq3000}) puis prolongeons analytiquement sur $z$ à $\mathbb{C}\setminus\mathbb{R}$; il y a deux termes entre les grands crochets de (\ref{eq3011}) donc deux contributions, dites de type I et de type II dans cet ordre:
\bea
\label{eq3012}
S_{P_i^{\rm I}}(z)&=&\frac{(k_B T)^{-2}s_1s_2s_4s_5(\bar{n}_{\veps_3}-\bar{n}_{-s_3s_5\veps_5})}{z_2-z}\left[
\frac{\bar{n}_{s_1\veps_1}}{(z_1-z)(z_1-z_4)}+\frac{\bar{n}_{-s_2\veps_2}}{(z-z_1)(z-z_4)}+\frac{\bar{n}_{s_3\veps_3+s_5\veps_5}}{(z_4-z_1)(z_4-z)}\right] \\
\label{eq3013}
S_{P_i^{\rm II}}(z)&=&\frac{(k_B T)^{-2}s_1s_2s_4s_5(\bar{n}_{-s_3s_5\veps_5}-\bar{n}_{-s_3s_4\veps_4})}{z_2-z}\left[
\frac{\bar{n}_{s_1\veps_1}}{(z_1-z)(z_3-z)}+\frac{\bar{n}_{-s_2\veps_2}}{(z-z_1)(z_3-z_1)}+\frac{\bar{n}_{s_5\veps_5-s_4\veps_4}}{(z-z_3)(z_1-z_3)}\right]
\eea
avec les énergies algébriques des états intermédiaires repérés par les droites rouges sur la partie droite de la figure \ref{fig7}, $z_1=s_1\veps_1+s_2\veps_2$ et $z_2=s_3\veps_3+s_4\veps_4$ communes aux deux types, $z_4=s_2\veps_2+s_3\veps_3+s_5\veps_5$ de type I et enfin $z_3=s_1\veps_1+s_4\veps_4-s_5\veps_5$ de type II. Ces variables ne sont pas indépendantes puisque $z_1+z_2=z_3+z_4$. Il reste à prendre la limite des petits angles $\epsilon\to 0$, dans laquelle {\yc $z=\veps_q+O(k_B T \epsilon^2)$ et} $z_j=\veps_q+O(k_B T \epsilon^2)$, $1\leq j\leq 4$. Alors $s_3\veps_3+s_5\veps_5=s_1\veps_1+O(k_BT\epsilon^2)$ et l'on peut, à l'ordre dominant, remplacer $\bar{n}_{s_3\veps_3+s_5\veps_5}$ par $\bar{n}_{s_1\veps_1}$ dans (\ref{eq3012}) ; en regroupant par nombre d'occupation et en simplifiant, on retrouve la première contribution au second membre de (\ref{eq233}). De même, on remplace $\bar{n}_{s_5\veps_5-s_4\veps_4}$ par $\bar{n}_{-s_2\veps_2}$ à l'ordre dominant dans (\ref{eq3013}), ce qui reproduit la seconde contribution à (\ref{eq233}).

\section{Une divergence linéaire de $\tilde{\Sigma}_{\qb}^{(4,2)}(\zetat)$ en dehors de l'origine}
\label{app4}

Comme nous le montrons dans cette annexe, la composante $\tilde{\Sigma}_{\qb}^{(4,2)}(\zetat)$ de la fonction énergie propre d'ordre quatre en $H_3$ et d'ordre deux en $\epsilon$, obtenue en sommant toutes les contributions (\ref{eq220}), (\ref{eq240}), (\ref{eq242}) des diagrammes d'ordre quatre de la section \ref{sec3.1}, diverge linéairement lorsque $\zetat\to\zetat_0\equiv -3\gamma\qb^3/32$ (ici par parties imaginaires positives): 
\be
\label{eq4000}
\tilde{\Sigma}_{\qb}^{(4,2)}(\zetat) \stackrel{\im\zetat>0}{\underset{\zetat\to \zetat_0}{\sim}} \frac{\qb^3(\bar{n}^{\rm lin}_{q/2}+{1}/{2})}{24\ii\pi^2\gamma(\zetat-\zetat_0)} \left(\frac{9(1+\Lambda)^2}{8\rho\xi^2}\right)^2\left[\int_0^{\qb/2} \dd\kb' \kb'(\kb'-\qb/2)(\bar{n}^{\rm lin}_{k'}-\bar{n}^{\rm lin}_{k'-{q}/{2}})+2\int_{\qb/2}^{\eta} \dd\kb'\kb'(\kb'-\qb/2)(\bar{n}^{\rm lin}_{k'}-\bar{n}^{\rm lin}_{k'-{q}/{2}})\right]
\ee
Ceci résulte de deux phénomènes. (i) À nombre d'onde réduit $\kb$ fixé, l'intégrale angulaire (\ref{eq221}) dans les diagrammes à boucle interne diverge en $\zetat=A_1^B$ avec un exposant $3/2$ car, pour cette valeur de $\zetat$, le dénominateur de l'intégrande admet $\thetat=0$ comme racine quadruple. Pour le voir quantitativement, on peut utiliser l'identité $\int_{\mathbb{R}}\dd\thetat/(z-\thetat^2)^2=-\ii\pi\,{\ycd\sig(\im z)}/(2 z^{3/2}), \forall z\in\mathbb{C}\setminus\mathbb{R}$, avec $z=(\zetat-A_1^B)/B_1^B$. (ii) Cette divergence {\yc en} $(\zetat-A_1^B)^{-3/2}$ se produit à une position $A_1^B$ dépendant de $\kb$, comme le montre l'équation (\ref{eq222}), donc elle est en général effacée par l'intégration sur $\kb$ à $\zetat$ fixé dans (\ref{eq220}), sauf si $\zetat$ coïncide avec un extrémum parabolique de $A_1^B(\kb)$ sur le domaine d'intégration $\mathcal{D}_1^B$. Pour le voir, choisissons un nombre d'onde réduit $\kb_0$ arbitraire dans $\mathcal{D}_1^B$, posons $\zetat=w+A_1^B(\kb_0)$ et passons à la limite $w\to 0$ par parties imaginaires positives dans (\ref{eq220}): le résultat ne peut diverger que si la valeur de l'intégrale 
\be
\label{eq4005}
\sigma_{\kb_0}(w)=\int_{\kb_0-\veps}^{\kb_0+\veps} \frac{\dd\kb}{[w+A_1^B(\kb_0)-A_1^B(\kb)]^{3/2}}
\ee
diverge dans cette limite, où $\veps>0$ est fixé. Comme on peut prendre $\veps$ aussi petit que l'on veut sans changer la conclusion, on peut remplacer $A_1^B(\kb)-A_1^B(\kb_0)$ dans (\ref{eq4005}) par son développement limité en $\kb_0$ à l'ordre dominant non nul. Si $(\dd A_1^B/\dd \kb)(\kb=\kb_0)\neq 0$, un développement à l'ordre un suffit, et il n'y a pas de divergence car $\int_{\mathbb{R}} \dd \kappa/(w-\kappa)^{3/2}=0$.\footnote{Si $\kb_0$ se trouve au bord de $\mathcal{D}_1^B$, le voisinage de $\kb_0$ se réduit à $[\kb_0,\kb_0+\veps]$ ou $[\kb_0-\veps,\kb_0]$ et il y a divergence en racine carrée car $\int_0^{+\infty} \dd \kappa/(w-\kappa)^{3/2}=-2/w^{1/2}$. Cette divergence est sous-dominante et ignorée ici.} Sinon, il faut développer à l'ordre deux, et il y a divergence linéaire $[\zetat-A_1^B(\kb_0)]^{-1}$ comme l'annonçait (\ref{eq4000}) car
\be
\label{eq4001}
\int_{\mathbb{R}} \frac{\dd \kappa}{(w-\kappa^2)^{3/2}} = \frac{2}{\ii w} \quad\mbox{et}\quad \int_{\mathbb{R}} \frac{\dd \kappa}{(w+\kappa^2)^{3/2}} = \frac{2}{w}\quad\forall w\in \mathbb{C}, \im w>0
\ee

Donnons plus de détails sur l'exemple du diagramme $B_1$ de la figure \ref{fig8}, dans lequel les signes $s_i$ des orientations des lignes internes valent tous $1$, le domaine d'intégration sur $\kb$ vaut $\mathcal{D}_1^B=[0,\qb]$ et celui sur $\kb'$ à $\kb$ fixé vaut $\mathcal{D}_2^B(\kb)=[0,\kb]$. On remarque que $A_1^B(\kb)$ atteint son minimum en $\kb_0=\qb/2$, à l'intérieur de $\mathcal{D}_1^B$ donc avec une dérivée nulle, et que ce minimum vaut justement $\zetat_0=-3\gamma\qb^3/32$ comme dans (\ref{eq4000}). Pour obtenir un équivalent de $\tilde{\Sigma}_{B_1}^{(4,2)}(\qb,\zetat)$ lorsque $\zetat\to\zetat_0$, nous pouvons traiter $\thetat^2$, $(\kb-\qb/2)^2$ et $w=\zetat-\zetat_0$ comme des infiniment petits du même ordre. Ainsi, dans l'intégrale angulaire (\ref{eq221}), nous pouvons remplacer $\zetat$ par $\zetat_0+\ii 0^+$ et $\thetat^2$ par $0$ dans le second facteur au dénominateur, et remplacer partout $\kb$ par $\kb_0=\qb/2$ sauf dans le coefficient $A_1^B(\kb)$ que l'on écrit $A_1^B(\kb)=\tilde{\zeta}_0+(3\gamma\qb/8)(\kb-\qb/2)^2$. Après calcul de l'intégrale sur $\phi$ puis sur $\thetat$ par le théorème des résidus, il reste
\be
\label{eq4002}
I(\kb,\kb') \sim \frac{-8\pi^2}{(6\gamma\qb^3)^{1/2} \kb'}\frac{1}{[w-(3\gamma/8)\qb(\kb-\qb/2)^2]^{3/2}}
\ee
ce que l'on peut reporter dans l'intégrande de (\ref{eq220}), en remplaçant de plus $\kb$ par $\kb_0=\qb/2$ dans tous les autres facteurs et dans la borne supérieure d'intégration sur $\kb'$. {\yc On change ensuite l'ordre d'intégration.} Dans l'intégrale {\yc maintenant intérieure} sur $\kb$ à $\kb'$ fixé, il ne reste plus que
\be
\label{eq4003}
\int_0^{\qb} \frac{\dd \kb}{[w-(3\gamma/8)\qb(\kb-\qb/2)^2]^{3/2}}\sim \left(\frac{8}{3\gamma\qb}\right)^{1/2}\int_{-\infty}^{+\infty} \frac{\dd\kappa}{(w-\kappa^2)^{3/2}}=\left(\frac{8}{3\gamma\qb}\right)^{1/2} \frac{2}{\ii w}
\ee
où nous avons effectué le changement de variable $\kb=\qb/2+[8/(3\gamma\qb)]^{1/2}\kappa$, étendu les bornes d'intégration sur $\kappa$ à $\pm\infty$ sans changer le comportement dominant et utilisé l'équation (\ref{eq4001}). Nous obtenons finalement la partie divergente du diagramme $B_1$:
\be
\label{eq4004}
\tilde{\Sigma}_{B_1}^{(4,2)}(\qb,\zetat) \stackrel{\im\zetat>0}{\underset{\zetat\to \zetat_0}{\sim}} \frac{\qb^3(\bar{n}^{\rm lin}_{q/2}+{1}/{2})}{24\ii\pi^2\gamma(\zetat-\zetat_0)} \left(\frac{9(1+\Lambda)^2}{8\rho\xi^2}\right)^2\int_0^{\qb/2} \dd\kb' \kb'(\kb'-\qb/2)(\bar{n}^{\rm lin}_{k'}-\bar{n}^{\rm lin}_{k'-{q}/{2}})
\ee
Le diagramme $B_2$ se traite de la même façon et donne une contribution divergente similaire (avec un préfacteur double et l'intégrale sur $\kb'$ allant de $\qb/2$ à $\eta$). Les diagrammes $B_3$ et $B_4$ sont sous-dominants en $\epsilon$ car $s_5\neq s_2$ et doivent être omis ici. Les diagrammes $B_5$ et $B_6$ conduisent à un coefficient $A_1^B(\kb)=(3\gamma/8)\qb\kb(\kb+\qb)$ strictement croissant sur l'intervalle d'intégration $\kb\in[0,\eta-\qb]$ donc ne peuvent contribuer à la divergence linéaire (\ref{eq4000}). Les diagrammes $B_7$ et $B_8$ ont le même problème, avec $A_1^B(\kb)=(3\gamma/8)\qb\kb(\kb-\qb)$ et $\mathcal{D}_1^B=[\qb,\eta]>\qb/2$. Il reste à regrouper les contributions de $B_1$ et $B_2$ pour obtenir (\ref{eq4000}), dont l'expression entre crochets dans le cas du champ quantique tend vers $-\qb(16\pi^2+\qb^2)/48$ lorsque $\eta\to+\infty$. 

Pour terminer, expliquons pourquoi les diagrammes à pont ne contribuent pas à la divergence linéaire (\ref{eq4000}) sur l'exemple du diagramme $P_1$ (tous les $s_i=1$, $\mathcal{D}_1^P=[0,\qb]$ et $\mathcal{D}_2^P(\kb)=[0,\kb]$, type II). Comme l'avait remarqué la section \ref{sec3.3}, l'intégrale angulaire (\ref{eq243}) diverge linéairement en {\yc un certain} $\zetat=X_0(\kb,\kb')$ si les lieux d'annulation des trois facteurs au dénominateur de (\ref{eq243}) se croisent en un même point dans le plan $(\tilde{\theta},\tilde{\theta}')$. La condition de triple intersection {\ycd dans le cas général $X_0\neq 0$} s'obtient en remplaçant tous les coefficients $A_i^P$ par $A_i^P-X_0$ dans (\ref{eq281}); l'équation du second degré sur $X_0$ qui en résulte admet la solution nulle évidente, à laquelle s'était implicitement limitée la section \ref{sec3.3}, et une deuxième racine réelle, $X_0(\kb,\kb')=-3\gamma \kb(\kb-\qb)\kb'(\kb'-\qb)/(2\qb)$ ; cette dernière conduit à une triple intersection dans le plan $(\tilde{\theta},\tilde{\theta}')$, aux points de coordonnées $\pm[\sqrt{3\gamma}/(2\qb)]((\qb-\kb)(\qb-2\kb'),-(\qb-2\kb)(\qb-\kb'))$, si on ajoute les contraintes $\kb>\qb/2$ et $\kb'<\qb/2$ pour avoir les bons signes dans les coefficients des petits écarts $(\tau,\tau')$ à ces points (sans quoi, comme dans (\ref{eq282}), il n'y a pas de divergence). Alors, en procédant comme dans (\ref{eq282}), on obtient la partie divergente de l'intégrale angulaire (pour $\zetat\to X_0$ par parties imaginaires positives),
\be
\label{eq4006}
K_{\rm div}(\kb,\kb') = \frac{-32\pi^2}{3\gamma\qb\kb\kb'(\kb-\kb')[\zetat-X_0(\kb,\kb')]}
\ee
à intégrer sur $(\kb,\kb')\in[\qb/2,\qb]\times[0,\qb/2]$ après multiplication par la même fonction de $(\kb,\kb')$ que dans (\ref{eq242}). Or $X_0(\kb,\kb')$ admet un minimum parabolique, situé en $(\kb,\kb')=(\qb/2,\qb/2)$ et valant précisément $\zetat_0$ ! La singularité (\ref{eq4006}) survivra donc à l'intégration sur les nombres d'onde si $\zetat\to \zetat_0$ comme dans (\ref{eq4000}). Effectuons alors le changement de variables $(\kb=\kappa+\qb/2,\kb'=-\kappa'+\qb/2)$ et traitons $w=\zetat-\zetat_0$, $\kappa^2$ et $\kappa'^2$ comme des infiniment petits du même ordre. En écrivant la fonction en facteur de $[\zetat-X_0(\kb,\kb')]^{-1}$ à l'ordre dominant en $\kappa$ et $\kappa'$ et en remplaçant le pôle $X_0(\kb,\kb')$ par son approximation quadratique en $\kappa$ et $\kappa'$ près du minimum, nous obtenons l'équivalent
\be
\label{eq4008}
\tilde{\Sigma}^{(4,2)}_{P_1^{\rm II}}(\qb,\zetat)\stackrel{\im w>0}{\underset{w\to 0}{\sim}} (\qb/2)^{\ycd 4} \left(\frac{9(1+\Lambda)^2}{8\rho\xi^2}\right)^2 (2\bar{n}^{\rm lin}_{q/2}+1)\frac{(-32\pi^2)}{3\gamma(2\pi)^4} \int_0^{\qb/2}\dd\kappa \int_0^{\qb/2} \dd\kappa' \frac{1}{\kappa+\kappa'} \frac{1}{w-(3\gamma\qb/8)(\kappa^2+\kappa'^2)}
\ee
On peut étendre les bornes d'intégration $\qb/2$ à $+\infty$ sans perdre l'équivalence. L'intégrale double se calcule alors facilement en coordonnées polaires $(K,\alpha)$ où $K=(\kappa^2+\kappa'^2)^{1/2}$ décrit $\mathbb{R}^+$ et $\alpha$ décrit $[0,\pi/2]$. On trouve finalement une divergence en racine carrée, $\propto (\zetat-\zetat_0)^{-1/2}$, sous-dominante dans (\ref{eq4000}).

\section{Le signal aux temps longs dans la limite $(\rho\xi^2)^{-1}\to 0$ des théories non perturbatives}
\label{app5}

Nous obtenons ici le comportement asymptotique (\ref{eq314}) du signal associé à la fonction énergie propre $\check{\Sigma}_{{\rm corr}\,\infty}^{(2,2)}(\qb,\check{\zeta})$ de l'équation (\ref{eq311}). À cette fin, nous appliquons la technique de l'\ref{app2} en rabattant le chemin d'intégration $C_+$ dans (\ref{eq012}) sur la ligne de coupure de $\check{\Sigma}_{{\rm corr}\,\infty}^{(2,2)}(\qb,\check{\zeta})$ située sur $\ii\mathbb{R}^{-}$ et décrite paramétriquement par $\zeta=-\ii\check{y}$, $\check{y}\geq 0$ (cette ligne de coupure est l'ensemble des valeurs de $\zetac$ annulant le dénominateur de l'intégrande dans (\ref{eq311})). En transposant l'équation (\ref{eq2001}) au présent cas, nous obtenons 
\be
\label{eq5001}
s^{(2,2)}_{{\rm corr}\,\infty}(t)\underset{\check{t}\to+\infty}{\sim}\int_0^{+\infty}\frac{\dd\check{y}}{2\pi}\frac{\delta\sigma_{\qb}(\check{y})\exp(-\check{y}\check{t})}{\left[\check{\Sigma}_{{\rm corr}\,\infty}^{(2,2)}(\qb,0)\right]^2}\quad\mbox{où }\quad
\delta\sigma_{\qb}(\check{y})=\check{\Sigma}_{{\rm corr}\,\infty}^{(2,2)}(\qb,0^+-\ii\check{y})-\check{\Sigma}_{{\rm corr}\,\infty}^{(2,2)}(\qb,0^--\ii\check{y})
\ee
est la discontinuité de la fonction énergie propre à travers sa ligne de coupure à la distance $\check{y}$ de l'origine. L'identité au sens des distributions $1/(\check{y}+\ii 0^+)=\mathrm{v.p.}(1/\check{y})-\ii\pi\delta(\check{y})$ conduit à 
\be
\label{eq5002}
\delta\sigma_{\qb}(\check{y})=-\frac{4\pi}{\qb^2}\int_{\qb}^{\eta}\dd \kb \int_0^{\eta-\kb}\dd\kb' \kb \kb' (\kb-\qb)(\kb'+\qb) (\bar{n}_{k'+q}^{\rm lin}-\bar{n}_{k'+k}^{\rm lin})(\bar{n}_{k-q}^{\rm lin}-\bar{n}_k^{\rm lin})
\delta\left(\check{y}-\kb'(\kb-\qb)\frac{\check{\Gamma}_{\kb-\qb}+\check{\Gamma}_{\kb'}}{2\qb(\kb+\kb')}\right)
\ee
Dans la limite $\check{t}\to+\infty$, la fonction exponentielle dans (\ref{eq5001}) limite l'intégrale à des valeurs arbitrairement faibles de $\check{y}$ et il suffit de connaître le comportement dominant de $\delta\sigma_{\qb}(\check{y})$ lorsque $\check{y}\to 0^+$. Comme les taux réduits $\check{\Gamma}_{\bar{K}}$ ne peuvent pas tendre vers zéro, l'argument du Dirac dans (\ref{eq5002}) ne peut s'annuler dans la limite $\check{y}\to 0$ que si $\kb'$ ou $\kb-\qb$ est infinitésimal. Pour un choix arbitraire d'une borne $\veps$ non infinitésimale mais $\ll 1$, on peut donc limiter le domaine d'intégration sur $\kb$ et $\kb'$ aux deux parties, (i) $\kb\in[\qb,\qb+\veps]$, $\kb'\in[0,\eta-\kb]$ et (ii) $\kb\in [\qb+\veps,\eta-\veps]$, $\kb'\in[0,\veps]$.\footnote{On remarque dans la partie (ii) que, si $\kb\leq\eta-\veps$, on a $\veps\leq\eta-\kb$ donc $\kb'$, qui doit rester $\leq \eta-\kb$, peut effectivement aller jusqu'à $\veps$. On vérifie par ailleurs que le \g{bout manquant} $\kb\in[\eta-\veps,\eta]$, $\kb'\in[0,\veps]$ donne une contribution en $O(\check{y}\veps)$ négligeable à $\delta\sigma_{\qb}(\check{y})$.} Dans la partie (i) du domaine d'intégration, on a $\kb-\qb\leq\veps\ll 1$ donc on peut remplacer les différentes quantités dépendant de $\kb$ dans (\ref{eq5002}) par leur développement limité en $\kb=\qb$ à l'ordre dominant, par exemple $(\kb-\qb)(\bar{n}^{\rm lin}_{k-q}-\bar{n}^{\rm lin}_{k})\simeq 1$, $\bar{n}_{k'+q}^{\rm lin}-\bar{n}_{k'+k}^{\rm lin}\simeq(\kb-\qb)(-\dd/\dd\kb')\bar{n}_{k'+q}^{\rm lin}$ {\yct ou} $\check{\Gamma}_{\kb-\qb}\simeq \check{\Gamma}_0$ et la borne {\yc supérieure} d'intégration sur $\kb'$ peut être remplacée par $\eta-\qb$. L'argument du Dirac dans (\ref{eq5002}) devient une fonction affine de $\kb-\qb$ ce qui permet d'intégrer facilement sur $\kb$:
\be
\label{eq5003}
\delta\sigma_{\qb}^{\rm (i)}(\check{y}) \simeq -16\pi\qb\check{y}\int_0^{\eta-\bar{q}} \dd\kb'\frac{(\kb'+\qb)^3}{(\check{\Gamma}_0+\check{\Gamma}_{\kb'})^2}\left(-\frac{\dd}{\dd\kb'}\bar{n}_{k'+q}^{\rm lin}\right) Y\left(\veps-\frac{2(\kb'+\qb)\qb\check{y}}{\kb'(\check{\Gamma}_0+\check{\Gamma}_{\kb'})}\right) \simeq -16\pi\qb\check{y}\int_{\qb^2\check{y}/\veps\check{\Gamma}_0}^{\eta-\qb}\frac{\dd\kb'}{\kb'} \frac{(\kb'+\qb)^3}{(\check{\Gamma}_0+\check{\Gamma}_{\kb'})^2} \left(-\frac{\dd}{\dd\kb'}\bar{n}_{k'+q}^{\rm lin}\right)
\ee
La forme du troisième membre de (\ref{eq5003}) repose sur le raisonnement suivant: si $\check{y}\to 0$ à $\veps$ fixé, l'argument de la fonction de Heaviside $Y$ au second membre est strictement positif, sauf si $\kb'\to 0$ ; au bord du support de cette fonction $Y$, on a donc $\kb'\ll 1$, ce qui autorise à remplacer dans son argument le deuxième terme par son approximation d'ordre dominant en $\kb'$ ; ceci fait apparaître $Y(\veps-\qb^2\check{y}/\kb'\check{\Gamma}_0)$, d'où la borne inférieure annoncée dans l'intégration sur $\kb'$. Dans la partie (ii) du domaine d'intégration, on procède de même: comme $\kb'\leq\veps\ll 1$, on remplace les différentes quantités dans (\ref{eq5002}) par leur ordre dominant en $\kb'$. L'argument du Dirac devenant une fonction affine de $\kb'$, on intègre sur cette variable ; la fonction $Y(\veps-\frac{2\qb\kb\check{y}}{(\kb-\qb)(\check{\Gamma}_{\kb-\qb}+\check{\Gamma}_0)})$ qui en résulte est inopérante lorsque $\check{y}\to 0$ car $\kb-\qb$ est ici $\geq\veps$, d'où
\be
\label{eq5004}
\delta\sigma_{\qb}^{\rm (ii)}(\check{y}) \simeq -16\pi\qb\check{y} \int_{\qb+\veps}^{\eta-\veps} \frac{\dd\kb}{\kb-\qb}\frac{\kb^3}{(\check{\Gamma}_{\kb-\qb}+\check{\Gamma}_0)^2} (\bar{n}_q^{\rm lin}-\bar{n}_k^{\rm lin})(\bar{n}_{k-q}^{\rm lin}-\bar{n}_k^{\rm lin})
\ee
Il reste à passer à la limite $\veps\to 0$ dans les résultats (\ref{eq5003},\ref{eq5004}). Pour cela, on sépare chaque intégrande en sa partie divergente infrarouge $\propto 1/\kb'$ ou $\propto 1/(\kb-\qb)$ et sa partie régulière.\footnote{Pour calculer le coefficient de la partie divergente $1/(\kb-\qb)$, on utilise le fait que $\lim_{\kb\to\qb}(\bar{n}_q^{\rm lin}-\bar{n}_k^{\rm lin})(\bar{n}_{k-q}^{\rm lin}-\bar{n}_k^{\rm lin})=(-\dd/\dd\qb)\bar{n}_q^{\rm lin}.$} Dans l'intégrale des parties régulières, on peut faire tendre sans dommage $\veps$ vers zéro dans les bornes. Dans l'intégrale des parties divergentes, faire de même donne naissance à des contributions dangereuses $\pm\ln\veps$, mais qui ont le bon goût de se compenser exactement dans la somme des parties (i) et (ii). Finalement
\be
\label{eq5005}
\delta\sigma_{\qb}(\check{y})\underset{\check{y}\to 0^+}{=}\frac{4\pi\qb^4}{\check{\Gamma}_0^2} \left(-\frac{\dd}{\dd\qb}\bar{n}^{\rm lin}_{{\yct q}}\right) \check{y}\ln\check{y} + C_{\qb} \check{y} + O(\check{y}^2\ln\check{y})
\ee
Pour alléger, nous ne donnons pas l'expression du coefficient $C_{\qb}$ de la partie linéaire dans (\ref{eq5005}), auquel notre méthode donne pourtant accès. La forme du $O(\ldots)$ dans (\ref{eq5005}) a en revanche été obtenue numériquement. Le report du terme dominant $\propto\check{y}\ln\check{y}$ dans (\ref{eq5001}) conduit au résultat annoncé (\ref{eq314}).
